\documentclass[pdftex,twoside,a4paper]{book}
\usepackage{amsmath}
\usepackage{amsthm}
\usepackage{amssymb}
\usepackage{graphicx}
\usepackage{braket}
\usepackage{verbatim}
\usepackage[margin=1in, top=1.2in, bottom=1.2in]{geometry}
\usepackage{leftidx}
\usepackage{wrapfig}
\usepackage{multicol}
\usepackage{array}
\usepackage[T1]{fontenc}
\usepackage{fancyhdr}
\usepackage{dsfont}
\usepackage{bbold}
\usepackage{caption}
\usepackage{subcaption}
\usepackage{multirow}
\usepackage{tikz}
\usepackage{mathtools}
\usepackage{booktabs}
\usepackage{bigstrut}
\usepackage{tabularx}
\usepackage{float}
\usepackage{url}
\usepackage{dcolumn}
\usepackage{makecell}
\usepackage[figuresright]{rotating}
\usetikzlibrary{positioning,shapes}
\usetikzlibrary{arrows}
\usetikzlibrary{calc}
\usetikzlibrary{fit,decorations.pathreplacing}

\title{Four dimensional supersymmetric Yang-Mills quantum mechanics with three colors}
\author{Zbigniew Ambrozi\'nski}
\date{\parbox{\linewidth}{\centering%
  Krakow, June 2014\endgraf\vspace{5cm}
  \flushright
\begin{tabular}{rl}
\multicolumn{2}{l}{PhD thesis at Jagiellonian University in Krakow}\\[1cm]
Supervisor: &  prof. Jacek Wosiek\\
Auxiliary supervisor:&dr Piotr Korcyl
\end{tabular}
  }}

\begin{document}
\maketitle

\large

\mbox{}\newpage

\chapter*{Abstract}
\addcontentsline{toc}{chapter}{Abstract}

The $D=4$ supersymmetric Yang-Mills quantum mechanics with $SU(2)$ and $SU(3)$ gauge symmetry groups is studied. A numerical method to find finite matrix representation of the Hamiltonian is presented in detail. It is used to find spectrum of the theory. In the $SU(2)$ case there are bound states in all channels with definite total number of fermions and angular momentum. For 2,3,4 fermions continuous and discrete spectra coexist in the same range of energies. These results are confirmation of earlier studies. With $SU(3)$ gauge group, the continuous spectrum is moved to sectors with more fermions. Supersymmetry generators are used to identify supermultiplets and determine the level of restoration of supersymmetry for a finite cutoff. For both theories, with $SU(2)$ and $SU(3)$ symmetry, wavefunctions are studied and different behavior of bound and scattering states is observed.

\newpage
\null
\vfill
\begin{flushright}
\emph{\Large To my wife.}
\end{flushright}
\vspace{3cm}
\chapter*{Acknowledgements}
I would like to express my gratitude to my supervisor prof. Jacek Wosiek for sharing his ideas and giving me inspiration. I am grateful for his assistance throughout my PhD studies and teaching me the physicist's point of view. Moreover, I thank for his thorough and critical reading of this thesis.

Secondly, I thank dr Piotr Korcyl for valuable discussions on SYMQM, his enthusiasm and friendship.

I am grateful to prof. Hermann Nicolai for hospitality during my visit in Max Planck Institute in Potsdam.

Finally, I would like to thank my wife Gosia for her patience and support.

\vspace{1cm}

This work was supported by Foundation for Polish Science MPD Programme co--financed by the European Regional Development Fund, agreement no. MPD/2009/6.

The research was carried out with the supercomputer ,,Deszno'' purchased thanks to the financial support of the European Regional Development Fund in the framework of the Polish Innovation Economy Operational Program (contract no. POIG. 02.01.00-12-023/08).
\tableofcontents

\chapter{Introduction}
The widely accepted model which describes physics in microscale is the Standard Model. Recent experiments (in particular at LHC) show an astonishing agreement with the Standard Model (SM). Nevertheless, it leaves some fundamental problems unresolved. This leads to the belief that the Standard Model is only an effective theory and needs to be extended. An example of such issue is the hierarchy problem \cite{Weinberg_hierarchy}. A possible solution proposed in \cite{Dimopoulos} is extending SM to a Minimal Supersymmetric Standard Model (MSSM). Another motivation for MSSM is that it ensures that the coupling constants of strong, electromagnetic and weak interactions meet at one point at the energy scale of grand unification \cite{Raby}. Finally, the Standard Model does not provide a good candidate for dark matter. On the other hand, certain models propose the lightest supersymmetric particle (LSP) as a possible constituent of the cold dark matter \cite{Ellis}. Although there are certain theoretical motivations for the supersymmetry to be a true symmetry of nature, no evidence for it was yet found in experiment. Instead, it starts to be more constrained and pushed away to higher energies. Nevertheless, supersymmetry remains at the center of interest of theoretical physics. One of the supersymmetric models is the supersymmetric Yang-Mills gauge theory \cite{Ferrara,Brink}.

The supersymmetric Yang-Mills gauge theory was studied in \cite{Brink}. It was found that it can be supersymmetric only in certain dimensions, namely $D=2,4,6,10$. The gauge theory can be used to construct supersymmetric Yang-Mills quantum mechanics (SYMQM). By definition, $D$--dimensional SYMQM is obtained from the $\mathcal N=1$ supersymmetric Yang-Mills field theory in $D=d+1$ dimensions by reducing the whole space to a single point. The number of supersymmetry generators in the resulting quantum mechanics is $\mathcal N=2,4,8,16$ respectively for the given dimensions. In \cite{Claudson} the authors gave complete solution for the case of $D=2$ and $SU(2)$ gauge group. Their result was then generalized and vacuum wavefunctions were found for arbitrary $SU(N)$ \cite{Samuel}.

Dimensional reduction is not only a simplification of the field theory, but SYMQM has its own motivations. In \cite{Hoppe} the purely bosonic part of SYMQM was proposed as a regularized description of a membrane. More precisely, coordinates of a membrane are first expanded in terms of an orthogonal set of functions $Y^a$ on the membrane. Then, a cutoff to the expansion is introduced. Finally, it was shown that for each $N\in\mathds N$ the algebra of the first $N^2-1$ functions $Y^a$ is closed and isomorphic to $su(N)$. A membrane is represented by an $N\times N$ matrix in the fundamental representation of $SU(N)$. These matrices are the bosonic degrees of freedom of the Yang-Mills theory. The parameter $N$ is a regulator and the continuum limit is recovered for $N\to\infty$. This correspondence can be extended to the full SYMQM describing the dynamics of supermembranes \cite{Bergshoeff,deWit}. It was later shown in \cite{Nicolai} that the spectrum of SYMQM is continuous and thus the supermembrane is unstable. This gave rise to the interpretation that a theory quantum supermembranes is second--quantized from the very beginning \cite{Helling}.

In \cite{BFSS} an M-theory compactified on a circle is considered in an infinite momentum frame \cite{Kogut}. It is argued that the dynamics of the M-theory are given by D0 branes. Those are described precisely by the large $N$ limit of SYMQM in $D=10$. This is called the BFSS conjecture (named after Banks, Fischler, Shenker and Susskind) and is perhaps the most remarkable application of SYMQM. The continuous spectrum of SYMQM turned to be a virtue and corresponds to scattering states in the M-theory. A necessary condition for the BFSS hypothesis to be true is the existence of a normalizable massless state in SYMQM. It would correspond to the graviton one the side of M--theory. It was shown that such state cannot exist in dimension lower than $D=10$ \cite{Moore}. Other papers suggest that there may be such state for $D=10$ \cite{Staudacher,Fischbacher}. More details can be found e.g. in a review article \cite{Taylor}. The BFSS conjecture aroused a large interest in this model \cite{Becker,Porrati,Taylor}. Asymptotic ground states for the $SU(2)$ case were found in \cite{Danielsson,Halpern}. The lattice methods were applied in \cite{Catterall,Wiseman}. SYMQM was also studied with a hybrid Monte Carlo approach \cite{Anagnostopoulos,Hanada,Nishimura}.

A different motivation for the Yang-Mills quantum mechanics is given in \cite{Bjorken}. It was advocated that the dynamics of a QCD gauge field can be understood by analyzing the theory in a small volume, of order $10^{-14}$cm. An approximate description is then given by constant fields. These are described by the dimensionally reduced theory, i.e. the purely bosonic part of four dimensional SYMQM.

A more complete study was proposed in \cite{Luscher} where a small volume (or equivalently weak coupling) expansion is considered. The zeroth order approximation is the Yang-Mills quantum mechanics. In \cite{Luscher,Munster,Weisz} this system was studied for $SU(2)$ and $SU(3)$ gauge groups with a numerical Rayleigh-Ritz technique. In \cite{Ziemann_phd} also higher groups, up to $SU(6)$ were considered. Further analysis, in medium volume was performed in \cite{Koller1,Koller2}.

A program to study the whole family of SYMQM with various dimensions and gauge groups was proposed in \cite{Janik}. The cut Fock space method, which will be our tool was used in  a set of papers. In  \cite{Wosiek,Campostrini} the four dimensional model with $SU(2)$ symmetry group was addressed. The $D=2$ case was studied and eventually a complete solution for SU(N) gauge group was found in \cite{Trzetrzelewski_susyd2,Korcyl,KorcylN}. There was yet another study based on the Schr\"odinger equation \cite{vanBaal,Kotanski,Kotanski2}. With this method the energies and eigenstates in 0-angular momentum sector of $D=4$ model with $SU(2)$ gauge group were found with great precision.

This thesis is a continuation of the program. The primary goal is to analyze the supersymmetric Yang-Mills quantum mechanics in four dimensions with $SU(3)$ gauge group. Analysis of this model is based on numerical results obtained using the cut Fock space. The main result is the spectrum of the theory. It is calculated in channels with definite number of fermions and spin. Eigenstates of the Hamiltonian with equal energies form supermultiplets. A way to identify members of these supermultiplets is given. An important part of this thesis is devoted to studying wavefunctions of the bound states and scattering states. Analysis of the wavefunctions delivers important information, in particular how the wavefunctions penetrate valleys of the potential. All results are given for $SU(2)$ as well as for $SU(3)$ gauge group. Construction of wavefunctions for this class of systems is a novelty. It was addressed before in \cite{vanBaal,Kotanski} but only for vanishing angular momentum and only for $SU(2)$. The method presented in this thesis applies to all angular momenta and in principle to all gauge groups.

Outline of the dissertation is following.

In Chapter 2 we provide an introduction to SYMQM with $D=4$ and $SU(N)$ gauge group symmetry. The dimensional reduction of the quantum field theory is performed. The Majorana condition is imposed so that the model is supersymmetric. Next, the symmetries of the Hamiltonian are identified. The two basic properties are conservation of the total angular momentum and number of fermions. It is also invariant under the particle--hole transformation. Thanks to the last symmetry, there is degeneracy between channels with $n_F$ and $n_F'=2(N^2-1)-n_F$ fermions and therefore only half of the sectors with given number of fermions need to be analyzed. Finally, the model is supersymmetric. The matrix notation is introduced. It is a common tool for simplifying the notation and plays an important role in the algorithm for constructing matrices of physical observables.

In Chapter 3 the cut Fock space method is presented. The idea of this approach is to represent the Hamiltonian as an infinite matrix and to introduce a cutoff. Here, as in many other cases, it is most useful to use the Fock basis. The cutoff limits the total occupation number of bosons. The maximal number of fermions is $2(N^2-1)$ because of the Pauli exclusion principle, so there is no need to limit the number of fermions. The basis of the gauge invariant space is generated by bricks which are traces of products of creation operators. Thanks to conservation of number of fermions, the Hamiltonian can be constructed in each fermionic sector separately. Matrix of the Hamiltonian could be constructed independently also in channels with definite angular momentum, as in \cite{Campostrini}. However, this procedure would involve constructing bricks with definite angular momentum which is inefficient for the $SU(3)$ gauge group. Instead, bricks with definite 3 bosonic and 2 fermionic occupation numbers (corresponding to three spatial and two spinor indices) are used. Construction of the matrix is performed using a recursive algorithm. This is the most numerically involved part of calculations. Once the matrix is created, it is diagonalized to obtain its eigenvectors and eigenvalues.

In Chapter 4 we present a method based on group theory to obtain dimensions of all sectors with definite set of quantum numbers of the cut Fock space. It was first introduced in \cite{Trzetrzelewski_number} to calculate dimensions of spaces with definite fermionic and bosonic occupation numbers $n_F$ and $n_B$ and also with definite angular momentum $j$. It is modified in order to calculate dimensions of spaces with more occupation numbers specified, i.e. $n_F^+$, $n_F^-$, $n_B^x$, $n_B^y$ and $n_B^z$. Dimensions of such subspaces can be also obtained numerically from the rank of Gramm matrix for the overcomplete set of basis vectors. At the end of the chapter a cross--check of both approaches is provided. The knowledge of dimensions of certain subspaces is also useful for constructing matrix elements of the Hamiltonian.

In Chapter 5 results for eigenenergies of the system are given. They are obtained in channels with definite number of fermions and angular momentum. An important question is to distinguish the discrete spectrum from the continuous one. This analysis is based on dependence of energies on the cutoff. Thanks to the particle--hole symmetry, only sectors with $n_F\leq N^2-1$ need to be examined in the theory with $SU(N)$ gauge group. It turns out that the spectrum in the few lowest fermionic channels is discrete, while the continuum spectrum appears for many fermions. Because of the supersymmetry, energies from neighboring fermionic sectors are degenerate. For this reason, for some $n_F$ there are discrete energy levels immersed in the continuous spectrum. It was found for the case of $SU(2)$ and is now confirmed also for $SU(3)$. We present results for both, $SU(2)$ and $SU(3)$ cases for comparison with each other and with earlier papers.

In Chapter 6 the supersymmetry multiplets are identified and discussed. Supermultiplets contain $SO(3)$ multiplets of states with equal energies and different numbers of fermions and angular momentum. Existence of supersymmetric multiplets is a direct consequence of the fact that supercharges conserve energy. On the other hand, the supercharge operators do not conserve bosonic (and fermionic) occupation number. Therefore, supersymmetry is broken for finite cutoff. In order to identify states which form multiplets in the continuous limit, we define so called supersymmetry fractions. They measure how supercharges map one energy states into other. Values of supersymmetry fractions in the continuum limit are known. Therefore, they are useful to analyze the degree of breaking of supersymmetry. Moreover, they help to identify SUSY partners.

In Chapter 7 we present a method to construct wavefunctions of energy eigenstates. These distributions are used to directly illustrate the effect of nonabelian interactions. First, given an angular momentum multiplet of states in the Fock space a single rotationally invariant function is introduced. It is an analogue of square of radial part of the wavefunction in the case of an ordinary three dimensional problem. Then, the structure of the configuration space in both cases, $SU(2)$ and $SU(3)$ is investigated. The flat valleys of the potential and directions in which the potential grows fastest are identified. They are essential in distinguishing between behavior of bound and scattering states. Finally, behavior of wavefunctions of several energy eigenstates along the flat valleys and steep directions is studied.

Finally, in Chapter 8 a summary of our results is given.

Some less significant and lengthy calculations are put in appendices. In Appendix \ref{ap:symmetries} symmetries of the SYMQM Hamiltonian are discussed. In \ref{ap:no_j1} we prove after \cite{Munster} that there are no $SU(2)$--invariant bosonic vector states. Appendix \ref{ap:finite_proof} contains a proof that the recursive algorithm for constructing matrix elements introduced in Chapter \ref{ch:Fock_space_method} is finite. Finally, Appendices \ref{ap:rotation_invariance} nad \ref{ap:fixing_gauge} are related to Chapter \ref{ch:wavefunctions}.

\chapter{The model}\label{ch:model}

The Yang Mills quantum mechanics is obtained by dimensional reduction \cite{Claudson} of a four dimensional gauge theory given by the action
\begin{align}
S=\int d^4x\mathcal L=\int d^4x\left(-\frac{1}{4}F_{\mu\nu}^aF^{\mu\nu a}+\frac{i}{2}\bar\psi^a\gamma^\mu D_\mu\psi^a\right)
\end{align}
The gauge group is $SU(N)$ and $a$ is the group index in the adjoint representation. The strength tensor and covariant derivative are given by
\begin{align}\begin{split}
F_{\mu\nu}^a&=\partial_\mu A_\nu^a-\partial_\nu A_\mu^a+gf^{abc}A_\mu^bA_\nu^c,\\
D_\mu\psi^a&=\partial_\mu\psi^a-gf^{abc}A^c_\mu\psi^b,
\end{split}\end{align}
where $f^{abc}$ are the structure constants of $SU(N)$.

The Lagrangian contains no time derivative of the field $A^a_0$. Therefore, $A^a_0$ is not a dynamical field and the Euler--Lagrange equations for $A^a_0$ are constraints of the system:
\begin{align}\label{eq:Gauss_law}
G^a&\equiv\frac{\delta \mathcal L}{\delta A^a_0}-\partial_i\frac{\delta\mathcal L}{\delta(\partial_i A^a_0)}=0.
\end{align}
$G^a$ is generator of the gauge symmetry and (\ref{eq:Gauss_law}) is the Gauss law. As $A^a_0$ is not dynamical, it can have an arbitrary value. One can use the gauge freedom to eliminate it. In the following, we work in the temporal gauge, with $A^a_0=0$. In the quantized theory, the Gauss law \cite{Bjorken} is imposed on the states of the Hilbert space. That is, the physical Hilbert space is composed of states annihilated by $G^a$.

In the dimensional reduction procedure one removes dependence of fields on spatial coordinates, so that $A^a_i(t,\vec x)=A^a_i(t)$. In the following we will adopt the notation of quantum mechanics $A^a_i\rightarrow x^a_i$. The resulting dimensionally reduced Lagrangian reads
\begin{align}
\mathcal L=\frac{1}{2}(\dot x_i^a)^2+\frac{1}{2}\bar\psi^a\gamma^0\partial_0\psi^a-\frac{g^2}{4}(f^{abc}x_i^bx_j^c)^2-\frac{i}{2}gf^{abc}\bar\psi^a\gamma^kx^c_k\psi^b.
\end{align}
The corresponding Hamiltonian is \cite{Claudson}
\begin{align}\label{eq:hamiltonian}\begin{split}
H&=H_K+g^2H_V+gH_F,\\
H_K&=\frac{1}{2}p^a_ip^a_i,\\
H_V&=\frac{1}{4}f^{abc}f^{ade}x^b_ix^c_jx^d_ix^e_j,\\
H_F&=\frac{i}{2}f^{abc}\psi^{a\dagger}\Gamma_k\psi^bx^c_k.
\end{split}\end{align}
$\Gamma_k$ are the alpha matrices, i.e. $\Gamma_k=\gamma^0\gamma^k$ where $\gamma^\mu$ are Dirac matrices. $H_K,\ H_V$ and $H_F$ are kinetic, potential and fermionic parts of the Hamiltonian respectively. Spatial indices $i,j,k$ take values $1,2,3$ and color indices $a,b,\ldots$ range from $1$ to $N^2-1$. Therefore there are 24 bosonic degrees of freedom in model for the $SU(3)$ group.

Once spatial dependence of $A^a_i$ is removed, the gauge group generator $G^a$ has particularly simple form
\begin{align}\label{eq:gauge_generator}
G^a&\equiv f^{abc}(x^b_ip^c_i-\frac{i}{2}\psi^{b^\dagger}_\alpha\psi^c_\alpha).
\end{align}
This is nothing but the total angular momentum in the color space.

In what follows in all explicit calculations the Weyl representation of Dirac matrices from \cite{Itzykson} will be used. The Dirac matrices are
\begin{align}
\gamma^0&=\left(\begin{array}{cc}0&\mathds 1\\\mathds 1&0\end{array}\right),&
\gamma^k&=\left(\begin{array}{cc}0&\sigma_k\\-\sigma_k&0\end{array}\right),&
\gamma^5&=\left(\begin{array}{cc}-\mathds 1&0\\0&\mathds1\end{array}\right).
\end{align}

This theory is not yet supersymmetric. Indeed, in SUSY models the number of bosonic and fermionic degrees of freedom must match. If $\psi$ is a Dirac fermion, then there are $4(N^2-1)$ fermionic d.o.f. On the other hand, the number of bosonic d.o.f. is reduced by the Gauss law from $3(N^2-1)$ to $2(N^2-1)$. The supersymmetry can be achieved by imposing Weyl or Majorana condition on the spinor $\psi$. In what follows we choose the Majorana condition. Then, $\psi$ must satisfy $\psi_C=\psi$ where for the charge conjugation we use the phase convention
\begin{align}
\psi^{a}_C&=-iC(\bar{\psi}^a)^T,\\
C&=\left(\begin{array}{cc}-i\sigma_2&0\\0&i\sigma_2\end{array}\right).
\end{align}

Finally, we proceed to quantizing the system. The quantization procedure gives algebra of operators
\begin{align}
[x^a_i,p^b_j]&=\delta^{ab}\delta_{ij},\\
\{\psi^a_\alpha,\psi^{b\dagger}_\beta\}&=\delta^{ab}\delta_{\alpha\beta}.\label{eq:psi_anticommutations}
\end{align}
Because of the Majorana condition, components $\psi^a_\alpha$ cannot be independent. Instead, $\psi$ can be constructed from a two--component Weyl spinor $f^a_\alpha$ ($\alpha=1,2$). By definition, the Weyl spinor satisfies anticommutation relations
\begin{align}\label{eq:fermionic_operators}
\{f^a_\alpha,f^{b\dagger}_\beta\}=\delta^{ab}\delta_{\alpha\beta}.
\end{align}
The Weyl spinor can be written in a four component notation
\begin{align}
\psi_W^a=-e^{i\pi/4}\left(\begin{array}{c}f^{a}_1\\f^{a}_2\\0\\0\end{array}\right).
\end{align}
$\psi_W$ satisfies $\gamma^5\psi_W=-\psi_W$, so it is indeed a left--handed spinor. The phase in front is chosen such that all components of spinor in the next formula (\ref{eq:Weyl_spinor}) have common phase. The Majorana spinor is then constructed from the Weyl spinor and its charge conjugate \cite{Weinberg}:
\begin{align}\label{eq:Weyl_spinor}
\psi^a=\psi_W^a+(\psi_W)_C^a=e^{i\pi/4}\left(\begin{array}{c}-f^{a}_1\\-f^{a}_2\\f^{a\dagger}_2\\-f^{a\dagger}_1\end{array}\right).
\end{align}
The spinor $\psi^a$ automatically satisfies the Majorana condition because charge conjugation is an involution, i.e. for any $\chi$ there is $(\chi_C)_C=\chi$.

\section{Symmetries}
Now we come to discussion of symmetries of the Hamiltonian. An extension of this paragraph can be found in Appendix \ref{ap:symmetries}. In particular, Appendix \ref{ap:symmetries} contains proofs that generators of the symmetries commute with the Hamiltonian.

The supercharges arising from invariance under the supersymmetry transformation are
\begin{align}\label{eq:supercharges}
Q_\alpha=(\Gamma_k\psi^a)_\alpha p^a_k+igf^{abc}(\Sigma_{jk}\psi^a)_\alpha x^b_jx^c_k,
\end{align}
where $\Sigma_{jk}=-\frac{i}{4}[\Gamma_j,\Gamma_k]$.
They obey anticommutation relations
\begin{align}\label{eq:anticommutation_relations}
\begin{split}
\{Q_\alpha,Q^\dagger_\beta\}&=2\delta_{\alpha\beta}H+g\Gamma_{k\alpha\beta}x^a_kG^a.
\end{split}
\end{align}
In the space of physical states, i.e. when $G_a$ vanishes, the anticommutator $\{Q_\alpha,Q^\dagger_\beta\}$ is proportional to the Hamiltonian. By acting with $Q_\alpha$ and $Q_\alpha^\dagger$ on eigenstates of the Hamiltonian one generates a supersymmetry multiplet. The supercharge operator $Q_\alpha$ commutes with the Hamiltonian, so all states in the supermultiplet have identical energy. Structure of these multiplets is analyzed in detail in Chapter \ref{ch:supersymmetric_multiplets}.

In the dimensional reduction procedure the full configuration space is reduced to a single point. Still, the resulting Hamiltonian exhibits an inherited rotational symmetry. The angular momentum operators are
\begin{align}\label{eq:angular_momentum}
J_i=L_i+S_i=\epsilon_{ijk}\left(x^a_jp^a_k+\frac{1}{4}\psi^{a\dagger}\Sigma_{jk}\psi^a\right).
\end{align}
Commutators of $H$ and $J_i$ vanish and thus the angular momentum is conserved.

A remarkable feature of the four--dimensional theory is conservation of the total number of fermions $n_F=\sum_{a\alpha}f^{a\dagger}_\alpha f^{a\dagger}_\alpha$. It is not true e.g. in 10 dimensions which is interesting due to the BFSS conjecture.

Furthermore, there is a particle--hole symmetry:
\begin{align}\label{eq:ph_symmetry}
f^a_1&\rightarrow -f^{a\dagger}_2,&f^{a\dagger}_1&\rightarrow -f^{a}_2,\nonumber\\
f^a_2&\rightarrow f^{a\dagger}_1,&f^{a\dagger}_2&\rightarrow f^{a}_1,\\
x_i^a&\rightarrow -x_i^a,&p_i^a&\rightarrow -p_i^a.\nonumber
\end{align}
A natural consequence of this symmetry is that one can find eigenstates of the Hamiltonian which are even or odd under (\ref{eq:ph_symmetry}). However, conservation of $n_F$ is a stronger symmetry. It implies that the full Hilbert space splits into independent sectors with definite number of fermions. The particle--hole symmetry relates these sectors. For each state with $n_F$ fermions there is a state with $2(N^2-1)-n_F$ fermions and the same energy. Construction of the other state is given explicitly so all its properties can be easily inferred. In particular, the particle--hole transformation preserves the total angular momentum.

Moreover, the Hamiltonian has a scaling property
\begin{align}
\begin{split}
x^a_i&\rightarrow g^{-1/3} x^a_i,\\
p^a_i&\rightarrow g^{1/3} p^a_i,\\
f^a_\alpha&\rightarrow f^a_\alpha,\\
H=H_K+g^2H_V+gH_F&\rightarrow g^{2/3}(H_K+H_V+H_K).
\end{split}
\end{align}
In what follows $g$ will be set to $1$. For any other finite $g$ the spectrum is merely scaled by $g^{2/3}$. Eigenstates are stretched by $g^{-1/3}$ and all their quantum numbers do not change.

Finally, consider parity. The operators transform under parity in the following way:
\begin{align}\label{eq:parity}\begin{split}
p^a_i&\rightarrow-p^a_i,\\
x^a_i&\rightarrow-x^a_i,\\
\psi_\alpha^a&\rightarrow(\psi_P)^{a}_\alpha\equiv\gamma^0_{\alpha\beta}\psi_\beta^a.
\end{split}\end{align}
One can check that the Hamiltonian is invariant under the transformation (\ref{eq:parity}). However, the spinor $\psi_P$ does not satisfy the Majorana condition. It follows that states generated by $\psi_P$ are not invariant under the charge conjugation. Therefore, the Hilbert space breaks the parity symmetry. A more detailed discussion of parity breaking can be found in Appendix \ref{ap:symmetries}.

\section{Matrix notation}

It is convenient to use the so called matrix notation. Let $T^a$ be the $N^2-1$ generators of $SU(N)$ in the fundamental representation. They are hermitean traceless $N\times N$ matrices and satisfy multiplication law
\begin{align}
T^aT^b=\frac{1}{2N}\delta^{ab}\mathbb 1_N+\frac{1}{2}(if^{abc}+d^{abc})T^c,
\end{align}
where $f^{abc}$ and $d^{abc}$ are antisymmetric and symmetric $SU(N)$ group structure constants respectively. For $N=2$ the generators are simply Pauli matrices $T^a=\frac{1}{2}\sigma^a$, $a=1,2,3$. The antisymmetric structure constant is $f^{abc}=\epsilon^{abc}$ and the symmetric $d^{abc}$ vanishes. For $N=3$ the generators are Gell--Mann matrices
\begin{align}
T^{1}&=\frac{1}{2}\left(\begin{array}{ccc}0&1&0\\1&0&0\\0&0&0\end{array}\right),&
T^{2}&=\frac{1}{2}\left(\begin{array}{ccc}0&-i&0\\i&0&0\\0&0&0\end{array}\right),\nonumber\\
T^{3}&=\frac{1}{2}\left(\begin{array}{ccc}1&0&0\\0&-1&0\\0&0&0\end{array}\right),&
T^{4}&=\frac{1}{2}\left(\begin{array}{ccc}0&0&1\\0&0&0\\1&0&0\end{array}\right),\nonumber\\
T^{5}&=\frac{1}{2}\left(\begin{array}{ccc}0&0&-i\\0&0&0\\i&0&0\end{array}\right),&
T^{6}&=\frac{1}{2}\left(\begin{array}{ccc}0&0&0\\0&0&1\\0&1&0\end{array}\right),\nonumber\\
T^{7}&=\frac{1}{2}\left(\begin{array}{ccc}0&0&0\\0&0&-i\\0&i&0\end{array}\right),&
T^{8}&=\frac{1}{2\sqrt3}\left(\begin{array}{ccc}1&0&0\\0&1&0\\0&0&-2\end{array}\right).
\end{align}
Consider an operator with an adjoint color index  $A^a$. Then $A=A^aT^a$ is a $N\times N$ matrix with operator--valued matrix elements. In this notation $x_i=x_i^aT^a$, $p_i=p_i^aT^a$, $\psi_\alpha=\psi_\alpha^aT^a$, etc. In the matrix notation the Hamiltonian, angular momentum and supersymmetry generators are
\begin{align}\label{eq:trace_ops}\begin{split}
H_K&=Tr(p_ip_i),\\
H_V&=-\sum_{i<j}Tr([x_i,x_j]^2),\\
H_F&=\sum_{k\alpha\beta}\Gamma_{k\alpha\beta} Tr(\psi^\dagger_\alpha[\psi_\beta,x^k]),\\
J_i&=\epsilon_{ijk}\left(2Tr(x_jp_k)+\frac{1}{2}(\Sigma_{jk})_{\alpha\beta}Tr(\psi^\dagger_\alpha\psi_\beta)\right),\\
Q_\alpha&=2\Gamma_{k\alpha\beta}Tr(\psi_\beta p_k)+\frac{i}{4}\Sigma_{jk\alpha\beta}Tr(\psi_\beta x_j x_k),\\
G&=-i[x_i,p_i]-\frac{1}{2}\{\psi^\dagger_\alpha,\psi_\alpha\}+(1+3i)\frac{N^2-1}{2N}\mathds 1_N.
\end{split}\end{align}
The first five operators in (\ref{eq:trace_ops}) are color singlets and are given in terms of traces. $G^a$ which is a vector in the color space is represented by a matrix. The last term in $G$ may look strange, since $G$ must be traceless. However, one has to remember that the two operators $[x_i,p_i]$, $\{\psi^\dagger_\alpha,\psi_\alpha\}$ have non--vanishing trace. The whole expression is indeed traceless.

The matrix notation in only a way to rewrite the same operators. However, it is much more convenient to manipulate with traces of matrix operators rather than to keep the color indices explicit.

\chapter{The Cut Fock space method}\label{ch:Fock_space_method}
In this chapter we discuss a numerical technique, called cut Fock space method, which is used to solve our model. It originates from the variational Tamm--Dancoff method \cite{Dancoff} where one uses a small set of trial states to construct an approximate ground state of a theory. Quality of the variational approach relies on how many parameters one uses to approximate the ground state and how clever one is in choosing these parameters. The ground state energy is always approximated from above.

The idea of this method is to first choose a basis in the Hilbert space and then to introduce a cutoff. The cutoff has to be such that dimension of the cut space is finite. Then one constructs and diagonalizes matrix of the Hamiltonian in the cut space. The lowest eigenvalue is an approximation to the ground state energy. It is the energy obtained with variational technique where all basis states in the cut space are trial states. It is usually most convenient to work in the Fock space (hence the name) with the standard occupation number basis. The cutoff is then the maximal occupation number $N_B$. As the cutoff increases, energies converge to their exact values.

One of the virtues of this approach is that one can construct not only Hamiltonian, but also other observables. Having computed eigenvectors of the Hamiltonian, one can easily obtain expectation value of other observable in these states. Another asset is that one has a direct access to wavefunctions of energy eigenstates. They will be studied in following chapters.

Wavefunctions $\braket{x|n}$ of basis states in the configuration space are simply Hermite functions. It is known that $\braket{x|n}$ in practice have finite support, i.e. they are exponentially suppressed for $\left|x\right|\gtrsim\sqrt{2n+1}$. That is, for a finite cutoff $n\leq N_B$ all states are localized inside a finite box of size $L=\sqrt{2N_B+1}$. Wavefunctions in the momentum space $\braket{p|n}$ share the same behavior and thus the momenta are also limited. One can then take a point of view that the cut Fock space is a regularized Hilbert space. The cutoff $N_B$ plays a role of both, IR and UV cutoffs which are approximately $(2N_B)^{-1/2}$ and $(2N_B)^{1/2}$.

The cut Fock space method was already applied with success to other simpler models. Properties of the cut Fock space were studied for one dimensional quantum mechanics \cite{Trzetrzelewski_spectra}. The technique was used for computations with high precision for the double well potential \cite{Ambrozinski} and multiple wells with periodic boundary conditions \cite{Ambrozinski_cosine}. Then it was also applied to SYMQM in two dimensions \cite{Trzetrzelewski_susyd2,Korcyl,KorcylN} and finally to four dimensional theory with $SU(2)$ gauge group \cite{Wosiek,Campostrini}.

\section{The cut Fock space}
In order to construct the Fock space we introduce creation and annihilation operators, which satisfy the usual commutation rules
\begin{align}\label{eq:bosonic_ca_ops}
a^a_i&=\frac{1}{\sqrt2}(x^a_i+ip^a_i),&a^{a\dagger}_i=\frac{1}{\sqrt 2}(x^a_i-ip^a_i),
\end{align}
\begin{align}
[a^a_i,a^{b\dagger}_j]=\delta^{ab}\delta_{ij}.
\end{align}
Fermionic creation and annihilation operators $f^{a\dagger}_\alpha$ and $f^a_\alpha$ were already introduced in (\ref{eq:fermionic_operators}). The Fock vacuum is defined as usual by
\begin{align}
a^a_i\ket{0}&=0,&f^a_\alpha\ket{0}&=0.
\end{align}
All other states are generated by acting with bosonic and fermionic creation operators on $\ket{0}$.

Since we would like to construct a Hilbert space consisting only of gauge invariant states, we use only specific combinations of creation operators. Take a set of matrix operators $A_1,\ldots,A_n$ where $A_k=A_k^aT^a$. For each $k$ operators $A_k^a$ are in adjoint $SU(N)$ representation. The lower index $k$ is not related to any symmetry and $A_k$ can be any operators. In the following the object
\begin{align}
(A_1\dotsm A_n)\equiv Tr(A_1\dotsm A_n)=A^{a_1}_1\dotsm A^{a_n}_nTr(T^{a_1}\dotsm T^{a_n})
\end{align}
is called a \emph{trace operator}. If all operators $A_k$ are bosonic or fermionic creation operators then $(A_1\dotsm A_n)$ is called a \emph{brick}. For the rest of the thesis the round bracket $(\cdot)$ is used for a short notation of the trace. It was shown \cite{Trzetrzelewski_trees} that the space of gauge invariant states is spanned by states obtained by repeatedly acting with bricks on the Fock vacuum. A product of bricks is called a \emph{composite brick}. A state generated by a composite brick has a definite number of bosons $n_B$ and fermions $n_F$
\begin{align}\label{eq:number_of_particles_operators}
n_B&=\sum a^{a\dagger}_ia^a_i,&n_F&=\sum f^{a\dagger}_\alpha f^a_\alpha.
\end{align}
There are only $2(N^2-1)$ fermionic creator operators $f^{a\dagger}_\alpha$, so the Pauli exclusion principle implies $n_F\leq 2(N^2-1)$.

Because the Hamiltonian conserves the fermion number $n_F$, it is convenient to work with subspaces with fixed number of fermions $\mathcal H_{n_F}$. For each $n_F$ we introduce a different cutoff $N_B$. The cut Fock space $\mathcal H_{n_F,N_B}$ is then the space of all states which contain precisely $n_F$ fermions and at most $N_B$ bosons.

The cut Fock space method will be used to construct matrices for several operators. The angular momentum operators conserve $n_F$ and $n_B$, so the cut matrices of these operators are $(J_i)_{n_F,N_B}:\mathcal H_{n_F,N_B}\rightarrow \mathcal H_{n_F,N_B}$. Eigenvalues of these matrices are exact eigenvalues of $J_i$. Therefore, we may us eigenvectors of $(J^2)_{n_F,N_B}$ and $(J_3)_{n_F,N_B}$ to construct Hilbert spaces $\mathcal H_{n_F,N_B,j,m}$ with definite quantum numbers $j$ and $m$.

The Hamiltonian conserves $n_F$ but not $n_B$. The matrix $H_{n_F,N_B}:\mathcal H_{n_F,N_B}\rightarrow \mathcal H_{n_F,N_B}$ has eigenvalues which approximate energy levels of the Hamiltonian in the $N_B\to\infty$ limit. In principle, one can construct $H$ directly on subspaces $\mathcal H_{n_F,N_B,j,m}$. This procedure would generate smaller matrices which are easier to diagonalize. However, then one needs to use bricks with definite angular momentum. The computational cost of this procedure would exceed the gain from smaller matrices.

Finally, matrices of $Q_\alpha$ will be constructed. Supercharges do not conserve $n_F$. In Chapter \ref{ch:supersymmetric_multiplets} we introduce operators $\mathcal Q_\pm$ which are closely related to the supercharges $Q_\alpha$ and play exactly the same role. $\mathcal Q_\pm$ do not conserve $n_F$. Instead, they decrease the number of fermions by $1$. Therefore, We generate matrices $(\mathcal Q_{\pm})_{n_F,N_B,N_B'}:\mathcal H_{n_F,N_B}\rightarrow \mathcal H_{n_F-1,N_B'}$. In practice the two cutoffs $N_B$ and $N_B'$ are always different.

\section{Relations between bricks for $SU(3)$}\label{sec:brick_relations}
The full Fock space of gauge invariant states is spanned by all possible composite brick acting on the Fock vacuum. However, these states are not linearly independent. For optimization reasons it is necessary to have as few states as possible. Therefore, we try to identify and remove those bricks which can be expressed in terms of other bricks. The more complicated case of $SU(3)$ gauge group is addressed in this subchapter.

The Cayley Hamilton theorem states that a matrix is a root of its characteristic polynomial. Let $M$ be a square traceless matrix of size 3. Then, the theorem implies that
\begin{align}\label{eq:CH}
M^3&=(M)M^2+\frac{1}{2}\left((M^2)-(M)^2\right)M+\frac{1}{6}\left((M)^3-3(M^2)(M)+2(M^3)\right)\mathds{1}_3.
\end{align}
Recall that $(\cdot)$ is a short notation for a trace. This theorem holds if the matrix is operator--valued, i.e. its matrix elements are operators, as long as the matrix elements commute. Multiply the above equation by another operator--valued matrix $O$ and take a trace. Then,
\begin{align}\label{eq:CH_trace}\begin{split}
(M^3O)&=(M)(M^2O)+\frac{1}{2}\left((M^2)-(M)^2\right)(MO)\\
&\quad+\frac{1}{6}\left((M)^3-3(M^2)(M)+2(M^3)\right)(O).
\end{split}\end{align}
It follows that if a brick contains an expression which is repeated three times and at least one more operator, then it can be written in terms of shorter bricks, e.g. for $M=a_1^{\dagger}a_2^{\dagger}$ and $O=f_1^{\dagger}a_1^{\dagger}$ there is
\begin{align}
\begin{split}
(a_1^{\dagger}a_2^{\dagger}a_1^{\dagger}a_2^{\dagger}a_1^{\dagger}a_2^{\dagger}f_1^{\dagger}a_1^{\dagger})&=
(a_1^{\dagger}a_2^{\dagger})(a_1^{\dagger}a_2^{\dagger}a_1^{\dagger}a_2^{\dagger}f_1^{\dagger}a_1^{\dagger})
+\frac{1}{2}(a_1^{\dagger}a_2^{\dagger}a_1^{\dagger}a_2^{\dagger})(a_1^{\dagger}a_2^{\dagger}f_1^{\dagger}a_1^{\dagger})\\
&-\frac{1}{2}(a_1^{\dagger}a_2^{\dagger})^2(a_1^{\dagger}a_2^{\dagger}f_1^{\dagger}a_1^{\dagger})
+\frac{1}{6}(a_1^{\dagger}a_2^{\dagger})^3(f_1^{\dagger}a_1^{\dagger})\\
&-\frac{1}{2}(a_1^{\dagger}a_2^{\dagger}a_1^{\dagger}a_2^{\dagger})(a_1^{\dagger}a_2^{\dagger})(f_1^{\dagger}a_1^{\dagger})
+\frac{1}{3}(a_1^{\dagger}a_2^{\dagger}a_1^{\dagger}a_2^{\dagger}a_1^{\dagger}a_2^{\dagger})(f_1^{\dagger}a_1^{\dagger}).
\end{split}
\end{align}
Because the trace of a single matrix operator is zero, i.e. $(a_i^\dagger)=(f^\dagger_\alpha)=0$, this formula simplifies if $M$ or $O$ in (\ref{eq:CH_trace}) are single creation operators. For $M=a_1^\dagger$ and $O=f_1^{\dagger}a_3^{\dagger}$ it reads
\begin{align}
\begin{split}
(a_1^{\dagger}a_1^{\dagger}a_1^{\dagger}f_1^{\dagger}a_3^{\dagger})&=
\frac{1}{2}(a_1^{\dagger}a_1^{\dagger})(a_1^{\dagger}f_1^{\dagger}a_3^{\dagger})
+\frac{1}{3}(a_1^{\dagger}a_1^{\dagger}a_1^{\dagger})(f_1^{\dagger}a_3^{\dagger}).
\end{split}
\end{align}

There are other relations for fermionic operators. The simplest example is vanishing of a square of single fermionic creation operator.
\begin{align}\label{eq:fermion_sq_vanishes}
(f^\dagger_1f^{\dagger}_1)=f^{a_1\dagger}_{1}f^{a_2\dagger}_{1}(T^{a_1}T^{a_2})=-f^{a_2\dagger}_{1}f^{a_1\dagger}_{1}(T^{a_2}T^{a_1})=-(f_1^{\dagger}f_1^{\dagger}),
\end{align}
so $(f^\dagger_1f^{\dagger}_1)=0$. This can be easy generalized. Let $M$ be a operator valued matrix with anticommuting matrix elements. Then,
\begin{align}\label{eq:vanishing_fermion_matrices}
(M^2)=(M^4)&=0,&M^6&=0.
\end{align}
The proof of the first two relations is analogous to (\ref{eq:fermion_sq_vanishes}). The last equality can be easily shown by plugging $M^2$ into (\ref{eq:CH}) and using the first two identities from (\ref{eq:vanishing_fermion_matrices}). There is also a Cayley--Hamilton relation for matrix with anticommuting matrix elements:
\begin{align}\label{eq:CH_anticommutation}
M^5&=\frac{1}{3}(M^3)M^2+\frac{1}{3}(M^5).
\end{align}
It can be proven by a direct calculation. In order to use it for eliminating bricks, one has to multiply (\ref{eq:CH_anticommutation}) by an arbitrary operator $O$ and take a trace. Then,
\begin{align}
(M^5O)&=\frac{1}{3}(M^3)(M^2O)+\frac{1}{3}(M^5)(O).
\end{align}
They Cayley-Hamilton relation for Grassmann--valued matrices can be also extended to any $SU(N)$ group \cite{Procesi}. For an $N\times N$ matrix $M$ whose entries anticommute it reads
\begin{align}
M^{2N-1}=\frac{1}{N}(M^{2N-1})+\frac{1}{N}\sum_{i=1}^{N-1}M^{2i}(M^{2N-2i-1}).
\end{align}

There is one more identity for $SU(3)$ which is used in eliminating bricks, namely \cite{Macfarlane}
\begin{align}\label{eq:symmetric_reduction}
T^{\{a}T^bT^{c\}}=\frac{1}{4}\delta^{\{ab}T^{c\}}+\frac{1}{3}(T^{\{a}T^bT^{c\}})\mathds 1_3.
\end{align}
Curly brackets denote symmetrization without additional coefficient $\frac{1}{3!}$. Formulas presented earlier (\ref{eq:CH}) -- (\ref{eq:CH_anticommutation}) allow to  discard some bricks because they are expressed by shorter bricks. (\ref{eq:symmetric_reduction}) is different, because it relates bricks of the same length. For any six bricks that differ by permutations of three operators then one of them can be eliminated.

Finally, the procedure of eliminating dependent bricks is the following. For each $n_B$ and $n_F$ generate all possible bricks with appropriate occupation numbers. Then, eliminate as many bricks as possible using relations (\ref{eq:CH_trace}), (\ref{eq:CH_anticommutation}), (\ref{eq:vanishing_fermion_matrices}) and (\ref{eq:symmetric_reduction}). All composite bricks with fermionic and bosonic occupation numbers $n_F$ and $n_B$ form an overcomplete basis in sector $(n_F,n_B)$. This overcompleteness cannot be eliminated entirely and is taken into account in the program for generating matrices.

There is yet another way to eliminate superfluous bricks. It involves objects introduced below and will be presented in \ref{sec:Gauss_bricks}.

\section{The algorithm}\label{sec:algorithm}
In this part we present the algorithm for constructing matrix elements of interesting operators. The same algorithm is used for both cases, SYMQM with $SU(2)$ and $SU(3)$ group, so it is presented in a general form for a group $SU(N)$.

First we introduce the notion of a trace operator. Take $n$ operators $A^a_k$, $k=1,\ldots,m$, where $a$ is the color index. Each $A^a_k$ is a fermionic of bosonic creation or annihilation operator. Then, $(A_1\dotsm A_m)$ given by the formula
\begin{align}\label{eq:general_trace}
(A_1\dotsm A_m)\equiv A_1^{a_1}\dotsm A_m^{a_m}(T^{a_1}\dotsm T^{a_m})
\end{align}
will be called a \emph{trace operator}. The number $m$ of operators inside the trace (\ref{eq:general_trace}) is called the \emph{length} of the trace. In particular, identity is a trace operator of length $0$, i.e. $id=\frac{1}{N}Tr(\mathds 1_N)$. A product of trace operators is called a \emph{composite trace}. Length of a composite trace is the total of all lengths of single traces in the product. All operators of our interest, i.e. angular momentum, Hamiltonian and supercharges can be expressed by traces of creation and annihilation operators. The angular momentum has a particularly simple form
\begin{align}\label{eq:angular_traces}\begin{split}
J_1&=2i(a_3^\dagger a_2)-2i(a_2^\dagger a_3)+(f_1^\dagger f_2)+(f_2^\dagger f_1),\\
J_2&=2i(a_1^\dagger a_3)-2i(a_3^\dagger a_1)-i(f_1^\dagger f_2)+i(f_2^\dagger f_1),\\
J_3&=2i(a_2^\dagger a_1)-2i(a_1^\dagger a_2)+(f_1^\dagger f_1)-(f_2^\dagger f_2).
\end{split}\end{align}
The explicit expression for Hamiltonian is much more complex and is not given here.

The cut Fock space can be decomposed into subspaces with definite occupation numbers
\begin{align}\label{eq:Hilbert_space_decomposition}
\mathcal H_{n_F,N_B}=\bigoplus_\mathbf n\mathcal H_\mathbf n,
\end{align}
where the sum runs over $\mathbf n=(n^1_F,n^2_F,n^1_B,n^2_B,n^3_B)\geq0$ which satisfy $\sum_\alpha n^\alpha_F=n_F$ and $\sum_i n^i_B\leq N_B$.

Given a composite brick $\mathcal P$ one can associate the occupation labels $\mathbf n$ with $\mathcal P$. Each component of $\mathbf n$ is the total number of corresponding creation operators in $\mathcal P$. A subspace $\mathcal H_{\mathbf n}$ is spanned by all composite bricks $\mathcal P$ with occupation labels $\mathbf n$ acting on the Fock vacuum.

\subsection{Maximally annihilating form}

The maximally annihilating form is an analog of the Wick's expansion. Consider an operator $(a_ia_j^\dagger)$. The Wick's expansion is
\begin{align}
(a_ia_j^\dagger)=a^a_ia_j^{b\dagger}(T^aT^b)=[a_j^{b\dagger} a^a_i+\delta^{ab}_{ij}](T^aT^b)=(a_j^\dagger a_i)+\frac{1}{2}\delta_{ij}.
\end{align}
The operator on the right is normal ordered and each term is a trace operator. However, not always all operators in the Wick's expansion are trace operators. For example, in the Wick expansion of $(a_i a_j^\dagger a_k a_l^\dagger)$ there is a term $a_j^{b\dagger} a_l^{d\dagger} a^a_i a^c_k(T^aT^bT^cT^d)$. Note that the order of color indices is different for creation/annihilation operators and $SU(N)$ generators. We would like to use the matrix notation introduced in Chapter \ref{ch:model}, so the trace--operator structure has to be preserved. The maximally annihilating form is defined in a way that all operators are trace operators and it is as close to the Wick's expansion as possible.

Consider a trace operator $A=(A_1\dotsm A_m)$ with $m\geq 1$. With operator $A$ we associate \emph{annihilation rank} $\boldsymbol \nu=(\nu^1_F, \nu^2_F, \nu^1_B, \nu^2_B, \nu^3_B)$. It is defined in the following way. Let $\mu_i$ be the number of operators $f_1$ minus the number of operators $f_1^\dagger$ in the set $\{A_{i+1},\ldots,A_m\}$ for $i=0,\ldots,m$. Then $\nu^1_F$ is defined as the maximum of all $\mu_i$'s. That means that if a state has quantum numbers $\mathbf n$ and $n^1_F<\nu^1_F$, then it is annihilated by $A$. The other components of $\boldsymbol\nu$ are defined in analogous way. Note that different operators with the same quantum numbers can have different annihilation ranks. For instance, a trace operator $(a_1^\dagger a_1)$ has annihilation rank $\boldsymbol\nu=(0,0,1,0,0)$ while the operator $(a_1 a_1^\dagger)$ has $\boldsymbol\nu=(0,0,0,0,0)$.

A cyclic permutation of operator $A$ is understood as $(A_i\ldots A_mA_1\ldots A_{i-1})$. Consider now the set $\mathcal N$ of annihilation ranks $\boldsymbol\nu_i$ of all cyclic permutations of $A$ and annihilation ranks $\bar{\boldsymbol\nu}_i$ of cyclic permutations of $A^\dagger=(A_m^\dagger\ldots A_1^\dagger)$. We say that $(A_i\ldots A_mA_1\ldots A_{i-1})$ is maximally annihilating rotation of $A$ if $\boldsymbol\nu_i$ or $\bar{\boldsymbol\nu}_i$ maximizes $\mathcal N$ (in lexicographical order). The crucial observation is that either $\boldsymbol\nu_i$ or $\bar{\boldsymbol\nu}_i$ is nonzero for the maximally annihilation rotation. For instance, the maximally annihilating rotation of $(a_1a_1^\dagger a_2)$ is $(a_1^\dagger a_2 a_1)$.

Let $\tilde A$ be the maximally annihilating rotation of $A$. We now show how to transform $A$ into the following expression:
\begin{align}\label{eq:max_annihilating_form}
A=\tilde A+\mathcal R,
\end{align}
where $\mathcal R$ is a linear combination of products of trace operators. Each product of trace operators in $\mathcal R$ has total length smaller than $m$. Operator $A$ can be written in the following way:
\begin{align}\begin{split}
(A_1\dotsm A_m)&=A_1^{a_1}\dotsm A_m^{a_m}(T^{a_1}\dotsm T^{a_m})\\
&=A_2^{a_2}\dotsm A_m^{a_m}A_1^{a_1}(T^{a_2}\dotsm T^{a_m}T^{a_1})+\mathcal R\\
&=(A_{2}\dotsm A_{m}A_{1})+\mathcal R.
\end{split}\end{align}
The remainder $\mathcal R$ is
\begin{align}
\mathcal R&=\sum_i\epsilon_i A_2^{a_2}\dotsm A_{i-1}^{a_{i-1}}[A_1^{a_1},A_i^{a_i}]_{\pm}A_{i+1}^{a_{i+1}}\dotsm A_m^{a_m}(T^{a_1}\dotsm T^{a_m}),
\end{align}
where $[\cdot,\cdot]_\pm$ is a commutator or anticommutator, depending on whether both, $A_1^{a_1}$ and $A_i^{a_i}$ are fermionic or not. Coefficients $\epsilon_i$ are $(-1)^k$ where $k$ is the number of $j\in\{1,\ldots,i-1\}$ for which $[A_1^{a_1},A_j^{a_j}]_{\pm}$ is an anticommutator.
Because all operators $A_i^a$ are creation or annihilation operators, each (anti)commutator $[A_1^{a_1},A_i^{a_i}]_{\pm}$ is zero or $\pm\delta^{a_1,a_i}$. We take the sign into account by changing $\epsilon_i$, which now can be $\pm1$ or $0$. Now,
\begin{align}\label{eq:remainder}
\mathcal R&=\sum_{i=2}^m\epsilon_i \delta^{a_1a_i} A_2^{a_2}\dotsm A_{i-1}^{a_{i-1}}A_{i+1}^{a_{i+1}}\dotsm A_m^{a_m}(T^{a_1}\dotsm T^{a_m}).
\end{align}
In order to perform sum over $a_1$ we use the identity $T^a_{ij}T^a_{kl}=\frac{1}{2}\delta_{il}\delta_{kj}-\frac{1}{2N}\delta_{ij}\delta_{kl}$. This is in fact the only place where the value of $N$ is used. For $i=3,\ldots,m-1$ the trace in formula (\ref{eq:remainder}) is
\begin{align}
\delta^{a_1a_i}(T^{a_1}\dotsm T^{a_m})&=\frac{1}{2}(T^{a_2}\dotsm T^{a_{i-1}})(T^{a_{i+1}}\dotsm T^{a_m})-\frac{1}{2N}(T^{a_2}\dotsm T^{a_{i-1}}T^{a_{i+1}}\dotsm T^{a_m}).
\end{align}
For $i=2,m$ it is
\begin{align}\begin{split}
\delta^{a_1a_2}(T^{a_1}\dotsm T^{a_m})&=\frac{N^2-1}{2N}(T^{a_3}\dotsm T^{a_m}),\ m>2,\\
\delta^{a_1a_n}(T^{a_1}\dotsm T^{a_m})&=\frac{N^2-1}{2N}(T^{a_2}\dotsm T^{a_{m-1}}),\ m>2,\\
\delta^{a_1a_2}(T^{a_1}T^{a_2})&=\frac{N^2-1}{2}.
\end{split}\end{align}
In each case $\mathcal R$ can be written in terms of products of traces. The total length of traces in each product is equal to $m-2$. Up to this point $A$ was rotated by one position and an additional remainder was produced. This procedure is continued until $A$ is turned into its maximally annihilating rotation. Then it is applied to each trace operator in $\mathcal R$ to turn it into its maximally annihilating rotation.

Now, take a composite trace operator $B=B_1\dotsm B_p$ where each $B_j$ is in its maximally annihilating rotation. We associate annihilation rank $\boldsymbol\nu$ with $B$ in the following way. Let $\mu_i$ be the difference of numbers of operators $f_1$ and $f_1^\dagger$ in trace operators $B_{i+1}\dotsm B_p$. Then $\nu^1_F$ is maximum of all $\mu_i$. Let now $\mathcal N$ be the set of all annihilation ranks $\boldsymbol\nu_\sigma$ and $\bar{\boldsymbol\nu}_\sigma$ which correspond to permutations $B_{\sigma(1)}\dotsm B_{\sigma(p)}$ and their conjugates $B^\dagger_{\sigma(p)}\dotsm B^\dagger_{\sigma(1)}$. We say that $B$ is in the most annihilating permutation if its annihilation rank or annihilation rank of its hermitean conjugate maximizes $\mathcal N$ (in lexicographical order). The maximally annihilating form of $B$ will be denoted by $\tilde B$.

We now commute trace operators in $B$ in such way that it transforms into maximally annihilating permutation. Commuting trace operators reduces to commuting creating and annihilation operators between different traces. This gives contractions of color indices between traces. Such contractions are eliminated by the following relation:
\begin{align}\label{eq:color_contractions}\begin{split}
\delta^{a_ib_j}(T^{a_1}\dotsm T^{a_m})(T^{b_1}\dotsm T^{b_k})&=\frac{1}{2}(T^{a_1}\dotsm T^{a_{i-1}} T^{b_{j+1}}\dotsm T^{b_k} T^{b_1}\dotsm T^{b_{j-1}}T^{a_{i+1}}\dotsm T^{a_m})\\
&-\frac{1}{2N}(T^{a_1}\dotsm T^{a_{i-1}}T^{a_{i+1}}\dotsm T^{a_m})(T^{b_1}\dotsm T^{b_{j-1}}T^{b_{j+1}}\dotsm T^{b_k}).
\end{split}\end{align}
Finally, commuting traces produces only more products of trace operators. Therefore, $B$ can be written as
\begin{align}
B=\tilde B+\mathcal R,
\end{align}
where $\mathcal R$ is a combination of products of trace operators. Length of each composite trace in $\mathcal R$ is smaller than length of $B$. Next, we turn all trace operators in $\mathcal R$ into their maximally annihilating rotations and all composite traces into their maximally annihilating permutations.

Finally, take an arbitrary linear combination of products of trace operators $C$. Transform all trace operators in $C$ into their maximally annihilating rotations and all products of these into their maximally annihilating permutations. The result of this procedure is called \emph{maximally annihilating form} of $C$. What is important, all trace operators and thus also products of these have positive annihilation rank $\boldsymbol \nu$ (or $\bar{\boldsymbol\nu}$) in the sense that it has a nonzero component and the first nonzero component is positive. The only exception is the identity for which $\boldsymbol \nu=\bar{\boldsymbol\nu}=0$

\subsection{Recursion}\label{sec:recursion}

We now present the algorithm of computing the matrix of a given operator $A$. Assume that $A$ is given by sum of composite traces: $A=\sum_i \alpha_i \prod_j O_{ij}$ where $O_{ij}$ are trace operators and $\alpha_i$ are constants. Assume that each $O_{ij}$ is given in maximally annihilating form.

Recall that the full Hilbert space is decomposed into orthogonal subspaces (\ref{eq:Hilbert_space_decomposition}) and each subspace $\mathcal H_\mathbf n$ is spanned by all composite bricks with occupation numbers $\mathbf n$ acting on the Fock vacuum. The matrix of operator $A$ can be written in a block form with blocks $A|_{\mathbf n'\mathbf n}:\mathcal H_\mathbf n\rightarrow\mathcal H_{\mathbf n'}$. Dimension $D_\mathbf n$ of $\mathcal H_\mathbf n$ is finite for each $\mathbf n$. It can be determined numerically as it will be shown below or calculated from the character method which is presented in Chapter \ref{ch:characters}. Clearly, each block $A|_{\mathbf n'\mathbf n}$ can be expressed as a combination of products of block of components of $A$: $O_{ij}|_{\mathbf n''\mathbf n'''}$. Therefore, we assume that $A$ is a trace operator itself: $A=(A_1\ldots A_m)$.

Let $\hat n^\alpha_F$ be the number of creation operators $f_\alpha^\dagger$ in $A$ minus the number of annihilation operators $f_\alpha$. Let $\hat n^i_B$ be defined in analogous way. Then $\hat{\mathbf n}=(\hat n^1_F,\hat n^2_F,\hat n^1_B,\hat n^2_B,\hat n^3_B)$ is called \emph{creation labels} for $A$.

The block $A|_{\mathbf n'\mathbf n}$ is known a priori in several cases. If $A$ is an identity operator then $A|_{\mathbf n'\mathbf n}$ is an identity matrix for $\mathbf n=\mathbf n'$ and vanished otherwise. The block vanishes for any $A$ if $\mathbf n'\neq\mathbf n+\hat{ \mathbf n}$. It is also true when the annihilation rank $\boldsymbol\nu$ of $A$ is greater than $\mathbf n$ for at least one of five components. Similarly, the block vanishes when annihilation rank $\bar{\boldsymbol\nu}$ of $A^\dagger$ is greater than $\mathbf n'$ for at least one component. From now on we assume that none of these is true. Moreover, if $\boldsymbol \nu$ is greater than $\bar{\boldsymbol\nu}$ then we calculate $A^\dagger|_{\mathbf n\mathbf n'}$ and write $A|_{\mathbf n'\mathbf n}=\left(A^\dagger|_{\mathbf n\mathbf n'}\right)^\dagger$.

Let $\{\mathcal B_i\}$ be the set of all bricks that contain at least one creation operator corresponding to the first not vanishing component of $\mathbf n$ (e.g. if $n_F^1\neq0$ it is $f^\dagger_1$ and if $n_F^1=0\neq n_F^2$ then it is $f^\dagger_2$, etc.). Next, remove all bricks for which $\mathcal H_{\mathbf n_i}$, where $\mathbf n_i=\mathbf n-\hat{\mathbf n}_i$, is empty. In particular, all bricks for which at least one component of $\mathbf n_i$ is negative have to be removed. $\mathcal H_{\mathbf n_i}$ is then spanned by composite bricks acting on Fock vacuum, each containing at least one brick from the set $\{\mathcal B_i\}$. Now, take an orthonormal basis $\ket{e^{\mathbf n_i}_j}$ in the sector $\mathcal H_{\mathbf n_i}$. Then, vectors $\ket{v_k}=\mathcal B_i\ket{e^{\mathbf n_i}_j}$ span $\mathcal H_\mathbf n$. The index $k$ enumerates all pairs $(i,j)$ on the right hand side. Vectors $\ket{v_k}$ may be not orthogonal and the number of them can be larger than the dimension $D_{\mathbf n}$ of $\mathcal H_\mathbf n$. This will be taken into account by orthogonalization matrix.

The block element of $A$ is constructed in two steps. First, a block in the overcomplete basis is built:
\begin{align}
(\bar{A}|_{\mathbf n'\mathbf n})_{lk}=\braket{e^{\mathbf n'}_l|A|v_k}=\braket{e^{\mathbf n'}_l|A\mathcal B_i|e^{\mathbf n_i}_j}.
\end{align}
It can be written in a block form:
\begin{align}
\bar{A}|_{\mathbf n'\mathbf n}=
\left(\begin{array}{c}
(A\mathcal B_1)|_{\mathbf n'\mathbf n_1}\\
\vdots\\
(A\mathcal B_q)|_{\mathbf n'\mathbf n_q}
\end{array}\right).
\label{eq:block_matrix}
\end{align}
For each block, one finds first the maximally annihilating form of $A\mathcal B_i$ and then constructs $(A\mathcal B_i)|_{\mathbf n'\mathbf n_i}$. After the whole block $\bar{A}|_{\mathbf n'\mathbf n}$ is constructed, we orthogonalize basis of $\mathcal H_\mathbf n$. Let $S_\mathbf n$ be the Gramm matrix in the sector $\mathcal H_\mathbf n$:
\begin{align}
S_{\mathbf n}&=
\left(\begin{array}{ccc}
(\mathcal B_1^\dagger \mathcal B_1)|_{\mathbf n_1\mathbf n_1}&\ldots& (\mathcal B_1^\dagger \mathcal B_q)|_{\mathbf n_1\mathbf n_q}\\
\vdots&\ddots&\vdots\\
(\mathcal B_q^\dagger \mathcal B_1)|_{\mathbf n_q\mathbf n_1}&\ldots& (\mathcal B_q^\dagger \mathcal B_q)|_{\mathbf n_q\mathbf n_q}\\
\end{array}\right).
\label{eq:gramm_matrix}
\end{align}
Matrix $S_\mathbf n$ has exactly $D_\mathbf n$ nonzero eigenvalues $\lambda_l$. Eigenvectors $w^l_k$ of $S_{\mathbf n}$ which correspond to nonzero eigenvalues are used to construct basis of $\mathcal H_\mathbf n$. More precisely, the orthonormal basis is given by
\begin{align}
\ket{e_l^{\mathbf n}}=\frac{1}{\sqrt{\lambda_l}}\sum_k w^l_k\ket{v_k}.
\end{align}
These states are indeed orthonormal:
\begin{align}\begin{split}
\braket{e_m^{\mathbf n}|e_l^{\mathbf n}}&=\frac{1}{\sqrt{\lambda_m\lambda_l}}\sum_{kj}w^{m*}_jw^l_k\braket{v_j|v_k}=\frac{1}{\sqrt{\lambda_m\lambda_l}}\sum_{kj}w^{m*}_jw^l_k(S_\mathbf n)^j_{\phantom j k}\\
&=\frac{1}{\sqrt{\lambda_m\lambda_l}}\sum_{j}\lambda_l w^{m*}_jw^l_j=\delta_{ml}.
\end{split}\end{align}
Finally, we construct the orthogonalization matrix $R_\mathbf n$ by setting its matrix elements to $(R_\mathbf n)_{kl}=\frac{1}{\sqrt\lambda_l}w^l_k$. Then, $A|_{\mathbf n'\mathbf n}$ is the product of the block in overcomplete basis and the orthogonalization matrix:
\begin{align}
(\bar{A}|_{\mathbf n'\mathbf n}R_\mathbf n)_{ij}=\sum_k(\bar{A}|_{\mathbf n'\mathbf n})_{ik}(R_\mathbf n)_{kj}=\sum_k\braket{e^{\mathbf n'}_i|A|v_k}\frac{1}{\sqrt\lambda_j}w^j_k=\braket{e^{\mathbf n'}_i|A|e_j^{\mathbf n}}.
\end{align}

The block $A|_{\mathbf n'\mathbf n}$ is constructed. We give now summary of the algorithm.
\begin{enumerate}
\item Find all bricks $\mathcal B_i$ with creation labels $\hat{\mathbf n}_i$ such that the first nonzero component of $\hat{\mathbf n}_i$ is at the same position as for $\mathbf n$ and $\mathbf n_i=\mathbf n-\hat{\mathbf n}_i$ is non negative.
\item For each brick $\mathcal B_i$ write the operator $A\mathcal B_i$ in its maximally annihilating form.
\item Find all blocks $(A\mathcal B_i)|_{\mathbf n'\mathbf n_i}$ and use them to construct $\bar{A}|_{\mathbf n'\mathbf n}$.
\item Write all operators $\mathcal B_i^\dagger \mathcal B_j$ in maximally annihilating form.
\item Find all blocks $(\mathcal B_i^\dagger \mathcal B_j)|_{\mathbf n_i\mathbf n_j}$ and construct Gramm matrix $S_\mathbf n$.
\item Diagonalize $S_\mathbf n$ and construct the orthogonalization matrix $R_\mathbf n$.
\item The matrix block of $A$ is $A|_{\mathbf n'\mathbf n}=\bar{A}|_{\mathbf n'\mathbf n}R_\mathbf n$.
\end{enumerate}

The algorithm is recursive, so it requires a proof that it is finite. The proof is given in Appendix \ref{ap:finite_proof}.

\subsection{Diagonalization}\label{subsec:diagonalization}

Some remarks concerning diagonalization of matrices are in place. Assume that we constructed matrices of the Hamiltonian $H$, square of total angular momentum $J^2$ and the third component of angular momentum $J_3$ in a sector with $n_F$ fermions and at most $N_B$ bosons. In order to obtain energies one can diagonalize $H$ and then eventually act with $J^2$ and $J_3$ on eigenvectors to check what are their quantum numbers. This procedure is however ineffective.

Recall that $J^2$ and $J_3$ conserve the number of bosons $n_B$ (c.f. \ref{eq:angular_traces}) while $H$ does not. Therefore, matrices of angular momentum decompose into smaller matrices on subspaces with fixed $n_B$:

\begin{align}
J^2_{N_B}&=\left(
\begin{array}{cccc}
J^2_{n_B=1}&&\multicolumn{2}{c}{\multirow{2}{*}{\LARGE 0}}\\
&J^2_{n_B=2}&\multicolumn{2}{c}{}\\
\multicolumn{2}{c}{\multirow{2}{*}{\LARGE 0}}&\ddots&\\
\multicolumn{2}{c}{}&&J^2_{n_B=N_B}
\end{array}\right)\\
J_{3,N_B}&=\left(\begin{array}{cccc}
J_{3,n_B=1}&&\multicolumn{2}{c}{\multirow{2}{*}{\LARGE 0}}\\
&J_{3,n_B=2}&\multicolumn{2}{c}{}\\
\multicolumn{2}{c}{\multirow{2}{*}{\LARGE 0}}&\ddots&\\
\multicolumn{2}{c}{}&&J_{3,n_B=N_B}
\end{array}\right)
\end{align}
We diagonalize $J^2_{n_B}$ for each $n_B$. Since these matrices are smaller, it is much faster to diagonalize them. For each value of angular momentum $j$ we construct a projection matrix $P^{n_B}_j$. It maps the sector $\mathcal H_{n_B}$ onto a subspace $\mathcal H_{n_B,j}$ corresponding to given $j$. The projection matrix is composed of eigenvectors of $J^2$. Then $J_3$ on these small subspaces are
\begin{align}
J_{3,n_B,j}=P^\dagger_{n_Bj}J_{3,n_B}P_{n_Bj}.
\end{align}
These matrices are yet smaller and can be diagonalized for each $n_B$ and $j$ separately. The projection matrices $P_{n_B,j,m}$ from $\mathcal H_{n_B,j}$ to $\mathcal H_{n_B,j,m}$ are given by eigenvectors of $J_{3,n_B,j}$. Then the projection matrices from $\mathcal H_{n_B}$ to $\mathcal H_{n_B,j,m}$ are $\mathcal P_{n_B,j,m}=P_{n_B,j,m}P_{n_B,j}$. We construct a transition matrix
\begin{align}
\mathcal P_{N_B,j,m}&=\left(
\begin{array}{cccc}
\mathcal P_{n_B=1,j,m}&&\multicolumn{2}{c}{\multirow{2}{*}{\LARGE 0}}\\
&\mathcal P_{n_B=2,j,m}&\multicolumn{2}{c}{}\\
\multicolumn{2}{c}{\multirow{2}{*}{\LARGE 0}}&\ddots&\\
\multicolumn{2}{c}{}&&\mathcal P_{n_B=N_B,j,m}
\end{array}\right).
\end{align}
Finally, we construct the matrix of Hamiltonian in a channel with given $(j,m)$: $H_{N_B,j,m}=\mathcal P_{N_B,j,m}^\dagger H_{N_B}\mathcal P_{N_B,j,m}$. It is much smaller than the initial full matrix $H_{N_B}$ and thus diagonalization is faster.

\section{Gauss elimination for bricks}\label{sec:Gauss_bricks}
It was mentioned at the end of subchapter \ref{sec:brick_relations} that there is one more way to eliminate superfluous bricks. Here this method is presented.

Gauss elimination for bricks is based on an observation that the usual Gauss elimination which can be used to identify linearly dependent vectors, works also for bricks. If a set of composite bricks $\mathfrak B_i$ acting on the Fock vacuum gives linearly dependent states, i.e.
\begin{align}\label{eq:dependent_paths}
\sum_i\alpha_i\mathfrak B_i\ket{0}&=0,&\alpha_i\neq0,
\end{align}
then the composite bricks themselves are linearly dependent:
\begin{align}
\sum_i\alpha_i\mathfrak B_i&=0.
\end{align}
This is not true for general operators. However, it holds for composite bricks because they consist only of creation operators. The idea is to identify linearly dependent states (\ref{eq:dependent_paths}) and based on this knowledge eliminate unnecessary bricks. The procedure is the following.
\begin{enumerate}
\item Choose a sector with occupation numbers $\mathbf n$ and assume that the Gauss elimination was already performed for all $\mathbf n'$ such that $\mathbf n-\mathbf n'\geq0$.
\item Take all bricks $\mathcal B_i$ ($i=1,\ldots,q$) with occupation numbers $\hat{\mathbf n}_i$ such that $\mathbf n_i=\mathbf n-\hat{\mathbf n}_i$ is nonnegative.
\item Order bricks $\mathcal B_i$ such that $\mathbf n_i$ is positive for $i=1,\ldots,l$ and vanishes for $i=l+1,\ldots,q$.
That means that the first $l$ bricks give rise to composite bricks. The last $q-l$ bricks $\mathcal B_i$ acting on the Fock vacuum give a state in $\mathcal H_{\mathbf n}$.
\item Construct the Gramm matrix $S_{\mathbf n}$. Note that matrix elements $(\mathcal B_i^\dagger \mathcal B_i')|_{\mathbf n_i\mathbf n_{i'}}$ have size $1\times1$ for $i,i'>l$.
\item Perform Gauss elimination for $S_{\mathbf n}$. It has to be done in such way that first row $i=1,\ldots,l$ are used to eliminate elements in other columns. That is, in the first $l$ steps the pivot element is chosen from the first $l$ rows. In the remaining $q-l$ steps the pivot elements are chosen from the last $l-q$ rows.
\item Bricks $\mathcal B_i$ with $i>l$ that correspond to rows which vanish can be removed. They are linearly dependent with other composite bricks.
\end{enumerate}

Using the Gauss elimination one removes all dependent bricks. Still, bases generated in sectors $\mathcal H_\mathbf n$ are overcomplete. It happens for the following reason. Consider occupation numbers $\mathbf n=(0,0,3,1,0)$. If one constructs a block matrix element of an operator $A$ in this sector, then two bricks, $\mathcal B_1=(a_1^\dagger a_1^\dagger)$ and $\mathcal B_2=(a_1^\dagger a_2^\dagger)$ are taken into account. Then,
\begin{align}
\bar{A}|_{\mathbf n'\mathbf n}=
\left(\begin{array}{c}
(A\mathcal B_1)|_{\mathbf n'\mathbf n_1}\\
(A\mathcal B_2)|_{\mathbf n'\mathbf n_2}
\end{array}\right).
\end{align}
The occupation labels in the smaller sectors are $\mathbf n_1=(0,0,1,1,0)$, $\mathbf n=(0,0,2,0,0)$. Each of them is one--dimensional. The single basis vector in $\mathcal H_{\mathbf n_1}$ is $\mathcal B_2\ket{0}$ and the basis vector in $\mathcal H_{\mathbf n_2}$ is $\mathcal B_1\ket{0}$. The matrix $\bar{A}|_{\mathbf n'\mathbf n}$ has then two columns. One corresponds to $\mathcal B_1\mathcal B_2\ket{0}$ and the other to $\mathcal B_2\mathcal B_1\ket{0}$. This is in fact the same vector, so the basis generated in $\mathcal H_\mathbf n$ is overcomplete. This overcompleteness cannot be removed if one would like to preserve gains comming from the recursive nature of the algorithm.

It turned out that with this algorithm all bricks with more than six creation operators were eliminated. That means that all simple bricks which are longer than $6$ are expressed in terms of composite bricks.

Motivated by this observation, we found a relation between generators of $SU(3)$. First we assumed a general form
\begin{align}\label{eq:su3_relation}\begin{split}
T^{a_1}T^{a_2}T^{a_3}T^{a_4}T^{a_5}T^{a_6}&=\frac{1}{3}(T^{a_1}T^{a_2}T^{a_3}T^{a_4}T^{a_5}T^{a_6})\mathds 1_3\\
&\quad+A_{i_1i_2i_3i_4i_5i_6}(T^{a_{i_1}}T^{a_{i_2}}T^{a_{i_3}}T^{a_{i_4}}T^{a_{i_5}})T^{a_{i_6}}\\
&\quad+B_{i_1i_2i_3i_4i_5i_6}(T^{a_{i_1}}T^{a_{i_2}}T^{a_{i_3}}T^{a_{i_4}})[T^{a_{i_5}}T^{a_{i_6}}]\\
&\quad+C_{i_1i_2i_3i_4i_5i_6}(T^{a_{i_1}}T^{a_{i_2}}T^{a_{i_3}})[T^{a_{i_4}}T^{a_{i_5}}T^{a_{i_6}}]\\
&\quad+D_{i_1i_2i_3i_4i_5i_6}(T^{a_{i_1}}T^{a_{i_2}}T^{a_{i_3}})(T^{a_{i_4}}T^{a_{i_5}})T^{a_{i_6}}\\
&\quad+E_{i_1i_2i_3i_4i_5i_6}(T^{a_{i_1}}T^{a_{i_2}})[T^{a_{i_3}}T^{a_{i_4}}T^{a_{i_5}}T^{a_{i_6}}]\\
&\quad+F_{i_1i_2i_3i_4i_5i_6}(T^{a_{i_1}}T^{a_{i_2}})(T^{a_{i_3}}T^{a_{i_4}})[T^{a_{i_5}}T^{a_{i_6}}],
\end{split}\end{align}
where $[\cdot]$ is the traceless part of a matrix, i.e. $[M]=M-\frac{1}{3}(M)\mathds 1_3$. The equation (\ref{eq:su3_relation}) can be regarded as a set of equations with unknown coefficients $A,\ B,\ C,\ D,\ E,\ F$. There are over 260 thousand equations labeled by different sets of indices $(a_1,a_2,a_3,a_4,a_5,a_6)$. Indices $(i_1,i_2,i_3,i_4,i_5,i_6)$ are permutations of $(1,2,3,4,5,6)$. This gives 4320 free coefficients. Their number can be reduced to 1134 because of symmetry of the trace under cyclic permutations. This set of equations can be solved e.g. with Mathematica\footnote{A Mathematica notebook for solving (\ref{eq:su3_relation}) is available at \url{http://th.if.uj.edu.pl/~ambrozinski/SU3_relation.zip}}. The essential statement is that a solution exists. Because of certain relations between $SU(3)$ generators the solution is not unique. There are in fact 432 free parameters. We do not give explicit form of the solution because it is rather complicated. Nevertheless, existence of relation (\ref{eq:su3_relation}) implies that there are no bricks that are longer than 6.

Finally, we found that there are 786 bricks for the $SU(3)$ symmetry group.

\section{Eigenvectors in terms of composite bricks}

Recall that we are interested not only in values of energies of the Hamiltonian but also in wavefunctions of the bound states. Here we reconstruct states in the Fock space from eigenvectors of the Hamiltonian in the matrix representation.

Once the Hamiltonian $H_{N_B,j,m}$ is diagonalized, its eigenvectors $\{c^\mathbf q_l\}$ are known. Here, $\mathbf q=(n_F,j,m,E)$ are quantum numbers of the corresponding state and $l$ labels elements of the vector. We assume that $\{c^\mathbf q_l\}$ are given with respect to the orthogonal basis in $\mathcal H_{n_F,N_B}$. That is, once the Hamiltonian was diagonalized on the subspace $\mathcal H_{n_F,N_B,j,m}$ the eigenvectors are transformed back to $\mathcal H_{n_FN_B}$ with the transition matrix. In what follows the quantum numbers $\mathbf q$ are omitted for simpler notation.

Sectors corresponding to occupation numbers $\mathbf n$ are sorted in lexicographical order. It is then easy to identify which coefficients $c_l$ correspond to which $\mathcal H_\mathbf n$. The eigenvector $\{c_l\}$ can be given with two labels $\{c^{\mathbf n}_l\}$. In each subspace $\mathcal H_\mathbf n$ there is a basis $\ket{e^{\mathbf n}_l}$. The ket $\ket{\psi}$ corresponding to $\{c^{\mathbf n}_l\}$ is then
\begin{align}
\ket{\psi}=\sum_{\mathbf n,l}c^{\mathbf n}_l \ket{e^{\mathbf n}_l}.
\end{align}
The orthogonal basis is given in terms of the orthogonalization matrix $R_{\mathbf n}$ and bricks acting on basis in lower sectors. Therefore,
\begin{align}\begin{split}
\ket{\psi}&=c^{\mathbf 0} \ket{0}+\sum_{\mathbf n>0,l}c^{\mathbf n}_l \ket{e^{\mathbf n}_l}
=c^{\mathbf 0} \ket{0}+\sum_{\mathbf n,l}c^{\mathbf n}_l \sum_k(R_{\mathbf n})_{kl}\ket{v_k}\\
&=c^{\mathbf 0} \ket{0}+\sum_{\mathbf n,l}c^{\mathbf n}_l \sum_{i,p}(R_{\mathbf n})_{k(i,p)l}\mathcal B_i\ket{e^{\mathbf n_i}_p}.
\end{split}\end{align}
We continue to express $\ket{e^{\mathbf n_i}_p}$ in terms of basis of yet lower sectors until $\psi$ is given in terms of composite bricks acting on the Fock vacuum
\begin{align}\label{eq:state_as_path}
\ket{\psi}&=\sum_l\alpha_l\left(\prod_i\mathcal B_{il}\right)\ket{0}.
\end{align}

This form of state $\ket{\psi}$ can be then used to construct the wavefunction. Wavefunctions are discussed in Chapter \ref{ch:wavefunctions}.

\section{Additional variational parameter}\label{sec:omega}
The cut Fock space method is effectively a variational method for finding the lowest energies of a system. For a given $n_F$ and $N_B$ the Hilbert space has a finite number of basis states $\ket{e_l}$. A general state is given by a combination $\ket{\psi}=\sum_l b_l\ket{e_l}$. The lowest energy is approximated by the minimum of $\braket{\psi|H|\psi}$ with respect to coefficients $\{b_l\}$ with the constraint $\braket{\psi|\psi}=1$. This is precisely the lowest eigenvalue of the matrix $\left(\braket{e_l|H|e_m}\right)_{lm}$. The second eigenvalue is the minimum of $\braket{\psi'|H|\psi'}$ with normalization $\braket{\psi'|\psi'}=1$ and orthogonality condition $\braket{\psi'|\psi}=0$, etc. With increasing $N_B$ the number of variational parameters $b_l$ grows and thus the approximation is more accurate. We now present a way to introduce another variational parameter to the method and show its influence on the final result.

Basis of the Fock space are states
\begin{align}\label{eq:general_Fock_state}
\ket{\boldsymbol n, \boldsymbol \chi}=\ket{\boldsymbol n}_B\otimes\ket{\boldsymbol \chi}_F=\prod_{b,\alpha}(f^{b\dagger}_\alpha)^{\chi^b_\alpha}\prod_{a,i}\frac{1}{\sqrt{(n^a_i)!}}(a^{a\dagger}_i)^{n^a_i}\ket{0},
\end{align}
where $\boldsymbol n=(n^a_i)$ and $\boldsymbol \chi=(\chi^b_\alpha)$ denote bosonic and fermionic occupation numbers respectively. These states are not gauge singlets. An orthogonal basis $\{\ket{e_l}\}$ of the space of gauge invariant states $\mathcal H_{N_B}$ for given $n_F$ was introduced earlier in this chapter. It can be expressed as linear combinations of states (\ref{eq:general_Fock_state}):
\begin{align}
\ket{e_l}=\sum_{\boldsymbol n \boldsymbol \chi}c_l^{\boldsymbol n\boldsymbol\chi}\ket{\boldsymbol n, \boldsymbol \chi}.
\end{align}

Wavefunctions of bosonic states in the configuration representation $\prescript{}{B}{\braket{\boldsymbol x|\boldsymbol n}}_B$ are given by Hermite functions. The basis can be modified by introducing a parameter $\omega>0$ in the following way:
\begin{align}\begin{split}
\prescript{}{B}{\braket{\boldsymbol x|\boldsymbol n}}_{B,\omega}&\equiv\omega^{d/2}\prescript{}{B}{\braket{\omega\boldsymbol x|\boldsymbol n}}_B,\\
\ket{e_l}_\omega&=\sum_{\boldsymbol n \boldsymbol \chi}c_l^{\boldsymbol n\boldsymbol\chi}\ket{\boldsymbol n, \boldsymbol \chi}_\omega=\sum_{\boldsymbol n \boldsymbol \chi}c_l^{\boldsymbol n\boldsymbol\chi}\ket{\boldsymbol n}_{B,\omega}\otimes\ket{\boldsymbol \chi}_F,
\end{split}\end{align}
where $d=3(N^2-1)$ is dimension of the bosonic configuration space. States $\ket{\boldsymbol n, \boldsymbol \chi}_\omega$ satisfy proper orthogonality relations and therefore states $\ket{e_l}_\omega$ are orthogonal.

Matrix elements of the position operator in the modified basis $\ket{\boldsymbol n, \boldsymbol \chi}_\omega$ satisfy
\begin{align}\begin{split}
\prescript{}{B,\omega}{\braket{\boldsymbol m|x^a_i|\boldsymbol n}}_{B,\omega}
&=\int d^d\boldsymbol x \prescript{}{B,\omega}{\braket{\boldsymbol m|\boldsymbol x}}x^a_i\braket{\boldsymbol x|\boldsymbol n}_{B,\omega}\\
&=\omega^{-1}\int d^d (\omega \boldsymbol x) \prescript{}{B}{\braket{\boldsymbol m|\omega\boldsymbol x}}\omega x^a_i\braket{\omega\boldsymbol x|\boldsymbol n}_B\\
&=\omega^{-1}\prescript{}{B}{\braket{\boldsymbol m|x^a_i|\boldsymbol n}}_B.
\end{split}\end{align}
A similar proof shows that $\prescript{}{B,\omega}{\braket{\boldsymbol m|p^a_i|\boldsymbol n}}_{B,\omega}=\omega\prescript{}{B}{\braket{\boldsymbol m|p^a_i|\boldsymbol n}}_B$. Finally, it can be read from (\ref{eq:angular_momentum}), (\ref{eq:gauge_generator}) and (\ref{eq:hamiltonian}) that matrix elements of other operators in the orthogonal basis $\ket{e_l}$ yield
\begin{align}\begin{split}
\prescript{}{\omega}{\braket{e_l|x^a_i|e_m}}_\omega&=\omega^{-1}\prescript{}{}{\braket{e_l|x^a_i|e_m}},\\
\prescript{}{\omega}{\braket{e_l|p^a_i|e_m}}_\omega&=\omega\prescript{}{}{\braket{e_l|p^a_i|e_m}},\\
\prescript{}{\omega}{\braket{e_l|J_i|e_m}}_\omega&=\prescript{}{}{\braket{e_l|J_i|e_m}},\\
\prescript{}{\omega}{\braket{e_l|G^a|e_m}}_\omega&=\prescript{}{}{\braket{e_l|G^a|e_m}},\\
\prescript{}{\omega}{\braket{e_l|H_K|e_m}}_\omega&=\omega^2\prescript{}{}{\braket{e_l|H_K|e_m}},\\
\prescript{}{\omega}{\braket{e_l|H_V|e_m}}_\omega&=\omega^{-4}\prescript{}{}{\braket{e_l|H_V|e_m}},\\
\prescript{}{\omega}{\braket{e_l|H_F|e_m}}_\omega&=\omega^{-1}\prescript{}{}{\braket{e_l|H_F|e_m}}.
\end{split}\label{eq:omega_dependence}\end{align}

It is significant that matrix elements of $J_i$ and $G^a$ are invariant under  $\omega$. It follows that $\ket{e_l}_\omega$ are gauge invariant because $G^a\ket{e_l}=0$. Therefore, states $\ket{e_l}_\omega$ form an orthogonal basis of the Hilbert space of gauge singlets. Moreover, given an angular momentum multiplet $\ket{j,m}=\sum_l c_l^{j,m}\ket{e_l}$, the scaled combination $\ket{j,m}_\omega=\sum_l c_l^{j,m}\ket{e_l}_\omega$ is also an angular momentum multiplet with the same quantum numbers $(j,m)$.

We now recall how the energies are calculated to show how $\omega$ can be included. First, for given $n_F$ matrices of operators $H_K,\ H_V,\ H_F, J_i$ are constructed in the cut space $\mathcal H_{N_B}$. Then, $J^2$ and $J_3$ are diagonalized. They are $\omega$--independent. Next, the three matrices $(H_K)_{N_B},\ (H_C)_{N_B},\ (H_F)_{N_B}$ are projected to the subspaces $\mathcal H_{N_B,j,m}$. Their dependence on $\omega$ is given by (\ref{eq:omega_dependence}). Therefore, $\omega$ can be taken into account be constructing
\begin{align}
(H_\omega)_{N_B,j,m}=\omega^2(H_K)_{N_B,j,m}+\omega^{-4}(H_V)_{N_B,j,m}+\omega^{-1} (H_F)_{N_B,j,m},
\end{align}
where all matrices on the right hand side are constructed for $\omega=1$. Finally, $\omega$ is chosen such that the lowest eigenvalue of $(H_\omega)_{N_B,j,m}$ is minimized. This requires diagonalizing $(H_\omega)_{N_B,j,m}$ multiple times. However, subspaces $\mathcal H_{N_B,j,m}$ are reasonably small, especially for small $j$, and this minimization does not consume significant computer resources.

We come to discussing the effect of including $\omega$. Results concerning the energies are given in Tab. \ref{tab:energies_omega_min}. In the sector $(n_F,j)=(0,0)$ the smallest eigenvalue with cutoff $N_B$ minimized with respect to $\omega$ is smaller then the the lowest eigenvalues for cutoff $N_B+2$ with $\omega=1$. That is, including the parameter $\omega$ effectively increases the cutoff by $2$. The effect is similar for $(n_F,j)=(2,1)$. In the singlet channel for six fermions inclusion of $\omega$ effectively rises $N_B$ by almost $1$.

\newcolumntype{d}[1]{D{.}{.}{#1}}
\def\arraystretch{1.3}
\begin{table}\centering
\begin{tabular}{|c||d{-1}|d{-1}||d{-1}|d{-1}||d{-1}|d{-1}|}
\hline
\multirow{2}{*}{$N_B$}&\multicolumn{6}{c|}{lowest eigenvalues in sector with quantum numbers $(n_F,j)$}\\
\cline{2-7}&\multicolumn{1}{c|}{$(0,0)$, no $\omega$}&\multicolumn{1}{c||}{$(0,0)$ with $\omega$}&\multicolumn{1}{c|}{$(2,1)$, no $\omega$}&\multicolumn{1}{c||}{$(2,1)$ with $\omega$}&\multicolumn{1}{c|}{$(6,0)$, no $\omega$}&\multicolumn{1}{c|}{$(6,0)$ with $\omega$}\\\hline
 0&15       &12.9919    &       &           &15        &12.99\\
 1&15       &12.9919    &17     &14.41      &9.06      &7.83\\
 2&13.327   &12.9517    &14.35  &12.15      &6.12      &5.10\\
 3&13.327   &12.9517    &13.03  &11.47      &4.59      &3.74\\
 4&12.8821  &12.6322    &12.05  &11.25      &          &\\
 5&12.8821  &12.6322    &11.54  &11.04      &          &\\
 6&12.7123  &12.6203    &11.20  &10.79      &          &\\
 7&12.7123  &12.6203    &10.97  &10.64      &          &\\
 8&12.6339  &12.5911    &       &           &          &\\
 9&12.6339  &12.5911    &       &           &          &\\
10&12.6038  &12.5889    &       &           &          &\\\hline
\end{tabular}
\caption{The lowest energies in selected sectors. In each case the energy is given with keeping $\omega=1$ and minimizing the energy with respect to $\omega$.}
\label{tab:energies_omega_min}
\end{table}

In Tab. \ref{tab:omega_min} optimal values of $\omega$ for different channels are given. It can be seen that all values of $\omega$ are similar. They are slightly larger for $n_F\geq 4$. This means that the functions are in general wider for high $n_F$ which may be a slight indication that the spectrum is continuous in these sectors.

\def\arraystretch{1.2}
\begin{table}\centering
\begin{tabular}{|c|d{-1}|d{-1}|d{-1}|d{-1}|d{-1}|d{-1}|d{-1}|d{-1}|d{-1}|}
\hline
\diaghead{\theadfont Diag Colu}{$j$}{$n_F$}&
0&1&2&3&4&5&6&7&8\\
\hline
$0$             &1.22&&1.21&&1.15&&1.15&&1.14\\
$1/2$   &&1.31&&1.23&&1.15&&1.16&\\
$1$             &1.25&&1.22&&1.15&&1.15&&1.15\\
$3/2$   &&1.27&&1.22&&1.15&&1.15&\\
$2$             &1.25&&1.20&&1.15&&1.16&&1.15\\\hline
\end{tabular}
\caption{Optimal values of $\omega$ in each $(n_F,j)$ channel. The value depends on the cutoff and is given for the maximal cutoff available.}
\label{tab:omega_min}
\end{table}
\def\arraystretch{1}

\section{Modifications for SU(2)}\label{sec:su2_modifications}
Up to this point the cut Fock space method was discussed in reference to the $SU(3)$ theory. The case of $SU(2)$ is simpler. Some changes can be done to the algorithm so that a higher cutoff can be reached. In this part we briefly discuss these modifications.

For $SU(2)$ there is a relation similar to (\ref{eq:su3_relation}), but it decomposes a product of three $SU(2)$ generators:
\begin{align}\label{eq:su2_relation}
T^aT^bT^c&=\frac{1}{2}(T^aT^b)T^c-\frac{1}{2}(T^aT^c)T^b+\frac{1}{2}(T^bT^c)T^a+\frac{1}{2}(T^aT^bT^c)\mathds 1_2.
\end{align}
This implies that there are no bricks longer than three. Some bricks can be eliminated because of invariance of trace under cyclic permutations as for the $SU(3)$ case. The Cayley--Hamilton theorem applied to a matrix $M$ of size $2\times2$ implies that $M^2$ can be expressed in terms of $(M^2)$ and $(M)$. Therefore, there are no bricks of the form $(a_i^\dagger a_i^\dagger b^\dagger)$, where $b^\dagger$ is any creation operator. All independent bricks can be easily identified and there are $35$ of them.

The relation (\ref{eq:su2_relation}) has further consequences. It implies that there are no single--trace operators of length higher than three. One can calculate all commutators of single--trace operators beforehand and then use the in the program for calculating matrix elements. Because of this, one does not have to calculate the maximally annihilating form of an operator during a run. Instead, one uses the commutators to write $A\mathcal B_i=\mathcal B_iA+ [A,\mathcal B_i]$ (cf. (\ref{eq:block_matrix})). This is in principle possible also for $SU(3)$. However, the formula (\ref{eq:su3_relation}) limits the single--trace operators to length $6$ and in practice there are too many commutators to be calculated and memorized.

Secondly, it is relatively easy to construct operators with definite angular momentum from single--trace operators. Using operators with definite angular momentum allows one to construct matrix elements of the Hamiltonian directly in channels $\mathcal H_{n_FN_Bjm}$. Then one does not have to diagonalize the matrix representations of $J^2$ and $J_3$ but directly the Hamiltonian $H$. This significantly reduces the time needed for diagonalization and allows one to reach higher cutoffs.

Both of above simplifications are taken into account in \cite{Campostrini}. This allowed the authors to reach a cutoff $N_B=18$ in all fermionic sectors. In this thesis $SU(2)$ case is used only as a guidance for analysis of $SU(3)$ and we are interested in small $N_B$ behavior. Therefore, we do not incorporate modifications discussed above.

\section{Summary}
In this chapter the cut Fock space method was presented. The bosonic and fermionic creation and annihilation operators and the Fock vacuum state were introduced. Because of the Gauss law, only gauge invariant states and operators are relevant. In the matrix notation, creation and annihilation operators are given by operator--valued matrices. Gauge invariant operators are traces of products of such matrices. They are called trace operators. A special class of trace operators, namely those which consist only of creation operators are called bricks. Composite bricks are products of bricks. The whole physical Fock space is generated by composite bricks acting on the empty state.

In the Hilbert space we introduce a cutoff $N_B$ for the maximal number of bosonic excitations. In this way the full space is reduced to a finite dimensional cut Fock space. In this subspace operators are represented by matrices. Because angular momentum and Hamiltonian conserve the total number of fermions, it is convenient to work with subspaces with definite $n_F$. It is argued that the spectrum of angular momentum operators is exact for finite cutoff because the angular momentum operators conserve the total number of bosons $n_B$. The spectrum of the Hamiltonian is recovered only in the $N_B\to\infty$ limit.

All possible composite bricks acting on the empty generate an overcomplete basis. In subchapters \ref{sec:brick_relations} and \ref{sec:Gauss_bricks} we have shown how to reduce the number of bricks to minimum. The three basic relations are
\begin{itemize}
\item invariance of trace under cyclic permutations,
\item Cayley-Hamilton theorem for matrices with commuting and anticommuting  matrix elements (\ref{eq:CH}) and (\ref{eq:CH_anticommutation}),
\item linear dependence of products of three $SU(3)$ generators $T^a$ (\ref{eq:symmetric_reduction}).
\end{itemize}
We argued that more bricks can be removed with Gauss elimination. We found that there are no bricks that are longer than 6. Finally, we explained this by showing existence of a decomposition formula for product of six generators $T^a$ (\ref{eq:su3_relation}).

In \ref{sec:algorithm} we proposed an algorithm for computing matrix of an operator. The key point is transforming an operator to its maximally annihilating form. For an operator $A$ and a brick $\mathcal B$ the maximally annihilating form of $A\mathcal B$ (\ref{eq:block_matrix}) is given by $\mathcal B A+\mathcal R$ where $\mathcal R$ is the maximally annihilating form of $[A,\mathcal B]$. Therefore, calculating the maximally annihilating form of $A\mathcal B$ is the same as calculating commutator of $A$ and $\mathcal B$ and giving it in a form convenient for further computations. Overcompleteness of basis is taken into account by introducing orthogonalization matrices $R_{\mathbf n}$. The algorithm is recursive. Proof of its finiteness is given in Appendix \ref{ap:finite_proof}.

We presented an effective way to diagonalize the Hamiltonian. It is efficient first to diagonalize the matrix of angular momentum $J^2$ and $J_3$. This is because these operators conserve the total number of bosons. Therefore, they can be diagonalized on smaller subspaces which is much faster.

Finally, it was shown how to reconstruct a state in terms of composite bricks acting on the Fock vacuum from eigenvectors of operators in matrix representation. This form of states is used in Chapter \ref{ch:wavefunctions} where we construct and study wavefunctions.

\chapter{Number of gauge invariant states}\label{ch:characters}
In the previous Chapter we introduced a set of vectors $\ket{v^k}$ which span a sector $\mathcal H_\mathbf n$ for given occupation numbers $\mathbf n$. It was noted that $\ket{v^k}$ form an overcomplete basis. The proper dimension of $\mathcal H_\mathbf n$ can be read out from the number of nonzero eigenvalues of the Gramm matrix $S_\mathbf n$. However, these eigenvalues bare some numerical errors. Therefore, one would like to cross check if there is a clear distinction between zero and nonzero eigenvalues and if the dimension of $\mathcal H_\mathbf n$ obtained in this way is correct. Secondly, the knowledge of dimensions of sectors a priori provides a check for completeness of the set of bricks. To this end an analytic way of calculating the number of independent $SU(3)$ gauge invariant states in a given sector is presented. It is called character method because the number of states is given by an integral of characters over the group as it is shown below. The analysis in this chapter is based on \cite{Trzetrzelewski_number} and modified. Introduction to the group theory and definitions of objects used in this chapter can be found e.g. in \cite{DiFrancesco}.

\section{The character method}
A general gauge invariant state in the Fock space can be written in the following form:
\begin{align}\label{eq:general_state}
\ket{s}=t^{a_1\ldots a_n b_1\ldots b_m}a_{i_1}^{a_1\dagger}\dotsm a_{i_{n}}^{a_n\dagger}f_{\alpha_1}^{b_1\dagger}\dotsm f_{\alpha_m}^{b_m\dagger}\ket{0}.
\end{align}
All bosonic and fermionic creation operators are in adjoint SU(3) representation. In this chapter the adjoint representation is denoted by $R$. Assume that  in the combination (\ref{eq:general_state}) there are $n_B^i$ bosonic creation operators $a^{a\dagger}_i$ for each $i$ and $n_F^\alpha$ fermionic creation operator $f^{b\dagger}_\alpha$ for each $\alpha$. The state $\ket{s}$ has occupation numbers $\mathbf n=(n_B^1,n_B^2,n_B^2,n_F^1,n_F^2)$. In order to simplify notation we introduce
\begin{align}
\mathbf n_B&=(n_B^1,n_B^2,n_B^3),&n_B&=n_B^1+n_B^2+n_B^3,\\
\mathbf n_F&=(n_F^1,n_F^2),&n_F&=n_F^1+n_F^2.
\end{align}

The state $\ket{s}$ is in the representation
\begin{align}\label{eq:product}
Sym(\otimes^{n_B^1}R)\times Sym(\otimes^{n_B^2}R)\times Sym(\otimes^{n_B^3}R)\times Alt(\otimes^{n_F^1}R)\times Alt(\otimes^{n_F^2}R),
\end{align}
where $Sym$ and $Alt$ are symmetrization and antisymmetrization of the tensor product. Let $D_\mathbf n$ be the number of independent singlet states in sector $\mathcal H_\mathbf n$. $D_\mathbf n$ is equal to the number of trivial representations that appear in the product (\ref{eq:product}). From orthogonality properties of characters it follows \cite{Hamermesh} that $D_\mathbf n$ is given by an integral of character over the group $SU(3)$:
\begin{align}\label{eq:character_integral}
D_{\mathbf n}=\int d\mu\chi^{n_B^1}_{Sym}(R)\chi^{n_B^2}_{Sym}(R)\chi^{n_B^3}_{Sym}(R)\chi^{n_F^1}_{Alt}(R)\chi^{n_F^2}_{Alt}(R),
\end{align}
where $\mu$ is the Haar measure of SU(3). Characters of symmetric and antisymmetric powers of representation $R$ are constructed with Frobenius formula later in this chapter. In order to make the integral (\ref{eq:character_integral}) simpler, we use the generating function
\begin{align}
G(\boldsymbol a, \boldsymbol b)=\sum_{n_B^i=0}^\infty\sum_{n_F^\alpha=0}^\infty a_1^{n_B^1}a_2^{n_B^2}a_3^{n_B^3}(-b_1)^{n_F^1}(-b_2)^{n_F^2} D_{\mathbf n}.
\end{align}
Symbols $\boldsymbol a=(a_1,a_2,a_3)$ and $\boldsymbol b=(b_1,b_2)$ are introduced for shorter notation.

Let $U$ be an element of $SU(3)$ in the fundamental representation. $U$ can be decomposed as $U=W\Lambda W^\dagger$ where $W$ is unitary and $\Lambda$ is diagonal with $\det(\Lambda)=1$. Then $\Lambda=diag(z_1,z_2,z_3)$ depends on two parameters because $z_1z_2z_3=1$. We write $z_j=e^{i\phi_j}$. The character $\chi(R)$ depends on $\Lambda$ and does not depend on parameters of $W$. Therefore, it would be convenient to have a Haar measure $\mu$ (cf. (\ref{eq:character_integral})) integrated over parameters of $W$. This measure is given by the formula (cf. \cite{Drouffe})
\begin{align}
d\mu=\frac{1}{6}\prod_{i=1}^3\frac{d\phi_i}{2\pi}\delta_P(\sum_{i=1}^3\phi_i)\prod_{i<j}\left|z_i-z_j\right|^2,
\end{align}
where $\phi\in[0,2\pi]$ and $\delta_P(x)=\sum_{k\in\mathbf Z} \delta(x-2\pi k)$. In this parametrization character of an element $R$ in adjoint representation is given by Schur function
\begin{align}
\chi(R)&=\chi(\{z_j\})=\det({z^{l_i+3-i}_j})/\det(\{z_j^{3-i}\}),
\end{align}
where $l_i=3-i$ is partition of the adjoint representation of $SU(3)$. This is exactly the Weyl determinant formula \cite{Weyl}. The determinants can be calculated explicitly and the character is
\begin{align}
\begin{split}
\chi(\{z_i\})&=\det\left(\left[\begin{array}{ccc}1&1&1\\z_1^2&z_2^2&z_3^2\\z_1^4&z_2^4&z_3^4\end{array}\right]\right)/\det\left(\left[\begin{array}{ccc}1&1&1\\z_1&z_2&z_3\\z_1^2&z_2^2&z_3^2\end{array}\right]\right)\\
&=\prod_{i<j}(z_i+z_j)=\sum_{i,j}\frac{z_i}{z_j}-1.
\end{split}
\end{align}
The character of $R^k$ is given simply by
\begin{align}
\chi(R^k)&=\chi(\{z_i^k\})=\sum_{i,j}\frac{z_i^k}{z_j^k}-1.
\end{align}
The character of symmetric and antisymmetric powers of $\chi(R)$ is given by the Frobenius formula \cite{Hamermesh}:
\begin{align}
\chi_{Sym}^{(n)}(R)&=\sum_{\sum_k k i_k=n}\prod_{k=1}^{n}\frac{1}{i_k!}\frac{\chi^{i_k}(R^k)}{k^{i_k}},\\
\chi_{Alt}^{(n)}(R)&=\sum_{\sum_k k i_k=n}(-1)^{\sum_k i_k}\prod_{k=1}^{n}\frac{1}{i_k!}\frac{\chi^{i_k}(R^k)}{k^{i_k}}.
\end{align}
Finally, we introduce auxiliary functions
\begin{align}
F_{Sym}(a,\{\phi_i\})&=\sum_{n=0}^\infty a^{n}\chi^{(n)}_{Sym}(R)\\
F_{Alt}(b,\{\phi_i\})&=\sum_{n=0}^\infty (-b)^{n}\chi^{(n)}_{Alt}(R).
\end{align}
Functions $F_{Sym}(a,\{\phi_i\})$ and $F_{Alt}(b,\{\phi_i\})$ can be simplified in the following way:
\begin{align}
\begin{split}
F_{Sym}(a,\{\phi_i\})&=\sum_{n=0}^\infty a^{n}\sum_{\sum_k k i_k=n}\prod_{k=1}^{n}\frac{1}{i_k!}\frac{\chi^{i_k}(R^k)}{k^{i_k}}
=\sum_{n=0}^\infty \sum_{\sum_k k i_k=n}\prod_{k=1}^{n}\frac{1}{i_k!}\left(\frac{a^k\chi(R^k)}{k}\right)^{i_k}\\
&=\sum_{n=0}^\infty \frac{1}{n!}\left(\sum_k\frac{a^k\chi(R^k)}{k}\right)^n
=\exp\left(\sum_{ij}\sum_k\frac{1}{k}\frac{a^k z_i^k}{z_j^k}-\sum_k\frac{a^k}{k}\right)\\
&=\exp\left(-\sum_{ij}\ln\left(1-\frac{a z_i}{z_j}\right)+\ln(1-a)\right)=(1-a)\prod_{ij}\left(1-\frac{a z_i}{z_j}\right)^{-1}.
\end{split}\end{align}
Similarly, one can show that
\begin{align}
\begin{split}
F_{Alt}(b,\{\phi_i\})=(1-b)^{-1}\prod_{ij}\left(1-b\frac{z_i}{z_j}\right).
\end{split}\end{align}
Then the generating function $G$ has the form
\begin{align}
G(\boldsymbol a,\boldsymbol b)&=\int_{[0,2\pi]^3}d\mu \prod_j F_{Sym}(a_j,\{\phi_i\})\prod_\alpha F_{Alt}(b_\alpha,\{\phi_i\}).
\end{align}
We change the variables
\begin{align}
\phi_i&\rightarrow -\sum_{j\neq i}\phi_j,&\phi_N&\rightarrow \sum_{j}\phi_j.
\end{align}
The jacobian is $1/3$. Integration over the last variable is over $[0,6\pi]$ giving $3$ identical integrals (from the $\delta_P$ function). This cancels with the jacobian and $\delta_P(\phi_3)$ is replaced by $\delta(\phi_3)$. The generating function is
\begin{align}\begin{split}
G(\boldsymbol a,\boldsymbol b)&= \frac{1}{6}\left(\frac{(1-b_1)(1-b_2)}{(1-a_1)(1-a_2)(1-a_3)}\right)^{2}\int_0^{2\pi}\prod_{i=1}^3\frac{d\phi_i}{2\pi}\delta(\phi_3)\times\\
&\quad\times\prod_{i\neq j}(1-\frac{z_i}{z_j})\frac{(1-b_1\frac{z_i}{z_j})(1-b_2\frac{z_i}{z_j})}{(1-a_1\frac{z_i}{z_j})(1-a_2\frac{z_i}{z_j})(1-a_3\frac{z_i}{z_j})}.
\end{split}\end{align}

\section{Calculation of the generating function}
The integrals are calculated one by one:
\begin{align}\label{eq:G_decomposition}\begin{split}
G(\boldsymbol a,\boldsymbol b)&=\frac{1}{6}\left(\frac{(1-b_1)(1-b_2)}{(1-a_1)(1-a_2)(1-a_3)}\right)^{2}\int_{\mathcal C}dz_2 G_1(\boldsymbol a,\boldsymbol b,z_2),\\
G_1(\boldsymbol a,\boldsymbol b,z_2)&=\int_{\mathcal C} dz_1 G_2(\boldsymbol a,\boldsymbol b,z_1,z_2),\\
G_2(\boldsymbol a,\boldsymbol b,z_1,z_2)&=\frac{1}{z_1z_2}\prod_{i\neq j}\frac{(z_j-z_i)(z_j-b_1 z_i)(z_j-b_2z_i)}{(z_j-a_1z_i)(z_j-a_2z_i)(z_j-a_3z_i)},
\end{split}\end{align}
where $\mathcal C$ is the unit circle on the complex plane. The range of indices $i,j$ is $1,2,3$. Remember that $z_3=1$. The integral is done by residues. After the integrals are calculated, the dimensions $D_\mathbf n$ of sectors $\mathcal H_\mathbf n$ are read from the Taylor expansion of $G(\boldsymbol a,\boldsymbol b)$ around 0. Therefore, in all calculations it is assumed that $a_i$ and $b_\alpha$ are small.

First we perform the integral over $z_1$ by residues. There are seven simple poles in $z_1$ variable: $z_1=0,a_k,a_k z_2$. The residues are
\begin{align}
res_0 G_2=\frac{(1-z_2)^2\prod_{i\neq j}\prod_\alpha (z_j-b_\alpha z_i)}{z_2\prod_ka_k(1-a_k z_2)(z_2-a_k)},
\end{align}
where $z_1=0$, $z_3=1$,
\begin{align}
res_{a_k} G_2=\frac{(z_2-1)(z_2-a_k)(1-a_k)}{a_k^2 z_2(z_2-a_k^2)(1+a_k)(1-a_kz_2)}\prod_{i\neq j}\frac{(z_j-b_1 z_i)(z_j-b_2z_i)}{\prod_{n\neq k}(z_j-a_nz_i)},
\end{align}
where $z_1=a_k,\ z_3=1$ and
\begin{align}
res_{a_k z_2} G_2=\frac{(1-a_kz_2)(1-a_k)(1-z_2)}{a_k^2z_2(1+a_k)(z_2-a_k)(1-a_k^2z_2)}\prod_{i\neq j}\frac{(z_j-b_1 z_i)(z_j-b_2z_i)}{\prod_{n\neq k}(z_j-a_nz_i)},
\end{align}
with $z_1=a_kz_2$, $z_3=1$.

$res_0 G_2$ has four simple poles, $z_2=0$ and $z_2=a_k$. There are six simple poles of $res_{a_k} G_2$, namely $z_2=0$, $z_2=a_n$ for $n\neq k$ and $z_2=a_na_k$. Finally, $res_{a_k z_2} G_2$ has four simple poles $z_2=0$ and $z_2=a_n$. None of these functions has poles of higher order.

Now the function $G_1(\boldsymbol a, \boldsymbol b,z_2)$ from (\ref{eq:G_decomposition}) is
\begin{align}
G_1(\boldsymbol a, \boldsymbol b,z_2)=\int_{\mathcal C}d z_1 G_2(\boldsymbol a,\boldsymbol b,z_1,z_2)=res_0G_2+\sum_k(res_{a_k}G_2+res_{a_k z_2}G_2)
\end{align}
Finally, $G(\boldsymbol a,\boldsymbol b)$ is an integral of $G_1(\boldsymbol a, \boldsymbol b,z_2)$. It is again calculated by residues. The form of the integrals becomes somewhat more complicated:
\begin{align*}
res_{0} G_1&=\frac{\prod_\alpha b_\alpha^3}{\prod_i a_i^3}+\frac{2 \prod_\alpha b_\alpha^2}{\prod_{i}a_i^2\prod_{j<k}(a_j a_k-1)(a_j-a_k)}\\
&\quad\times\sum_i\frac{(a_i-1)\prod_\alpha(a_i-b_\alpha)(a_ib_\alpha-1)(a_m a_n-1)(a_m-a_n)}{a_i(a_i+1)}
\end{align*}
where $m$ and $n$ are such that $(i,m,n)$ is a cyclic permutation of $(1,2,3)$.

\begin{align*}
res_{a_k} G_1&=\frac{1}{\sum_{i\neq j}(a_j a_j-1)}\frac{\prod_\alpha (a_k-b_\alpha)(a_k b_\alpha-1)}{a_k^2(1+a_k)\prod_{i\neq k}(a_k-a_i)}\Bigg(
\frac{(a_k-1)(\prod_{i\neq k} a_i-1)\prod_\alpha b_\alpha^2}{a_k\prod_{i\neq k a_i^2}}\\
&+\frac{(a_k-1)^2(a_k+1)(\prod_{i\neq k}a_i-1)\prod_\alpha(a_k-b_\alpha)(a_k^2-b_\alpha)(a_k b_\alpha-1)(a_k^2 b_\alpha-1)}{a_k(1+a_k+a_k^2)\prod_{i\neq k}(a_k-a_i)(a_k^2-a_i)(a_k a_i-1)(a_k^2 a_i-1)}\\
&+\sum_{i\neq k}\Big(
\frac{(a_i-a_k)\prod_\alpha(a_i-b_\alpha)(a_i-a_k b_\alpha)(a_k-a_i b_\alpha)(a_i b_\alpha-1)}{a_i^2(a_i-a_k^2)(a_k-a_i^2)(1+a_i)(a_ia_k-1)(a_i-a_m)(a_i-a_ka_m)(a_k-a_ia_m)}\\
&+\frac{(a_ka_i-1)\prod_\alpha(a_i-b_\alpha)(a_ia_k-b_\alpha)(a_ka_ib_\alpha-1)(a_ib_\alpha-1)}{a_i^2(a_k-a_i)(a_i-a_m)(a_m-a_ia_k)(1+a_i)(a_i^2a_k-1)(a_ia_k^2-1)(a_ia_ka_m-1)}
\Big)\Bigg)
\end{align*}
where $m$ is assumed to be different then $i$ and $k$, i.e. $(i,k,m)$ is a permutation of $(1,2,3)$.

\begin{align*}
res_{a_k^2}G_1&=\frac{(-1+a_k)^2\prod_\alpha (a_k-b_\alpha)^2(a_k^2-b_\alpha)(-1+a_k b_\alpha)^2(-1+a_k^2b_\alpha)}{a_k^3(1+a_ka_k^2)\prod_{i\neq k}(a_k-a_i)^2(a_k^2-a_i)(-1+a_ka_i)^2(-1+a_k^2a_i)}.
\end{align*}

\begin{align*}
res_{a_k a_n}G_1&=-2\frac{\prod_\alpha(a_k-b_\alpha)(a_n-b_\alpha)(a_ka_n-b_\alpha)(-1+a_kb_\alpha)(-1+a_nb_\alpha)(-1+a_ka_nb_\alpha)}
{(a_k-a_n)^2(a_k a_n-a_l)(-1+a_l a_k a_n)}\\
&\quad\times\prod_{j=k,n}\frac{1}{a_j^2(1+a_j)(a_j-a_l)(-1+a_la_j)(-1+a_ja_ka_n)},
\end{align*}
where $l\neq k,n$.

Finally,
\begin{align}\label{eq:number_of_states}
\begin{split}
G(\boldsymbol a,\boldsymbol b)&=\frac{1}{6}\left(\frac{(1-b_1)(1-b_2)}{(1-a_1)(1-a_2)(1-a_3)}\right)^{2}\\
&\quad\times\Big(res_0 G_1+\sum_k res_{a_k}G_1+res_{a_k^2}G_1+\sum_{k<n}res_{a_ka_n}G_1\Big).
\end{split}
\end{align}

The dimension of $\mathcal H_\mathbf n$ can be read out from Taylor expansion of $G$. The explicit expressions for residues are rather involved, so it is convenient to use a program for symbolic manipulations, e.g. Mathematica to do the Taylor expansion. The dimension of sectors with given $n_B,n_F$ is given in Table \ref{tab:number_of_states}.

\begin{table}\centering
\begin{tabular}{r|rrrrrrrrr|}
$n_B$&\multicolumn{9}{c}{$n_F$}\\
&0&1&2&3&4&5&6&7&8\\\hline
0& 1& 0& 1& 4& 5& 12& 11& 16&  28\\
1& 0& 6& 12& 30& 72& 114& 180& 234& 240\\
2& 6& 18& 66& 198&  444& 810& 1212& 1590& 1776\\
3& 11& 70& 292& 870& 2048& 3926& 6135&   8010& 8780\\
4& 30& 228& 969& 3192& 7941& 15396& 24549& 32520&  35694\\
5& 75& 624& 3042& 10224& 26010& 52260& 84669& 112632&  123888\\
6& 186& 1632& 8328& 29376& 77030& 157348& 258150& 346292&  381908\\
7& 381& 4008& 21366& 77448& 207684& 431712& 717150&  967104& 1067382\\
8& 885& 9156& 51078& 190500& 520782& 1097892&  1839834& 2495124& 2760186\\
9& 1785& 20108& 115894& 441212&  1225864& 2617004& 4423238& 6024572& 6672230\\
10& 3618& 42300& 249780& 971028& 2737962& 5903544& 10048302& 13743936& 15242292
\end{tabular}
\caption{Total number of states in sectors with given $n_F$ and $n_B$.}
\label{tab:number_of_states}
\end{table}

\section{Comparison with the numerical approach}

In this chapter we have shown a method to calculate dimensions $D_\mathbf n$ of sectors $\mathcal H_\mathbf n$ analytically. A direct, numerical way to find $D_\mathbf n$ was presented in Chapter \ref{ch:Fock_space_method}. First, one constructs the Gramm matrix of scalar products of all vectors that span $\mathcal H_\mathbf n$. Then, $D_\mathbf n$ is equal to the number of nonzero eigenvalues of the Gramm matrix.

\begin{figure}
\centering
\begin{subfigure}{.48\textwidth}
\includegraphics[width=\textwidth]{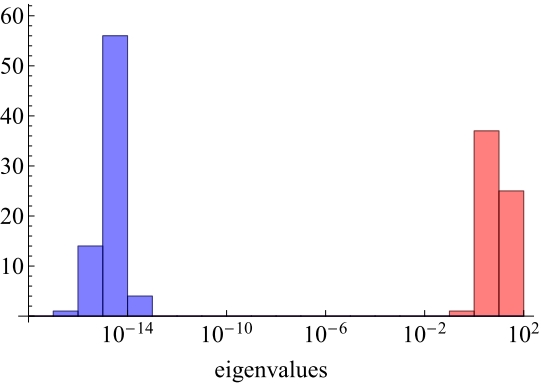}
\caption{$\mathbf n=(0,0,3,3,2)$}
\end{subfigure}
\begin{subfigure}{.48\textwidth}
\includegraphics[width=\textwidth]{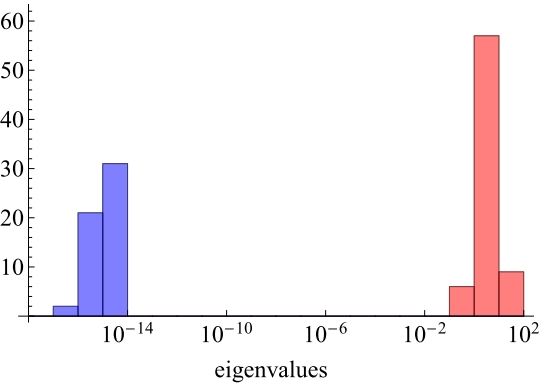}
\caption{$\mathbf n=(1,0,3,2,2)$}
\end{subfigure}
\begin{subfigure}{.48\textwidth}
\includegraphics[width=\textwidth]{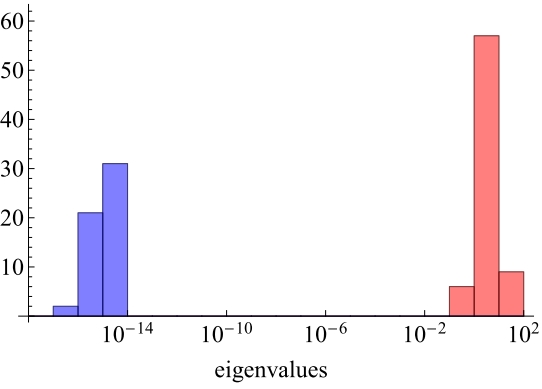}
\caption{$\mathbf n=(2,1,1,1,1)$}
\end{subfigure}
\caption{Histograms of eigenvalues of the Gram matrices $S_\mathbf n$ in three different sectors with quantum numbers $\mathbf n$. Red bars correspond to $D_{\mathbf n}$ largest eigenvalues where $D_{\mathbf n}$ is the dimension of $\mathcal H_\mathbf n$ which follows from the character method.}
\label{fig:histograms}
\end{figure}

For a cross check of these two approaches we present histograms of eigenvalues in illustrative sectors in Fig. \ref{fig:histograms}. The $D_{\mathbf n}$ largest eigenvalues (where numerical value of $D_\mathbf n$ follows from the character method) are marked with red bars. Bars which correspond to smaller eigenvalues are blue. One can see that the large and small eigenvalues are well separated. Also, the number of large eigenvalues agrees with the dimension $D_\mathbf n$. This holds also in all other sectors. More precisely, the 'zero' eigenvalues are always smaller than $10^{-12}$ and the 'nonzero' are larger than $10^{-3}$.

The agreement also means that all relevant bricks are included. If that was not the case, that there would be not enough large eigenvalues.

\chapter{The spectrum}\label{ch:spectrum}

Spectrum of a Hamiltonian is a basic characteristic of a model. The main goal of the cut Fock space method is to find eigenvalues of a given operator. Here, we apply this algorithm to find energies of the supersymmetric Yang-Mills quantum mechanics. In particular, it is interesting to identify type of the spectrum, whether it is continuous or discrete. The nature of the spectrum can be identified from behavior of energy levels for growing cutoff \cite{Trzetrzelewski1}. If a given eigenvalue converges fast, then it corresponds to a discrete energy in the continuum limit. A signature of a continuous spectrum are many eigenvalues which decrease slowly (to zero if there is no mass gap) as the cutoff increases. As it was mentioned before, the fermion number and angular momentum are conserved. Therefore, we study the spectrum in channels with definite $n_F$ and $j$.

First we present the results for the $SU(2)$ theory. We give a short review of other methods and then proceed to detailed analysis of the spectrum obtained with our approach. This part serves as a check of our program and a guide to understand the results for the more complicated model with $SU(3)$ gauge group. Finally, we proceed to analysis of the spectrum of SYMQM with $SU(3)$.

\section{SYMQM with SU(2) gauge group}
\subsection{Review of other approaches}
The $SU(2)$ model has been already extensively studied in \cite{Campostrini}. The energies were obtained in all fermonic sectors and all angular momenta for cutoff up to $N_B=18$. The method used was analogous to our approach, but it included modifications discussed in \ref{sec:su2_modifications}. These modifications incorporate computing matrix elements in channels with given angular momentum. This is possible for $SU(2)$ because the algebra of operators is much simpler than in the $SU(3)$ case. One can then rearrange all single--trace operators so that they have definite angular momentum and calculate all commutators.

There exists yet another way to obtain energies in the $SU(2)$ case \cite{vanBaal}. However, it is limited to channels with zero angular momentum and $n_F=0,2$. In these channels it is very effective and enables one to reach very large cutoff $N_B=60$ for $n_F=2$ \cite{Kotanski}. We give more details concerning this approach in Chapter \ref{ch:wavefunctions}.

On the other hand, the algorithm proposed in this thesis allows us to calculate energy levels in all $(n_F,j)$ channels. On the other hand, the cutoff is much smaller. It is $N_B=8-11$ depending on $n_F$. Results of the computations for all $n_F$, $N_B$ and several angular momenta are shown in Figs \ref{fig:su2_spectrum_nf0} - \ref{fig:su2_spectrum_nf3}. Values of the energies were checked against \cite{Campostrini} and perfect agreement was found.

An important characteristic of a spectrum is its type --  discrete or continuous. It was argued \cite{Trzetrzelewski_spectra} that the two types of spectra reveal a different behavior for a finite cutoff. Energy levels which belong to the discrete spectrum converge very fast, e.g. exponentially. On the other hand, eigenvalues of the Hamiltonian which correspond to the continuous spectrum decrease to $0$ like $1/N_B$. Identification of the type of spectrum is the main subject of following paragraphs.

\subsection{Spectra in sectors with given number of fermions}
\begin{figure}[H]\centering
\begin{subfigure}[b]{0.3\textwidth}\includegraphics[width=\textwidth]{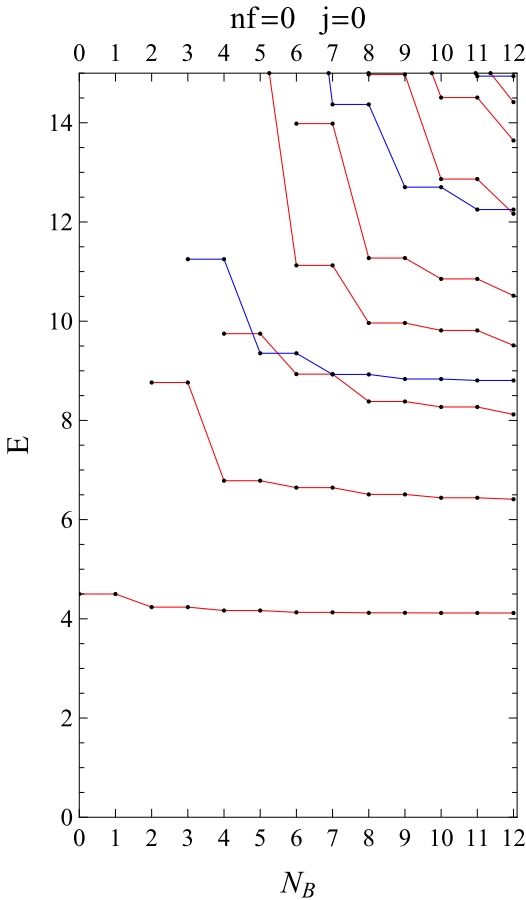}\end{subfigure}
\begin{subfigure}[b]{0.3\textwidth}\includegraphics[width=\textwidth]{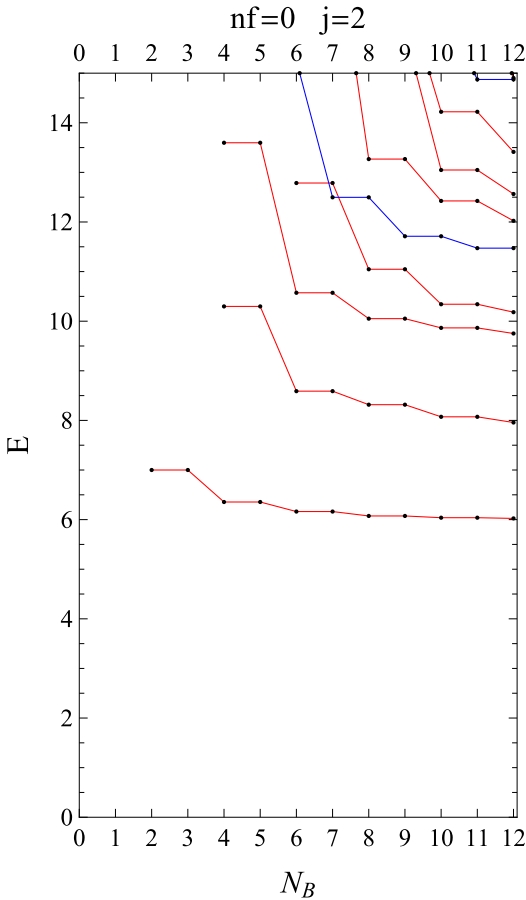}\end{subfigure}
\caption{Spectrum of the $SU(2)$ theory in the bosonic sector. The maximal cutoff is $N_B=12$. There are no states with angular momentum $j=1$ (cf. Appendix \ref{ap:no_j1}). Energies corresponding to even and odd states are marked with red and blue color respectively.}
\label{fig:su2_spectrum_nf0}
\end{figure}

We start discussing spectra for particular $n_F$ with $n_F=0$. We first comment on the 'stepwise' convergence of eigenvalues in Fig. \ref{fig:su2_spectrum_nf0}. In the bosonic channel the parity is conserved. The creation operator $a_i^\dagger$ has negative parity. Therefore, states with odd $n_B$ (see eq. (\ref{eq:number_of_particles_operators})) have negative parity. For a given $N_B$, all states with $n_B\leq N_B$ are included in the basis of the cut space. When $N_B$ changes from an odd to even value, the basis of odd states does not grow and eigenvalues corresponding to odd states remain constant. Similarly, energy levels of even states do not change when $N_B$ changes form an even to odd value.

\begin{figure}[H]\centering
\begin{subfigure}[b]{0.3\textwidth}\includegraphics[width=\textwidth]{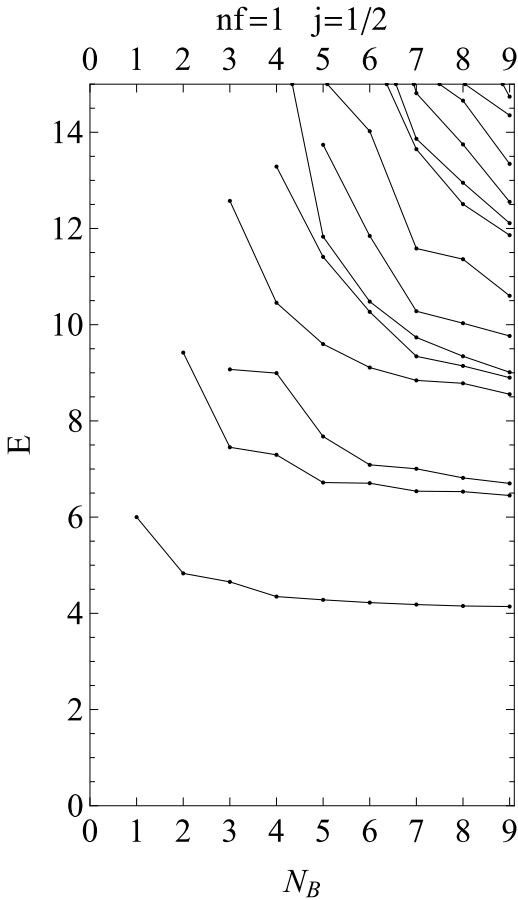}\end{subfigure}
\begin{subfigure}[b]{0.3\textwidth}\includegraphics[width=\textwidth]{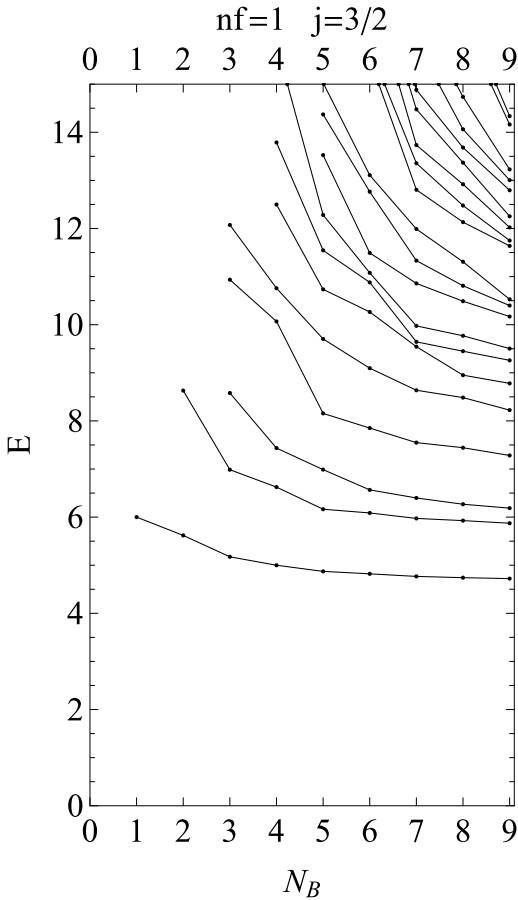}\end{subfigure}
\begin{subfigure}[b]{0.3\textwidth}\includegraphics[width=\textwidth]{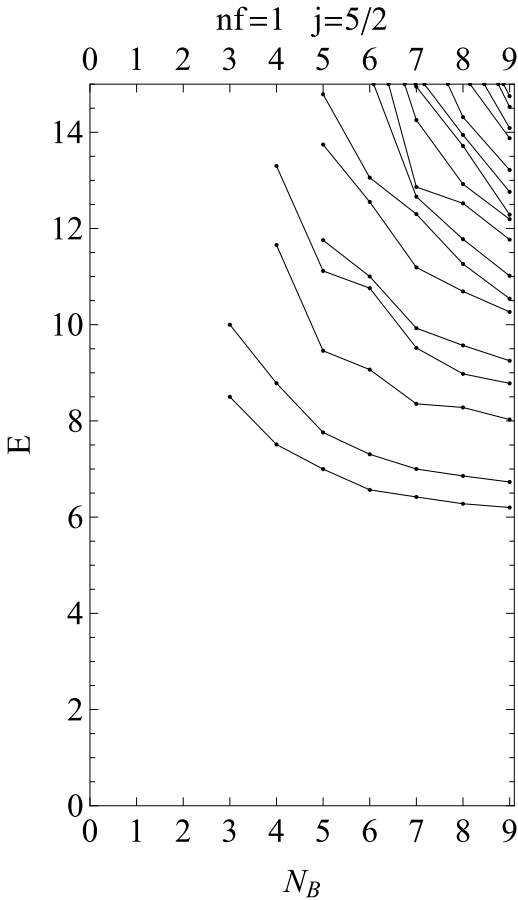}\end{subfigure}
\caption{Spectrum of the $SU(2)$ theory in sector with a single fermion.}
\label{fig:su2_spectrum_nf1}
\end{figure}

For $n_F=0$ (Fig. \ref{fig:su2_spectrum_nf0}) all energies converge fast with growing cutoff. This behavior is typical for discrete spectrum. The fact that the spectrum is discrete is contrary to what one may naively expect. Indeed, if the three matrices $X_i$ commute then the potential vanishes: $V\equiv-\sum_{i<j}Tr([X_i,X_j]^2)=0$. The region where the potential vanishes forms a vector subspace in the configuration space which we call flat valleys. Such situation usually generates a continuous spectrum. On the other hand, the potential becomes steeper in the transverse directions as one moves deeper inside the valleys. The transverse oscillations cost more energy and thus the effective potential inside the flat valleys grows. Therefore, the eigenstates of the Hamiltonian with $n_F=0$ are localized and the spectrum is discrete. A simpler system with similar properties was studied in \cite{Simon}. The sector with one fermion shares the same behavior (Fig. \ref{fig:su2_spectrum_nf1}).

\begin{figure}[H]\centering
\begin{subfigure}[b]{0.3\textwidth}\includegraphics[width=\textwidth]{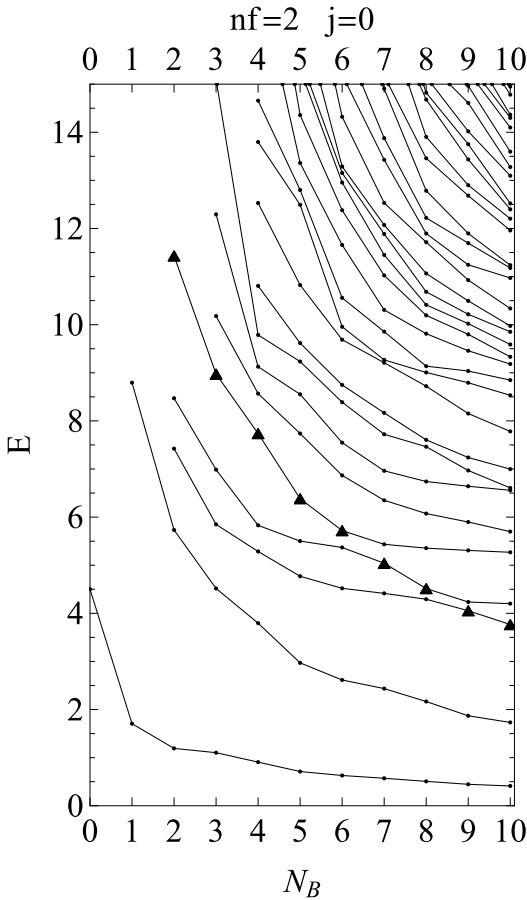}\end{subfigure}
\begin{subfigure}[b]{0.3\textwidth}\includegraphics[width=\textwidth]{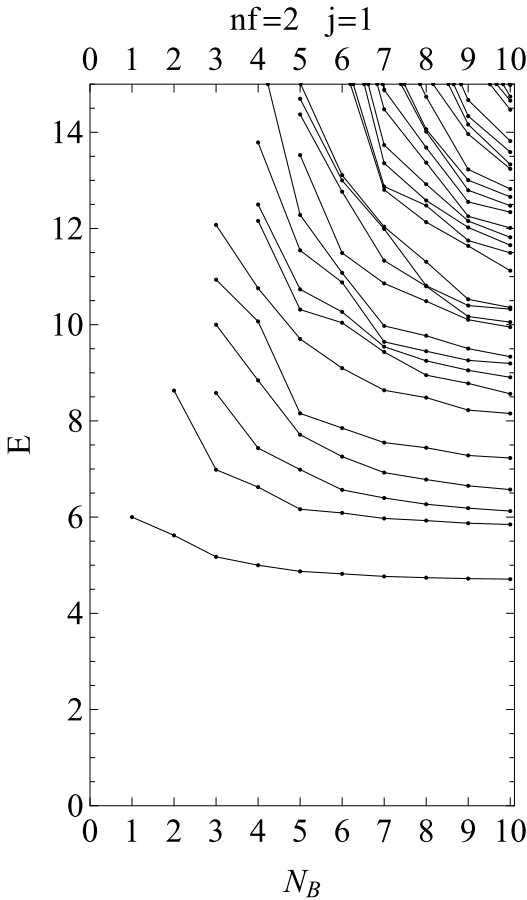}\end{subfigure}
\begin{subfigure}[b]{0.3\textwidth}\includegraphics[width=\textwidth]{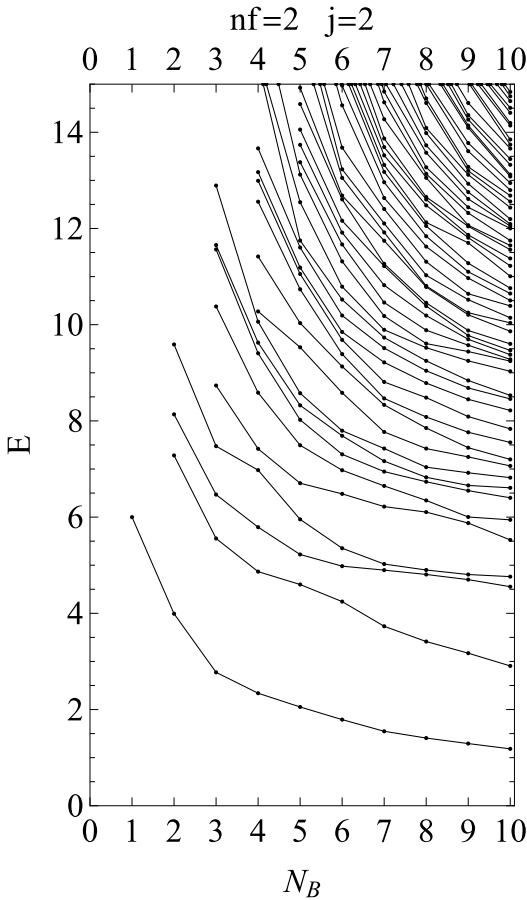}\end{subfigure}
\caption{Spectrum of the $SU(2)$ theory in sector with two fermions. For $j=0$ the two lowest energies correspond to continuous spectrum. The third line from the continuous spectrum crosses two discrete energy levels. This line is marked with triangles. Fourth energy from the continuous spectrum is probably the fifth excited energy.}
\label{fig:su2_spectrum_nf2}
\end{figure}

The picture is different with two fermions (Fig. \ref{fig:su2_spectrum_nf2}). In the $j=0$ channel the lowest energies approach $0$ and the convergence is much slower. This is a signature of a continuous spectrum. This is because inside the flat valley the contributions form interactions of bosonic and fermionic degrees of freedom with the potential have different sign. They cancel exactly causing the effective potential to vanish. The flat valleys are open and the spectrum is continuous. The eigenenergies corresponding to the continuous spectrum will be denoted by $E_k^c$ where $k=0,1,\ldots$ is the label of the energy level. The two lowest energies $E_k$ correspond to the continuous spectrum, so $E_k^c=E_k$ for $k=0,1$. The second excited energy which belongs to the continuous spectrum goes through discrete energy levels. It is marked with triangles in Fig. \ref{fig:su2_spectrum_nf2}. The third energy seems to be $E^c_3=E_5$.

Even with two fermions, the $j=1$ channel has discrete spectrum (Fig. \ref{fig:su2_spectrum_nf2}). In the next sector, with angular momentum $j=2$ the spectrum is again continuous. With our data we checked that the spectrum is continuous for all even $j$ and discrete for odd $j$ at least up to $j=6$. For three fermions the spectrum is continuous for all $j$.

All these effects, i.e. nonvanishing effective potential in a flat valley, cancelation between bosons and fermions and finally discrete spectrum inside continuous one were already observed for simpler supersymmetric systems \cite{Simon,Nicolai,Korcyl_effective}.

\subsection{Scaling relations}
\begin{figure}[H]\centering
\includegraphics[width=.7\textwidth]{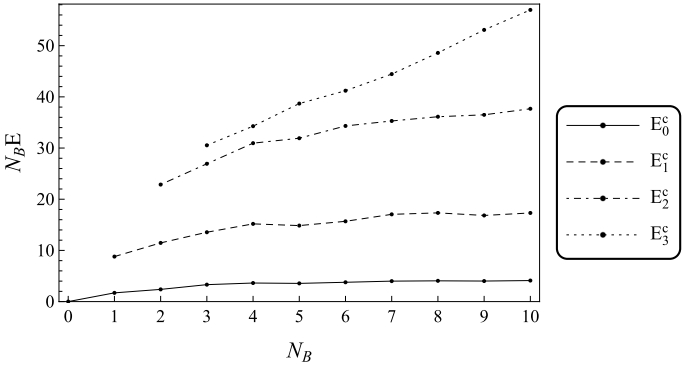}
\caption{Scaled energies for $SU(2)$ in channel with $n_F=2$, $j=0$. The energies corresponding to the continuous spectrum scaled by the cutoff $N_B$ converge to constant values. This is in agreement with (\ref{eq:continuous_scaling}). The convergence is best for lowest energies. The cutoff is too small for $N_B E^c_3$ to level off or it is misidentified as an energy from the continuous spectrum.}
\label{fig:su2_continuous_scaling}
\end{figure}

\begin{figure}[H]\centering
\includegraphics[width=.7\textwidth]{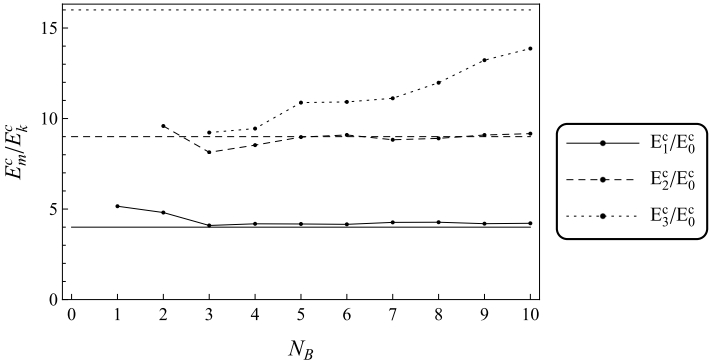}
\caption{Ratios of the energies for $SU(2)$ satisfy the relation $E^c_m/E^c_k\approx(m+1)^2/(k+1)^2$ (flat lines) in the continuum limit. The higher ratio $E_3/E_0$ has not yet converged for $N_B=10$.}
\label{fig:su2_continuous_ratios}
\end{figure}

There are interesting scaling relations for eigenvalues corresponding to the continuous spectrum. They originate from dispersion relations for energies. In \cite{Trzetrzelewski_spectra} the free Hamiltonian in a one dimensional quantum mechanics was studied with the cut Fock space method. For this system there is a standard dispersion relation $E(p)=\frac{1}{2}p^2$. It was shown that in the infinite $N$ limit momentum is related to the label of energy level by $p=\frac{\pi}{\sqrt{2N_B}}(m+1)$. It follows that the energies satisfy
\begin{align}\label{eq:continuous_scaling}
E^c_m&\approx\frac{\pi^2(m+1)^2}{4N_B}&m=0,1,\ldots
\end{align}
In the continuum limit the energies fill densely the positive real axis. In \cite{Campostrini,Kotanski} it was argued and checked numerically that the scaling of energies are similar for the continuous spectrum in SYMQM. Although the relation (\ref{eq:continuous_scaling}) does not hold precisely, it was shown that $N_BE_m$ converge to nonzero constants and ratios of the energies agree, i.e. $E^c_m/E^c_k\approx(m+1)^2/(k+1)^2$. Corresponding results based on our data are presented in Figs \ref{fig:su2_continuous_scaling} and \ref{fig:su2_continuous_ratios}. Both scaling relations are confirmed for the three lowest energies in the continuous spectrum $E^c_0,\ E^c_1$ and $E^c_2$. For the third excited continuous energy we took $E^c_3=E_5$. The scaling relations suggest that $E^c_3=E_5$ is smaller than the theoretical expectation. This can be an effect of too small cutoff or misidentification of the continuous energy.

\begin{figure}[H]\centering
\begin{subfigure}[b]{0.3\textwidth}\includegraphics[width=\textwidth]{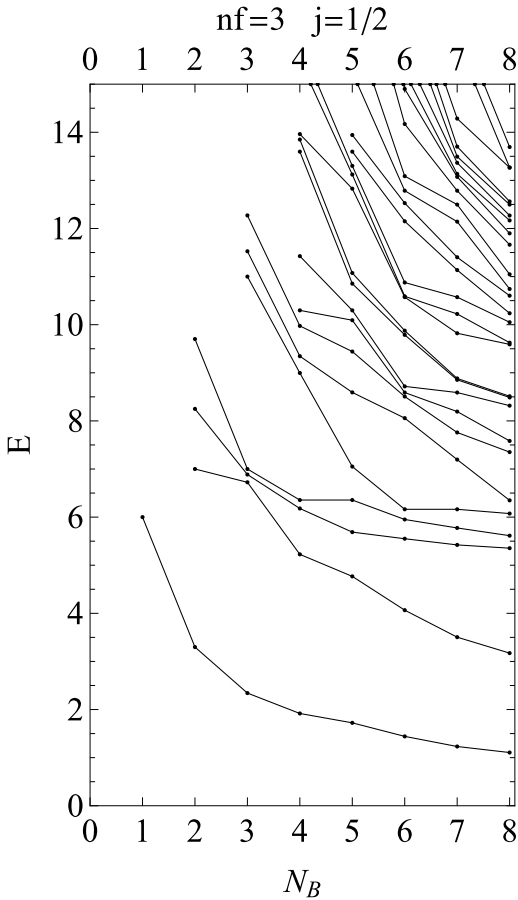}\end{subfigure}
\begin{subfigure}[b]{0.3\textwidth}\includegraphics[width=\textwidth]{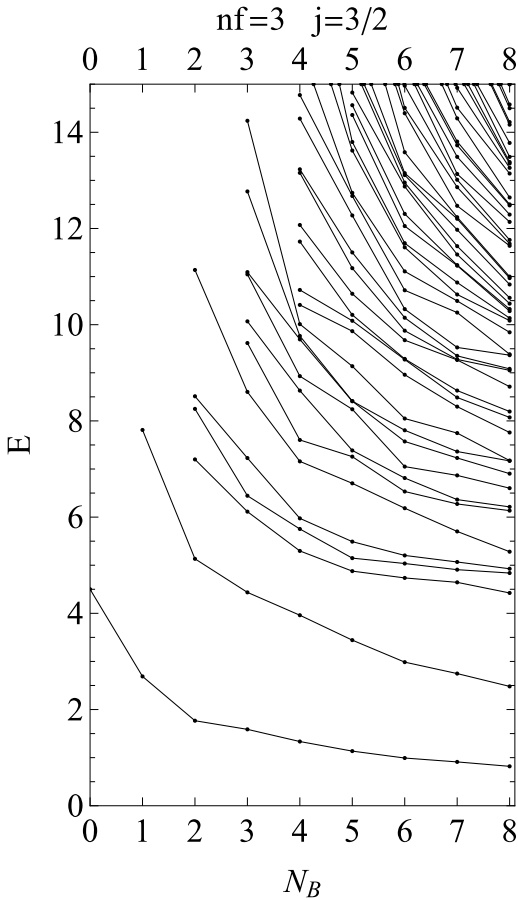}\end{subfigure}
\begin{subfigure}[b]{0.3\textwidth}\includegraphics[width=\textwidth]{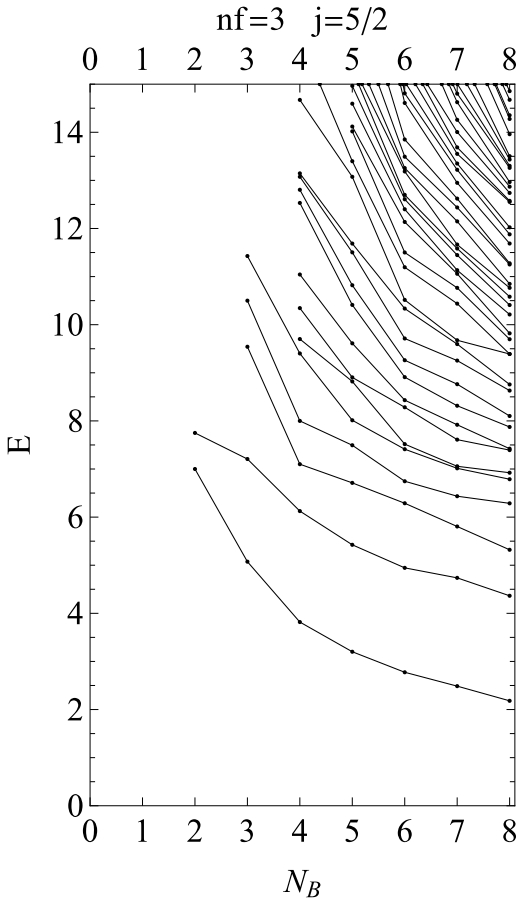}\end{subfigure}
\caption{Spectrum of the $SU(2)$ theory in sector with $n_F=3$.}
\label{fig:su2_spectrum_nf3}
\end{figure}

\section{The spectrum for the SU(3) gauge group}
Here our results for the $SU(3)$ symmetry group are presented. In order to obtain the spectrum we constructed matrices with sizes up to 36000. Computations were performed on a supercomputer Deszno located in the Institute of Physics at Jagiellonian University. 96 cores were used in parallel with the OpenMP interface using up to 256 GB shared memory.

\begin{figure}[H]\centering
\begin{subfigure}[b]{0.3\textwidth}\includegraphics[width=\textwidth]{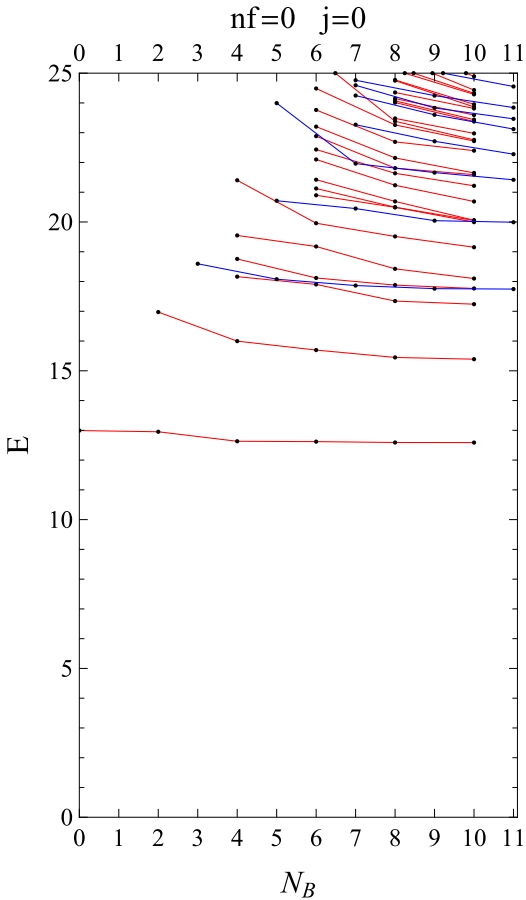}\end{subfigure}
\begin{subfigure}[b]{0.3\textwidth}\includegraphics[width=\textwidth]{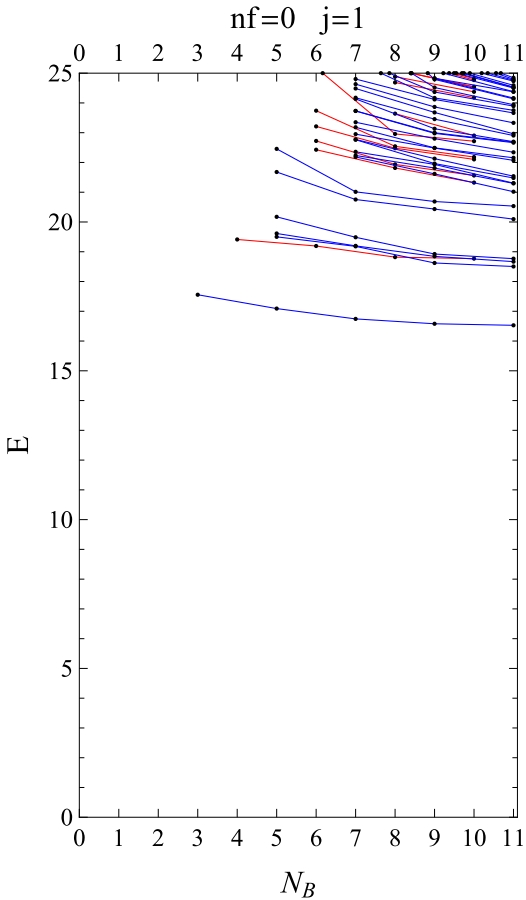}\end{subfigure}
\begin{subfigure}[b]{0.3\textwidth}\includegraphics[width=\textwidth]{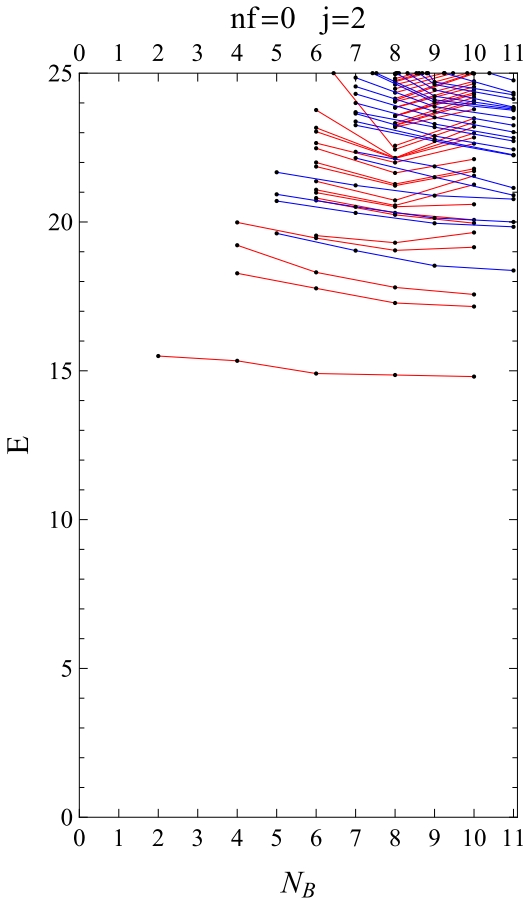}\end{subfigure}
\caption{Spectrum of the $SU(3)$ theory in the bosonic sector. The maximal cutoff is $N_B=11$. Energies marked with red correspond to states with positive parity and blue to negative parity. Even energies are marked only for even $N_B$ because they do not change when $N_B$ passes from even to odd value. Conversely, odd energies are marked only for odd $N_B$.}
\label{fig:su3_spectrum_nf0}
\end{figure}

\begin{figure}[H]\centering
\begin{subfigure}[b]{0.3\textwidth}\includegraphics[width=\textwidth]{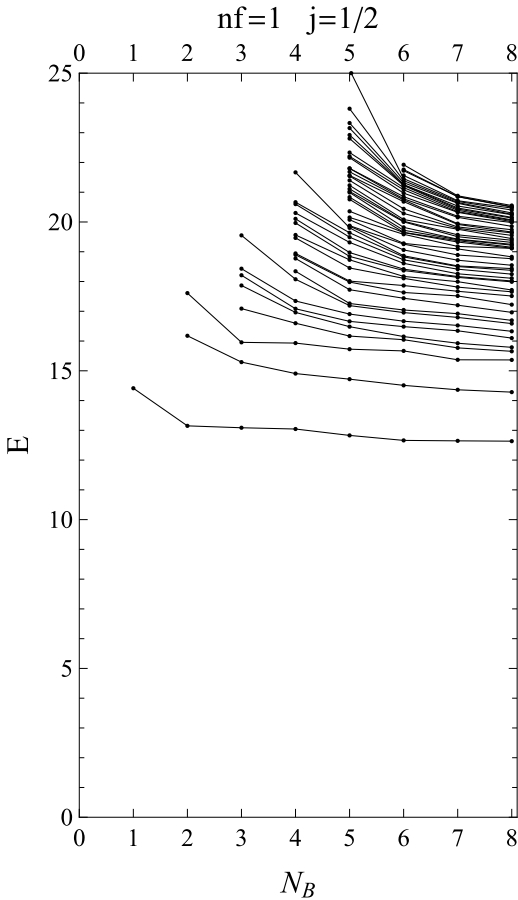}\end{subfigure}
\begin{subfigure}[b]{0.3\textwidth}\includegraphics[width=\textwidth]{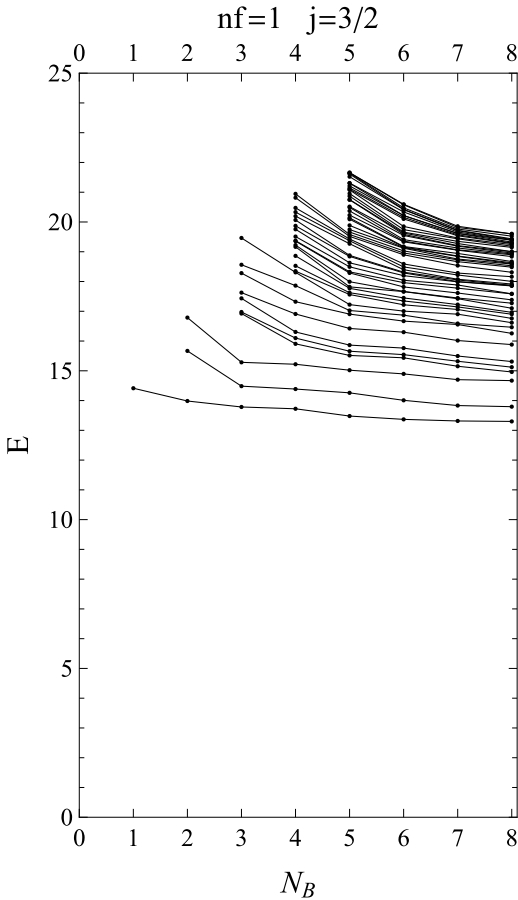}\end{subfigure}
\begin{subfigure}[b]{0.3\textwidth}\includegraphics[width=\textwidth]{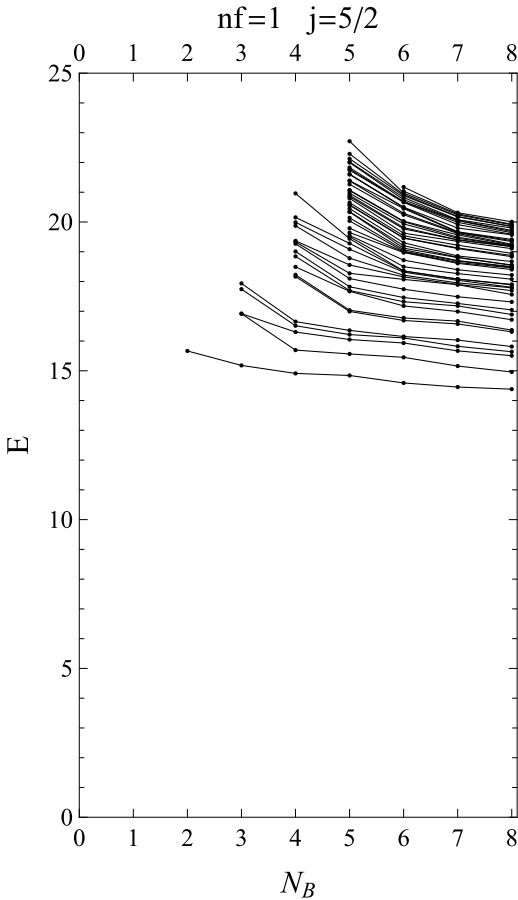}\end{subfigure}
\caption{Spectrum of the $SU(3)$ theory with a single fermion.}
\label{fig:su3_spectrum_nf1}
\end{figure}

\begin{figure}[H]\centering
\begin{subfigure}[b]{0.3\textwidth}\includegraphics[width=\textwidth]{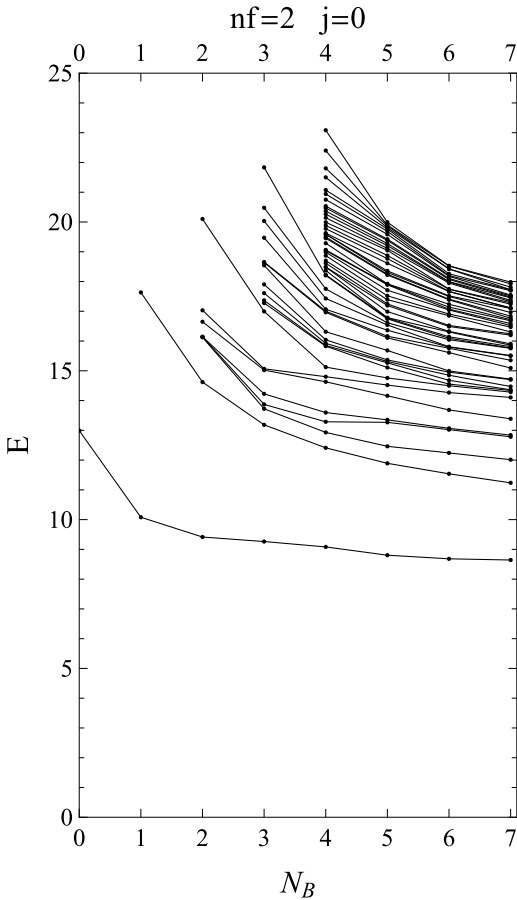}\end{subfigure}
\begin{subfigure}[b]{0.3\textwidth}\includegraphics[width=\textwidth]{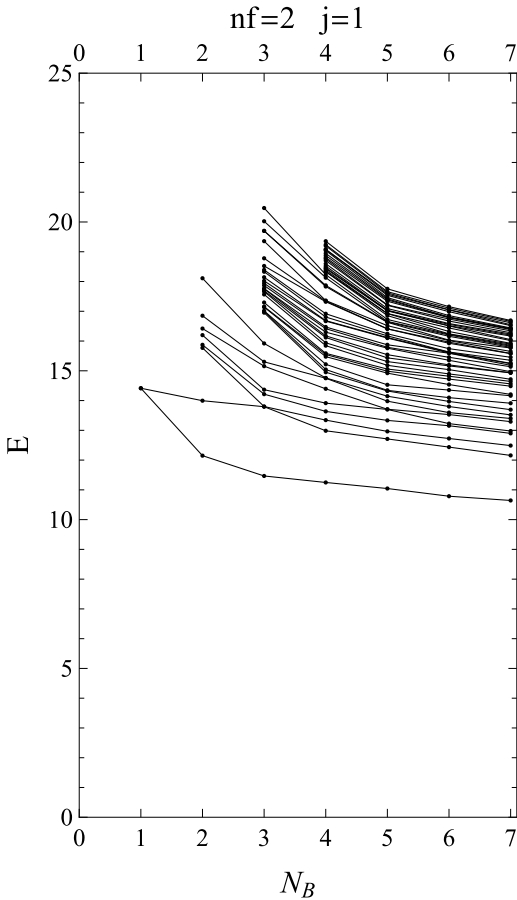}\end{subfigure}
\begin{subfigure}[b]{0.3\textwidth}\includegraphics[width=\textwidth]{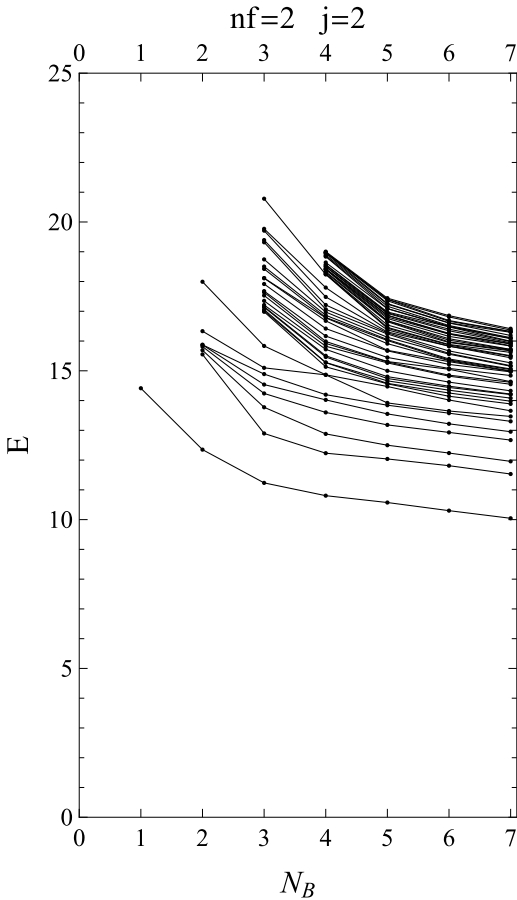}\end{subfigure}
\caption{Spectrum of the $SU(3)$ theory with two fermions.}
\label{fig:su3_spectrum_nf2}
\end{figure}

\begin{figure}[H]\centering
\begin{subfigure}[b]{0.3\textwidth}\includegraphics[width=\textwidth]{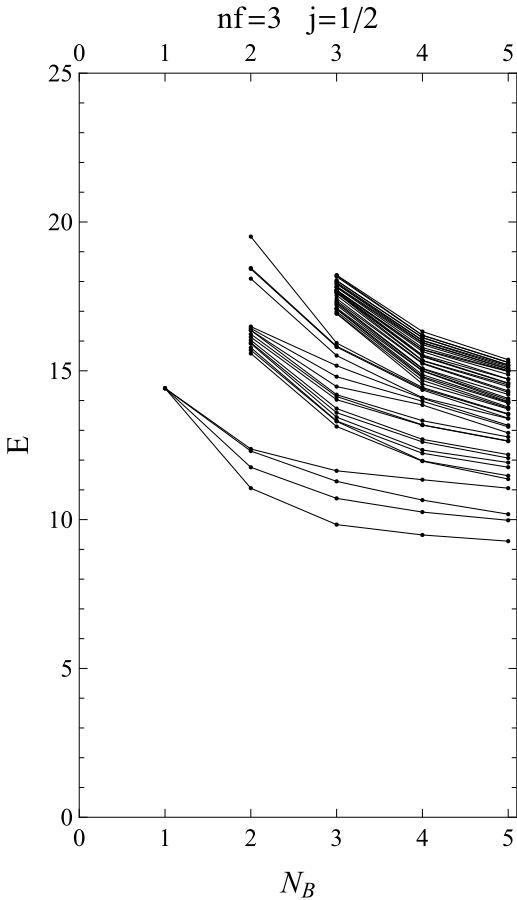}\end{subfigure}
\begin{subfigure}[b]{0.3\textwidth}\includegraphics[width=\textwidth]{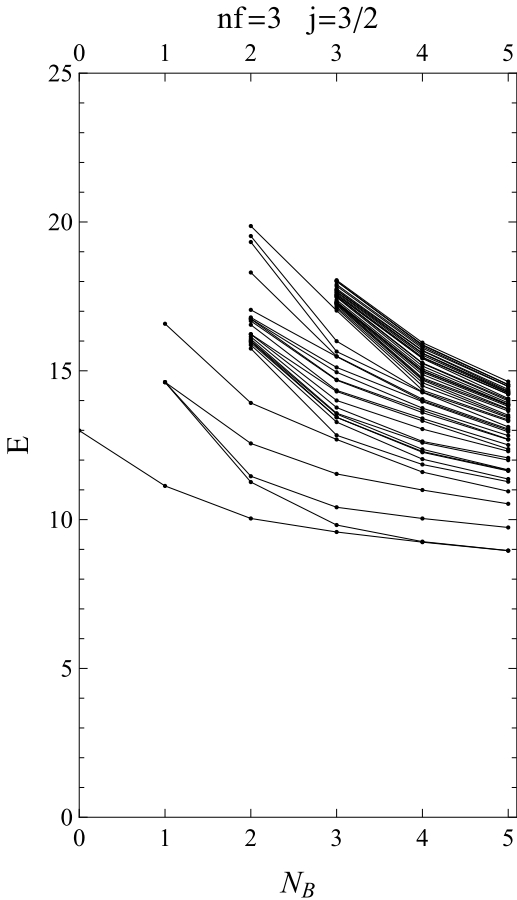}\end{subfigure}
\begin{subfigure}[b]{0.3\textwidth}\includegraphics[width=\textwidth]{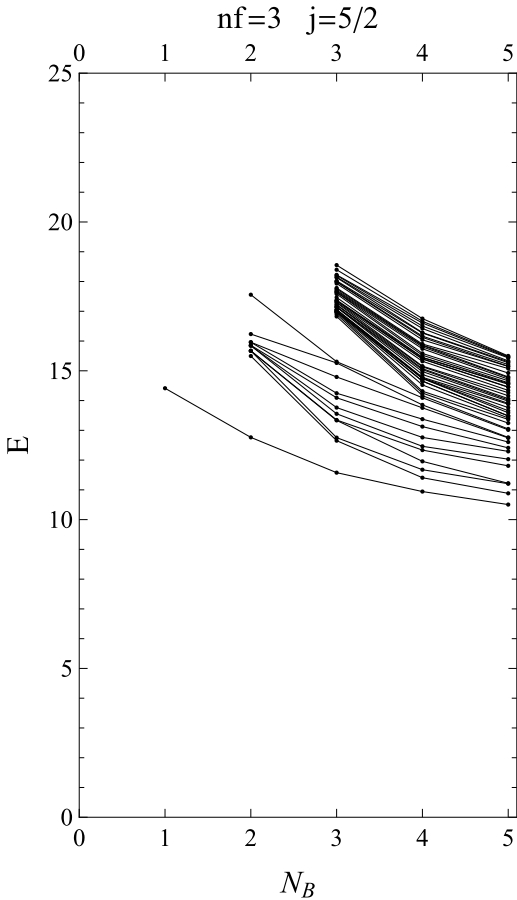}\end{subfigure}
\caption{Spectrum of the $SU(3)$ theory with $n_F=3$.}
\label{fig:su3_spectrum_nf3}
\end{figure}

\begin{figure}[H]\centering
\begin{subfigure}[b]{0.3\textwidth}\includegraphics[width=\textwidth]{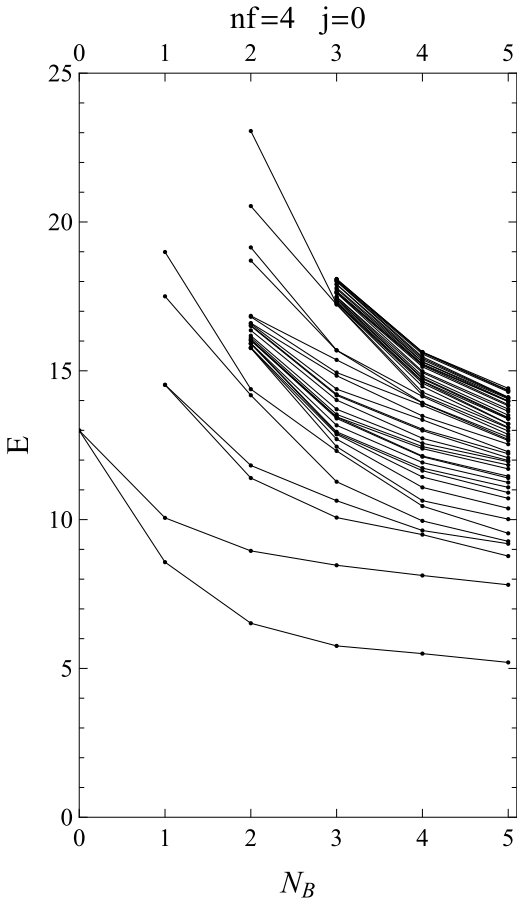}\end{subfigure}
\begin{subfigure}[b]{0.3\textwidth}\includegraphics[width=\textwidth]{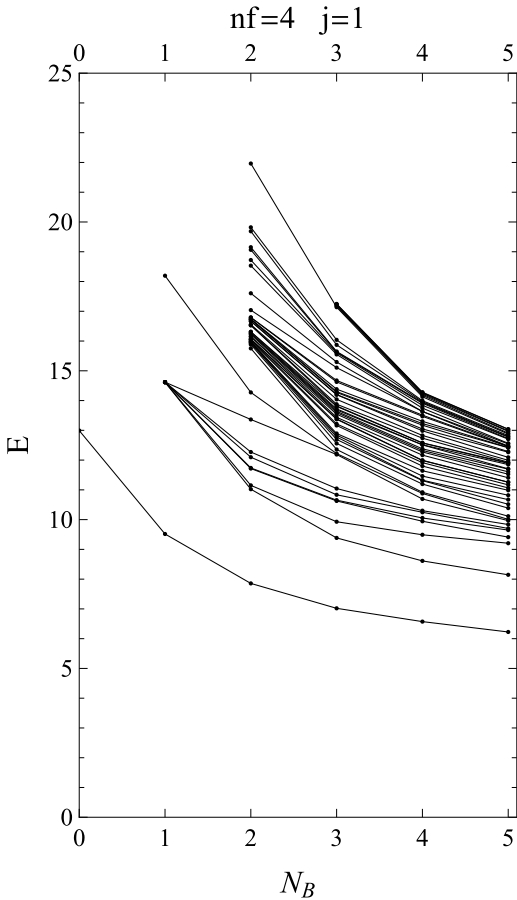}\end{subfigure}
\begin{subfigure}[b]{0.3\textwidth}\includegraphics[width=\textwidth]{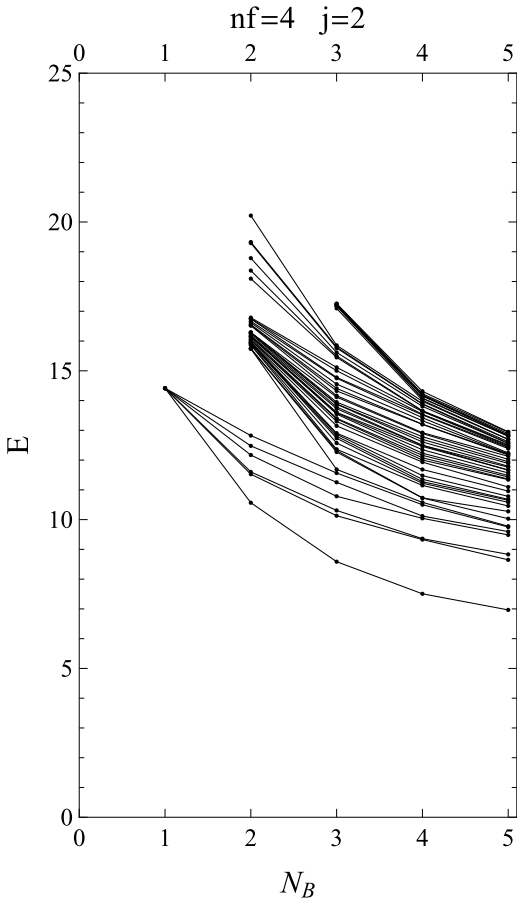}\end{subfigure}
\caption{Spectrum of the $SU(3)$ theory with $n_F=4$.}
\label{fig:su3_spectrum_nf4}
\end{figure}

\begin{figure}[H]\centering
\begin{subfigure}[b]{0.3\textwidth}\includegraphics[width=\textwidth]{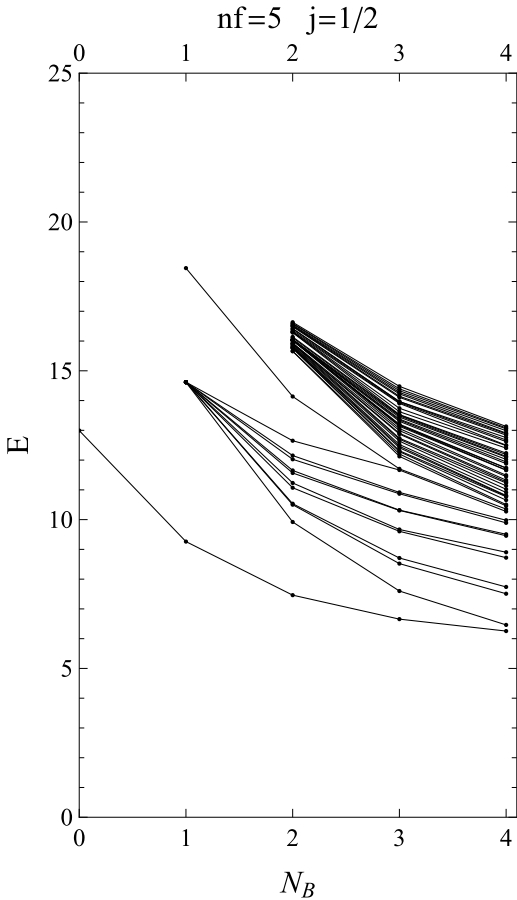}\end{subfigure}
\begin{subfigure}[b]{0.3\textwidth}\includegraphics[width=\textwidth]{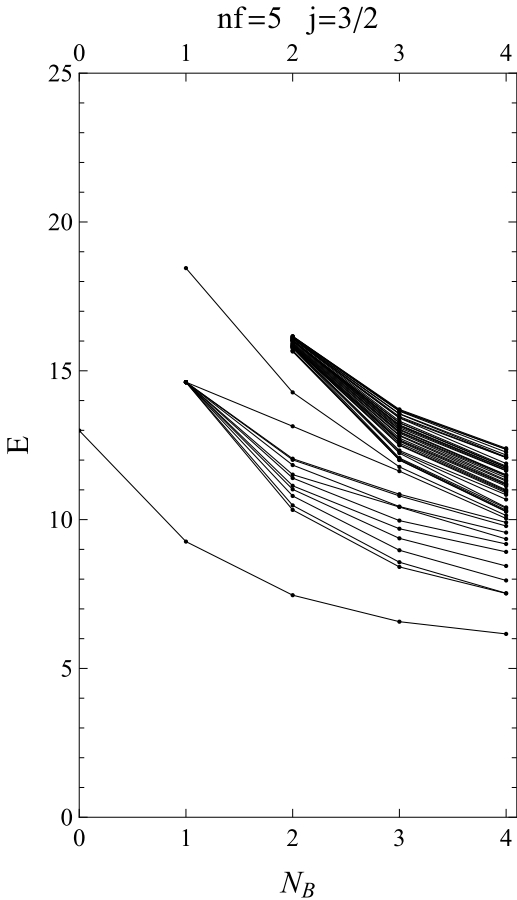}\end{subfigure}
\begin{subfigure}[b]{0.3\textwidth}\includegraphics[width=\textwidth]{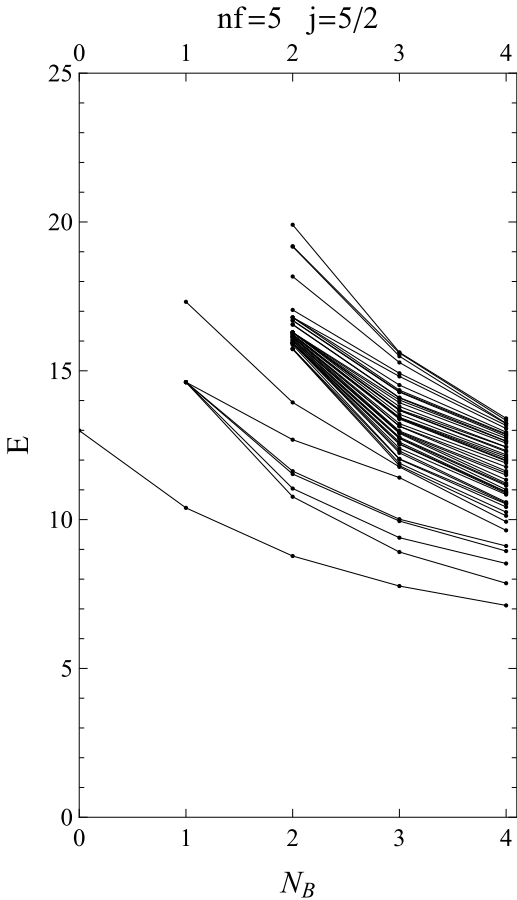}\end{subfigure}
\caption{Spectrum of the $SU(3)$ theory with $n_F=5$.}
\label{fig:su3_spectrum_nf5}
\end{figure}

\begin{figure}[H]\centering
\begin{subfigure}[b]{0.3\textwidth}\includegraphics[width=\textwidth]{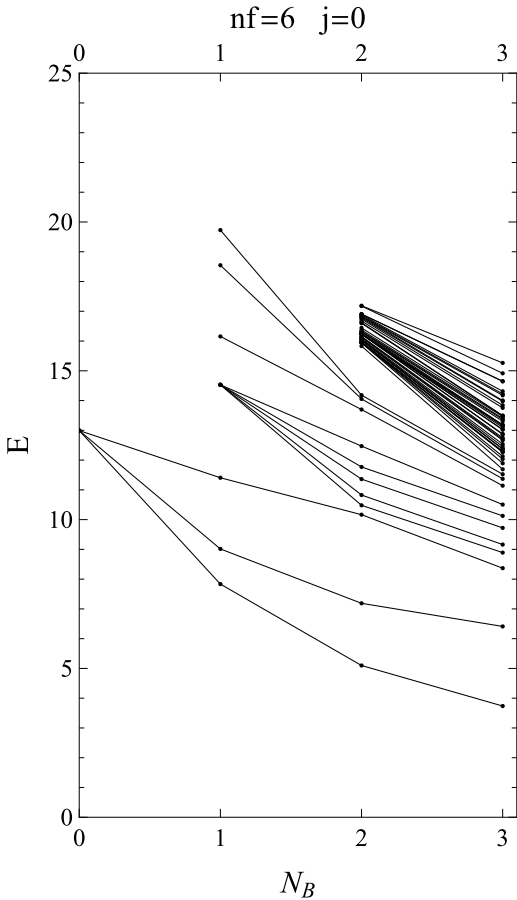}\end{subfigure}
\begin{subfigure}[b]{0.3\textwidth}\includegraphics[width=\textwidth]{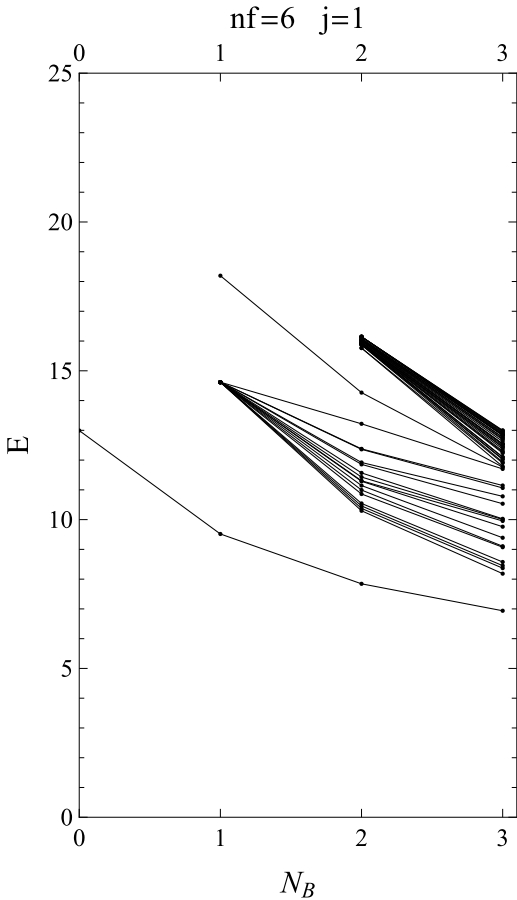}\end{subfigure}
\begin{subfigure}[b]{0.3\textwidth}\includegraphics[width=\textwidth]{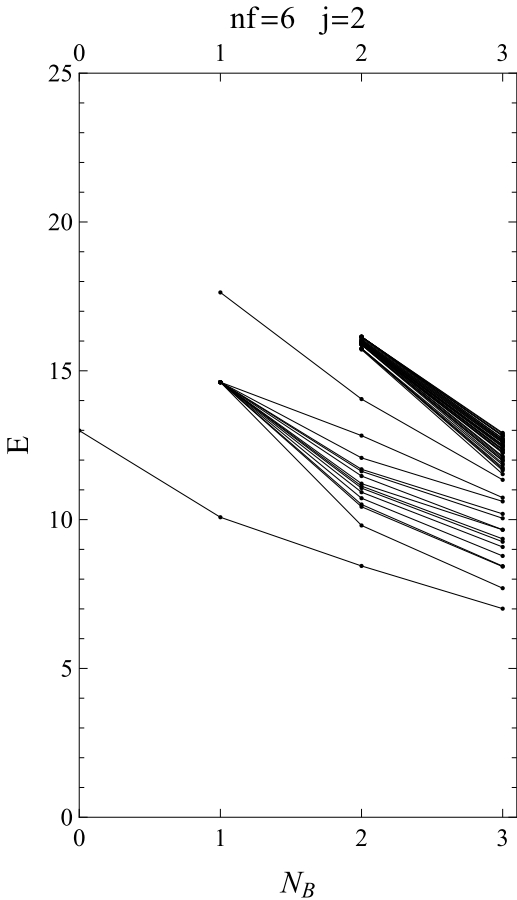}\end{subfigure}
\caption{Spectrum of the $SU(3)$ theory with $n_F=6$.}
\label{fig:su3_spectrum_nf6}
\end{figure}

\begin{figure}[H]\centering
\begin{subfigure}[b]{0.3\textwidth}\includegraphics[width=\textwidth]{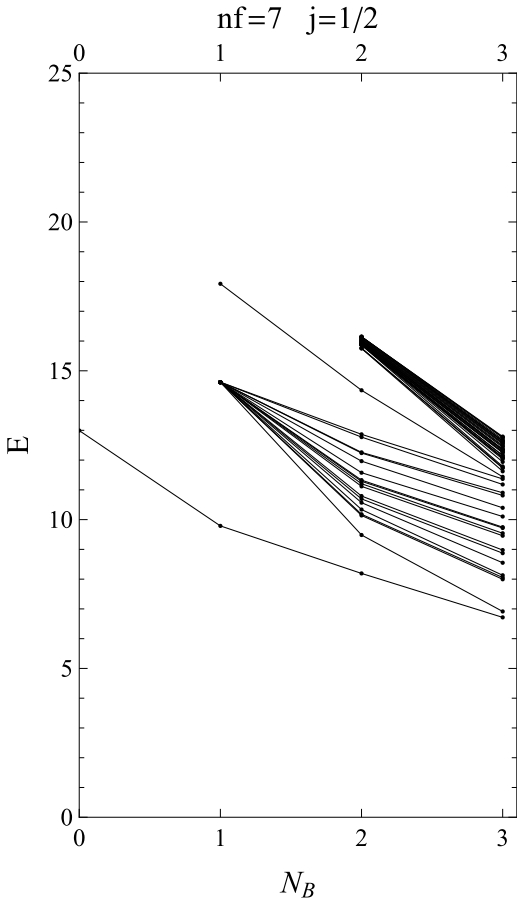}\end{subfigure}
\begin{subfigure}[b]{0.3\textwidth}\includegraphics[width=\textwidth]{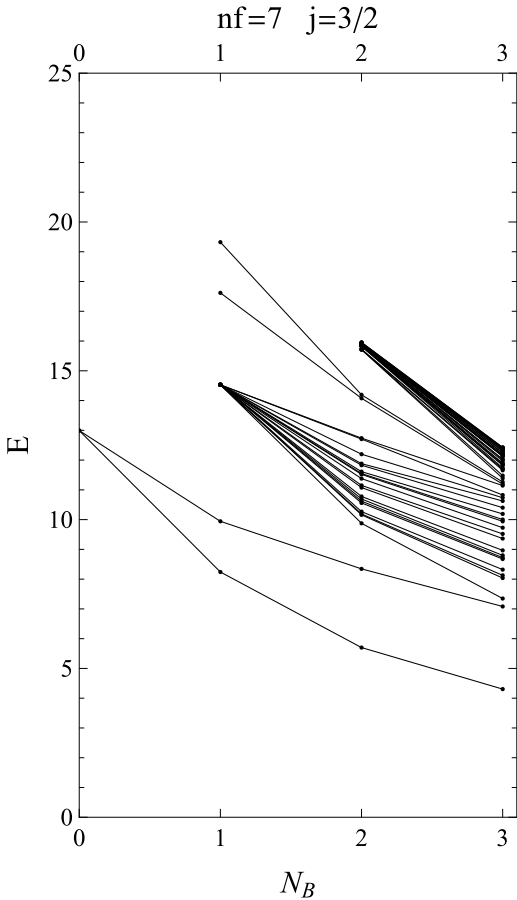}\end{subfigure}
\begin{subfigure}[b]{0.3\textwidth}\includegraphics[width=\textwidth]{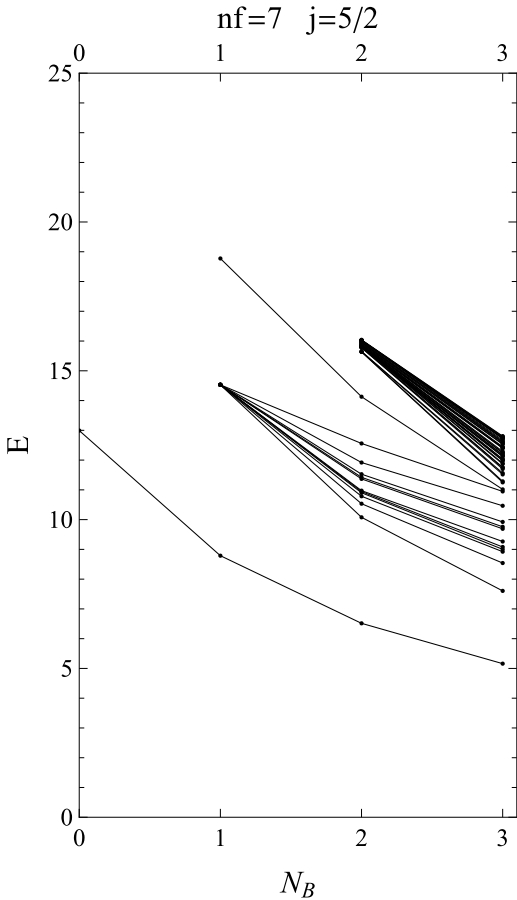}\end{subfigure}
\caption{Spectrum of the $SU(3)$ theory with $n_F=7$.}
\label{fig:su3_spectrum_nf7}
\end{figure}

\begin{figure}[H]\centering
\begin{subfigure}[b]{0.3\textwidth}\includegraphics[width=\textwidth]{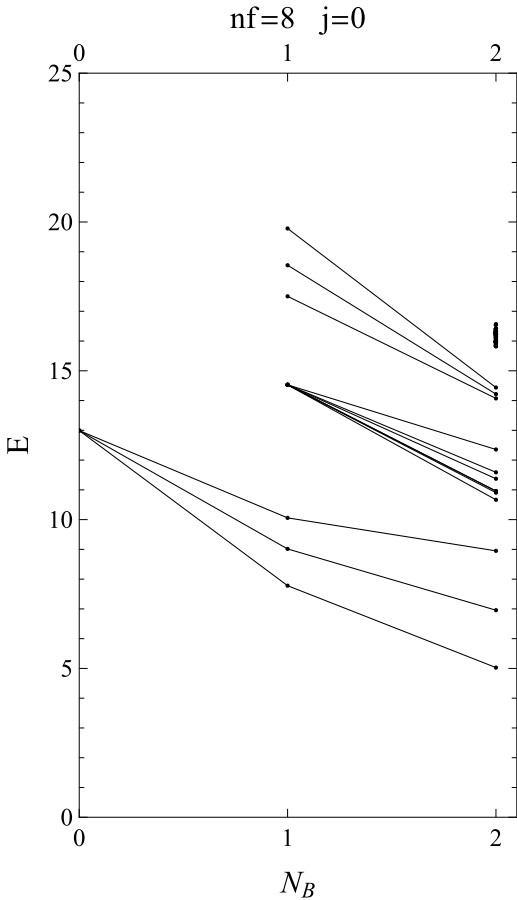}\end{subfigure}
\begin{subfigure}[b]{0.3\textwidth}\includegraphics[width=\textwidth]{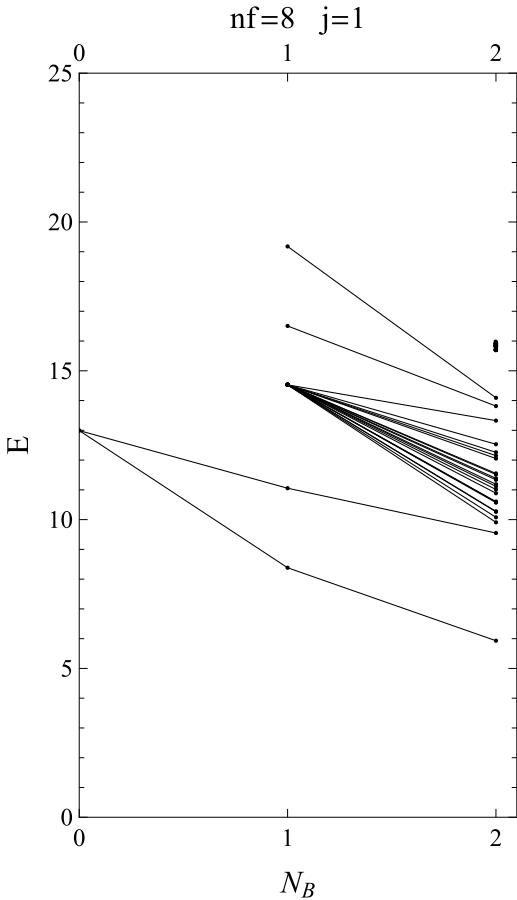}\end{subfigure}
\begin{subfigure}[b]{0.3\textwidth}\includegraphics[width=\textwidth]{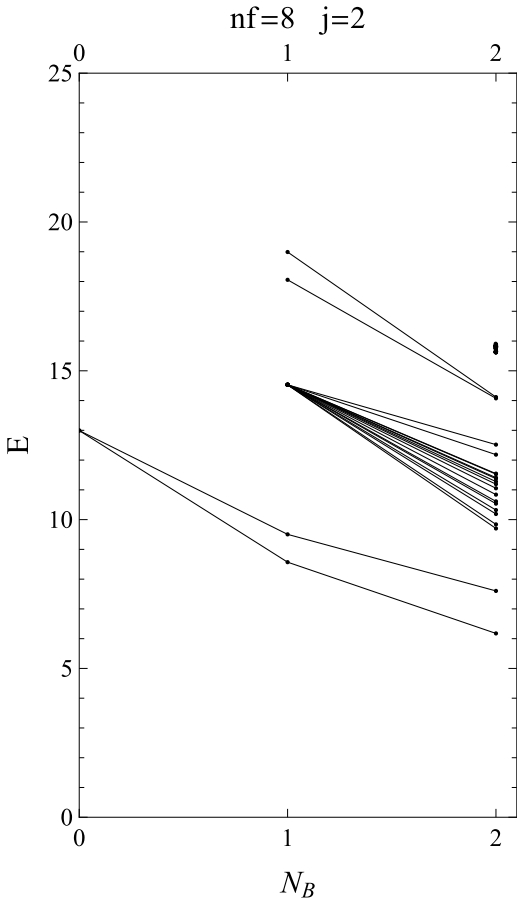}\end{subfigure}
\caption{Spectrum of the $SU(3)$ theory with $n_F=8$.}
\label{fig:su3_spectrum_nf8}
\end{figure}

The eigenenergies of the model are shown in Figs \ref{fig:su3_spectrum_nf0} - \ref{fig:su3_spectrum_nf8}. All energies are obtained in channels with definite total angular momentum $j$ (cf. \ref{subsec:diagonalization}). For $SU(3)$ there are more degrees of freedom and therefore there are many more states for a given $N_B$ compared to the $SU(2)$ case. This is reflected by a larger density of eigenvalues. Because there are more degrees of freedom, the energies are generally larger for $SU(3)$ than for $SU(2)$. As before, parity is conserved in the bosonic channel.

The general picture is similar to the case of $SU(2)$. For zero and one fermion the spectrum is purely discrete. Unlike in the $SU(2)$ case, the argument from Appendix \ref{ap:no_j1} doesn't work and there are states with $j=1$ for $n_F=0$.

\begin{figure}[H]\centering
\includegraphics[width=\textwidth]{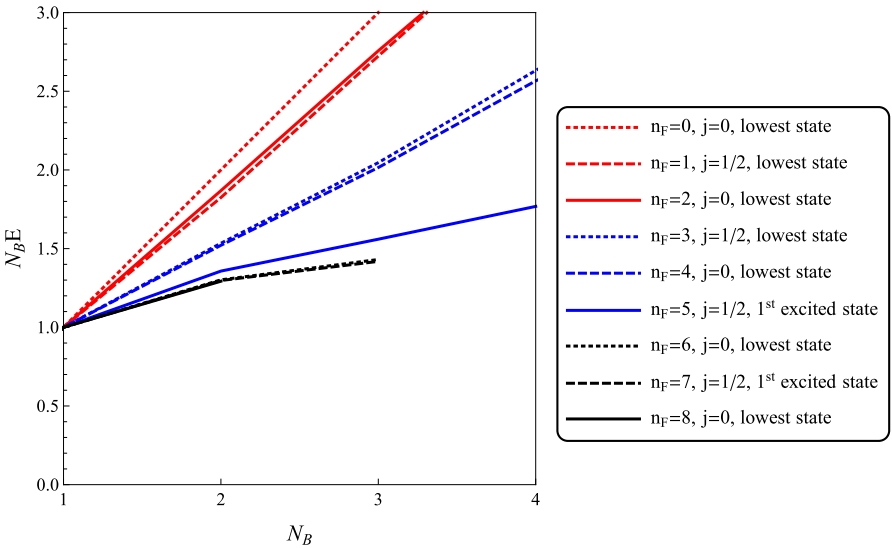}
\caption{Scaled energy $N_BE$ for $SU(3)$ in sectors with $n_F=3,\ldots,8$. All energies are divided by $E$ for $N_B=1$. In each channel the lowest $j$ was picked, i.e. $j=0$ for even $n_F$ and $j=1/2$ for odd $n_F$. In most sectors the lowest energy state was picked. For $n_F=5,7$ the first excited state was chosen because these seem to be better candidates for states in the continuum spectrum. For states corresponding to the continuum spectrum the scaled energy should be flat.}
\label{fig:su3_scaling}
\end{figure}

The main difference from $SU(2)$ is that in the sector with two fermions the spectrum is still discrete. It means that there are not enough fermions to cancel with bosonic degrees of freedom and the effective potential does not vanish. The flat valleys remain closed. For the cutoff we were able to reach the continuous spectrum is best manifested for the ground state of $n_F=6$ and $j=0$. We checked the scaling of the energies to confirm this conclusion. The result is shown in Fig. \ref{fig:su3_scaling}. The scaled energies are given for all $n_F$ in channels with the lowest angular momentum (i.e. $j=0$ for even $n_F$ and $j=1/2$ for odd $n_F$). For $n_F=0,1,2,3,4,6,8$ we picked the lowest energy. For $n_F=5,7$ we chose the first excited energy, because those seem to be better candidates for the continuous spectrum (cf. Figs. \ref{fig:su3_spectrum_nf5}, \ref{fig:su3_spectrum_nf7}). Scaled energies corresponding to the continuous spectrum should be flat for large $N_B$. All energies are divided by $E(N_B=1)$ so that the overall scale for each channel is removed. From Fig. \ref{fig:su3_scaling} one can see that the scaled energies become more flat as the number of fermions grows.

As it was mentioned in the Introduction, the Yang--Mills quantum mechanics without fermions was considered as a zeroth order approximation to pure gauge $SU(2)$ theory in a small volume in \cite{Luscher,Munster}. In \cite{Ziemann_phd,Weisz} authors give numerical results also for $SU(3)$. Their results are obtained numerically by a variational technique, which is essentially the same as our method (but restricted to $n_F=0$). In Table \ref{tab:Weisz_comparison} we present comparison of the lowest energies in several channels with given angular momentum and parity. The results are consistent. The relative difference usually does not exceed 4\%. Naturally, the results differ less in cases where both methods are more precise (i.e. for lowest energies).

\begin{table}[h]\centering
\begin{tabular}{|c|c|c|c|}
\hline
sector&cut Fock space&Rayleigh-Ritz&relative difference\\\hline
\multirow{3}{*}{$0^{+}$}&12.5889&12.5887&0.0016\%\\&15.39&15.38&0.07\%\\&17.24&17.23&0.06\%\\\hline
\multirow{1}{*}{$0^{-}$}&17.75&17.8&-0.28\%\\\hline
\multirow{2}{*}{$1^{+}$}&18.77&18&4.1\%\\&22.4&>20&\\\hline
\multirow{2}{*}{$1^{-}$}&16.52&17.05&-3.2\%\\&18.5&23&-24\%\\\hline
\multirow{2}{*}{$2^{+}$}&14.806&14.854&-3.2\%\\&17.159&17.26&-0.59\%\\\hline
\multirow{1}{*}{$2^{-}$}&18.37&>20&\\\hline
$3^{-}$&16.10&16.5&-2.5\%\\\hline
$4^{+}$&17.3&18&-4\%\\\hline
\end{tabular}
\caption{Comparison of lowest energies in the bosonic sector for the $SU(3)$ case obtained by our method and \cite{Weisz}. Since both approaches are variational, lower values of energies give better approximation to energies in the continuum limit. In the $0^+$ sector results are perfectly consistent - eigenvalues differ only at the last digit which was given in \cite{Weisz}. In other channels our results are usually slightly more accurate.}
\label{tab:Weisz_comparison}
\end{table}

Concluding, the spectrum of the $SU(3)$ theory has discrete and continuous part. In contrary to the $SU(2)$ case, for $SU(3)$ the spectrum is discrete for two fermions. The continuous spectrum is moved to channels with higher $n_F$ because more fermions are needed to have supersymmetric cancelations between fermionic and bosonic degrees of freedom. Only then the effective potential in the flat valleys vanishes. Our results indicate that this occurs for $n_F=6$.

\vfill
\newpage


\chapter{Supersymmetry multiplets}\label{ch:supersymmetric_multiplets}

Having computed spectrum of the model, we are now interested in its supersymmetric properties. It is known \cite{Wess} that energy eigenstates of a supersymmetric model form supersymmery multiplets (or supermultiplets). A supermultiplet is a representation of the supersymmetry algebra (\ref{eq:anticommutation_relations}). In each supermultiplet there is a Clifford vacuum $\ket{\Omega}$. It can have any angular momentum $j$. For $j>0$ it is degenerate and belongs to a $(2j+1)$--dimensional representation of rotational group. By acting with supercharges on $\ket{\Omega}$ one recovers the whole supermultiplet. Other elements of the supermultiplet also form multiplets of angular momentum. The supermultiplet has in total $8j+4$ states. Half of them has odd and half of them even $n_F$. In general, these representations are reducible. The only exception is $j=0$.

A powerful method to identify supermultiplets in organized way is to use supersymmetry fractions \cite{Campostrini}. One can check whether two states are elements of the same multiplet by acting with a supercharge on one of them and analyzing the overlap with the other state. A supersymmetry fraction is essentially such an overlap.

In this chapter the parameter $\omega$ (cf. Chapter \ref{ch:Fock_space_method}) is not included, i.e. here $\omega=1$.

\section{Supersymmetry multiplets in the continuum limit}
The four supersymmetry generators $Q_\alpha$ introduced in (\ref{eq:supercharges}) satisfy anticommutation relations (\ref{eq:anticommutation_relations}) which play a central role in supersymmetric theories. They are however not the only operators which fulfill the algebra (\ref{eq:anticommutation_relations}). Moreover, they are somewhat inconvenient because they do not form a spin multiplet. For this reasons we introduce operators $\mathcal Q_\alpha$ ($\alpha=\pm$):
\begin{align}
\begin{split}
\mathcal Q_+^\dagger&=-e^{i\pi/4}Q_4,\\
\mathcal Q_-^\dagger&=e^{i\pi/4}Q_3.
\end{split}
\end{align}
The overall phase is chosen such that the phase factors simplify when $\mathcal Q^\dagger_\pm$ are written in terms of traces. $\mathcal Q^\dagger_\alpha$ form a spin doublet. The magnetic quantum number of operator $\mathcal Q_\pm^\dagger$ is $m=\pm1/2$, i.e. $[J_3,\mathcal Q^\dagger_\pm]= \pm\frac{1}{2}\mathcal Q^\dagger_\pm$. Both $\mathcal Q^\dagger_\pm$ rise the fermionic number $n_F$ by $+1$. Their conjugates
\begin{align}
\begin{split}
\mathcal Q_+&=e^{i\pi/4}Q_1,\\
\mathcal Q_-&=e^{i\pi/4}Q_2
\end{split}
\end{align}
also form a spin doublet and have opposite quantum numbers. The new supersymmetry generators fulfill following commutation relations:
\begin{align}\label{eq:anticommutations_modified}
\begin{split}
\{\mathcal Q_\alpha,\mathcal Q_{\beta}\}=\{\mathcal Q_\alpha^\dagger,\mathcal Q_{\beta}^\dagger\}=0,\\
\{\mathcal Q_\alpha,\mathcal Q_{\beta}^\dagger\}=2\delta_{\alpha\beta}H,\\
[H,\mathcal Q_\alpha]=[H,\mathcal Q_\alpha^\dagger]=0.
\end{split}
\end{align}

Now we come to discussion of the structure of a supermultiplet. A Clifford vacuum $\ket{\Omega,m}$ is defined by
\begin{align}
\mathcal Q_\alpha\ket{\Omega,m}&=0&\alpha=\pm\frac{1}{2}.
\end{align}
Assume that it has $n_F$ fermions, angular momentum $j$ and energy $E$. The magnetic number $m$ is kept explicit. One has to keep in mind that the Clifford vacuum has nothing to do with the vacuum of the model, i.e. the state with the lowest energy. In order to avoid confusion, it will be always called a Clifford vacuum or a vacuum of a supermultiplet. If $E=0$ then $\mathcal Q^\dagger_\alpha\ket{\Omega,m}$ vanishes and $\ket{\Omega,m}$ is a supersymmetry singlet. Indeed,
\begin{align}
\left\|\mathcal Q^\dagger_\alpha\ket{\Omega,m}\right\|^2=\braket{\Omega,m|\mathcal Q_\alpha\mathcal Q^\dagger_\alpha|\Omega,m}=\braket{\Omega,m|2H-\mathcal Q^\dagger_\alpha\mathcal Q_\alpha|\Omega,m}=0.
\end{align}
If this is not the case, new states are created by acting with $\mathcal Q^\dagger_\alpha$ on $\ket{\Omega,m}$. With the aid of Clebsh-Gordan coefficients, these states can be arranged in spin multiplets
\begin{align}\label{eq:other_states}
\begin{split}
\begin{aligned}
\ket{\Omega_-,m}&=\frac{1}{\sqrt {2E}}\sum_\alpha\braket{j,m-\alpha,\frac{1}{2},\alpha|j-\frac{1}{2},m}\mathcal Q^\dagger_\alpha\ket{\Omega,m-\alpha},&& m=-j+\frac{1}{2},\ldots,j-\frac{1}{2},\\
\ket{\Omega_+,m}&=\frac{1}{\sqrt {2E}}\sum_\alpha\braket{j,m-\alpha,\frac{1}{2},\alpha|j+\frac{1}{2},m}\mathcal Q^\dagger_\alpha\ket{\Omega,m-\alpha},&& m=-j-\frac{1}{2},\ldots,j+\frac{1}{2},\\
\ket{\Omega_{0},m}&=\frac{1}{2E}\mathcal Q_-^\dagger \mathcal Q_+^\dagger\ket{\Omega,m},&&m=-j,\ldots,j.
\end{aligned}
\end{split}
\end{align}
If $j=0$ then $\ket{\Omega_-,m}=0$. Otherwise none of these states vanishes. All of them are normalized thanks to the relation of Clebsh-Gordan coefficients
\begin{align}\label{eq:CG_normalization}
\sum_\alpha\left|\braket{j\pm\frac{1}{2},m-\alpha,\frac{1}{2},\alpha|j,m}\right|^2&=1&\text{for all $j\geq\frac{1}{2}$ and $m=-j,\ldots,j$.}
\end{align}
Obviously, (\ref{eq:CG_normalization}) is not valid for $j=0$ when it is taken with the minus sign, because there is spin $-\frac{1}{2}$ on the left hand side. The three new states have energy $E$ because the supercharges commute with the Hamiltonian. $\ket{\Omega_-,m},\ket{\Omega_+,m},\ket{\Omega_{0},m}$ are spin multiplets with total angular momentum $j-\frac{1}{2},\ j+\frac{1}{2}$ and $j$ respectively. Together with the vacuum $\ket{\Omega,m}$ they form a \emph{supersymmetric multiplet}.

Several remarks concerning supermultiplets are in place. A supersymmetry multiplet is closed under the action of supercharges. That is, $\mathcal Q_\alpha$ or $\mathcal Q^\dagger_\alpha$ acting on an element of a supermultiplet gives zero or combination of other elements. It follows that none of the three states $\ket{\Omega_-,m},\ket{\Omega_+,m},\ket{\Omega_{0},m}$ can be a vacuum state of another supermultiplet.

The full supersymmetry multiplet forms a diamond in the $(n_F,j)$ plane (cf. Fig. \ref{fig:supermultiplet}). The exception is $j=0$ -- there is no state $\ket{\Omega_-,m}$, i.e. the lower node $(n_F+1,j-\frac{1}{2})$ in Fig. \ref{fig:supermultiplet} is not present. In the supermultiplet there are $4j+2$ bosonic states (i.e. states which contain an even number of fermions) and the same number of fermionic states. As a consequence, a supersymmetry multiplet with nonzero energy does not contribute to the Witten index.

\newlength{\vdiam}
\newlength{\hdiam}
\setlength{\vdiam}{1cm}
\setlength{\hdiam}{3cm}
\begin{figure}
\begin{center}
\begin{tikzpicture}
\node[draw,ellipse, minimum width = 3cm](swest){$n_F,j$};
\node[above right = \vdiam and \hdiam of swest,draw,ellipse, minimum width = 3cm](snorth){$n_F+1,j+\frac{1}{2}$};
\node[below right = \vdiam and \hdiam of swest,draw,ellipse, minimum width = 3cm](south){$n_F+1,j-\frac{1}{2}$};
\node[above right = \vdiam and \hdiam of south,draw,ellipse, minimum width = 3cm](seast){$n_F+2,j$};
\draw[-] (swest) -- (snorth) node [midway,fill=white,draw] {$q=j+1$};
\draw[-] (swest) -- (south) node [midway,fill=white,draw] {$q=j$};
\draw[-] (south) -- (seast) node [midway,fill=white,draw] {$q=j$};
\draw[-] (snorth) -- (seast) node [midway,fill=white,draw] {$q=j+1$};
\end{tikzpicture}
\end{center}
\caption{Structure of a supermultiplet together with supersymmetry fractions. For $j=0$ the bottom vertex and the two lower links are absent.}
\label{fig:supermultiplet}
\end{figure}
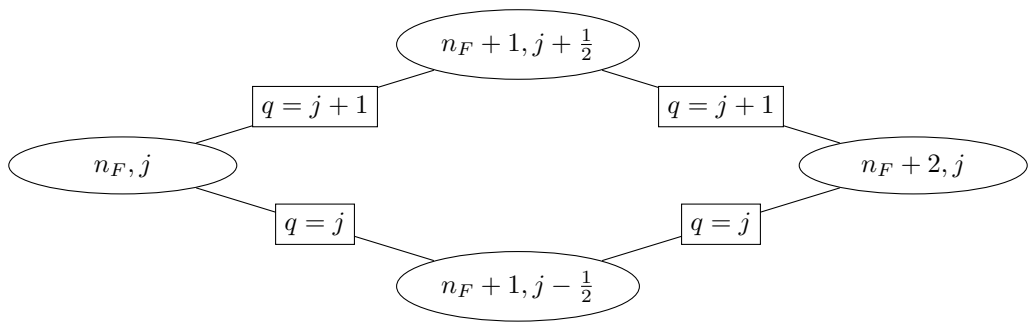

\begin{figure}
\centering
\includegraphics[width=.9\textwidth]{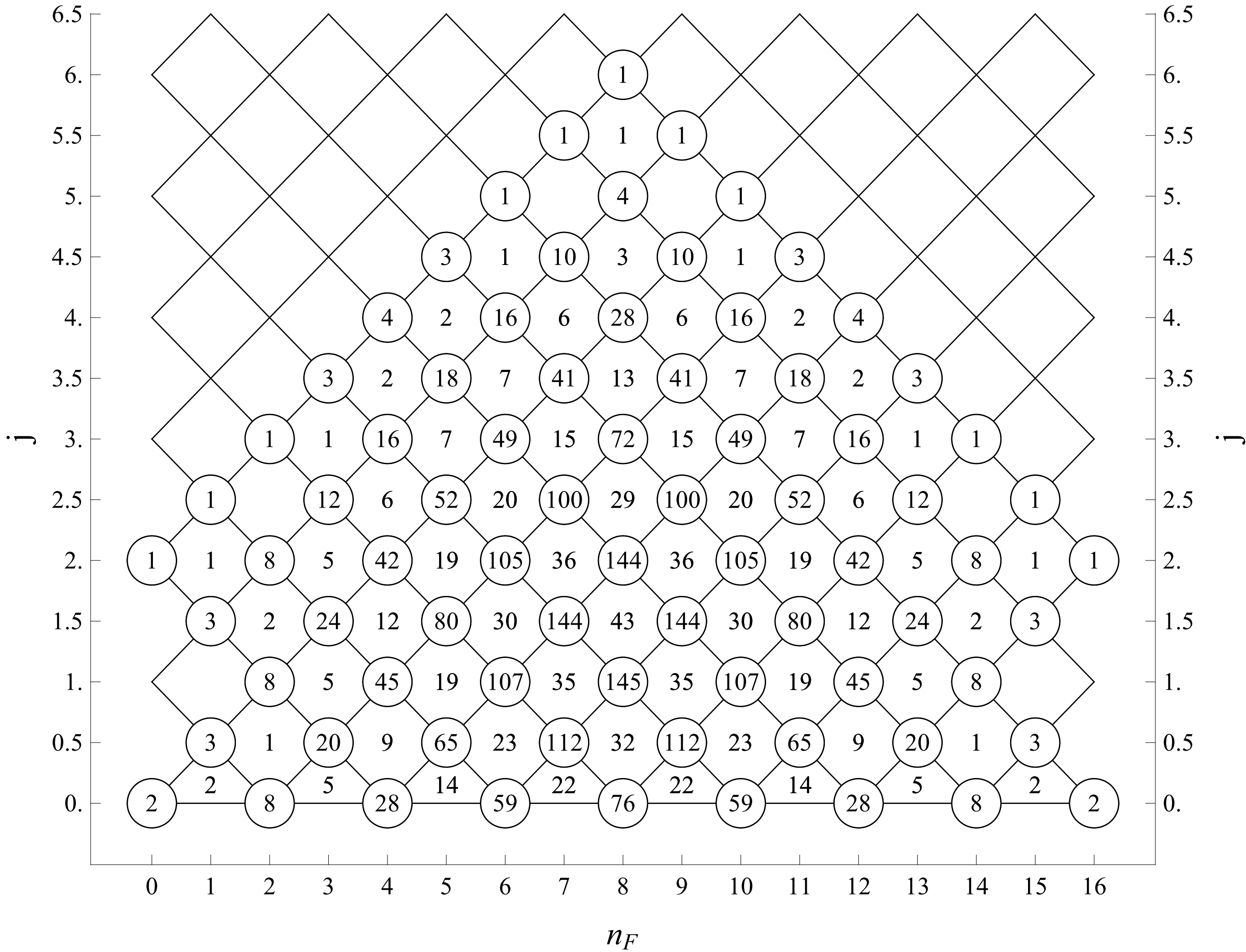}
\caption{Structure of the angular momentum multiplets and supersymmetry multiplets for SU(3) and $N_B=2$. Numbers in circles represent the number of angular momentum multiplets in a given channel. The numbers in between are the numbers of supermultiplets (the diamonds).}
\label{fig:su3structure}
\end{figure}

\begin{figure}
\centering
\includegraphics[width=.5\textwidth]{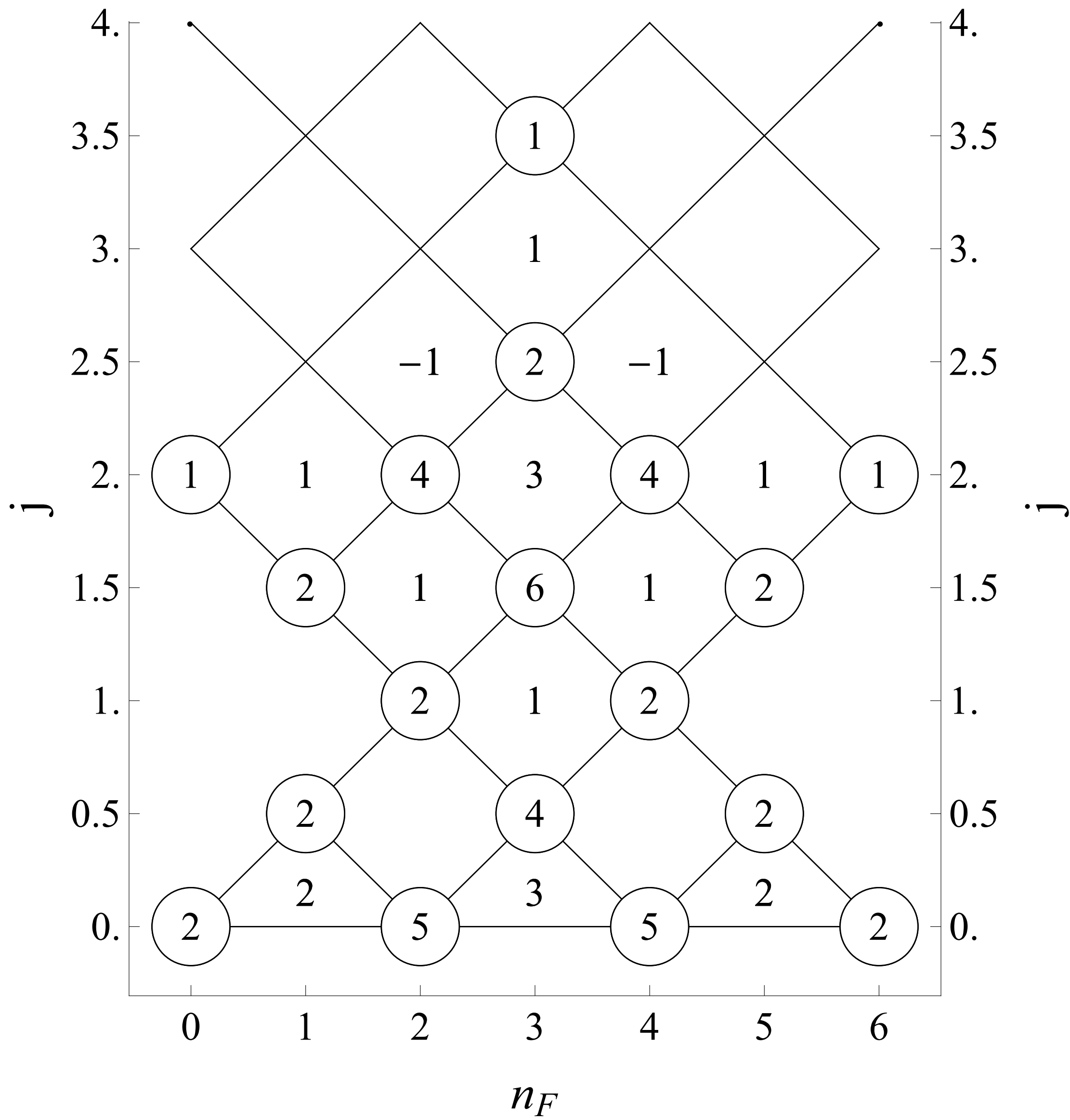}
\caption{Structure of the angular momentum multiplets and supersymmetry multiplets for SU(2) and $N_B=2$.}
\label{fig:su2structure}
\end{figure}

For finite $N_B$ there is only a finite number of states in each $(n_F,j)$ channel. In Fig. \ref{fig:su3structure} the total number of multiplets in each sector for the $SU(3)$ gauge group and $N_B=2$ is given. Each circle is a vertex of a diamond or a triangle. These figures are the same as the diamond in Fig. \ref{fig:supermultiplet} and they correspond to supermultiplets. The number inside each diamond or triangle is the number of according supermultiplets for finite $N_B$. These numbers are determined in such way that the total number of supermultiplets adjacent to a vertex $(n_F,j)$ is the same as the number of angular momentum multiplets in the sector $(n_F,j)$. This is motivated by the fact that each $SO(3)$ multiplet belongs to exactly one supermultiplet. However, one has to keep in mind that for a finite $N_B$ the supersymmetry is broken and there are actually no supermultiplets. Therefore, the number of supermultiplets determined in this way has no strict meaning. In particular, there is a negative number of $(1,5/2)$ supermultiplets for $SU(2)$ with $N_B=2$ (cf. Fig. \ref{fig:su2structure}).


For $j=0$ the supermultiplet is an irreducible representation of the algebra (\ref{eq:anticommutations_modified}). In this case is consists of four states. For $j>0$ the representation is contains $2j+1$ irreducible representations of the supersymmetry algebra. An irreducible representation can be constructed starting from $\ket{\Omega,m}$ with a single value of $m$. Then the other three states in the (small) supermultiplet are $\frac{1}{\sqrt{2E}}\mathcal Q^\dagger_\alpha\ket{\Omega,m}$ and $\frac{1}{2E}\mathcal Q_-^\dagger \mathcal Q_+^\dagger\ket{\Omega,m}$. Each of these states has definite magnetic number, but $\mathcal Q^\dagger_\alpha\ket{\Omega,m}$ do not have definite total angular momentum. For convenience, in this thesis we use full supermultiplets with complete angular momentum multiplets.

\section{Supersymmetry fractions}
A useful tool to analyze the supermultiplets are supersymmetry fractions. They serve two purposes. First is studying breaking of the supersymmetry for finite cutoff. Second is identifying supermultiplets. Consider two angular momentum multiplets $\ket{n_FjmE}$ and $\ket{n_F+1j'm'E'}$. A supercharge $\mathcal Q^\dagger_\alpha$ maps $\ket{n_FjmE}$ into $\ket{n_F+1j'm'E'}$ if they are in the same supermultiplet and $m'=m+\alpha$. A supersymmetry fraction measures the overlap. It is defined by
\begin{align}
q_{n_F}\left(j'E'|jE\right)&=\frac{1}{4E}\sum_{mm'\alpha}\left|\braket{n_F+1j'm'E'|\mathcal Q_\alpha^\dagger|n_FjmE}\right|^2.
\end{align}
Formulas (\ref{eq:other_states}) together with the anticommutation relations (\ref{eq:anticommutations_modified}) give straightforward
\begin{align}\label{eq:pre_fractions}
\begin{split}
\sum_{mm'\alpha}\left|\braket{\Omega,m|\mathcal Q_\alpha|\Omega_+,m'}\right|^2=\sum_{mm'\alpha}\left|\braket{\Omega_+,m|\mathcal Q_\alpha|\Omega_{0},m'}\right|^2&=4E(j+1),\\
\sum_{mm'\alpha}\left|\braket{\Omega,m|\mathcal Q_\alpha|\Omega_-,m'}\right|^2=\sum_{mm'\alpha}\left|\braket{\Omega_-,m|\mathcal Q_\alpha|\Omega_{0},m'}\right|^2&=4Ej.
\end{split}
\end{align}
These relations specify what are the supersymmetry fractions within a supermultiplet. If the Clifford vacuum $\ket{\Omega,m}$ has $n_F$ fermions, energy $E$ and angular momentum $j$, then the fractions are
\begin{align}\label{eq:supersummetry_fractions_exact}\begin{split}
q_{n_F}\left(j+\frac{1}{2},E\Big|j,E\right)=q_{n_F+1}\left(j,E\Big|j+\frac{1}{2},E\right)&=j+1,\\
q_{n_F}\left(j-\frac{1}{2},E\Big|j,E\right)=q_{n_F+1}\left(j,E\Big|j-\frac{1}{2},E\right)&=j.
\end{split}\end{align}
The supersymmetry fractions are denoted by rectangles in Fig. \ref{fig:supermultiplet}. The supermultiplet is closed under the action of $\mathcal Q^\dagger_\alpha$ so other fractions must necessarily vanish.

\begin{figure}
\centering
\includegraphics[width=.9\textwidth]{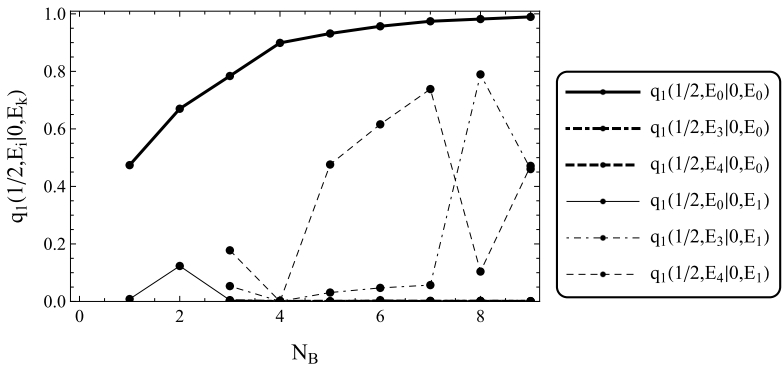}
\caption{Supersymmetry fractions as a function of cutoff. $N_B$ is the cutoff in the sector $n_F=1$. The cutoff for $n_F=0$ is higher by $2$.}
\label{fig:fractions_convergence}
\end{figure}

\section{Finite cutoff effects}

For a finite cutoff the supersymmetry is broken. Now we analyze restoration of the supersymmetry for growing $N_B$. For small cutoff the supersymmetry fractions are far from the values given by (\ref{eq:supersummetry_fractions_exact}). We say that the supersymmetry is recovered when the supersymmetry fractions are close to their exact values. Next we address the problem of identifying superpartners when two supermultiplets are almost degenerate.

Consider a supersymmetry fraction $q_{n_F}(j\pm\frac{1}{2},E_i|j,E_k)$. For a large cutoff it is expected that the fraction is zero unless corresponding states, labeled with $i$ and $k$, are in the same supermultiplet. In particular, $E_i$ and $E_k$ should be almost the same if the fraction is nonzero. In contrast, for a small cutoff eigenstates of the cut Hamiltonian do not give a good approximation to exact bound states. In this situation the supersymmetry fractions $q_{n_F}(j\pm\frac{1}{2},E_i|j,E_k)$ are small but nonzero for many different pairs $(i,k)$.

Analysis of supersymmetry fractions for growing cutoff provides us the information at which value of $N_B$ supersymmetry is approximately restored. In Fig. \ref{fig:fractions_convergence} dependence of supersymmetry fractions $q_1(\frac{1}{2},E_i|0,E_k)$ on the cutoff is shown for selected $i$ and $k$. All states with $(n_F,j)=(0,0)$ are Clifford vacua, so the exact values of shown supersymmetry fractions are $0$ and $1$. For the ground state in the bosonic sector there is only one fraction, $q_1(\frac{1}{2},E_0|0,E_0)$, which is significant. It approaches $1$ fast and we may say that the supersymmetry between the two corresponding states is essentially restored for $N_B\approx7$. In general higher cutoffs are required to regain supersymmetry for higher eigenstates. This is because the lowest energy states converge first.

Behavior of fractions $q_1(\frac{1}{2},E_i|0,E_1)$, $i=3,4$ in Fig. \ref{fig:fractions_convergence} is related to an approximate degeneracy of energies with $n_F=1$ and $N_B=9$ (cf. Fig. \ref{fig:su3_spectrum_nf1}). We address this case on a general ground. Suppose that with the available cutoff a good approximation of a bound state $\ket{n_FjmE}$ is obtained. Denote its (exact) superpartner a in channel $(n_F+1,j'=j\pm\frac{1}{2})$ by $\ket{n_F+1j'm'E}$. In general, for larger number of fermions cutoffs are smaller and thus the approximation is worse. Therefore, a good approximation of the state $\ket{n_F+1j'm'E}$ is given by some combination
\begin{align}\label{eq:superpartner_decomposition}
\ket{n_F+1j'm'E}\approx\sum_i c_i\ket{n_F+1j'm'E_i}_{N_B},
\end{align}
where $\ket{n_F+1j'm'E_i}_{N_B}$ are eigenstates of the Hamiltonian for cutoff $N_B$. The question is which states contribute to the sum (\ref{eq:superpartner_decomposition}). Each energy $E_i$ has an error $\Delta_i$ which is a finite cutoff effect. States which have energies equal within the error $\Delta$ mix for higher $N_B$ and give a better approximation of exact eigenstates. Therefore, only these $c_i$ are nonzero for which $|E_i-E|\lesssim \Delta_i$. If the cutoff is not very small and $\ket{n_F+1j'm'E}$ is not a highly excited state, then the errors $\Delta_i$ are small and the energies $E_i$ have to be almost degenerate. The supersymmetry fractions $q_{n_F}(j'E_i|jE)$ for the cutoff $N_B$ are fully determined by coefficients $c_i$. In order to calculate values of the fractions first assume that $\ket{n_FjmE}$ is vacuum of a supermultiplet. Then the supersymmetry fraction is
\begin{align}\begin{split}
&q_{n_F}(j'E_i|jE)=\\
&\frac{1}{4E}\sum_{mm'\alpha}\left|\prescript{}{N_B}{\braket{n_F+1j'm'E_i|Q_\alpha^\dagger|n_FjmE}}\right|^2\\
&=\frac{1}{4E}\sum_{mm'\alpha}\Big|\prescript{}{N_B}{\braket{n_F+1j'm'E_i|n_F+1j'm'E}}\braket{n_F+1j'm'E|Q_\alpha^\dagger|n_FjmE}\Big|^2\\
&=\frac{1}{4E}\sum_{mm'\alpha}\left|c_i\right|^2\Big|\frac{1}{\sqrt{2E}}\sum_\beta\braket{j,m'-\beta,\frac{1}{2},\beta|j'm'}\braket{n_Fjm'-\beta E|Q_\beta Q_\alpha^\dagger|n_FjmE}\Big|^2\\
&=\frac{1}{4E}\sum_{m\alpha}\left|c_i\right|^2\left|\sqrt{2E}\braket{j,m,\frac{1}{2},\alpha|j'm+\alpha}^*\right|^2.
\end{split}\end{align}
Finally, we use the following property of Clebsch-Gordan coefficients:
\begin{align}
\sum_{m\alpha}\left|\braket{j,m,\frac{1}{2},\alpha|j'm+\alpha}\right|^2=(2j'+1)
\end{align}
and obtain
\begin{align}\label{eq:split_fractions}
q_{n_F}(j'E_i|jE)&=\left(j'+\frac{1}{2}\right)\left|c_i\right|^2.
\end{align}

If $\ket{n_FjmE}$ is not a Clifford vacuum then it is $\ket{\Omega_+,m}$ or $\ket{\Omega_-,m}$ from (\ref{eq:other_states}). One can show by an analogous proof that then
\begin{align}
q_{n_F}(j'E_i|jE)&=\left(j+\frac{1}{2}\right)\left|c_i\right|^2.
\end{align}
In both cases $q_{n_F}(j'E_i|jE)$ is the exact supersymmetry fraction multiplied by $|c_i|^2$. Concluding, the supersymmetry fractions $q_{n_F}(j'E_i|jE)$ add up to the exact value. Therefore, if one finds several fractions $q_{n_F}(j'E_i|jE)$ which sum up almost to $j+\frac{1}{2}$ or $j'+\frac{1}{2}$ then the superpartner of $\ket{n_FjmE}$ is given by (\ref{eq:superpartner_decomposition}) where $c_i$ are determined up to phases.

Consider again the supersymmetry fractions $q_1(\frac{1}{2},E_i|0,E_1)$ (Fig. \ref{fig:fractions_convergence}). For $N_B$ up to $7$ the fraction $q_1(\frac{1}{2},E_4|0,E_1)$ grows while $q_1(\frac{1}{2},E_3|0,E_1)$ is close to $0$. For $N_B=8$ the two fractions interchange. This is because the energy of the state $\ket{1,\frac{1}{2},E_4}$ decreases faster than the energy of $\ket{1,\frac{1}{2},E_3}$ and labels of energies are swapped at $N_B=8$. For $N_B=9$ the fractions are $0.46$ and $0.47$ and sum up to $0.93$. Based on earlier analysis we conclude that a good approximation to the superpartner $\ket{\Omega_+,m}$ (cf. ( \ref{eq:other_states})) of the first excited state in the bosonic sector $\ket{0,0,E_1}$ is given by a combination of the third and fourth state:
\begin{align}
\ket{\Omega_+,m}&\approx\sqrt{0.46} e^{i\alpha_1}\ket{1,\frac{1}{2},m,E_3}_{N_B=9}+\sqrt{0.47} e^{i\alpha_2}\ket{1,\frac{1}{2},m,E_4}_{N_B=9}.
\end{align}

Such situation as described above, with several supersymmetry fractions which add up almost to an exact value is in fact very common. The main reason for this is that because large number of degrees of freedom density of eigenstates is large. Therefore, states that correspond to different but close energies mix easily for a finite cutoff. Obviously, these states disentangle for a cutoff which is high enough. Nevertheless, if energies of different supermultiplets are almost degenerate, the needed cutoff is very large.

All above considerations address only the case of discrete spectrum. This is because the continuous spectrum appears in channels with many fermions. There the density of energy states is higher while the available cutoff is small. This makes the analysis of supersymmetry fractions yet more demanding.

\section{Detailed analysis of supersymmetry fractions}

A general method for identifying superpartners was presented. Here more examples are considered. In particular, full identification of a supermultiplet with a Clifford vacuum in sector $(0,0)$ is carried out. In Tables \ref{tab:susyf00up} - \ref{tab:susyf01up} supersymmetry fractions for several $(n_F,j)$ channels are shown. A fraction $q_{n_F}(j+\frac{1}{2}E'|jE)$ can take values either $j+\frac{1}{2}$ or $j+1$ depending on whether corresponding states are in $(n_F-1,j+\frac{1}{2})$ or $(n_F,j)$ supermultiplet respectively. Analogously, $q_{n_F}(j-\frac{1}{2}E'|jE)=j+\frac{1}{2}$ or $j$ if the states are in $(n_F-1,j-\frac{1}{2})$ of $(n_F,j)$ supermultiplet. Because the cutoff we are able to reach are not very large, the supersymmetry fractions are in general significantly smaller than their values in the continuum limit. Still, they give valuable information on at least some supermultiplets.

In Tab. \ref{tab:susyf00up} there are shown supersymmetry fractions $q_1(\frac{1}{2}E_i|0,E_k)$. All states from the $(0,0)$ sector are vacua of a $(0,0,E)$ supermultiplet. Therefore, for all of them there should be a supersymmetry fraction equal to $1$ with some states with $(n_F,j)=(1,\frac{1}{2})$. The states from channel $(1,\frac{1}{2})$ belong to one of three supermultiplets, $(0,0,E),\ (0,1,E)$ or $(1,\frac{1}{2},E)$. Therefore, only some of them share nonzero supersymmetry fractions with states in the $(0,0)$ sector.

The first two columns of Tab. \ref{tab:susyf00up} were already analyzed in the previous part of this chapter. It was stated that the ground states in $(0,0)$ and $(1,\frac{1}{2})$ sectors are elements of the same multiplet. The superpartner of the first excited bosonic state, $\ket{0,0,15.48}$ is a combination of third and fourth excited states in the channel $(1,\frac{1}{2})$. The third column is related to the singlet $\ket{0,0,17.47}$. There are five fractions which correspond to doublets in single fermion sector that add up to $0.83$. Energies of these doublets range from $17.88$ to $18.85$. These states will possibly mix for higher cutoff and give the superpartner of $\ket{0,0,17.47}$. This is consistent with the energy dependence on the cutoff (cf. Fig. \ref{fig:su3_spectrum_nf1}). For $E\sim18$ all energies decrease by $\Delta E\sim 1$ when $N_B$ changes from $8$ to $9$. Therefore, the error bars to these energies are of order $1$. All five energies are the same within error bars and thus the states can mix before the continuum limit is reached.

Consider fractions $q_2(0,E_i|\frac{1}{2},E_k)$ (Tab. \ref{tab:susyf10do}). In the first column there are two significant fractions $q_2(0,13.34|\frac{1}{2},12.72)$ and $q_2(0,13.45|\frac{1}{2},12.72)$ which add up to $0.92$. This is evidence that a combination of states $\ket{2,0,13.34}$ and $\ket{2,0,13.45}$ gives the last state of the supermultiplet $(0,0,12.6)$. In the second column the largest element is $q_2(0,15.67|\frac{1}{2},14.48)=0.3$. If $\ket{1,\frac{1}{2},14.48}$ is vacuum of a supermultiplet, then the exact value of the fraction is $0.5$. If this is the case then there should be a fraction $q_2(1,E|\frac{1}{2},14.48)\approx 1.5$. Indeed, there is $q_2(1,15.62|\frac{1}{2},14.48)=0.65\times 1.5$.

Finally, we study supersymmetry fractions $q_0(j',E'|1,E)$ with $j'=\frac{1}{2}$ (Tab. \ref{tab:susyf01do}) and $j'=\frac{3}{2}$ (Tab. \ref{tab:susyf01up}). Since the bosonic state is always a vacuum of a supermultiplet, the exact values of fractions are $1$ and $2$ respectively. For the bosonic ground state with $j=1$ the relevant fractions are $q_0(\frac{1}{2},17.5|1,16.66)=0.77$ and $q_0(\frac{3}{2},17.44|1,16.66)=0.76\times 2$. It is remarkable that in both $(n_F=1,j')$ channels the approximation of the energies is of the same quality ($E\approx17.5$) as well as the fractions $q\approx 0.76\times q_{exact}$. For the first excited bosonic state the susy fractions are rather poor: $q_0(\frac{1}{2},20.75|1,18.79)=0.53$ and $q_0(\frac{3}{2},20.74|1,18.79)+q_0(\frac{3}{2},20.82|1,18.79)=0.52\times 2$. Again, the approximation of energies and fractions are similar for $\ket{1,\frac{1}{2},20.74}$ and $\ket{1,\frac{3}{2},20.82}$.

Analysis of all other sectors is similar. Summary of identified supermultiplets is given in Tab. \ref{tab:spectroscopy}. Also values of supersymmetry fractions are given to indicate how strong is the identification. Clearly, large difference of energies in a supermultiplet means that states did not yet converge and one should not expect large values of fractions. Supersymmetry fractions given in Tab. \ref{tab:spectroscopy} are normalized, i.e. divided by exact values.

\section{Summary}
\begin{figure}\centering
\includegraphics[width=.5\textwidth]{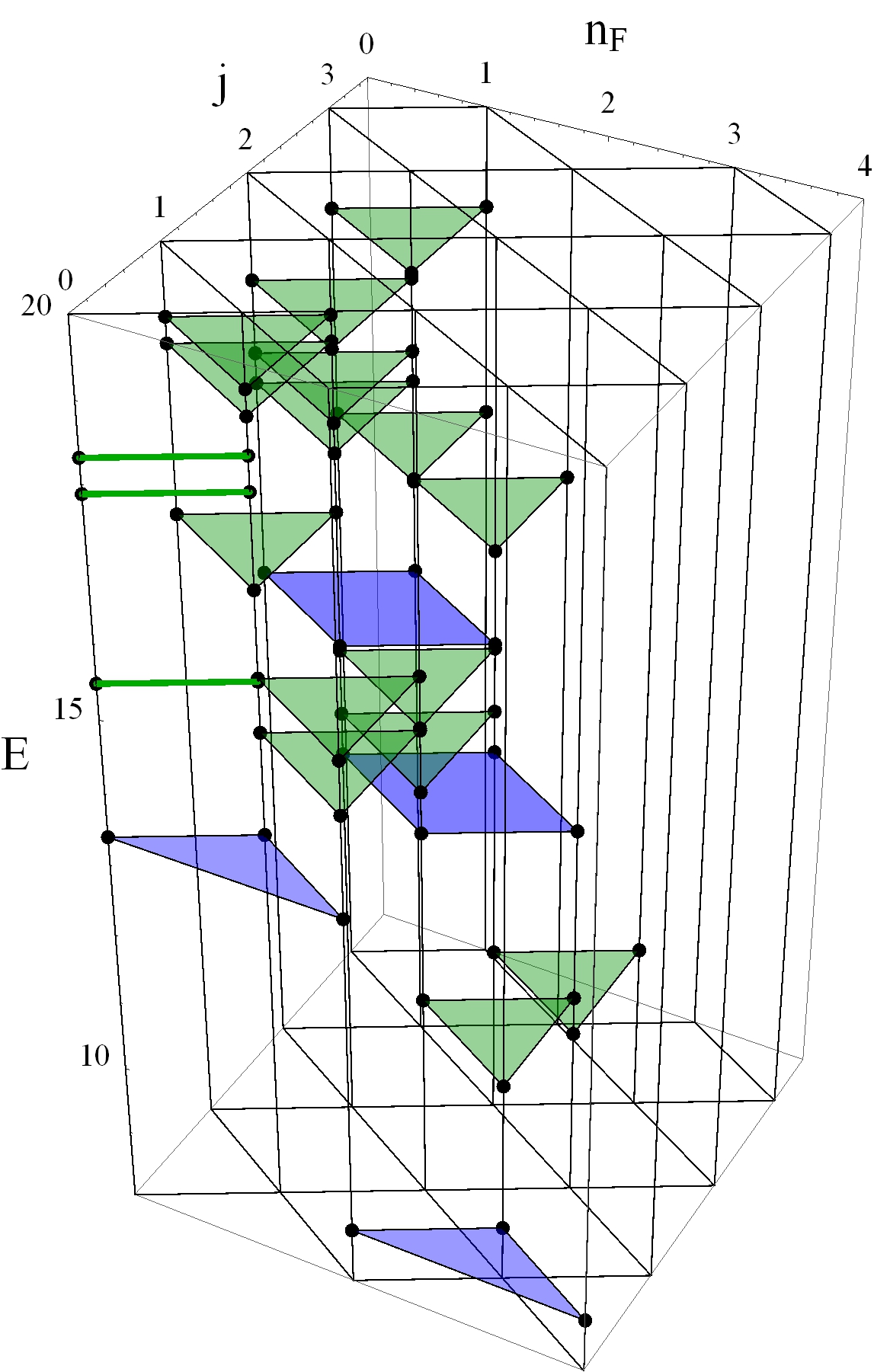}
\caption{All identified supermultiplets. The diamond and two triangles with two vertices at $j=0$ are completely identified multiplets. The other triangles correspond to diamonds with one state unidentified. Lines correspond to triangle multiplets with one vertex not identified.}
\label{fig:overall}
\end{figure}

In Fig. \ref{fig:overall} all identified supermultiplets are shown. Triangles with vertices in sectors $(n_F,0),\ (n_F+1,\frac{1}{2}),\ (n_F+2,0)$ and all diamonds are fully identified supermultiplets. They are marked with blue color. The triangles which have vertices at $(n_F,j)$, $(n_F+1,j-\frac{1}{2})$ and $(n_F+1,j+\frac{1}{2})$ are marked with green color and represent diamonds with one element not identified. Green lines correspond to two states out of three in a multiplet with Clifford vacuum in $(n_F,0)$. Not identified states are marked with a question mark in Tab. \ref{tab:spectroscopy}. There are also many energies in between for which there were no supersymmetry fractions much different from $0$. In the continuum limit all states should group in supersymmetric multiplets. The method which was used for finding the energies is variational, so all energies are approximated from above. For this reason all states in Fig. \ref{fig:overall} are shifted to the lowest energy in the corresponding multiplet.

We succeeded to fully identify four supermultiplets (cf. Tab. \ref{tab:spectroscopy}). In general, it is easier to identify states which 'open' the multiplet, i.e. the vacuum in the sector with $n_F$ fermions and two states with $n_F+1$ fermions. The remaining state which 'closes' the supermultiplet is significantly harder to find. This is because with more fermions the lowest energy state has smaller energy than in those with less fermions and the density of eigenenergies is higher (see Figs \ref{fig:su3_spectrum_nf0} - \ref{fig:su3_spectrum_nf8}). Therefore, even if the Clifford vacuum is a low excited state, the state which closes the supermultiplet has $n_F+2$ fermions and is highly excited. For excited states one needs a higher cutoff before the convergence is reached. On the other hand, the available cutoff gets smaller for more fermions.

\begin{table}\centering
\begin{tabular}{cc|cccccc}
&& \multicolumn{6}{c}{$n_F=0,\ j=0$}    \\
\multirow{27}{*}{\rotatebox[origin=c]{90}{$n_F=1,\ j={1/2}$}}&
$E$&12.6&15.48&17.47&17.88&18.32&19.76\\\hline&12.72&0.99&-&-&-&-&-\\&14.48&-
&-&-&-&-&-\\&15.52&-&0.01&-&-&-&-\\&15.88&-&0.46&-&-&-&-\\&15.96&-&0.47&0.01&-&-&-\\&16.46&-&-&-&-&-&-\\&16.69&-&-&-&-&-&-\\&16.97&-&-&-&-&
-&-\\&17.09&-&-&-&-&-&-\\&17.5&-&-&-&-&-&-\\&17.58&-&-&-&-&-&-\\&17.88&-&-&0.11&0.03&-&-\\&18.26&-&-&-&-&-&-\\&18.32&-&-&-&-&-&-\\&18.52&-
&-&0.2&0.14&0.04&0.02\\&18.62&-&-&0.42&0.03&0.05&-\\&18.73&-&-&0.06&0.02&0.01&-\\&18.85&-&-&0.05&0.61&0.01&-\\&18.96&-&-&-&-&-&-\\&19.08&-&-
&-&-&-&-\\&19.21&-&-&-&0.02&0.03&0.02\\&19.3&-&-&-&-&-&-\\&19.46&-&-&-
&-&-&-\\&19.66&-&-&0.01&0.03&0.11&0.06\\&19.73&-&-&-&-&0.52&-\\&19.75&
-&-&-&-&-&-\\&20.03&-&-&-&-&-&-\\&20.11&-&-&-&-&-&-\\&20.38&-&-&-&-&-&
-\\&20.47&-&-&-&-&-&-\\&20.57&-&-&-&-&-&-\\&20.74&-&-&-&-&-&0.03\\&20.75&-&-&-&-&-&-\\&20.79&-&-&-&-&-&-\\&20.92&-&-&-&-&0.02&0.01\\&20.99&-
&-&-&-&-&-\\&21.16&-&-&-&-&-&-\\&21.17&-&-&-&-&-&0.01\\&21.33&-&-&-&-&
-&-\\&21.34&-&-&-&-&-&0.01\\&21.49&-&-&-&-&-&-\\&21.5&-&-&-&-&-&0.1\\&21.57&-&-&-&-&-&-\\&21.64&-&-&-&-&-&0.3
\end{tabular}
\caption{Supersymmetry fractions between sectors $(n_F,j)=(0,0)$ and $(1,1/2)$.}
\label{tab:susyf00up}
\end{table}

\begin{table}\centering
\begin{tabular}{cc|cccc}
&& \multicolumn{4}{c}{$n_F=1,\ j=1/2$}    \\
\multirow{17}{*}{\rotatebox[origin=c]{90}{$n_F=2,\ j=0$}}& $E$&12.72&14.48&15.52&15.88\\
\hline
&8.8&-&-&-&-\\
&11.57&-&-&-&-\\
&12.47&-&-&-&-\\
&13.34&0.71&-&-&-\\
&13.45&0.21&-&-&-\\
&14.23&-&-&-&-\\
&14.64&-&-&-&-\\
&14.82&-&0.11&-&-\\
&15.09&-&-&-&-\\
&15.55&-&-&-&-\\
&15.67&-&0.3&-&-\\
&\vdots(4)\\
&16.61&-&-&0.09&0.02\\
&16.83&-&-&0.12&0.04\\
&17.05&-&-&0.14&0.01\\
&\vdots(7)\\
&18.05&-&-&0.04&0.28
\end{tabular}
\caption{Supersymmetry fractions between sectors with $(n_F,j)=(1,1/2)$ and $(2,0)$.}
\label{tab:susyf10do}
\end{table}

\begin{table}\centering
\begin{tabular}{cc|cccc}
&& \multicolumn{4}{c}{$n_F=0,\ j=1$}    \\
\multirow{12}{*}{\rotatebox[origin=c]{90}{$n_F=1,\ j=1/2$}}& $E$&16.66&18.79&19.02&19.11\\
\hline
&\vdots(9)\\
&17.5&0.77&-&-&-\\
&\vdots(12)\\
&19.46&-&0.11&-&0.14\\
&\vdots(4)\\
&20.11&-&-&0.02&0.14\\
&20.38&-&-&-&0.4\\
&\vdots(3)\\
&20.75&-&0.53&-&-\\
&20.79&-&0.1&0.04&0.04\\
&\vdots(6)\\
&21.49&-&-&0.4&-
\end{tabular}
\caption{Supersymmetry fractions between bosonic sector with $j=1$ and fermionic sector with $j=1/2$.}
\label{tab:susyf01do}
\end{table}

\begin{table}\centering
\begin{tabular}{cc|cccc}
&& \multicolumn{4}{c}{$n_F=0,\ j=1$}    \\
\multirow{22}{*}{\rotatebox[origin=c]{90}{$n_F=1,\ j=3/2$}}& $E$&16.66&18.79&19.02&19.11\\
\hline
&\vdots(12)&&&&\\
&17.44&0.14&-&-&-\\
&17.55&1.52&0.04&0.01&-\\
&\vdots(16)&&&&\\
&19.38&-&0.1&-&0.24\\
&\vdots(6)&&&&\\
&20.1&-&-&-&0.15\\
&\vdots(3)&&&&\\
&20.41&-&0.02&-&0.92\\
&20.54&-&0.04&-&-\\
&20.58&-&0.02&0.03&0.14\\
&20.66&-&-&-&-\\
&20.74&-&0.73&-&0.02\\
&20.75&-&-&-&-\\
&20.79&-&0.12&-&-\\
&20.82&-&0.31&-&-\\
&20.88&-&-&-&-\\
&20.93&-&0.11&-&-\\
&\vdots(9)&&&&\\
&21.47&-&-&0.38&-\\
&21.52&-&-&0.8&-
\end{tabular}
\caption{Supersymmetry fractions between bosonic sector with $j=1$ and fermionic sector with $j=3/2$. Rows with no supersymmetry fractions greater than 0.1 are omitted. Numbers of omitted states are given in brackets.}
\label{tab:susyf01up}
\end{table}

\begin{table}\centering
\begin{tabular}{c|cccc|cccc}
&\multicolumn{4}{c|}{energy}&\multicolumn{4}{c}{fractions}\\
\hline
$(n_F,j)$& \rotatebox[origin=c]{70}{$(n_F,j)$}& \rotatebox[origin=c]{70}{$(n_F+1,j-\frac{1}{2})$}& \rotatebox[origin=c]{70}{$(n_F+1,j+\frac{1}{2})$}& \rotatebox[origin=c]{70}{$(n_F+2,j)$}& \rotatebox[origin=c]{70}{$\tilde q_{n_F}(j-\frac{1}{2}|j)$}& \rotatebox[origin=c]{70}{$\tilde q_{n_F}(j+\frac{1}{2}|j)$}& \rotatebox[origin=c]{70}{$\tilde q_{n_F+1}(j|j-\frac{1}{2})$}& \rotatebox[origin=c]{70}{$\tilde q_{n_F+1}(j|j+\frac{1}{2})$}\\
\hline
(0,0)&12.6&-&12.72&13.4*&-&.99&-&.92*\\
(0,0)&15.48&-&15.92*&?&-&.93*&-&?\\
(0,0)&17.88&-&18.85&?&-&.61&-&?\\
(0,0)&18.32&-&19.7*&?&-&.63*&-&?\\
(0,1)&16.66&17.5&17.55&?&.77&.76&?&?\\
(0,1)&18.79&20.75&20.78*&?&.53&.52&?&?\\
(0,1)&19.11&20.38*&20.41&?&.54*&.53*&?&?\\
(0,2)&14.89&15.27&15.29&17.06&.9&.95&.75*&.67*\\
(0,2)&17.41&18.47&18.49&?&.72&.79&?&?\\
(0,2)&17.79&18.94&18.98*&?&.66&.71*&?&?\\
(0,2)&18.7&20.8&20.83*&?&.6&.63*&?&?\\
(0,3)&16.1&16.75&16.8&?&.86&.9&?&?\\
(0,3)&18.77&20.9*&21.02&?&.5*&.66&?&?\\
(1,1/2)&14.48&15.67&15.62&?&.6&.65&?&?\\
(1,1/2)&15.52&16.93*&16.72&?&.52&.75&?&?\\
(1,3/2)&13.39&13.86&13.97&14.96&.83*&.88*&.2*&.18\\
(1,3/2)&13.96&14.97&15.05&?&.73*&.8*&?&?\\
(1,3/2)&14.83&15.81&16.14&?&.71*&.71*&?&?\\
(1,5/2)&16.13&17.61&17.94&?&.49&.57*&?&?\\
(2,0)&8.8&-&9.84&12.74&-&.83&-&.58\\
(2,1)&10.97&13.&12.93&?&.62&.71*&?&?\\
(2,2)&10.43&12.68&12.72&?&.59*&.59&?&?
\end{tabular}
\caption{Energies and supersymmetry fractions in identified supermultiplets. Fractions are divided by the value they should take in continuum. The closer the normalized supersymmetry fraction is to one, the higher is the quality of identification of a supermultiplet. Stars mean that the fraction is summed over two states and energies are averaged. Question marks stand for unidentified states and corresponding supersymmetry fractions in a supermultiplet. Dashes stand for states which do not appear in a supermultiplet.}
\label{tab:spectroscopy}
\end{table}

\chapter{Wavefunctions}\label{ch:wavefunctions}
This chapter is dedicated to wavefunctions in coordinate representation for $SU(2)$ and $SU(3)$ gauge groups. The wavefunctions of supersymmetric Yang--Mills quantum mechanics in two dimensions are studied in great detail. The complete solution of the model, i.e. explicit formulae for wavefunctions and corresponding energies was given for $SU(2)$ gauge group \cite{Halpern}. Later, wavefunctions of the vacuum were found for general $SU(N)$ \cite{Samuel}. For $D=4$ only s--waves of $SU(2)$ were analyzed so far \cite{vanBaal}. In this thesis a new, more general approach is proposed. In principle, it can be used to study wavefunctions with arbitrary angular momentum and for any $SU(N)$ gauge group. It is used here to analyze the cases of $SU(2)$ and $SU(3)$.

We will now briefly review the approach of \cite{vanBaal}. It is a numerical technique to obtain energies and eigenstates of s--waves in $D=4$ for $SU(2)$. The idea of this method applied to the two--fermion sector is following. There is one symmetric and one antisymmetric combination of spinors that create two fermions. Each of these two can be contracted in three different ways to give a gauge and rotationally invariant operator. These six operators acting on the Fock vacuum generate a basis $\ket{e_m}$. A most general gauge invariant state with $j=0$ is a combination of these basis states, where the coefficients are functions $h_m(r,u,v)$ of three bosonic variables:
\begin{align}
\ket{\psi}=\sum_{m=1}^6h_m(r,u,v)\ket{e_m}.
\end{align}
The three invariants $(r,u,v)$ are discussed later in this chapter. The Schr\"odinger equation $H\ket{\psi}=E\ket{\psi}$ is a system of six coupled differential equations for functions $h_m$ in a three dimensional space. There exists an 'angular momentum' $L$ related to a differential operator in variables $(r,u,v)$. Eigenstates of the angular momentum are 6--component 'spherical harmonics'. They are functions of two angle variables $u$ and $v$. The spherical harmonics can be found analytically. Finally, one can introduce a basis for the radial part of functions with given $L$. There is a natural cutoff in this complete set of functions. It is equivalent to the cutoff for the number of bosonic excitations which is used in this paper. With this method one can reach $N_B=60$ \cite{Kotanski}. A limitation of this approach is that it addresses rotationally invariant states. In particular, it gives no information about sectors with odd number of fermions.

Here we propose a new approach. One starts with an angular momentum multiplet expressed in terms of composite bricks acting on the empty state. States in the Fock space are in one to one correspondence with wavefunctions in the configuration space. Because of fermions, the wavefunctions are multi--component. It is shown how to construct a rotationally invariant single--component function $\rho(x)$. This functions is an analogue to probability density of bosonic variables. Next, the structure of the potential is studied in a fixed gauge and some characteristic directions in the configuration space are chosen. These are the flat valley (i.e. a direction in which the potential vanishes) and two steep directions, where the potential grows fastest. Finally, sections of wavefunctions along these directions are studied.

The main purpose of analyzing the wavefunctions is exploring the difference between states corresponding to discrete and continuous spectra. In \ref{subsec:bound_and_scattering_states} we discuss the expected behavior of bound and scattering states. We argue that the localized states are well approximated by the cut Fock basis already for finite cutoff. On the other hand, the scattering states are non--localized and they can be reconstructed only when one goes to high $N_B$.

There are two ways to study scattering states. First is to inspect the limit of constant energy $E=E_*=const$ while $N_B\to\infty$. It was mentioned in Chapter \ref{ch:spectrum} that the eigenenergies corresponding to the continuous spectrum decrease slowly to $0$ as one increases $N_B$. Therefore, at every few values of the cutoff there is a state with energy $E\approx E_*$. Such limit was studied in \cite{Kotanski} for $SU(2)$. It was observed that a periodic function is recovered as the cutoff grows. When $N_B$ increases, new peaks are reconstructed. Period of the function agrees with the dispersion relation for energy $E_*$. For $SU(2)$ the constant energy analysis requires increasing cutoff by about $10$ at each step.

Another approach to address scattering states is to fix the energy level (e.g. the lowest state) and raise $N_B$ one by one. Then the energy converges to zero. One does not recover a periodic function. Instead, the probability density is smeared over an increasing range. Explanation of this behavior is given in detail later. This type of analysis is easier from the numerical point of view, because one does not need large cutoff.

\begin{figure}
\centering
\includegraphics[width=.6\textwidth]{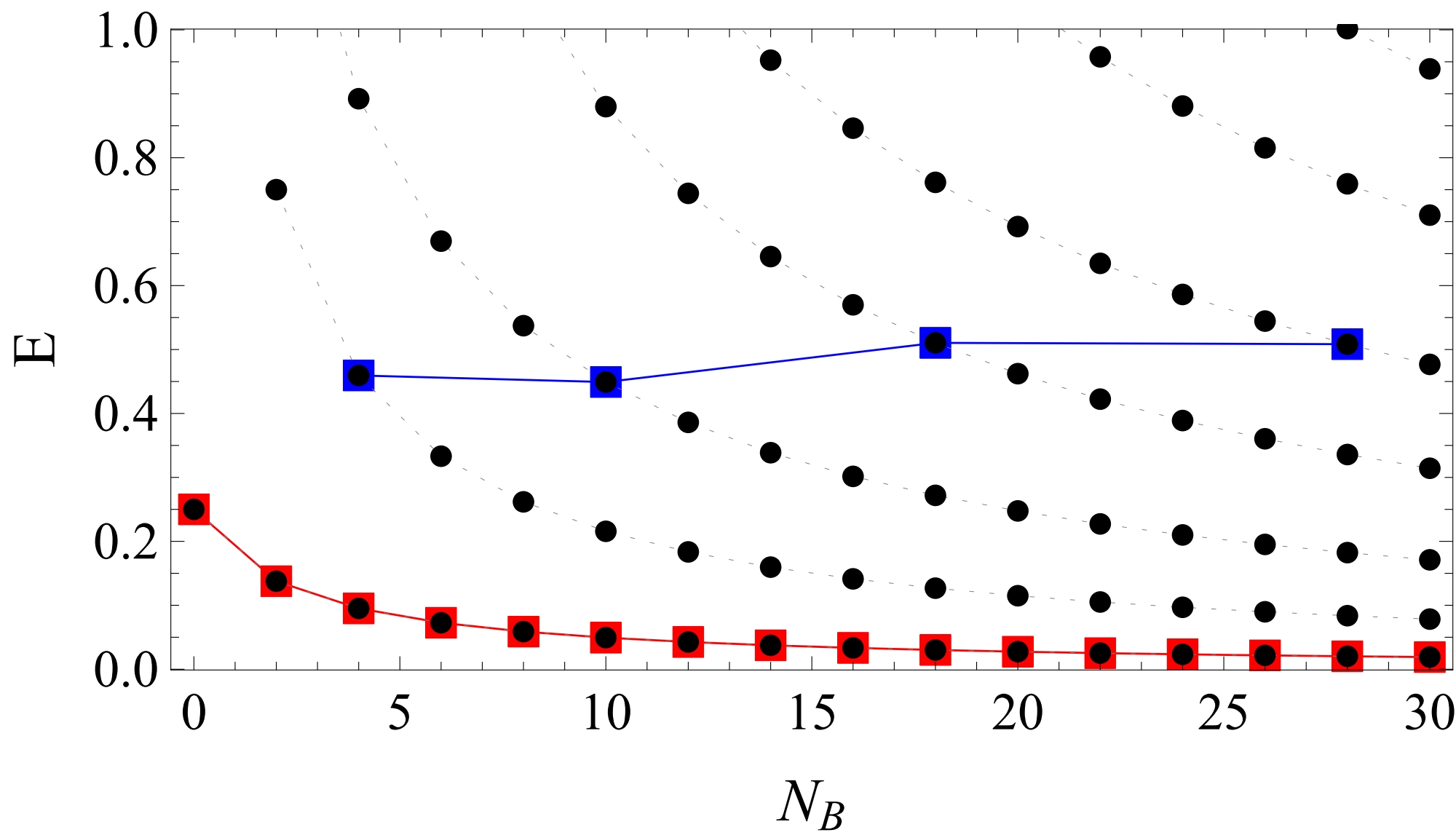}
\caption{Two ways to analyze non--localized energy eigenstates. Circles are energies of the free one--dimensional quantum mechanics. Blue line represents fixed energy limit with $E_*= 0.5$. Red line corresponds to the ground states for each $N_B$.}
\label{fig:energy_analysis}
\end{figure}

\section{Structure of fermionic wavefunctions}
In Chapter \ref{ch:Fock_space_method} it was shown how to obtain eigenstates of the Hamiltonian. These eigenstates are expressed in terms of creation operators acting on a Fock vacuum. In this part of the theses we study their wavefunctions in the configuration space.

Let $\boldsymbol n=(n^a_i)$ and $\boldsymbol \chi=(\chi^b_\alpha)$ denote bosonic and fermionic occupation numbers respectively. $n^a_i$ take integer, nonnegative values while $\chi^b_\alpha$ are $0$ or $1$. As before, $a,b=1,\ldots,N^2-1$ are color indices, $i=1,2,3$ is the spatial index and $\alpha=\pm\frac{1}{2}$ is the Weyl spinor index. Labels $\boldsymbol n$ and $\boldsymbol \chi$ have $3(N^2-1)$ and $2(N^2-1)$ components respectively.  The basis of the Fock space with no restriction to gauge--singlet states is
\begin{align}
\ket{\boldsymbol n, \boldsymbol \chi}=\ket{\boldsymbol n}_B\otimes\ket{\boldsymbol \chi}_F=\prod_{b,\alpha}(f^{b\dagger}_\alpha)^{\chi^b_\alpha}\prod_{a,i}\frac{1}{\sqrt{(n^a_i)!}}(a^{a\dagger}_i)^{n^a_i}\ket{0}.
\end{align}
These states satisfy relations
\begin{align}
a^{a\dagger}_ia^{a}_i\ket{\boldsymbol n, \boldsymbol \chi}&=n^a_i\ket{\boldsymbol n, \boldsymbol \chi},
&f^{b\dagger}_\alpha f^{b}_\alpha\ket{\boldsymbol n, \boldsymbol \chi}&=\chi^b_\alpha\ket{\boldsymbol n, \boldsymbol \chi},
\end{align}
where no sum over indices is performed. Bosonic states $\ket{\boldsymbol n}_B$, bosonic creation and annihilation operators $a^{a\dagger}_i$, $a^{a}_i$ and the Fock vacuum $\ket{0}$ depend on the parameter $\omega$ introduced in \ref{sec:omega}. In this chapter the index $\omega$ will remain implicit.

The coordinates in the bosonic $3(N^2-1)$--dimensional configuration space are $\boldsymbol x=(x^a_i)$. A ket $\ket{\boldsymbol x}_B$ is defined by the relation
\begin{align}
\hat{x}^a_i\ket{\boldsymbol x}_B=x^a_i\ket{\boldsymbol x}_B,
\end{align}
where $\hat{x}^a_i$ is the position operator. Wavefunction of bosonic states $\ket{\boldsymbol n}_B$ are simply wavefunctions of eigenstates of the harmonic oscillator
\begin{align}
\prescript{}{B}{\braket{\boldsymbol x|\boldsymbol n}}_B=\prod_{a,i}\sqrt\frac{\omega}{2^{n^a_i}(n^a_i)!}\pi^{-1/4}exp(-(\omega x^a_i)^2/2)H_{n^a_i}(\omega x^a_i),
\end{align}
where $H_n(x)$ is the Hermite polynomial.

In order to handle the fermionic states, a representation of $\ket{\boldsymbol\chi}_F$ is needed. In what follows we incorporate the matrix representation of fermions \cite{Salomonson}. In this notation fermionic creation and annihilation operators are represented by matrices and wavefunctions are columns. For the simplest case of one fermionic type there is a vacuum state $\ket{0}_F$ and a state with one fermion $\ket{1}_F$. The states and operators are represented as follows:
\begin{align}
\ket{0}_F&=\left(\begin{array}{c}0\\1\end{array}\right),&f^\dagger&=\left(\begin{array}{cc}0&1\\0&0\end{array}\right),\\
\ket{1}_F&=\left(\begin{array}{c}1\\0\end{array}\right),&f&=\left(\begin{array}{cc}0&0\\1&0\end{array}\right).
\end{align}
Then the action of creation and annihilation operators is
\begin{align}
f^\dagger\ket{0}_F&=\ket{1}_F,&f\ket{1}_F&=\ket{0}_F,&f\ket{0}_F=f^\dagger\ket{1}_F&=0.
\end{align}
A wavefunction of a general state $\ket{\phi}$ is a two--component function
\begin{align}
\phi(x)&=\left(\begin{array}{c}\phi_1(x)\\\phi_0(x)\end{array}\right).
\end{align}
Wavefunctions associated with the zero- and one-fermion part are given by scalar products
\begin{align}
\phi_\chi(x)&=\braket{x,\chi|\phi},& \chi=0,1.
\end{align}

This representation naturally generalizes to a larger number of fermionic types. Consider again SYMQM. The label $\boldsymbol\chi$ has $2(N^2-1)$ components, each taking values $0$ or $1$. Therefore, $\boldsymbol\chi$ has $2^{2(N^2-1)}$ possible values and the wavefunction is a column of size $2^{2(N^2-1)}$. Elements of the column are bosonic wavefunctions labeled by $\boldsymbol \chi$.

Consider a general state $\ket{\phi}$. Its expansion in basis states is given by
\begin{align}
\ket{\phi}&=\sum_{\boldsymbol n,\boldsymbol\chi}c(\boldsymbol n,\boldsymbol \chi)\ket{\boldsymbol n,\boldsymbol\chi}
\end{align}
with some complex coefficients $c(\boldsymbol n,\boldsymbol\chi)$. Then the $\boldsymbol \chi$--component of the wavefunction of $\ket{\phi}$ is the scalar product
\begin{align}
\phi_{\boldsymbol \chi}(\boldsymbol x)&\equiv \braket{\boldsymbol x,\boldsymbol \chi|\phi}=\sum_{\boldsymbol n^a_i}c(\boldsymbol n,\boldsymbol \chi)\prescript{}{B}{\braket{\boldsymbol x|\boldsymbol n}}_B.
\end{align}

The fact that a wavefunction has many components introduces a major difficulty to analyzing them in a direct way. For $SU(2)$ the column $\left(\phi_{\boldsymbol\chi}(\boldsymbol x)\right)$ has $2^{2\times 3}=64$ and for $SU(3)$ it has $2^{2\times 8}=65536$ elements. Fortunately, we consider only states in sectors with definite number of fermions. Then $\left(\phi_{\boldsymbol\chi}(\boldsymbol x)\right)$ has only $\left(\begin{array}{c}2(N^2-1)\\n_F\end{array}\right)$ elements that are nonzero. Still, the direct study of $\phi_{\boldsymbol\chi}(\boldsymbol x)$ is complicated. For example, consider a multiplet $\ket{\phi^m}$ with angular momentum $j$. Then $\ket{\phi^m}$ is represented by a $(2j+1)\times\binom{2(N^2-1)}{n_F}$--component complex--valued bosonic functions in a $3(N^2-1)$--dimensional bosonic space. For instance, for $SU(3)$, $j=1$ and $n_F=2$ these are 360 functions of 24 variables.

\section{Probability density from fermionic wavefunctions}\label{sec:probability_density}

A wavefunction carries full information about a state. In particular, it depends on fermionic occupation numbers and angular momentum. In order to simplify the wavefunctions, we would like to integrate out fermionic degrees of freedom and angles which correspond to rotations in the physical space. To this end we introduce the probability distribution $\rho(\boldsymbol x)$. Take an angular momentum multiplet $\ket{\phi^m}$. Then the probability density is defined as
\begin{align}\label{eq:square_wavefunction}
\rho(\boldsymbol x)=\sum_{m\boldsymbol \chi}|\phi^m_{\boldsymbol\chi}(\boldsymbol x)|^2,
\end{align}
which sums over fermionic occupation numbers and the magnetic quantum number.  Studying $\rho(\boldsymbol x)$ is easier since it is a single--component real function. Still, it depends on many variables. However, there are some simplifications. It is shown in Appendix \ref{ap:rotation_invariance} that $\rho(\boldsymbol x)$ is invariant under rotations. It is also gauge invariant because the original states $\ket{\phi^m}$ are such. Because of these symmetries $\rho(\boldsymbol x)$ depends only on $3(N^2-1)-(N^2-1)-3=2N^2-5$ invariants of gauge and rotation transformations.

One more step needs to be taken before $\rho(\boldsymbol x)$ can be considered as a probability density integrated over rotations and gauge transformations. Suppose we are interested in calculating the expectation value of an observable $A$ in a state given by a wavefunction $\psi(\boldsymbol x)$ and $A(\boldsymbol x)$ is invariant under rotations and action of gauge group. Rotations are parameterized by $3$ angles $\boldsymbol \theta=(\theta_1,\theta_2,\theta_3)$. Gauge transformations depend on $N^2-1$ angles $\boldsymbol\alpha=(\alpha_1,\ldots,\alpha_{N^2-1})$. Then there are $2N^2-5$ variables left. They will be denoted by $\boldsymbol y=(y_1,\ldots,y_{2N^2-5})$. $A$ depends only on $\boldsymbol y$. The mean value of $A$ is
\begin{align}
\begin{split}
\braket{\psi|A|\psi}&=\int d\boldsymbol x A(\boldsymbol x)\left|\psi(\boldsymbol x)\right|^2=\int d\boldsymbol{y} d\boldsymbol\alpha d\boldsymbol\theta A(\boldsymbol\alpha,\boldsymbol\theta,\boldsymbol{y})\left|\psi(\boldsymbol\alpha,\boldsymbol\theta,\boldsymbol{y})\right|^2|J(\boldsymbol\alpha,\boldsymbol\theta,\boldsymbol{y})|\\
&=\int d\boldsymbol{y} A(\boldsymbol{y})\left|\psi(\boldsymbol{y})\right|^2\int d\boldsymbol\alpha d\boldsymbol\theta |J(\boldsymbol{y},\boldsymbol\alpha,\boldsymbol\theta)|
=\int d\boldsymbol{y} A(\boldsymbol{y})\left|\psi(\boldsymbol{y})\right|^2 J(\boldsymbol{y}),
\end{split}
\end{align}
where $J$ is the Jacobian. The integral was reduced from $3(N^2-1)$ to $2N^2-5$ dimensional. $N^2+2$ variables were integrated out, so the Jacobian $J(\boldsymbol y)$ is proportional to $r^{N^2+2}$. In order to take the Jacobian into account, the function $\rho(\boldsymbol x)$ will be multiplied by $r^{N^2+2}$.

Concluding, $r^{N^2+2}\rho(\boldsymbol x)$ is the probability density, summed over fermionic degrees of freedom and integrated over angles which correspond to rotations and gauge transformations. The integration over rotational angles is performed by summing over members of the angular momentum multiplet. In order to further simplify the analysis we will choose some interesting directions in the configurations space.

\section{Wavefunctions for SU(2) gauge group}
\subsection{Invariants}
For the $SU(2)$ gauge group a general wavefunction depends on $9$ variables $x^a_i$. It was argued previously that $\rho(\boldsymbol x)$  defined in (\ref{eq:square_wavefunction}) depends on $2N^2-5=3$ gauge and rotation invariants. For $SU(2)$ they can be found explicitly. If $\boldsymbol x=(x^a_i)$ is treated as a $3\times 3$ matrix then it transforms under rotations and $SU(2)$ in the following way:
\begin{align}
\boldsymbol x\rightarrow \boldsymbol x'= U\boldsymbol xV^\dagger
\end{align}
where $U,V$ are both SO(3) matrices. $U$ corresponds to gauge transformation (in adjoint representation) and $V$ to rotation in the configuration space. It is known that one can diagonalize $\boldsymbol x$ with $U$ and $V$ \cite{Goldstone,Savvidy}. We define $\boldsymbol y=diag(y_1,y_2,y_3)$ by
\begin{align}\label{eq:UVdiagonalization}
\boldsymbol x=U\boldsymbol{y}V^\dagger.
\end{align}
The diagonal terms $y_i$ are defined uniquely up to permutations and reflection of two of them. These symmetries give 24--fold degeneracy. We impose additional constraint $y_1\geq y_2\geq |y_3|$ so that the diagonal elements are uniquely defined. Under this constraint $y_i$ are proper invariants of gauge transformations and rotations. It convenient to introduce other set of invariants \cite{vanBaal}:
\begin{align}\label{eq:su2_invariants}
\begin{split}
r^2&=y_1^2+y_2^2+y_3^2=\sum_{i,a}(x^a_i)^2,\\
r^3v&=y_1y_2y_3=det(x^a_i),\\
r^4u&=y_1^2y_2^2+y_2^2y_3^2+y_3^2y_1^2=\sum_{a,b,i<j}((x^a_i)^2(x^b_j)^2-x^a_ix^a_jx^b_ix^b_j).
\end{split}
\end{align}
Note that if $r$ vanishes, then the other two variables are undetermined. If $u=0$ then also $v=0$. Moreover, two of these variables are bounded, $0\leq u\leq1/3$, $|v|\leq 3^{-3/2}$. The potential takes the form $V=\frac{1}{2}r^4u$.

We will be interested in two directions in the configuration space, namely the flat valley and the steepest direction. Potential vanishes when $u=0$. Since then also $v$ vanishes, the flat valley is one--dimensional and can be parameterized by $r$. In terms of $y_i$ the potential vanishes when two variables vanish, i.e. only $y_1$ is nonzero.

The potential grows fastest for $u=1/3$ which implies $v=\pm 3^{-3/2}$. There are thus two steep directions and both are one dimensional. In terms of $y_i$ the steep direction is $y_1=y_2=r/\sqrt 3,\ y_3=\pm r/\sqrt 3$.

Although the wavefunction $\rho(\boldsymbol x)$ depends on $\boldsymbol y$, in practice it is given as an explicit function of $\boldsymbol x$ (\ref{eq:UVdiagonalization}). Therefore, we calculate $\rho(\boldsymbol x)$ along certain directions which correspond to values of invariants in the flat and steep directions. The directions are parameterized as $\boldsymbol x=r \boldsymbol{\hat x}$ where $|\boldsymbol{\hat x}|=1$. Versor of the flat direction is denoted by $\boldsymbol{\hat x}_{flat}$. Versors of the two steep directions are $\boldsymbol{\hat x}_{steep}$ and $\boldsymbol{\hat x'}_{steep}$.
\begin{align}\label{eq:directions}
\boldsymbol{\hat x}_{flat}&=\left(\begin{array}{ccc}1&0&0\\0&0&0\\0&0&0\end{array}\right),&
\boldsymbol{\hat x}_{steep}&=\frac{1}{\sqrt3}\left(\begin{array}{ccc}1&0&0\\0&1&0\\0&0&1\end{array}\right),&
\boldsymbol{\hat x}'_{steep}&=\frac{1}{\sqrt3}\left(\begin{array}{ccc}-1&0&0\\0&-1&0\\0&0&-1\end{array}\right).
\end{align}
For $\boldsymbol{\hat x}_{flat}$ and $\boldsymbol{\hat x}_{steep}$ the invariants $y_i$ are the diagonal elements. $\boldsymbol{\hat x}'_{steep}$ corresponds to $y_1=y_2=-y_3=1/\sqrt 3$. Note that $r \boldsymbol{\hat x}'_{steep}=-r\boldsymbol{\hat x}_{steep}$. In the following we will drop $\boldsymbol{\hat x}'_{steep}$ and slightly abuse the notation by taking $r \boldsymbol{\hat x}_{steep}$ with positive and negative $r$. This way, both steep directions are included. This is done merely for the convenience to include both steep directions on one plot.

\subsection{Bound and scattering states -- models and expectations}\label{subsec:bound_and_scattering_states}

Here we recall what is the characteristic behavior of states associated with discrete and continuous spectrum in general quantum mechanical models. It will serve us as a hint to analyze wavefunctions of SYMQM. The states which correspond to discrete spectrum are bound states, i.e. their wavefunctions are localized. That is, the probability density is centered at minimum of the potential and vanishes as one moves away to infinity. Because the basis of the Fock space consists of localized states, the bound states can be well approximated for finite $N_B$.

Bound states are typical for a potential which grows to infinity in all directions or at least $V(x)>V_0$ for $|x|>r_0$ (then there may be discrete energies smaller than $V_0$). In the considered model the potential has flat valleys and neither condition is fulfilled. However, although the potential has a flat valley, its width shrinks as one moves outwards. The transverse oscillations cost more energy as one moves along the valley and an effective potential is generated. The effective potential is unbounded from above and vanishes only at the origin. On the other hand, because of the supersymmetry, interactions of the potential with bosonic degrees of freedom may cancel and then the effective potential vanishes. However, in the purely bosonic channel such cancelations cannot take place simply because there are no fermions. Therefore, the effective potential is present for $n_F=0$ and the spectrum is discrete. Surprisingly, the effective potential does not vanish also for nonzero but small $n_F$. Only in sectors rich in fermions the supersymmetric cancelations are possible.

For $n_F$ large enough the continuous spectrum appears. The continuous spectrum is associated with scattering states. They are not localized and non--normalizable. The scattering states are concentrated in regions where the potential is flat. The probability density does not decay for $|x|\to\infty$. A scattering state cannot be well approximated for finite $N_B$ because the basis of cut Fock space consists of localized functions. Instead, such state is reconstructed step by step as the cutoff grows.

In order to better understand what is the behavior of scattering states, consider a three dimensional, quantum mechanical, free Hamiltonian $H=\frac{1}{2}\sum_{i=1}^3p_i^2$. This simple system can be solved analytically. Its spectrum in all channels with definite angular momentum is continuous and ranges from zero to infinity. For vanishing angular momentum the wavefunctions are just the spherical waves
\begin{align}\label{eq:0_momentum}
\braket{\vec x|E}=\sqrt\frac{2p}{\pi}\frac{\sin(p r)}{r},
\end{align}
where $p=\sqrt{2E}$ \cite{Sakurai}.

One way to analyze scattering states is to fix $E$ and observe how the functions approach the spherical wave. However, as it was mentioned in the introduction to this chapter, states with given $E$ appear at every few $N_B$. The cutoff that we are able to explore is too small to perform this analysis. Instead, we analyze the ground state and thus take the limit $E\to0$.

For $E\to 0$ the wavefunction (\ref{eq:0_momentum}) converges to a flat function. With the cut Fock space method one does not exactly recover a ground state of the theory, but rather an approximation which improves for growing cutoff. Therefore, one should expect that the ground state will be approximated by a roughly flat function. Because $N_B$ effectively gives an infrared cutoff (cf. Chapter \ref{ch:Fock_space_method}), the flat function is smeared over a finite region which grows with $N_B$.

\begin{figure}[H]\centering
\begin{subfigure}[b]{0.48\textwidth}\includegraphics[width=\textwidth]{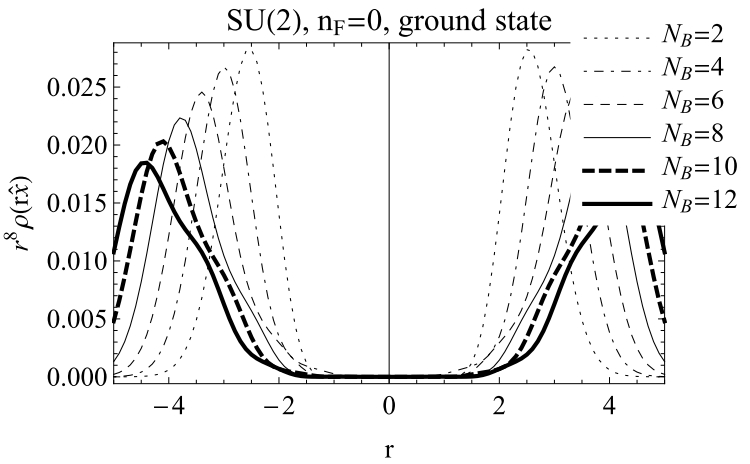}\end{subfigure}
\begin{subfigure}[b]{0.48\textwidth}\includegraphics[width=\textwidth]{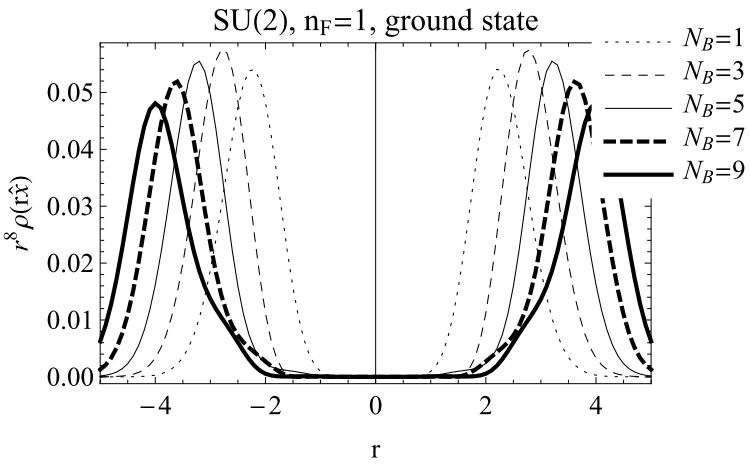}\end{subfigure}
\caption{Wavefunctions of lowest eigenstates for free $SU(2)$ with $n_F=0$ and $n_F=1$. As there is no potential, the wavefunctions depend only on the radius $r$. The angular momentum for $n_F=0$ is $j=0$ and $j=1/2$ for $n_F=1$. Wavefunction with $n_F=2$ is exactly the same as in the sector with $n_F=0$. It is obvious since the free Hamiltonian has no fermionic terms.
}\label{fig:su2_free}
\end{figure}

For comparison, consider the free Yang--Mills quantum mechanics with $SU(2)$ symmetry $H=Tr(p_ip_i)$. Because of symmetry of the Hamiltonian, the wavefunctions should depend only on one invariant $r$. Therefore, instead of $r^{N^2+2}=r^6$ we multiply the probability density $\rho(\boldsymbol x)$ by $r^8$. The function $r^8\rho(\boldsymbol x)$ is then the probability density integrated over all variables apart from the radius. Wavefunctions of ground state in channels with $n_F=0,1$ for finite cutoffs are shown in Fig. \ref{fig:su2_free}. One can see that the wavefunction move outside as the cutoff grows. This is in agreement with what was expected.

\subsection{Interacting case -- results}
\begin{figure}[H]\centering
\begin{subfigure}[b]{0.48\textwidth}\includegraphics[width=\textwidth]{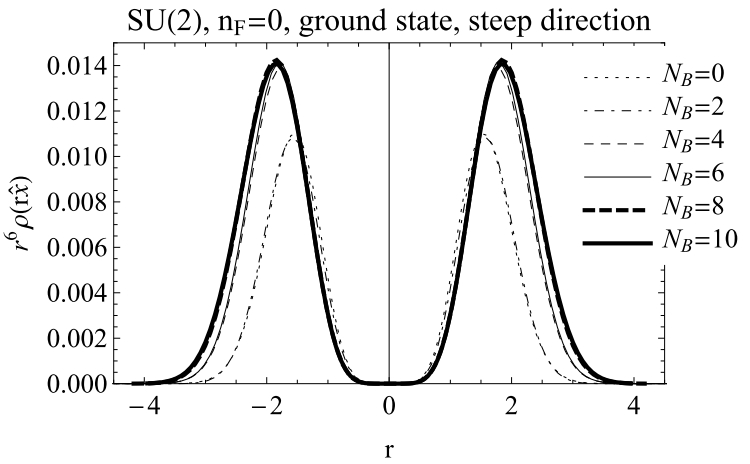}\end{subfigure}
\begin{subfigure}[b]{0.48\textwidth}\includegraphics[width=\textwidth]{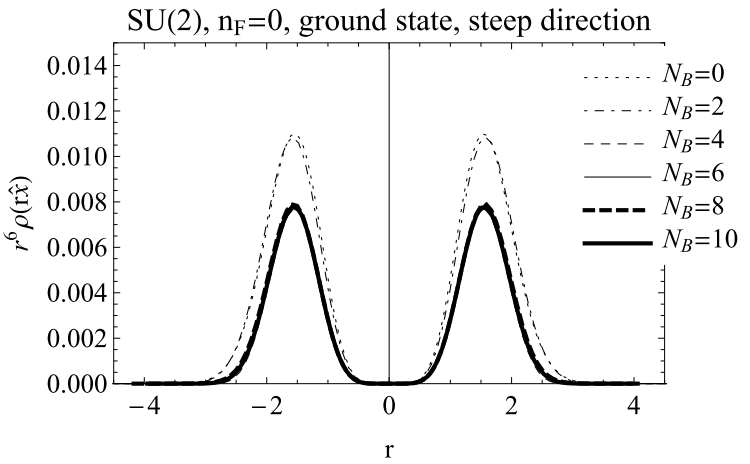}\end{subfigure}
\begin{subfigure}[b]{0.48\textwidth}\includegraphics[width=\textwidth]{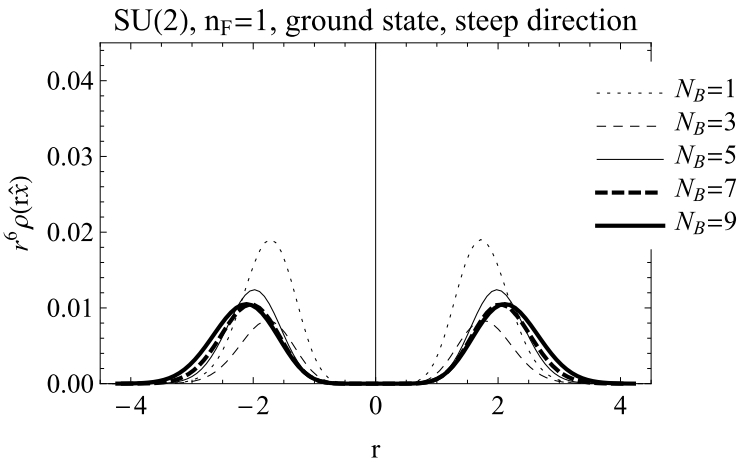}\end{subfigure}
\begin{subfigure}[b]{0.48\textwidth}\includegraphics[width=\textwidth]{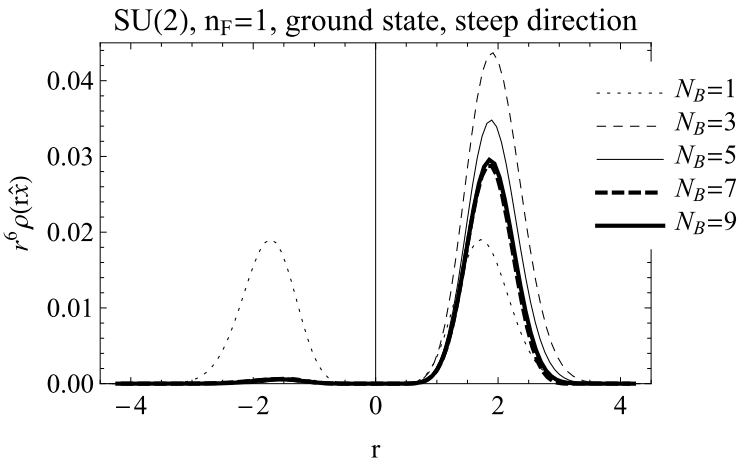}\end{subfigure}
\begin{subfigure}[b]{0.48\textwidth}\includegraphics[width=\textwidth]{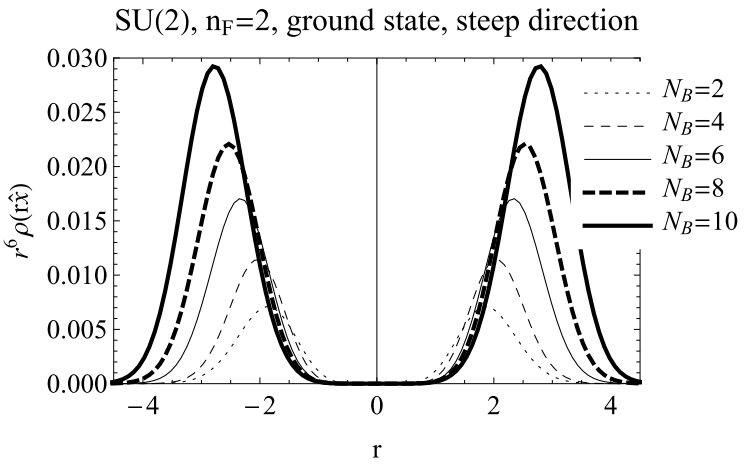}\end{subfigure}
\begin{subfigure}[b]{0.48\textwidth}\includegraphics[width=\textwidth]{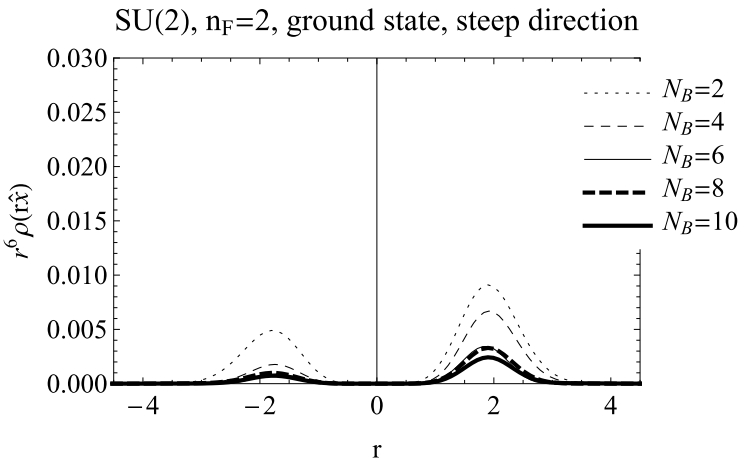}\end{subfigure}
\begin{subfigure}[b]{0.48\textwidth}\includegraphics[width=\textwidth]{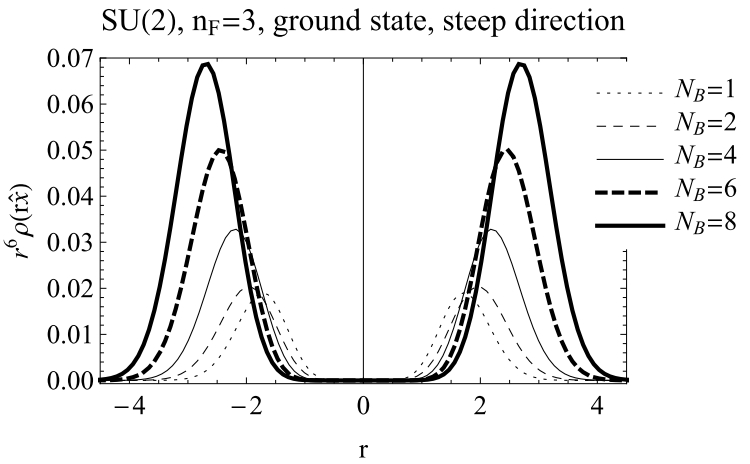}\end{subfigure}
\begin{subfigure}[b]{0.48\textwidth}\includegraphics[width=\textwidth]{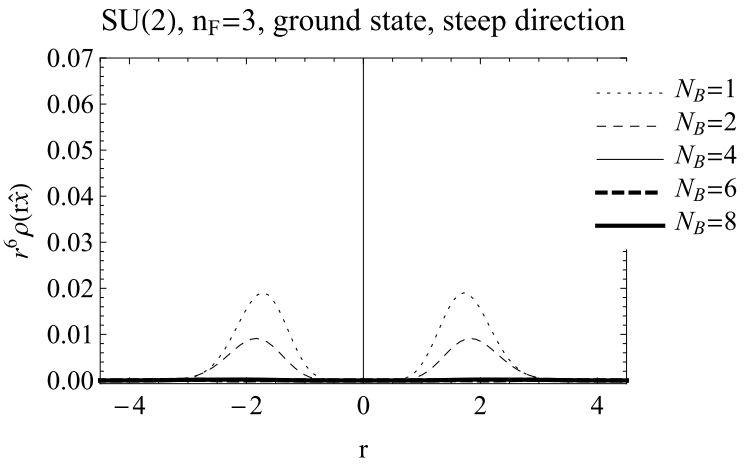}\end{subfigure}
\caption{Plots of wavefunctions of ground states in sectors $n_F=0,1,2,3$. The angular momentum is $j=0$ for $n_F=0,2$ and $j=1/2$ for $n_F=1,3$. The versor is $\hat x =\boldsymbol{\hat x}_{flat}$ for the flat direction and $\hat x=\boldsymbol{\hat x}_{steep}$ for the steep direction.
}\label{fig:su2_full}
\end{figure}

Finally, consider the interacting case. Profiles of wavefunctions with $n_F=0,1,2,3$ are presented in Fig. \ref{fig:su2_full}. They correspond to the ground states, which for $n_F=0,2$ and $n_F=1,3$ are in channels with $j=0$ and $j=1/2$ respectively.

For $n_F=0$ the wavefunction converges to a fixed shape. In fact, the wavefunction moves a little deeper into the flat valley. Convergence is seen for $N_B\approx4$. In the steep direction the wavefunction initially decreases with $N_B$ and finally converges for $N_B\approx4$. We shall also note that the values of wavefunction are of the same order inside and outside of the valley. The wavefunction in the steep direction is the same for positive and negative $r$. This is because for $n_F=0$ the parity is conserved. The two steep directions are related by the parity transformation. The behavior of wavefunction in $n_F=1$ sector is essentially the same. The wavefunction converges to a fixed shape although the rate of convergence is slower. For $n_F>0$ the parity symmetry is broken wavefunctions are different for $r>0$ and $r<0$.

Channels $n_F=2,3$ correspond to continuous spectrum and thus the behavior of wavefunctions is very different. In both cases, the probability density moves deeper into the flat valleys as the cutoff grows. In turn, in the steep directions the wavefunctions get significantly smaller. That is, the probability density concentrates along the flat valleys.

\begin{figure}[H]\centering
\begin{subfigure}[b]{0.48\textwidth}\includegraphics[width=\textwidth]{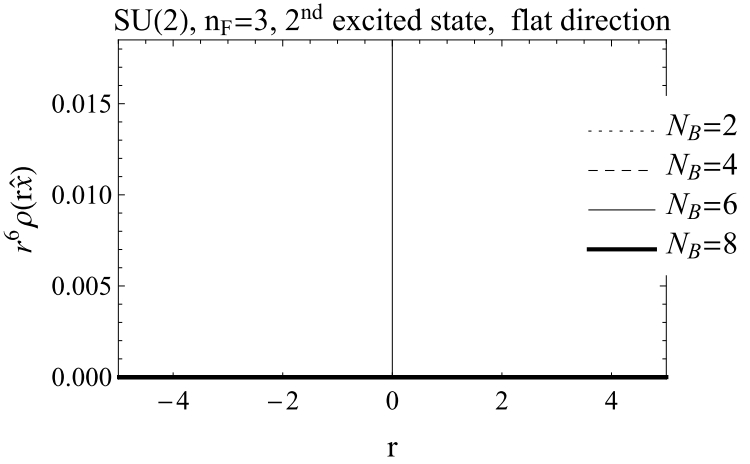}\end{subfigure}
\begin{subfigure}[b]{0.48\textwidth}\includegraphics[width=\textwidth]{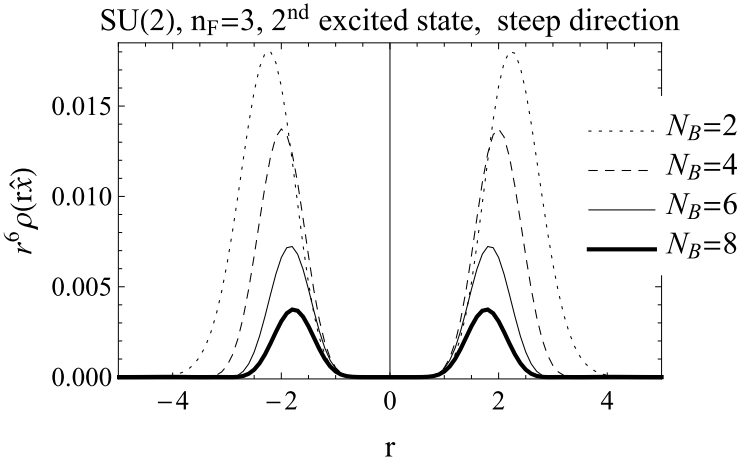}\end{subfigure}
\caption{Plots of wavefunctions of the second excited state in sector with three fermions and $j=1/2$.}
\label{fig:su2_bound}
\end{figure}

As it was observed in Chapter \ref{ch:spectrum}, there are discrete energies immersed in the continuous spectrum for $n_F=2,3$. The two plots in Fig. \ref{fig:su2_bound} show the second excited state of $(n_F,j)=(3,1/2)$. We see that the wavefunction vanishes along the whole the valley. Indeed, values of the wavefunction are of the order of numerical errors. In the steep direction the function decreases and approaches the origin as the cutoff grows. This behavior may be considered strange. However, this is merely a consequence of looking only along the steep and flat directions. A combined plot gives a more complete picture. First, we introduce another flat direction
\begin{align}
\boldsymbol {\hat x'}_{flat}=
\left(\begin{array}{ccc}
0&1&0\\0&0&0\\0&0&0
\end{array}\right).
\end{align}
It can be mapped into the flat direction $\boldsymbol {\hat x}_{flat}$ defined in (\ref{eq:directions}) with gauge transformations and rotations. On the other hand, it is orthogonal to $\boldsymbol {\hat x}_{steep}$. One can see that the probability density is concentrated around the origin with some preference of the flat direction. This is similar to the behavior of ground state in the purely bosonic sector (cf. Fig. \ref{fig:su2_full}). The difference is that the wavefunction vanishes along the bottom of the valley.

In conclusion, lowest energy states with $n_F=0,1$ fermions are localized near the origin and thus they are bound states. Ground states in channels with $n_F=2,3$ fermions penetrate the valley and are scattering states. There are also bound states for $n_F=2,3$. These observations are in agreement with conclusions of Chapter \ref{ch:spectrum} where the discrete spectrum in $n_F=0,1$ sectors and continuous spectrum for $n_F=2,3$ were identified.

\begin{figure}[H]\centering
\begin{subfigure}[b]{0.48\textwidth}\includegraphics[width=\textwidth]{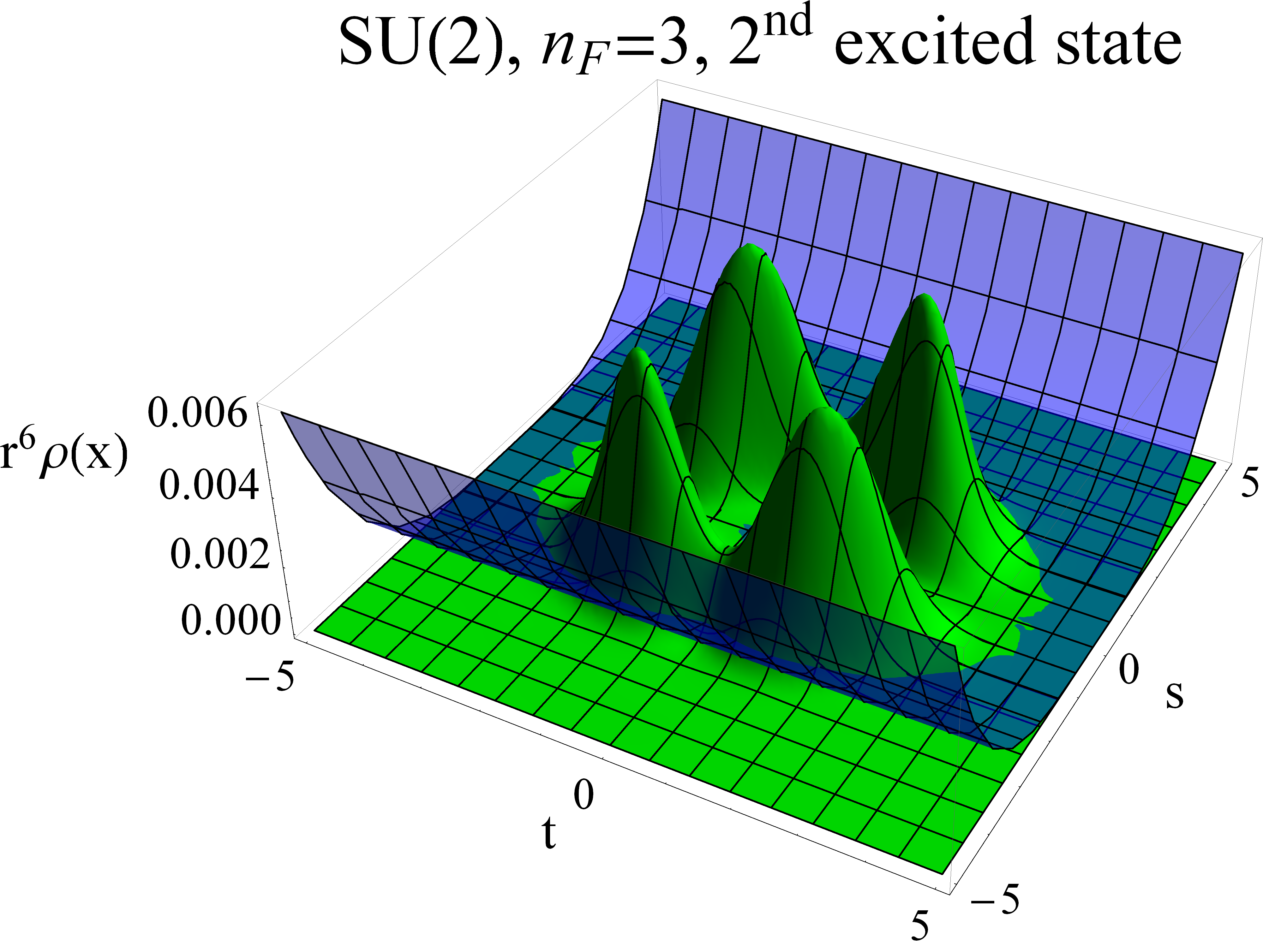}\end{subfigure}
\caption{Second excited states for $n_F=3$, $j=1/2$, $N_B=8$. The parametrization is $\boldsymbol x=t \boldsymbol{\hat x'}_{flat} + s\boldsymbol{\hat x}_{steep}$. The blue surface represents the potential.}
\label{fig:su2_full_bound}
\end{figure}

\section{Wavefunctions for SU(3) gauge group}
Now we analyze wavefunctions for the $SU(3)$ gauge group. There are 24 bosonic degrees of freedom $x_i^a$. Equivalently, they are expressed in terms of three hermitean traceless $3\times3$ matrices $X_i=x_i^aT^a$. The gauge transformation of $X_i$ is $X_i\rightarrow UX_iU^\dagger$, where $U\in SU(3)$.
The potential and wavefunctions are gauge invariant. Therefore, analyzing any of them along the orbits of gauge transformations gives no new information about dynamics of the system. One would like to eliminate these directions. This can be achieved in two ways. One is to find gauge invariants, as is was done for $SU(2)$. However, these are difficult to find for $SU(3)$. Another way is to fix the gauge. This procedure is straightforward and eliminates precisely $N^2-1=8$ degrees of freedom of $X_i$.

\subsection{Fixing the gauge}\label{sec:fixing_gauge}
The procedure of gauge fixing consists of several steps. One starts with a general $U$ in the fundamental representation of $SU(3)$ and a general set of three traceless, hermitean matrices $X_i$. In the first step $U$ is used to diagonalize $X_1$. This done, there is a residual gauge transformation that leaves $X_1$ unchanged. In following steps the gauge transformations are used to impose additional constraints on $X_2$ and $X_3$. The procedure continues until no more constraints on $X_i$ can be introduced.

Consider first a generic case. The gauge transformation $U$ can be chosen such that $X_1$ is diagonalized and the diagonal elements are in the increasing order. Once $X_1$ is diagonalized, the gauge is fixed up to transformations with $U=diag(e^{i\phi},e^{i\psi},e^{-i\phi-i\psi})$. Now the question is how this gauge freedom can be used to fix matrix elements of $X_2$. First, consider a general matrix $M=(m_{ij})$. The transformation applied to $M$ is
\begin{align}
UMU^\dagger=\left(\begin{array}{ccc}m_{11}&e^{i\phi-i\psi}m_{12}&e^{2i\phi+i\psi}m_{13}\\e^{-i\phi+i\psi}m_{21}&m_{22}&e^{i\phi+2i\psi}m_{23}\\e^{-2i\phi-i\psi}m_{31}&e^{-i\phi-2i\psi}m_{32}&m_{33}\end{array}\right).
\end{align}
One can see that $U$ can be used to set the off--diagonal matrix elements $(X_2)_{12}$ and $(X_2)_{23}$ to be real and positive. In the generic case this additional condition fixes the gauge completely. The conditions on matrices $X_i$ can be translated to constraints on coordinates $x^a_i$:
\begin{align}\label{eq:gauge_constraints}\begin{split}
& x_1^1=x_1^2=x_1^4=x_1^5=x_1^6=x_1^7=x_2^2=x_2^5=0,\\
&x_1^8\leq\frac{\sqrt 3}{2}x_1^3\leq0.
\end{split}\end{align}
The space of configurations in the fixed gauge is 16--dimensional.

There are many nongeneric cases in which the procedure presented above does not completely fix the gauge. For example, if $X_1=0$ then after the first step the residual gauge group is the full $SU(3)$. It is then used to diagonalize $X_2$ and set $(X_3)_{12}$ and $(X_3)_{23}$ positive.

Other special case is when $X_1$ is not zero, but the two lowest eigenvalues are equal. Then the remaining gauge freedom after the first step is
\begin{align}
U=\left(\begin{array}{ccc}\multicolumn{2}{c}{\multirow{2}{*}{$U_2$}}&0\\\multicolumn{2}{c}{}&0\\0&0&(\det U_2)^*\end{array}\right),
\end{align}
where $U_2$ is a general $U(2)$ matrix. Then it is used to diagonalize the upper--left $2\times 2$ block of $X_2$ with increasing order of eigenvalues. There still remains a gauge freedom with $U=diag(e^{i\phi},e^{i\psi},e^{-i\phi-i\psi})$. It is utilized to set the coefficients $(X_2)_{13}$ and $(X_2)_{23}$ positive.

A third degenerate case is when all three eigenvalues of $X_1$ are different but $(X_2)_{12}=0$. Then after fixing $(X_2)_{23}>0$ there is still a freedom of one phase in $U$. It is used to set $(X_2)_{13}$ to be positive.

These are not all special cases. Moreover, all of these three contain some degenerate 'subcases'. Therefore, some care has to be taken in order to completely fix the gauge. The complete procedure is presented in Appendix \ref{ap:fixing_gauge}. It is chosen in such a way that the conditions (\ref{eq:gauge_constraints}) are always fulfilled after the final step of gauge fixing.

One may question the necessity to completely fix the gauge rather then to restrict the gauge fixing to (\ref{eq:gauge_constraints}). However, when the gauge is only partially fixed, some gauge freedom still remains. This leads to 'false' flat valleys as will be shown later.

\subsection{Flat and steep directions}\label{sec:su3_directions}
After the gauge is fixed, a wavefunction depends on 16 coordinates, which are encoded in matrices $X_i$. This is still too many to analyze behavior of the wavefunctions. Again, we consider flat and steep directions. The potential is given by
\begin{align}
V=-\sum_{i<j}Tr([X_i,X_j]^2)=-\sum_{i<j}Tr([X_i,X_j]^2).
\end{align}
$V$ is nonnegative and has flat valleys. Indeed, each $i[X_i,X_j]$ is hermitean, so $-[X_i, X_j]^2$ is semipositive definite and so is the trace. The potential vanishes only if $[X_i,X_j]=0$ for all $i,j$. These equations together with constraints (\ref{eq:gauge_constraints}) give six different solutions. There is one six--parameter solution
\begin{align}\label{eq:valley}
 X_i^{val}= x_i^3 T^3+ x_i^8 T^8.
\end{align}
The other five solutions do not satisfy additional constraints which follow from fixing the gauge. More precisely, although they satisfy (\ref{eq:gauge_constraints}), they fall into the special cases mentioned before. The special cases impose additional constraints which are not fulfilled by these solutions. An example of such 'false valley' is
\begin{align}
X_1&=0,\nonumber\\
X_2&=\frac{1}{2}\left(\begin{array}{ccc}x^3_2+\frac{1}{\sqrt 3}x^8_2&x^1_2&0\\ x^1_2&-x^3_2+\frac{1}{\sqrt 3}x^8_2&0\\0&0&-\frac{2}{\sqrt 3}x^8_2\end{array}\right),\label{eq:false_valley}\\
X_3&=\frac{1}{2}\left(\begin{array}{ccc}x^3_3+\frac{1}{\sqrt 3}x^8_3&x^1_2x^1_3/x^3_2&0\\x^1_2x^1_3/x^3_2&-x^3_3+\frac{1}{\sqrt 3}x^8_3&0\\0&0&-\frac{2}{\sqrt 3}x^8_3\end{array}\right).\nonumber
\end{align}
One can easily check that it satisfies (\ref{eq:gauge_constraints}) and the potential vanishes on this configuration. However, if $X_1=0$ then the gauge symmetry is used to diagonalize $X_2$. The matrix $X_2$ in (\ref{eq:false_valley}) is not diagonal and thus this set of matrices does not fulfill all gauge constraints.

The probability density $\rho(\boldsymbol x)$ is invariant under rotations. Therefore, we would like to construct rotational invariants from (\ref{eq:valley}). The two vectors $\vec v_3=( x_1^3, x_2^3, x_3^3)$ and $\vec v_8=( x_1^8, x_2^8, x_3^8)$ transform under rotations like ordinary vectors. There are three rotation invariants, namely $\vec v_3^2$, $\vec v_8^2$, $\vec v_3\cdot \vec v_8$. They can be rewritten in terms of other parameters which will be more useful for later purposes:
\begin{align}\label{eq:valley_parametrization}\begin{split}
r^2&=\vec v_3^2+\vec v_8^2,\\
\tan\theta&=|\vec v_8|/|\vec v_3|,\\
\cos\phi&=\frac{\vec v_3\cdot \vec v_8}{|\vec v_8| |\vec v_3|}.
\end{split}\end{align}
This gives us the complete description of the flat valley.

Second interesting direction is the one in which the potential grows fastest. The scaling of the potential is known, $V(r\boldsymbol X)=r^4 V(\boldsymbol X)$. It is therefore enough to find such configuration $\boldsymbol X$, normalized by the condition $\sum_i Tr(X_i^2)=1$, which maximizes $V(\boldsymbol X)$. First observe that we can assume $Tr(X_i^2)=\frac{1}{3}$ for each $i$. Indeed, if this assumption is not fulfilled, there exists a rotation matrix $R_{ij}$ such that $X_i'=R_{ij}X_j$ where $R_{ij}$ satisfies $Tr(X_i'{}^{2})=\frac{1}{3}$. The potential is invariant under rotations, so $V(\boldsymbol X')=V(\boldsymbol X)$.

Consider one component of the potential, namely $V(X_1,X_2)=-Tr([X_1, X_2]^2)$. Maximizing this component with $Tr(X_i^2)=\frac{1}{3}$ and constraints from the gauge fixing gives a unique solution. It is
\begin{align}\label{eq:component_max}
X_1=\left(\begin{array}{ccc}-\frac{1}{\sqrt 6}&0&0\\0&0&0\\0&0&\frac{1}{\sqrt 6}\end{array}\right),\
X_2=\left(\begin{array}{ccc}0&0&\frac{1}{\sqrt 6}\\0&0&0\\\frac{1}{\sqrt 6}&0&0\end{array}\right).
\end{align}
Obviously, maximum of the potential is not greater than sum of maxima of the three components, i.e.
\begin{align}
max_{\boldsymbol X}V(\boldsymbol X)\leq-\sum_{i<j}max_{X_i,X_j}Tr([X_i X_j]^2).
\end{align}
Suppose that there exists a matrix $X_3$ such that $Tr(X_3^2)=\frac{1}{3}$ and $Tr([X_1,X_2]^2)=Tr([ X_2, X_3]^2)=Tr([ X_1, X_3]^2)$ with $X_1,\ X_2$ given by (\ref{eq:component_max}). If it is true, then the set $(X_1,X_2,X_3)$ necessarily maximizes the potential. There are two such matrices $ X_3$:
\begin{align}
X_3=\left(\begin{array}{ccc}0&0&-\frac{i}{\sqrt 6}\\0&0&0\\\frac{i}{\sqrt 6}&0&0\end{array}\right),\
X'_3=\left(\begin{array}{ccc}0&0&\frac{i}{\sqrt 6}\\0&0&0\\-\frac{i}{\sqrt 6}&0&0\end{array}\right).
\end{align}

Concluding, there are two one--dimensional steep directions, just as for the $SU(2)$ gauge group. In particular, these solutions are invariant under rotations. That is, for a given $SO(3)$ matrix $R_{ij}$ the configuration $X'_i=R_{ij} X_j$ is gauge--equivalent to $X_i$. In other words, rotations of the two steep directions are special cases of gauge transformations.

For consistency with previous paragraphs, we denote the versors of the two steep directions by $\boldsymbol{\hat x}_{steep}$ and $\boldsymbol{\hat x}'_{steep}$. As in the $SU(2)$ case, $-\boldsymbol{\hat x}'_{steep}$ can be transformed with the gauge transformation to $\boldsymbol{\hat x}_{steep}$. Therefore, in all profiles we use only one steep direction $\boldsymbol{\hat x}_{steep}$ and let $r$ to be positive and negative.

\subsection{Analysis of wavefunctions}
\begin{figure}\centering
\includegraphics[width=.5\textwidth]{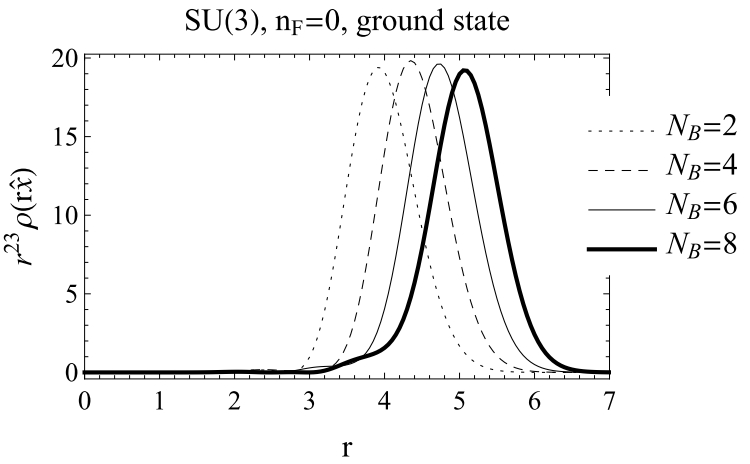}
\caption{Plot of the wavefunctions of ground state in the noninteracting case for $n_F=0$ and $j=0$.}
\label{fig:su3_free}
\end{figure}

Consider first the noninteracting case. The Hamiltonian has only the kinetic part, $H=Tr(p_ip_i)$. As in the $SU(2)$ case the wavefunction depends only on the radius. Therefore, the probability density has to be multiplied by $r^{23}$. The profile of the probability distribution in the noninteracting case is shown in Fig. \ref{fig:su3_free}. One can see that indeed the wavefunction moves outwards as the cutoff grows.

Let us now come to the interacting case. Previously it was shown that $\rho(\boldsymbol x)$ needs to be multiplied by a factor $r^{N^2+2}$ so that is can be considered as the probability density integrated over the gauge group and rotations. For $SU(3)$, $r^{N^2+2}=r^{11}$. However, additional factor has to be taken into account.

The flat valleys are three dimensional. They are parameterized by the radius $r$ and two angles $\theta,\phi$ (cf. (\ref{eq:valley_parametrization})). It turns out that $\rho(\boldsymbol x)$ does not depend on two angles, but only on the radius $r$. Therefore, it is convenient to consider probability density integrated over the gauge group, rotations and the two angles $\theta$ and $\phi$. Another $r^2$ has to be included and finally, $\rho(\boldsymbol x)$ has to be multiplied by $r^{13}$.

For the steep direction the situation is reversed. Spatial Rotations are included in gauge rotations. Thus, $\rho(\boldsymbol x)$ needs to be multiplied only by $r^9$ which comes from integration over the gauge group. However, for consistency $\rho(\boldsymbol x)$ in the steep direction is also multiplied by $r^{13}$. Only then one can compare wavefunctions in the two directions.

\begin{figure}\centering
\begin{subfigure}[b]{0.48\textwidth}\includegraphics[width=\textwidth]{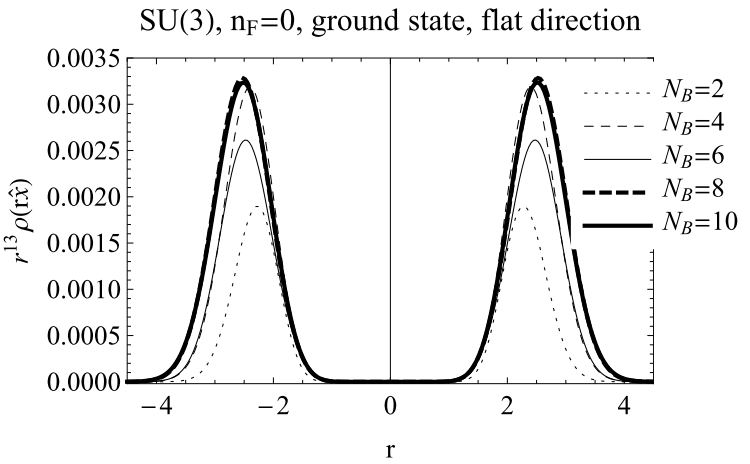}\end{subfigure}
\begin{subfigure}[b]{0.48\textwidth}\includegraphics[width=\textwidth]{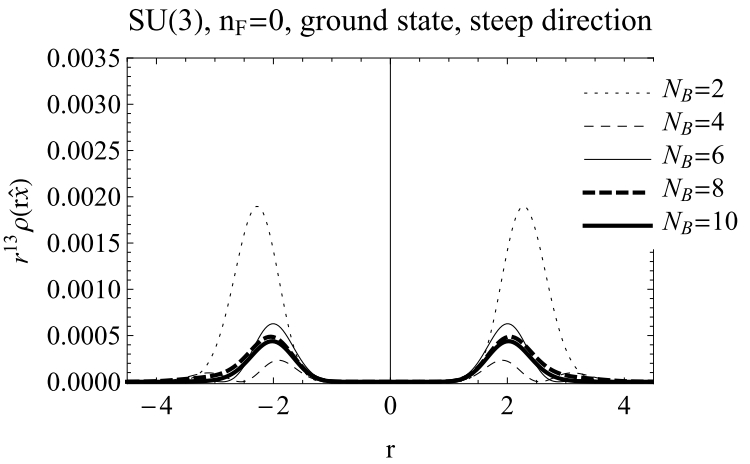}\end{subfigure}
\begin{subfigure}[b]{0.48\textwidth}\includegraphics[width=\textwidth]{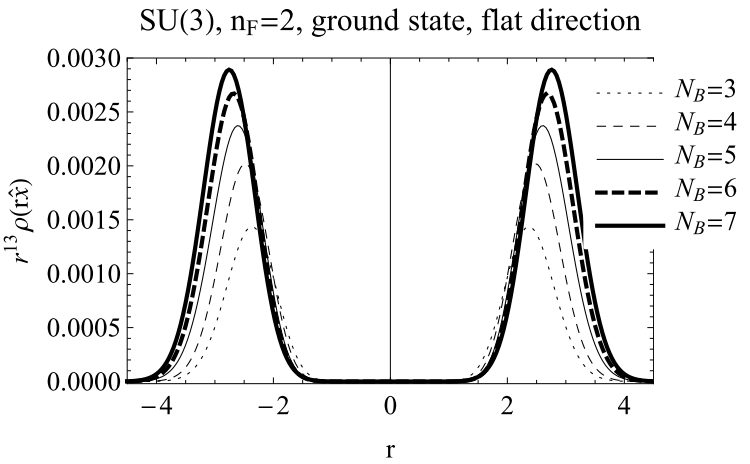}\end{subfigure}
\begin{subfigure}[b]{0.48\textwidth}\includegraphics[width=\textwidth]{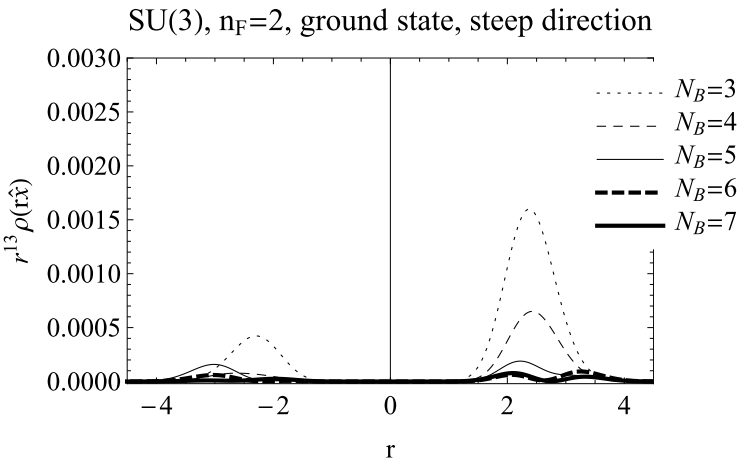}\end{subfigure}
\caption{Plots of wavefunctions of ground states in sectors $n_F=0$. The angular momentum is $j=0$.
}\label{fig:su3_discrete}
\end{figure}

First, consider the ground state in the purely bosonic sector (Fig. \ref{fig:su3_discrete}). One can see that the probability distribution converges for $N_B\approx 8$. The profile in the steep direction is suppressed compared to the flat direction. Moreover, it penetrates the valley deeper than the region where the potential is nonzero. All these effects are similar to the case of $SU(2)$, yet there are some differences. First, the convergence is slower. Secondly, the steep direction is suppressed stronger than for $SU(2)$. This may be caused by higher dimension of the configuration space. There are more allowed directions, so the steepest one is naturally more disfavored. For $n_F=2$ behavior of the probability distribution is very similar. The convergence is even slower, but rather clear.

Next, take the first excited state for $n_F=7$. One can see that in the flat direction the probability density grows while in the steep direction it is suppressed. This may be the first step of the characteristic behavior of wavefunctions of scattering states, which was observed for $SU(2)$ (cf. Fig. \ref{fig:su2_full}). On the other hand, bound states for $SU(3)$ share the same tendencies for small cutoffs. Therefore, a firm statement can be given only from analysis for higher $N_B$.

\begin{figure}\centering
\begin{subfigure}[b]{0.48\textwidth}\includegraphics[width=\textwidth]{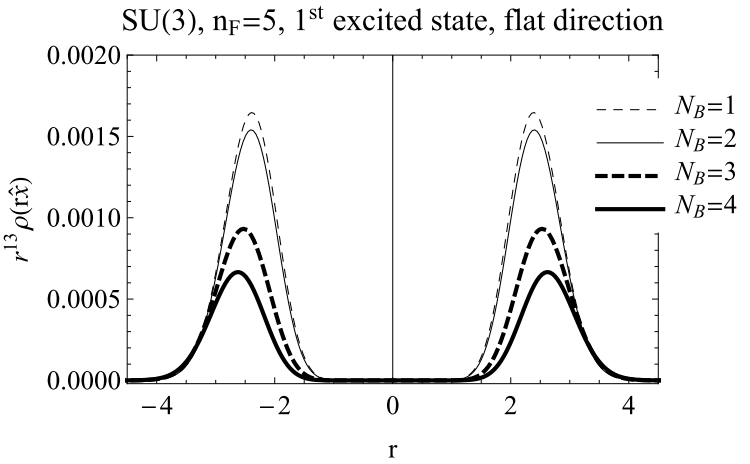}\end{subfigure}
\begin{subfigure}[b]{0.48\textwidth}\includegraphics[width=\textwidth]{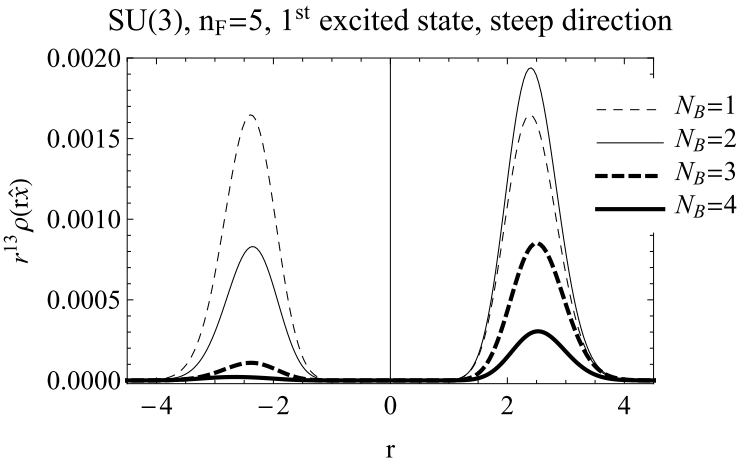}\end{subfigure}
\begin{subfigure}[b]{0.48\textwidth}\includegraphics[width=\textwidth]{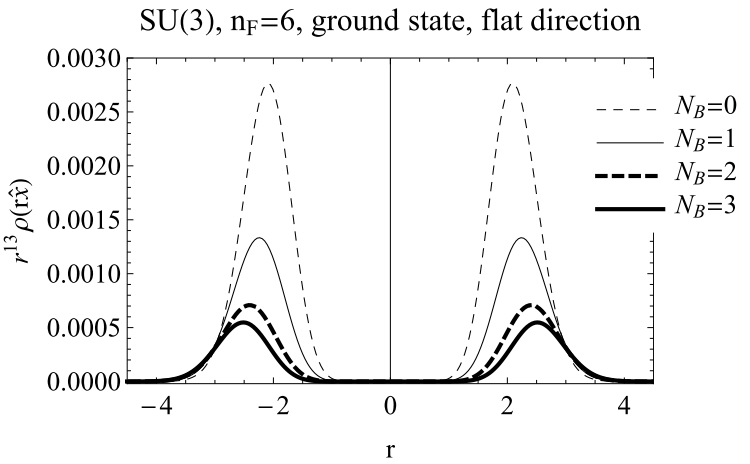}\end{subfigure}
\begin{subfigure}[b]{0.48\textwidth}\includegraphics[width=\textwidth]{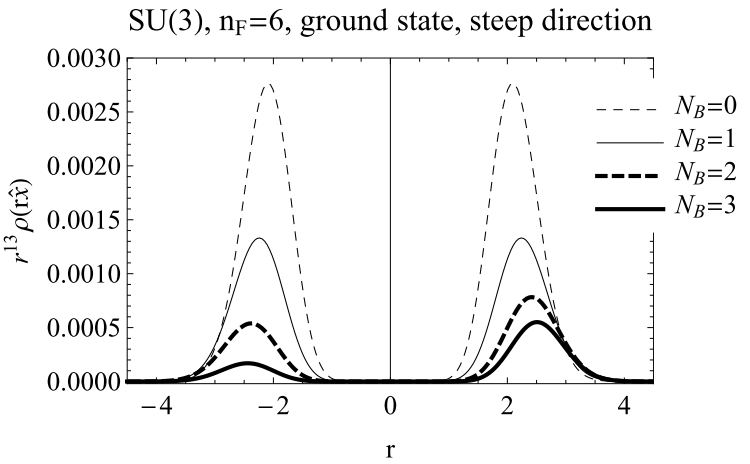}\end{subfigure}
\begin{subfigure}[b]{0.48\textwidth}\includegraphics[width=\textwidth]{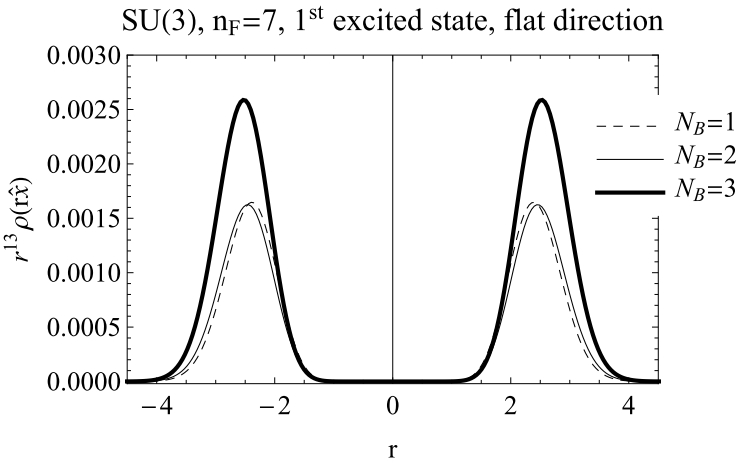}\end{subfigure}
\begin{subfigure}[b]{0.48\textwidth}\includegraphics[width=\textwidth]{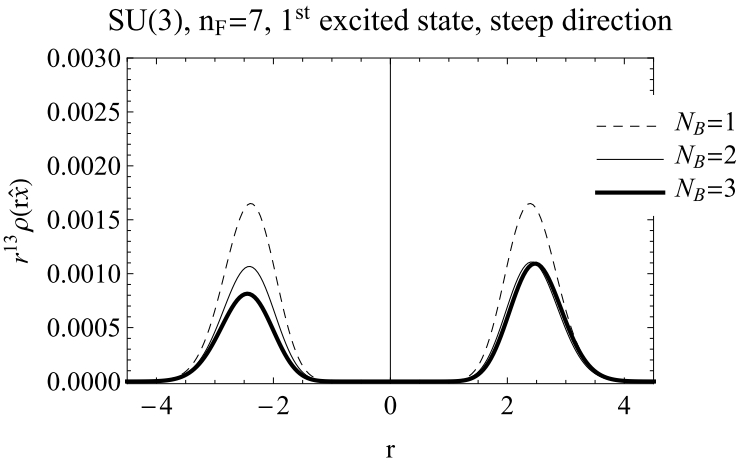}\end{subfigure}
\caption{Plots of wavefunctions of ground states in sectors $n_F=0$. The angular momentum is $j=0$.
}\label{fig:su3_continuous}
\end{figure}

Finally, consider the other states presented in Fig \ref{fig:su3_continuous} -- the first excited state for $n_F=5$ and the ground state for $n_F=6$. These are the candidates for the scattering states with minimal number of fermions (see Chapter \ref{ch:spectrum}). The probability distribution moves outwards as the cutoff grows. However, this effect is not stronger than for the lowest eigenstate for $n_F=2$ and $N_B\leq5$. Therefore, the asymptotic behavior of these profiles remains unclear.

Concluding, analysis of wavefunctions for the $SU(3)$ gauge group is more delicate than for $SU(2)$. In sectors with small number of fermions, the convergence is slower that for $SU(2)$, but is clearly seen in our data. Moreover, suppression of the steep direction is stronger. For five and six fermions the wavefunctions share some behavior of scattering states, but it could be a small--cutoff effect. Finally, the first excited state with seven fermions is a candidate for a scattering state. This is in agreement with scaling relations for continuous spectrum (cf. Chapter \ref{ch:spectrum}).


\chapter{Summary}
In this thesis the cut Fock space approach was used to study the supersymmetric Yang-Mills quantum mechanics in $D=4$. Central results concern the theory with $SU(3)$ gauge group. However, substantial part of the analysis is redone also for $SU(2)$, mainly for comparison. Our studies cover the following topics:
\begin{itemize}
\item construction of the basis of cut Fock space,
\item algorithm for constructing matrix representation of operators $H$, $J_i$ and $Q_\alpha$ in occupation number basis,
\item spectrum of the model obtained numerically,
\item restoration of supersymmetry in the continuum limit,
\item wavefunctions of energy eigenstates.
\end{itemize}
These points will be now discussed in order.

The Fock space consists of states which are singlets of the gauge group $SU(N)$. Basis of the Hilbert space is generated by so called bricks (cf. \cite{Trzetrzelewski_trees}). A brick is a trace of product of creation operators contracted with $SU(N)$ generators, e.g. $(a_1^\dagger a_2^\dagger) \equiv a_1^{a\dagger}a_2^{b\dagger} Tr(T^aT^b)$. The whole Hilbert space is spanned by composite bricks (i.e. products of bricks) acting on the empty state. The set of vectors generated in this way is overcomplete. This is because the set of all possible bricks is not independent. Eliminating dependent bricks is nontrivial but essential for performance of the numerical algorithm. In \ref{sec:brick_relations} we addressed this problem for $SU(3)$. Many bricks can be eliminated with the following relations:
\begin{itemize}
\item invariance of the trace under cyclic permutations,
\item Cayley-Hamilton theorem for matrices with commuting (\ref{eq:CH}) and anticommuting (\ref{eq:CH_anticommutation}) matrix elements,
\item symmetric combination of product of three $SU(3)$ generators reduces to shorter bricks.
\end{itemize}
After these relations are exploited to reduce the number of bricks, the Gauss elimination algorithm can be applied. With this method one eliminates all dependent bricks. It was found that all bricks consisting of more than six creation operators can be reduced to shorter ones. This fact was explained by showing that there exists a solution to (\ref{eq:su3_relation}). It follows from (\ref{eq:su3_relation}) that a trace of product of seven or more $SU(3)$ generators can be expressed by a combination of products of shorter traces. Finally, the number of independent bricks is finite and 786 bricks were found. The structure of $SU(2)$ group is simpler and there are only 35 independent bricks.

After construction of the occupation number basis, a cutoff $N_B$ for the maximal number of bosonic excitations is introduced. Because of the Pauli exclusion principle, the total number of fermions $n_F$ is limited by $2(N^2-1)$, so no cutoff for $n_F$ is needed. The resulting cut Fock space is spanned by composite bricks which consist of at most $N_B$ bosonic and any number of fermionic creation operators. In this finite--dimensional space operators are represented by matrices. In \ref{sec:algorithm} a recursive algorithm for calculating matrix elements of the Hamiltonian and other relevant operators is introduced. In Appendix \ref{ap:finite_proof} finiteness of this procedure is proven.
The Hamiltonian and angular momentum conserve the total number of fermions, so their matrix elements can be generated in each sector with given $n_F$ independently. Supersymmetry generators change the number of fermions by one, so they are represented by matrices which act between sectors.

The program was used to construct matrix representations of angular momentum, Hamiltonian and supersymmetry generators for certain cutoffs in (or between) sectors with given $n_F$. Construction of these matrices was performed with a parallel program on a supercomputer Deszno at the Jagiellonian University. The main limitation was the amount of available memory (256 GB). First, matrices of angular momentum were diagonalized. Then, the Hamiltonian was diagonalized in subspaces corresponding to given total angular momentum $j$ and the third component $m$.

The Hilbert space $\mathcal H$ can be split into subspaces $\mathcal H_{\mathbf n}$, $\mathbf n=(n_F^1,n_F^2,n_B^1,n_B^2,n_B^3)$ with definite occupation numbers of bosons and fermions for corresponding labels $\mathbf n$. In Chapter 4 a group theoretical approach was used to calculate dimensions $D_\mathbf n$ of $\mathcal H_\mathbf n$. Our analysis is an extension of a method discussed in \cite{Trzetrzelewski_number} where sectors with total number of bosons $n_B=n_B^1+n_B^2+n_B^3$ and total number of fermions $n_F=n_F^1+n_F^2$ are considered. An essential component of this technique is the generating function
\begin{align}
G(a_1,a_2,a_3,b_1,b_2)=\sum_{n^i_B=0}^\infty\sum_{n^\alpha_F=0}^\infty a_1^{n^1_B}a_2^{n^2_B}a_3^{n^3_B}(-b_1)^{n^1_F}(-b_2)^{n^2_F}D_{\mathbf n}.
\end{align}
$D_\mathbf n$ is expressed as an integral of character over the $SU(N)$ group. The generating function was calculated by residues and an explicit result was given in (\ref{eq:number_of_states}). Then, $D_\mathbf n$ were extracted from Taylor expansion of $G$. The dimensions were checked against the number of independent vectors determined numerically in the computer program and an agreement was found (cf. Fig. \ref{fig:histograms}). The dimensions of channels with definite total occupation numbers $n_B$ and $n_F$ were given in Tab. \ref{tab:number_of_states}. It is in perfect agreement with earlier results \cite{Trzetrzelewski_number}. As expected, dimensions $D_\mathbf n$ grow fast with $n_B$.

In Chapter \ref{ch:spectrum} we analyzed the spectrum for $SU(2)$ and $SU(3)$ cases. Dependence of the eigenenergies on the cutoff $N_B$ is crucial for identifying type of the spectrum. Discrete energy levels converge very fast to the exact value. On the contrary, if the continuous spectrum is present, then all eigenvalues $E_m$ converge slowly to zero for fixed $m$. From another point of view, one can recognize a continuous spectrum by a distribution of eigenenergies that fill densely a full interval as $N_B$ grows. The spectrum for $SU(2)$ was already found with a great precision in \cite{Campostrini} and served as a check of our program. A perfect agreement was established. The spectrum is purely discrete for $n_F=0,1$ (cf. Figs \ref{fig:su2_spectrum_nf0}--\ref{fig:su2_spectrum_nf1}). It is continuous for two and three fermions (cf. Figs \ref{fig:su2_spectrum_nf2}--\ref{fig:su2_spectrum_nf3}). Moreover, there are discrete energy levels immersed in the continuous part. For the $SU(3)$ case the cutoffs we were able to reach were significantly smaller than for $SU(2)$, especially for large $n_F$. However, a number of regularities already appear. Our results in the purely bosonic part of the Hilbert space were compared with \cite{Weisz} where no fermions were considered. Very good agreement was found (cf. Tab. \ref{tab:Weisz_comparison}). It was argued in \cite{Campostrini,Kotanski} that eigenenergies corresponding to the continuous spectrum should satisfy the scaling relation (\ref{eq:continuous_scaling}):
\begin{align}
E_m\propto \frac{(m+1)^2}{N_B},
\end{align}
where $m$ is the label of the energy. Therefore, the scaled energy $N_B\times E_m$ should approach a constant when $N_B\to\infty$. This effect is seen clearly for $SU(2)$ for at least the two lowest energies in the $(n_F,j)=(2,0)$ sector (cf. Fig. \ref{fig:su2_continuous_scaling}). Ratios of the lowest energies also agree with the prediction of (\ref{eq:continuous_scaling}), i.e. $E_m/E_0\approx (m+1)^2$ (cf. Fig. \ref{fig:su2_continuous_ratios}). For $SU(3)$ the scaled energies grow approximately linearly with increasing $N_B$ for $n_F\leq 4$ and seem to be approximately constant for $n_F=5,6,7,8$ (Fig. \ref{fig:su3_scaling}). Therefore, the spectrum is purely discrete for $n_F\leq 4$. It seems to be continuous for $n_F=6,7,8$. The intermediate sector $n_F=5$ is disputable. Because of the particle--hole symmetry, the energy levels for $8<n_F\leq16$ are the same as for $16-n_F$ fermions. Therefore, supersymmetric cancelations can happen only in the middle sectors. Only then there are enough fermionic degrees of freedom. For a more detailed picture concerning the type of spectrum a higher cutoff would be needed.

In Chapter \ref{ch:supersymmetric_multiplets} we studied supersymmetric structure of the model for a finite cutoff. Finite $N_B$ breaks the supersymmetry. This is because the supercharges $\mathcal Q_\alpha$ do not conserve the number od bosons. However, the supersymmetry is being restored as $N_B$ grows. For exact supersymmetry massive states form supersymmetry multiplets. For a given state one can act on it with $\mathcal Q_\alpha$ and  $\mathcal Q_\alpha^\dagger$ to recover the other states in its supermultiplet. To check how much the supersymmetry is broken for a given cutoff we used supersymmetry fractions
\begin{align}
q_{n_F}\left(j'E'|jE\right)&=\frac{1}{4E}\sum_{mm'\alpha}\left|\braket{n_F+1j'm'E'|\mathcal Q_\alpha^\dagger|n_FjmE}\right|^2,
\end{align}
where $\ket{n_FjmE}$ is a state with $n_F$ fermions, total angular momentum $j$, the third component of angular momentum $m$ and energy $E$. Values of supersymmetry fractions are known when the supersymmetry is exact. $q_{n_F}\left(j'E'|jE\right)$ is $j$ or $j+\frac{1}{2}$ if the two corresponding states are in the same multiplet and zero otherwise. Therefore, supersymmetry fractions can be used to measure the SUSY breaking. We found that at least for some low energy states the supersymmetry fractions converge to their exact values (cf. Fig. \ref{fig:fractions_convergence}). They were used to identify supermultiplets. Four complete supermultiplets were fully identified. More supermultiplets were found with one state in the channel with highest $n_F$ missing. Complete list of identified supermultiplets is given in Tab. \ref{tab:spectroscopy} and an overview picture is presented in Fig. \ref{fig:overall}. We concluded that within the available cutoff, some supersymmetry is already restored at low energies and small number of fermions.

Finally, in Chapter \ref{ch:wavefunctions} we studied the wavefunctions. For a given state $\ket{\phi}$ the probability density $\left|\phi_{\boldsymbol\chi}(\boldsymbol x)\right|^2$ in the configuration space is a multi--component function. The label $\boldsymbol\chi$ corresponds to occupation numbers of fermions and variable $\boldsymbol x$ is a $3(N^2-1)$--dimensional coordinate. As were are not interested in particular dependence on fermionic degrees of freedom, sum over $\boldsymbol\chi$ was performed. For an angular momentum multiplet $\ket{\phi^m}$ a sum over elements of the multiplet eliminates dependence on rotations. The probability distribution
\begin{align}
\rho(\boldsymbol x)\equiv\sum_{m\boldsymbol \chi}|\phi^m_{\boldsymbol\chi}(\boldsymbol x)|^2
\end{align}
is a real, single--component function. $\rho(\boldsymbol x)$ is gauge and rotationally invariant. This is proven in Appendix \ref{ap:rotation_invariance}.

We were interested in behavior of the wavefunctions inside and outside flat valleys of the potential. To this end we identified the directions in the configurations space in which the potential vanishes and grows most rapidly. For $SU(2)$ this was done with invariants (\ref{eq:su2_invariants}) introduced in \cite{vanBaal}. There is one direction in the space of gauge and rotation invariants where the potential is flat and there are two steep directions. We observed that the low energy wave functions converge to a fixed shape as the cutoff grows. They are a little stretched along the flat valleys, as one may expect. On the contrary, the scattering states penetrate the flat valleys deeper for increasing $N_B$ (Fig. \ref{fig:su2_full}). They remain localized in the steep direction.

For $SU(3)$ structure of the potential was studied in a fixed gauge. The full procedure of fixing the gauge is given in Appendix \ref{ap:fixing_gauge}. As for $SU(2)$, there are two one--dimensional steep directions. The flat direction is six dimensional. The flat valley can be parameterized by three angles which correspond to rotations and three rotation invariants. As it was pointed out, $\rho(\boldsymbol x)$ does not depend on rotations. Moreover, it turned out that in all studied cases $\rho(\boldsymbol x)$ depends only on the radius $r=\left(x_i^a x_i^a\right)^{1/2}$. For small $n_F$ the energy eigenstates correspond to discrete energy levels. Convergence of $\rho(\boldsymbol x)$ to a fixed shape is observed (Fig. \ref{fig:su3_discrete}). The available cutoff for large $n_F$ is small and the studies of wavefunctions in these sectors are preliminary.

To summarize, we analyzed the supersymmetric Yang-Mills quantum mechanics in four dimensions with $SU(2)$ and $SU(3)$ gauge groups. The case of $SU(2)$ was already considered before and served mainly as a check of the program for numerical calculations. Results for $SU(3)$ with fermions are original. Only the purely bosonic channel was addressed in earlier papers. The studies were performed with the cut Fock space method. On the technical side, we identified independent bricks which generate basis of the Hilbert space and developed a recursive algorithm for constructing matrices of relevant operators in the occupation number basis. The program worked nicely. Results are consistent with previous results for $SU(2)$ and the purely bosonic sector for $SU(3)$. Dimensions of $\mathcal H_\mathbf n$ calculated analytically and numerically agree.

The numerical results provide information about dynamics of zero modes of the supersymmetric Yang--Mills theory. In the sector with no fermions the spectrum is discrete. Although the potential has flat valleys, the transverse oscillations give rise to an effective potential. It prevents wavefunctions from propagating in these directions. The situation is identical for small number of fermions. Only in channels rich in fermions the supersymmetric cancelations of bosonic and fermionic degrees of freedom is possible. Then the effective potential vanishes and flat valleys are open. The spectrum becomes continuous. For $SU(2)$ this happens already for two fermions and for $SU(3)$ the smallest required number of fermions is $n_F=5$ or $6$. Because of the particle--hole symmetry, spectra for numer of fermions large enough are again discrete.

The supersymmetry is broken for finite cutoff $N_B$. It is recovered for $N_B\to\infty$ and then all states are grouped in supermultiplets. A full angular momentum and supersymmetry multiplet is labeled by $(n_F,j,E)$ where $(n_F,j)$ are the number of fermions and total angular momentum of the Clifford vacuum (cf. Chapter \ref{ch:supersymmetric_multiplets}). All $8j+4$ states of the multiplet are degenerate and have energy $E$. Restoration of the supersymmetric structure takes place first for low energy states and is best seen for small $n_F$. Four supermultiplets were completely identified and for eighteen more there is one state missing (due to the finite cutoff effects).

Finally, the wavefunctions were analyzed. In earlier papers \cite{vanBaal,Kotanski} only wavefunctions of s--waves were studied (and thus only channels with two fermions were available). In this thesis we developed a new method that gives access to all $(n_F,j)$ sectors. We have shown for $SU(2)$ that the bound states converge to a fixed shape for $N_B\to\infty$. On the other hand, the scattering states penetrate the flat valleys of the potential deeper as the cutoff grows. Analogous study of the $SU(3)$ case shows similar behavior of bound states.

\appendix

\chapter{Symmetries}\label{ap:symmetries}
In this chapter we discuss symmetries of the Hamiltonian. The Hamiltonian conserves angular momentum, total number of fermions and supercharge. Apart from these there is a particle--hole symmetry which relates sectors with $n_F$ and $2(N^2-1)-n_F$ fermions. All this symmetries were mentioned in Chapter \ref{ch:model}. Here they are analyzed in more detail. In the following part we use definitions of operators introduced in Chapter \ref{ch:model}. Recall definition of the Hamiltonian:
\begin{align}\begin{split}
H&=H_K+g^2H_V+gH_F,\\
H_K&=\frac{1}{2}p^a_ip^a_i,\\
H_V&=\frac{1}{4}f^{abc}f^{ade}x^b_ix^c_jx^d_ix^e_j,\\
H_F&=\frac{i}{2}f^{abc}\psi^{a\dagger}\gamma^0\gamma^k\psi^bx^c_k.
\end{split}\end{align}
The Majorana spinor $\psi$ is
\begin{align}\label{eq:spinor}
\psi^a=e^{i\pi/4}\left(\begin{array}{c}-f^{a}_1\\-f^{a}_2\\f^{a\dagger}_2\\-f^{a\dagger}_1\end{array}\right).
\end{align}

\section{Number of fermions}
The operator of the total number of fermions is
\begin{align}
n_F=\sum_{\alpha,a}f_\alpha^{a\dagger}f_\alpha^a.
\end{align}
It is more convenient to write it in terms of $\psi$. One can find that the matrix $\gamma^5$ is given by
\begin{align}
\gamma^5&=\left(\begin{array}{cccc}-1&0&0&0\\0&-1&0&0\\0&0&1&0\\0&0&0&1\end{array}\right).
\end{align}
Then, it is easy to check that
\begin{align}
n_F=8-\frac{1}{2}\psi^{a\dagger}\gamma^5\psi^a.
\end{align}
The kinetic part and potential obviously do not change the number of fermions. One needs to show that $H_F$ commutes with $\psi^{a\dagger}\gamma^5\psi^a$:
\begin{align}\begin{split}
[H_F,\psi^{a\dagger}\gamma^5\psi^a]&=\frac{i}{2}f^{abc}[\psi^{a\dagger}\gamma^0\gamma^k\psi^b,\psi^{d\dagger}\gamma^5\psi^d]x^c_k\\
&=\frac{i}{2}f^{abc}x^c_k\left(\psi^{a\dagger}\gamma^0\gamma^k\gamma^5\psi^b-\psi^{a\dagger}\gamma^5\gamma^0\gamma^k\psi^b\right)\\
&=\frac{i}{2}f^{abc}x^c_k\psi^{a\dagger}\left(\gamma^0\gamma^k\gamma^5-\gamma^5\gamma^0\gamma^k\right)\psi^b\\
&=0.
\end{split}\end{align}
Finally, conservation of the total number of fermions is proven.

This symmetry is in practice the most important one in calculating spectrum of SYMQM. One can consider each sector with definite $n_F$ separately and the full Hilbert space splits into $2(N^2-1)$ parts. Taking a channel with definite $n_F$ is easier than e.g. taking states with definite angular momentum.

\section{Parity}
In Chapter \ref{ch:model} we noted that the Hilbert space breaks the parity symmetry. It is a consequence of the condition that $\psi$ has to be a Majorana spinor. We will now show how this happens in detail. Color indices are irrelevant for this discussion and are implicit throughout this part.

First, take a Dirac spinor $\psi$. It can be parameterized in a convenient way by two $2$--component spinors $\zeta$ and $\chi$. These can be used to form left-- and right--handed spinors
\begin{align}
\psi_L&=\left(\begin{array}{c}\zeta\\0\end{array}\right),&\psi_R&=\left(\begin{array}{c}0\\\chi\end{array}\right).
\end{align}
Next, $\zeta$ is used to build a Majorana spinor $\psi_M$. In turn, $\chi$ is used to construct an 'anti-Majorana' spinor $\psi_{\bar M}$, i.e. $\psi_{\bar M}$ satisfies the Majorana condition with a '-' sign $(\psi_{\bar M})_C=-\psi_{\bar M}$. The explicit formulas are
\begin{align}
\psi_M&=\psi_L+(\psi_L)_C=\left(\begin{array}{c}\zeta\\\bar\zeta\end{array}\right),&\psi_{\bar M}&=\psi_R-(\psi_R)_C=\left(\begin{array}{c}\bar\chi\\\chi\end{array}\right),
\end{align}
where $\bar\zeta\equiv-\sigma_2\zeta^*$. A sum of $\psi_M$ and $\psi_{\bar M}$ gives a general, Dirac spinor
\begin{align}
\psi&=\frac{1}{\sqrt 2}(\psi_M+\psi_{\bar M})=\frac{1}{\sqrt2}\left(\begin{array}{c}\zeta+\bar\chi\\\bar\zeta+\chi\end{array}\right).
\end{align}
The parity transformation interchanges the left-- and right--handed spinors. Indeed,
\begin{align}
\psi\rightarrow\psi_P\equiv\frac{1}{\sqrt2}\left(\begin{array}{c}\zeta_P+\bar\chi_P\\\bar\zeta_P+\chi_P\end{array}\right)
=\gamma^0\psi=\frac{1}{\sqrt2}\left(\begin{array}{c}\bar\zeta+\chi\\\zeta+\bar\chi\end{array}\right).
\end{align}
It follows that $\zeta_P=\chi$ and $\chi_P=\zeta$. One can also see that $(\psi_{\bar M})_C$ is a Majorana spinor while $(\psi_M)_C$ is 'anti--Majorana'.

We quantize the system by imposing standard anticommutation relations on $\zeta$ and $\chi$:
\begin{align}
\{\zeta_\alpha,\zeta_\beta\}&=\{\zeta_\alpha^\dagger,\zeta_\beta^\dagger\}=0,&\{\zeta_\alpha,\zeta_\beta^\dagger\}&=\delta_{\alpha\beta},\nonumber\\
\{\chi_\alpha,\chi_\beta\}&=\{\chi_\alpha^\dagger,\chi_\beta^\dagger\}=0,&\{\chi_\alpha,\chi_\beta^\dagger\}&=\delta_{\alpha\beta}.
\end{align}
Then the Dirac spinor $\psi$ satisfies
\begin{align}
\{\psi_\alpha,\psi_\beta\}&=\{\psi_\alpha^\dagger,\psi_\beta^\dagger\}=0,&\{\psi_\alpha,\psi_\beta^\dagger\}=\delta_{\alpha\beta}.
\end{align}
The empty state $\ket{0}$ is defined by
\begin{align}
\zeta_\alpha\ket{0}=\chi_\alpha\ket{0}=0.
\end{align}
The other states in the Hilbert space are generated by repeatedly acting with $\zeta_\alpha^\dagger$ and $\chi_\alpha^\dagger$ on $\ket{0}$.

We are interested in parity--transformation of states. The parity--transformed empty state satisfies conditions
\begin{align}
(\zeta_P)_\alpha\ket{0}_P=(\chi_P)_\alpha\ket{0}_P=0.
\end{align}
We see that $\ket{0}_P=\ket{0}$, i.e. $\ket{0}$ is invariant under parity. Take now a different state, e.g. $\ket{s}=\zeta_1^\dagger\ket{0}$. The transformed state is
\begin{align}
\ket{s}_P=(\zeta_P)_1^\dagger\ket{0}=(\chi)_1^\dagger\ket{0}.
\end{align}

Recall that for the Yang--Mills quantum mechanics to be supersymmetric a condition $\psi=\psi_M$ has to be imposed. Therefore, $\psi$ contains only $\zeta$ and not $\chi$. The Hilbert space consists of the empty state $\ket{0}$ and all states obtained by acting with $\zeta_\alpha^\dagger$ on it. Therefore, $\ket{s}$ is an element of the Hilbert space, but $\ket{s}_P$ is not. In fact, only $\ket{0}$ survives the parity transformation. We can see that the parity is broken on the level of the Hilbert space.

\section{Particle -- hole symmetry}
The particle--hole transformation of operators is the following:
\begin{align}\begin{split}
\psi^a&\rightarrow \psi^a_{ph}=e^{i\pi/4}\left(\begin{array}{c}f_2^{a\dagger}\\-f_1^{a\dagger}\\f_1^{a}\\f_2^{a}\end{array}\right)=\gamma^0\gamma^5\psi^a,\\
x_i^a&\rightarrow-x_i^a.
\end{split}\end{align}
In the view of the discussion on parity, it is important to note that $\psi_{ph}$ is a Majorana spinor. Now, it has to be shown that it is a symmetry of the Hamiltonian. Obviously, the kinetic terms and potential do not change under the particle--hole transformation. Proof for the last term is following:
\begin{align}\begin{split}
H_F&\rightarrow-\frac{i}{2}f^{abc}\psi_{ph}^{a\dagger}\gamma^0\gamma^k\psi_{ph}^bx^c_k=-\frac{i}{2}f^{abc}\psi^{a\dagger}\gamma^5\gamma^0\gamma^0\gamma^k\gamma^0\gamma^5\psi^bx^c_k\\
&=\frac{i}{2}f^{abc}\psi^{a\dagger}\gamma^0\gamma^k\psi^bx^c_k=H_F.
\end{split}\end{align}
Therefore, this is a true symmetry.

The name of this symmetry lies in the transformation of states. The transformation of fermionic operators is
\begin{align}
f_1^a&\rightarrow -f_2^{a\dagger},&f_2^a&\rightarrow f_1^{a\dagger}.
\end{align}
Therefore, the definition of empty state transforms into
\begin{align}
f_\alpha^{a\dagger}\ket{0}_{ph}=0,
\end{align}
i.e. $\ket{0}_{ph}$ is the state with the maximal number of fermions
\begin{align}
\ket{0}_{ph}=\prod_{a,\alpha}f_\alpha^{a\dagger}\ket{0}.
\end{align}
One can readily see that a state $\ket{\phi}$ with $n_F$ fermions transforms to state $\ket{\phi}_{ph}$ with $2(N^2-1)-n_F$ fermions. It follows that the spectra in corresponding fermionic sectors are the same and one needs to study only $n_F\leq N^2-1$.

\section{Conservation of supercharge}
The proof of supercharge conservation is much simpler in the Majorana representation. The Gamma matrices in this representation are
\begin{align}
\gamma^0&=\left(\begin{array}{cc}0&\sigma_2\\\sigma_2&0\end{array}\right),&
\gamma^1&=\left(\begin{array}{cc}0&-\sigma_1\\-\sigma_1&0\end{array}\right),\nonumber\\
\gamma^2&=\left(\begin{array}{cc}\mathds 1&0\\0&-\mathds 1\end{array}\right),&
\gamma^3&=\left(\begin{array}{cc}0&-\sigma_3\\-\sigma_3&0\end{array}\right).
\end{align}
Note that $(\gamma^0)^T=-\gamma^0$, $(\gamma^i)^T=\gamma^i$. In Majorana representation the alpha matrices are real, i.e. $\Gamma_k=\Gamma_k^*$, where ${}^*$ denotes complex conjugation. It follows that $\Sigma_{jk}^*=-\Sigma_{jk}$.

The charge conjugation matrix is
\begin{align}
C&=\left(\begin{array}{cc}0&-i\sigma_2\\-i\sigma_2&0\end{array}\right).
\end{align}
The Majorana condition for spinor $\psi$ written explicitly reads
\begin{align}
\psi^a_\alpha=(\psi_C)^a_\alpha\equiv-iC_{\alpha\beta}(\gamma^0)_{\gamma\beta}\psi^{a\dagger}_\gamma=\psi^{a\dagger}_\alpha.
\end{align}
Supercharges are also self--conjugate. Indeed,
\begin{align}\begin{split}
Q_\alpha^\dagger&=(\Gamma_k^*\psi^{a\dagger})_\alpha p^a_k-igf^{abc}(\Sigma_{jk}^*\psi^{a\dagger})_\alpha x^b_jx^c_k\\
&=(\Gamma_k\psi^{a})_\alpha p^a_k-igf^{abc}(-\Sigma_{jk}\psi^{a})_\alpha x^b_jx^c_k=Q_\alpha.
\end{split}\end{align}
The Hamiltonian has a particularly simple form
\begin{align}
H=\frac{1}{2}\{Q_\alpha,Q_\alpha^\dagger\}=Q_\alpha Q_\alpha,
\end{align}
where no summation over $\alpha$ is performed. Obviously, it commutes with $Q_\alpha$.

Conservation of supercharge gives rise to supersymmetry multiplets. These multiplets consist of four states in different $n_F$ channels. One can combine multiplets of the supersymmetry and angular momentum. This gives rise to $8j+4$--dimensional representation of the two algebras. It consists of $2j+1$ states with angular momentum $j$ and $n_F$ fermions, angular momentum multiplets with $j'=j+1/2$ and $j'=j-1/2$ in $n_F+1$ channel and a multiplet with $j'=j$ with $n_F+2$. These multiplets are studied in detail in Chapter \ref{ch:supersymmetric_multiplets}.

\chapter{Absence of vector states in the bosonic sector for SU(2)}\label{ap:no_j1}
It is shown that for the $SU(2)$ gauge group there are no states with angular momentum $j=1$ and no fermions. This proof is done after \cite{Munster}. For a proof by contradiction, assume that $\psi_i(x)$ is a wavefunction with $j=1$. Variable $x^a_i$ has an adjoint color index $a=1,2,3$ and space index $i=1,2,3$. For convenience it will be treated as a matrix where $a$ stands for the row index and $i$ for the column index. Gauge transformations and rotations are represented by orthogonal matrices $R,S\in SO(3)$ and they act on $x^a_i$ as left and right multiplication respectively
\begin{align}
x^{a}_i\to x'^a_i=R^{ab}x^b_jS_{ji}.
\end{align}
The wavefunction is a gauge singlet and vector with respect to rotations:
\begin{align}
\psi_i(x')\equiv\psi_i(RxS)=\psi_j(x)S_{ji}
\end{align}
The matrix $x$ can be diagonalized with two matrices $R,S\in SO(3)$ \cite{Goldstone}, i.e. $x=RdS$ where $d$ is a diagonal matrix. Then
\begin{align}\label{eq:psi_transformation}
\psi_i(x)=\psi_j(d)S_{ji}.
\end{align}
Let us fix the index $i$. Consider a diagonal matrix $D\in SO(3)$ such that $D_{ii}=-1$ and the other two diagonal elements are $-1$ and $1$. Clearly, $d$ and $D$ commute. It follows that
\begin{align}
\psi_i(d)&=\psi_i(DDd)=\psi_i(DdD)=\psi_j(d)D_{ji}=-\psi_i(d).
\end{align}
This implies $\psi_i(d)=0$ and it holds for each $i=1,2,3$ and all $d$. From (\ref{eq:psi_transformation}) it follows that $\psi_i(x)=0$ for all $x$. The wavefunction must vanish identically and therefore a vector state cannot exist.

\chapter{Finiteness of the recursive algorithm}\label{ap:finite_proof}

In \ref{sec:algorithm} we presented a recursive algorithm for finding matrix elements of operators in the Fock space representation. In each step of this algorithm one needs to find matrix elements of other operators (cf. formula (\ref{eq:block_matrix})). It needs to be shown that such procedure is finite. This is done in this chapter. All symbols defined here are introduced in \ref{sec:algorithm}.

In a given step of the algorithm a matrix element $A|_{\mathbf n'\mathbf n}$ is calculated. Assume that the annihilation rank $\boldsymbol\nu$ of $A$ is not smaller than the annihilation rank $\bar{\boldsymbol\nu}$ of $A^\dagger$. Otherwise, one would construct matrix element of $A^\dagger$ and use the relation $A|_{\mathbf n'\mathbf n}=(A^\dagger|_{\mathbf n\mathbf n'})^\dagger$.

Let $L(A)$ be the length of the trace operator $A$ and $|\mathbf n|=\sum_{\alpha} n_F^\alpha+\sum_i n_B^i$ be a norm of the occupation number $\mathbf n$. We define \emph{recursion level} of the block $A|_{\mathbf n'\mathbf n}$ by
\begin{align}
R(A,\mathbf n',\mathbf n)=L(A)+|\mathbf n|+|\mathbf n'|.
\end{align}
It will be shown that one needs only matrix elements with smaller recursion level to construct the block $A|_{\mathbf n'\mathbf n}$.

There is a triangle inequality for $L(A)$, $|\mathbf n|$ an $|\mathbf n'|$. Namely, each quantity is not greater than the sum of the two others. In particular, $|\mathbf n|\leq L(A)+|\mathbf n'|$. If $A$ consisted only of creation operators, then we would have $|\mathbf n'|=L(A)+|\mathbf n|$. However, we assumed that $\boldsymbol\nu\geq\bar{\boldsymbol\nu}$, so $A$ contains at least one annihilation operator. Therefore, $|\mathbf n'|<L(A)+|\mathbf n|$.

First we show that $(A\mathcal B_i)|_{\mathbf n'\mathbf n_i}$, which are building blocks of $\bar{A}|_{\mathbf n'\mathbf n}$ (formula (\ref{eq:block_matrix})) have recursion levels smaller than $R(A,\mathbf n',\mathbf n)$. The operator $\mathcal B_i$ is a brick, so it is composed only of creation operators and therefore $L(\mathcal B_i)+|\mathbf n_i|=|\mathbf n|$. It is easy to check that the maximally annihilating permutation of $A\mathcal B_i$ is $\mathcal B_i A$. Therefore, the maximally annihilating form of $A\mathcal B_i$ is $\pm\mathcal B_i A+\mathcal R$. The sign is '-' if both, $A$ and $\mathcal B_i$ are fermionic and '+' otherwise. The remainder $\mathcal R=\sum_k \alpha_k\prod_l C_{kl}$ is a combination of products of trace operators. The length of operator $\mathcal R$ is defined by $L(\mathcal R)=\max_k\sum_l L(C_{kl})$.

$\mathcal R$ originates from commutators of operators in the product $A\mathcal B_i$. A commutator of creation and annihilation operator is proportional to identity. Therefore, each commutation removes two operators from the product, so $L(\mathcal R) < L(A)+L(\mathcal B_i)$. The block $\mathcal R|_{\mathbf n'\mathbf n_i}$ is a combination of products of blocks $C_{kl}|_{\mathbf n'_{kl}\mathbf n_{kl}}$. For each $k,\ l$
\begin{align}\begin{split}
R(C_{kl},\mathbf n'_{kl},\mathbf n_{kl})&=L(C_{kl})+|\mathbf n'_{kl}|+|\mathbf n_{kl}|\\
&\leq L(C_{kl})+\left(|\mathbf n'|+\sum_{i<l}L(C_{ki})\right)+\left(|\mathbf n_i|+\sum_{i>l}L(C_{ki})\right)\\
&\leq L(\mathcal R)+|\mathbf n'|+|\mathbf n_i|<L(A)+L(\mathcal B)+|\mathbf n'|+|\mathbf n_i|\\
&=L(A)+|\mathbf n'|+|\mathbf n|=R(A,\mathbf n,\mathbf n').
\end{split}\end{align}

The block $(\mathcal B_iA)|_{\mathbf n'\mathbf n}$ is calculated as a product of $\mathcal B_i|_{\mathbf n'\mathbf n'_i}$ and $A|_{\mathbf n_i'\mathbf n}$ where $\mathbf n'_i=\mathbf n'-\hat{\mathbf n}_i$ and $\hat{\mathbf n}_i$ is the creation label of $\mathcal B_i$. The brick $\mathcal B_i$ contains only creation operators, so $|\mathbf n'_i|=|\mathbf n'|-|\hat{\mathbf n}_i|$ and $L(\mathcal B_i)=|\hat{\mathbf n}_i|$.
Recursion levels of these blocks are
\begin{align}\begin{split}
R(\mathcal B_i,\mathbf n',\mathbf n'_i)&=L(\mathcal B_i)+|\mathbf n'|+|\mathbf n'_i|=L(\mathcal B_i)+2|\mathbf n'|-|\hat{\mathbf n}_i|=2|\mathbf n'| < R(A,\mathbf n',\mathbf n),\\
R(A,\mathbf n_i',\mathbf n)&=L(A)+|\mathbf n'_i|+|\mathbf n|=L(A)+|\mathbf n'|+|\mathbf n|-L(\mathcal B_i)< R(A,\mathbf n',\mathbf n).
\end{split}\end{align}
We have shown that each element in the block matrix (\ref{eq:block_matrix}) has recursion level smaller than $R(A,\mathbf n',\mathbf n)$.

It remains to calculate the recursion levels of block in the Gramm matrix. The product of two bricks $\mathcal B_i^\dagger \mathcal B_j$ transformed into its maximally annihilating form is $\mathcal B_j\mathcal B_i^\dagger+\mathcal R'$. The block $\mathcal B_i^\dagger \mathcal B_j|_{\mathbf n_i\mathbf n_j}$ is a sum of $\mathcal B_j\mathcal B_i^\dagger |_{\mathbf n_i\mathbf n_j}=\mathcal B_j|_{\mathbf n_i\mathbf n_{ij}}\times\mathcal B_i^\dagger |_{\mathbf n_{ij}\mathbf n_j}$ and $\mathcal R'|_{\mathbf n_i\mathbf n_j}$. Here, $\mathbf n_{ij}\equiv\mathbf n-\hat{\mathbf n}_i-\hat{\mathbf n}_j$ and we know that $|\mathbf n_{ij}|=|\mathbf n|-|\hat{\mathbf n}_i|-|\hat{\mathbf n}_j|=|\mathbf n|-L(\mathcal B_i)-L(\mathcal B_j)$. Recursion levels of these three blocks are following
\begin{align}\begin{split}
R(\mathcal B_j,\mathbf n_i,\mathbf n_{ij})&=2|\mathbf n|-2L(\mathcal B_i)\leq R(A,\mathbf n',\mathbf n)-2L(\mathcal B_i),\\
R(\mathcal B_i^\dagger,\mathbf n_{ij},\mathbf n_j)&=2|\mathbf n|-2L(\mathcal B_j)\leq R(A,\mathbf n',\mathbf n)-2L(\mathcal B_j),\\
R(\mathcal R',\mathbf n_i,\mathbf n_j)&=L(\mathcal R')+|\mathbf n_i|+|\mathbf n_j|< L(\mathcal B_i)+L(\mathcal B_j)+|\mathbf n_i|+|\mathbf n_j|\\
&=2|\mathbf n|\leq R(A,\mathbf n',\mathbf n)-2L(\mathcal B_j).
\end{split}
\end{align}

We have shown that the recursion levels of all blocks that we need to construct $A|_{\mathbf n'\mathbf n}$ are smaller than the recursion level of the block itself. It was assumed that $|\mathbf n|\neq0$. If $|\mathbf n|$ is zero, then the block is calculated in the sector containing only the empty state. Since we know that the first component of annihilation rank of $A$ is positive, it annihilates Fock vacuum and the block vanishes. The recursion level is always nonnegative and it is zero only for identity operator in the Fock vacuum state. Therefore, one needs only a finite number of recursive steps to construct a given block. In order to construct $A|_{\mathbf n'\mathbf n}$ one needs at most $R(A,\mathbf n',\mathbf n)$ steps.

\chapter{Construction of zero angular momentum functions}\label{ap:rotation_invariance}
In \ref{sec:probability_density} we constructed the probability density $\rho(\boldsymbol x)$. Given an angular momentum multiplet $\ket{\phi^m}$ with total angular momentu $j$ we constructed a wavefunction $\phi_{\boldsymbol\chi}^m(\boldsymbol x)$, where $\boldsymbol\chi$ labels occupation numbers of fermions. Then the probability distribution is given by
\begin{align}\begin{split}
\rho(\boldsymbol x)&=\sum_{m\boldsymbol\chi}\left|\phi^m_{\boldsymbol\chi}(\boldsymbol x)\right|^2\\
&=\sum_{m\boldsymbol\chi}\braket{\boldsymbol x\boldsymbol\chi|\phi^m}\braket{\phi^m|\boldsymbol x\boldsymbol\chi}
\end{split}\end{align}
$\rho(\boldsymbol x)$ is gauge and rotationally invariant. Invariance under gauge transformation is obvious since the wavefunction $\phi^m_{\boldsymbol\chi}(\boldsymbol x)$ is gauge invariant. Here we prove invariance under rotations, i.e.
\begin{align}
L^2\rho(\boldsymbol x)=0,
\end{align}
where $L_i=\epsilon_{ijk}x^a_jp^a_k$ is an orbital momentum operator.

First, we introduce some useful notation. The symbol $\boldsymbol \chi$ has $2(N^2-1)$ components $\chi^a_\alpha$. In a sector with $n_F$ fermions there are $n_F$ components of $\boldsymbol \chi$ which have value $1$ and the other elements are zeros. $\boldsymbol\chi$ can be alternatively written as a pair $(\boldsymbol a,\boldsymbol\alpha)$ where $\boldsymbol a=(a_1,\ldots,a_{n_F})$, $\boldsymbol \alpha=(\alpha_1,\ldots,\alpha_{n_F})$ are color and spinor indices for which $\chi^{a_i}_{\alpha_i}=1$. The range of indices is $a_i=1,\ldots,8$, $\alpha_i=\pm\frac{1}{2}$. We assume that pairs $(a_i,\alpha_i)$ are ordered in some way. Then the wavefunction of $\ket{\phi^m}$ can be written as $\braket{\boldsymbol x\boldsymbol a\boldsymbol \alpha|\phi^m}$.

Complex conjugation of the wavefunction can be transformed by using properties of time reversal operation
\begin{align}\begin{split}
\braket{\phi^m|\boldsymbol x\boldsymbol a\boldsymbol\alpha}&={}_T\braket{\boldsymbol x\boldsymbol a\boldsymbol\alpha|\phi^m}_T,\\
\ket{\phi^m}_T&=(-1)^m\ket{\phi^{-m}},\\
\ket{\boldsymbol x\boldsymbol a\boldsymbol\alpha}_T&=(-1)^{m_{\boldsymbol\alpha}}\ket{\boldsymbol x\boldsymbol a-\boldsymbol\alpha}.
\end{split}\end{align}
where $m_{\boldsymbol\alpha}=\sum_i\alpha_i$. Then, the probability density $\rho(\boldsymbol x)$ can be expressed as
\begin{align}\label{eq:density}
\rho(\boldsymbol x)\equiv\sum_{\boldsymbol a\boldsymbol\alpha m}\left|\braket{\boldsymbol x\boldsymbol a\boldsymbol\alpha|\phi^m}\right|^2
&=\sum_{\boldsymbol a\boldsymbol\alpha m}(-1)^{m+m_{\boldsymbol\alpha}}\braket{\boldsymbol x\boldsymbol a\boldsymbol\alpha|\phi^m}\braket{\boldsymbol x\boldsymbol a-\boldsymbol\alpha|\phi^{-m}}.
\end{align}

The orbital angular momentum operator can be written as
\begin{align}
L^2=L_3^2+\frac{1}{2}L_-L_++\frac{1}{2}L_+L_-,
\end{align}
where $L_\pm=L_1\pm iL_2$. From now on indices $i,j$ will take values $\{+,-,3\}$. The angular momentum $L_i$ is a differential operator and it satisfies the Leibniz rule. For any two different states $\ket{\psi_1}$ and $\ket{\psi_2}$ there is
\begin{align}\label{eq:Leibnitz}\begin{split}
L_i(\braket{\boldsymbol x|\psi_1}\braket{\boldsymbol x|\psi_2})&=(L_i\braket{\boldsymbol x|\psi_1})\braket{\boldsymbol x|\psi_2}+\braket{\boldsymbol x|\psi_1}(L_i\braket{\boldsymbol x|\psi_2})\\
&=\braket{\boldsymbol x|L_i|\psi_1}\braket{\boldsymbol x|\psi_2}+\braket{\boldsymbol x|\psi_1}\braket{\boldsymbol x|L_i|\psi_2}.
\end{split}\end{align}
Relation (\ref{eq:Leibnitz}) applied to (\ref{eq:density}) gives
\begin{align}\label{eq:expanding_rho}
\begin{split}
L_iL_j\rho(\boldsymbol x)&=\sum_{\boldsymbol a\boldsymbol\alpha m}(-1)^{m+m_{\boldsymbol\alpha}}\Big(\\&\quad
\braket{\boldsymbol x\boldsymbol a\boldsymbol\alpha|L_iL_j|\phi^m}\braket{\boldsymbol x\boldsymbol a-\boldsymbol\alpha|\phi^{-m}}+
\braket{\boldsymbol x\boldsymbol a\boldsymbol\alpha|L_i|\phi^m}\braket{\boldsymbol x\boldsymbol a-\boldsymbol\alpha|L_j|\phi^{-m}}+\\&\quad
\braket{\boldsymbol x\boldsymbol a\boldsymbol\alpha|L_j|\phi^m}\braket{\boldsymbol x\boldsymbol a-\boldsymbol\alpha|L_i|j-m}+
\braket{\boldsymbol x\boldsymbol a\boldsymbol\alpha|\phi^m}\braket{\boldsymbol x\boldsymbol a-\boldsymbol\alpha|L_iL_j|\phi^{-m}}\Big)
\end{split}
\end{align}
We plug the relation between total and orbital angular momentum, $L_i=J_i-S_i$, into (\ref{eq:expanding_rho}) and gather terms with double $J_i$, double $S_i$ and mixed.
\begin{align}\label{eq:LiLj_big_expression}
\begin{split}
L_iL_j\rho(\boldsymbol x)&=\sum_{\boldsymbol a\boldsymbol\alpha m}(-1)^{m+m_{\boldsymbol\alpha}}\left((\mathcal J\mathcal J)_{ij}+(\mathcal J\mathcal S)_{ij}+(\mathcal S\mathcal S)_{ij}\right).
\end{split}
\end{align}
The corresponding terms are
\begin{align}\begin{split}
(\mathcal J\mathcal J)_{ij}&=\braket{\boldsymbol x\boldsymbol a\boldsymbol\alpha|J_iJ_j|\phi^m}\braket{\boldsymbol x\boldsymbol a-\boldsymbol\alpha|\phi^{-m}}+
\braket{\boldsymbol x\boldsymbol a\boldsymbol\alpha|J_i|\phi^m}\braket{\boldsymbol x\boldsymbol a-\boldsymbol\alpha|J_j|\phi^{-m}}\\&\quad+
\braket{\boldsymbol x\boldsymbol a\boldsymbol\alpha|J_j|\phi^m}\braket{\boldsymbol x\boldsymbol a-\boldsymbol\alpha|J_i|\phi^{-m}}+
\braket{\boldsymbol x\boldsymbol a\boldsymbol\alpha|\phi^m}\braket{\boldsymbol x\boldsymbol a-\boldsymbol\alpha|J_iJ_j|\phi^{-m}},
\end{split}\end{align}
\begin{align}\begin{split}
(\mathcal J\mathcal S)_{ij}&=-\braket{\boldsymbol x\boldsymbol a\boldsymbol\alpha|S_iJ_j|\phi^m}\braket{\boldsymbol x\boldsymbol a-\boldsymbol\alpha|\phi^{-m}}-
\braket{\boldsymbol x\boldsymbol a\boldsymbol\alpha|J_iS_j|\phi^m}\braket{\boldsymbol x\boldsymbol a-\boldsymbol\alpha|\phi^{-m}}\\&\quad-
\braket{\boldsymbol x\boldsymbol a\boldsymbol\alpha|S_i|\phi^m}\braket{\boldsymbol x\boldsymbol a-\boldsymbol\alpha|J_j|\phi^{-m}}-
\braket{\boldsymbol x\boldsymbol a\boldsymbol\alpha|J_i|\phi^m}\braket{\boldsymbol x\boldsymbol a-\boldsymbol\alpha|S_j|\phi^{-m}}\\&\quad-
\braket{\boldsymbol x\boldsymbol a\boldsymbol\alpha|S_j|\phi^m}\braket{\boldsymbol x\boldsymbol a-\boldsymbol\alpha|J_i|\phi^{-m}}-
\braket{\boldsymbol x\boldsymbol a\boldsymbol\alpha|J_j|\phi^m}\braket{\boldsymbol x\boldsymbol a-\boldsymbol\alpha|S_i|\phi^{-m}}\\&\quad-
\braket{\boldsymbol x\boldsymbol a\boldsymbol\alpha|\phi^m}\braket{\boldsymbol x\boldsymbol a-\boldsymbol\alpha|S_iJ_j|\phi^{-m}}-
\braket{\boldsymbol x\boldsymbol a\boldsymbol\alpha|\phi^m}\braket{\boldsymbol x\boldsymbol a-\boldsymbol\alpha|J_iS_j|\phi^{-m}},
\end{split}\end{align}
\begin{align}\begin{split}
(\mathcal S\mathcal S)_{ij}&=\braket{\boldsymbol x\boldsymbol a\boldsymbol\alpha|S_iS_j|\phi^m}\braket{\boldsymbol x\boldsymbol a-\boldsymbol\alpha|\phi^{-m}}+
\braket{\boldsymbol x\boldsymbol a\boldsymbol\alpha|S_i|\phi^m}\braket{\boldsymbol x\boldsymbol a-\boldsymbol\alpha|S_j|\phi^{-m}}\\&\quad+
\braket{\boldsymbol x\boldsymbol a\boldsymbol\alpha|S_j|\phi^m}\braket{\boldsymbol x\boldsymbol a-\boldsymbol\alpha|S_i|\phi^{-m}}+
\braket{\boldsymbol x\boldsymbol a\boldsymbol\alpha|\phi^m}\braket{\boldsymbol x\boldsymbol a-\boldsymbol\alpha|S_iS_j|\phi^{-m}}.
\end{split}\end{align}

Action of spin operators on kets $\ket{\boldsymbol x\boldsymbol a\boldsymbol\alpha}$ can be easily read from (\ref{eq:angular_momentum})
\begin{align}\label{eq:spin_action}
S_3\ket{\boldsymbol x\boldsymbol a\boldsymbol\alpha}&=\frac{1}{2}\sum_i\alpha_i\ket{\boldsymbol x\boldsymbol a\boldsymbol\alpha},\\
S_+\ket{\boldsymbol x\boldsymbol a\boldsymbol\alpha}&=\sum_k \delta_{\alpha_k,-1/2}\ket{\boldsymbol x\boldsymbol a\boldsymbol\alpha_+^k},\\
S_-\ket{\boldsymbol x\boldsymbol a\boldsymbol\alpha}&=\sum_k \delta_{\alpha_k,1/2}\ket{\boldsymbol x\boldsymbol a\boldsymbol\alpha_-^k}.
\end{align}
where
\begin{align}
\boldsymbol \alpha_\pm^k=(\alpha_1,\ldots,\alpha_k\pm\frac{1}{2},\ldots,\alpha_{n_F}).
\end{align}

In the following we consider $(\mathcal J\mathcal J)_{ij}$, $(\mathcal J\mathcal S)_{ij}$ and $(\mathcal S\mathcal S)_{ij}$ separately. Moreover, in each case the sum over $(i,j)$ splits into $(i,j)=(3,3)$ and $(i,j)=(+,-),(-,+)$ and these vanish independently. For clarity, further considerations are organized in paragraphs.

\paragraph{\large Terms with double $J_i$.}
\begin{align}
&\sum_{m}(-1)^{m+m_{\boldsymbol\alpha}}(\mathcal J\mathcal J)_{33}\\
&=\left(m^2+m(-m)+m(-m)+(-m)^2\right)\braket{\boldsymbol x\boldsymbol a\boldsymbol\alpha|\phi^{m}}\braket{\boldsymbol x\boldsymbol a-\boldsymbol\alpha|\phi^{-m}}=0.
\end{align}
Before we consider $(\mathcal J\mathcal J)_{ij}$ with $i,j=\pm$, recall that
\begin{align}\begin{split}
J_{\pm}\ket{j,m}=\sqrt{j(j+1)-m(m\pm1)}\ket{j,m\pm1}.
\end{split}\end{align}
\begin{align}
\begin{split}
&(\mathcal J\mathcal J)_{-+}+(\mathcal J\mathcal J)_{+-}\\
&=\sum_{m}(-1)^{m}\Big(\\
&\quad\frac{1}{2}(j(j+1)-m(m-1)+j(j+1)-m(m+1))\braket{\boldsymbol x\boldsymbol a\boldsymbol\alpha|\phi^m}\braket{\boldsymbol x\boldsymbol a-\boldsymbol\alpha|\phi^{-m}}\\&\quad+
(j(j+1)-m(m+1))\braket{\boldsymbol x\boldsymbol a\boldsymbol\alpha|\phi^{m+1}}\braket{\boldsymbol x\boldsymbol a-\boldsymbol\alpha|\phi^{-(m+1)}}\\&\quad+
(j(j+1)-m(m-1))\braket{\boldsymbol x\boldsymbol a\boldsymbol\alpha|\phi^{m-1}}\braket{\boldsymbol x\boldsymbol a-\boldsymbol\alpha|\phi^{-(m-1)}}\\&\quad+
\frac{1}{2}(j(j+1)-m(m+1)+j(j+1)-m(m-1))\braket{\boldsymbol x\boldsymbol a\boldsymbol\alpha|jm}\braket{\boldsymbol x\boldsymbol a-\boldsymbol\alpha|\phi^{-m}}\Big)\\
&=\sum_{m}(-1)^{m}\Big(
2(j(j+1)-m^2)\braket{\boldsymbol x\boldsymbol a\boldsymbol\alpha|\phi^m}\braket{\boldsymbol x\boldsymbol a-\boldsymbol\alpha|\phi^{-m}}\\&\quad+
(j(j+1)-m(m+1))\braket{\boldsymbol x\boldsymbol a\boldsymbol\alpha|\phi^{m+1}}\braket{\boldsymbol x\boldsymbol a-\boldsymbol\alpha|\phi^{-(m+1)}}\\&\quad+
(j(j+1)-m(m-1))\braket{\boldsymbol x\boldsymbol a\boldsymbol\alpha|\phi^{m-1}}\braket{\boldsymbol x\boldsymbol a-\boldsymbol\alpha|\phi^{-(m-1)}}\Big)\\
&=\sum_{m}(-1)^{m}\Big(
2(j(j+1)-m^2)\braket{\boldsymbol x\boldsymbol a\boldsymbol\alpha|\phi^m}\braket{\boldsymbol x\boldsymbol a-\boldsymbol\alpha|\phi^{-m}}\\&\quad-
(j(j+1)-(m-1)m)\braket{\boldsymbol x\boldsymbol a\boldsymbol\alpha|\phi^m}\braket{\boldsymbol x\boldsymbol a-\boldsymbol\alpha|\phi^{-m}}\\&\quad-
(j(j+1)-(m+1)m)\braket{\boldsymbol x\boldsymbol a\boldsymbol\alpha|\phi^m}\braket{\boldsymbol x\boldsymbol a-\boldsymbol\alpha|\phi^{-m})}\Big)\\
&=0.
\end{split}
\end{align}
In the third equality terms with the same $m$ were gathered, hence additional minus sign from $(-1)^m$.

\paragraph{\large Mixed terms.}

Note that $[J_3,S_3]=0$.
\begin{align}\begin{split}
(\mathcal J\mathcal S)_{33}
&=-m\braket{\boldsymbol x\boldsymbol a\boldsymbol\alpha|S_3|\phi^m}\braket{\boldsymbol x\boldsymbol a-\boldsymbol\alpha|\phi^{-m}}
-m\braket{\boldsymbol x\boldsymbol a\boldsymbol\alpha|S_3|\phi^m}\braket{\boldsymbol x\boldsymbol a-\boldsymbol\alpha|\phi^{-m}}\\&\quad
+m\braket{\boldsymbol x\boldsymbol a\boldsymbol\alpha|S_3|\phi^m}\braket{\boldsymbol x\boldsymbol a-\boldsymbol\alpha|\phi^{-m}}
-m\braket{\boldsymbol x\boldsymbol a\boldsymbol\alpha|\phi^m}\braket{\boldsymbol x\boldsymbol a-\boldsymbol\alpha|S_3|\phi^{-m}}\\&\quad
+m\braket{\boldsymbol x\boldsymbol a\boldsymbol\alpha|S_3|\phi^m}\braket{\boldsymbol x\boldsymbol a-\boldsymbol\alpha|\phi^{-m}}
-m\braket{\boldsymbol x\boldsymbol a\boldsymbol\alpha|\phi^m}\braket{\boldsymbol x\boldsymbol a-\boldsymbol\alpha|S_3|\phi^{-m}}\\&\quad
+m\braket{\boldsymbol x\boldsymbol a\boldsymbol\alpha|\phi^m}\braket{\boldsymbol x\boldsymbol a-\boldsymbol\alpha|S_3|\phi^{-m}}
+m\braket{\boldsymbol x\boldsymbol a\boldsymbol\alpha|\phi^m}\braket{\boldsymbol x\boldsymbol a-\boldsymbol\alpha|S_3|\phi^{-m}}\\
&=0
\end{split}\end{align}
For $(i,j)=(+-),(-,+)$ in all brackets of the form $\braket{\boldsymbol x\boldsymbol a\boldsymbol\alpha|J_iS_j|\psi}$ we would like to move $S_j$ to the left and use (\ref{eq:spin_action}). Therefore, we need commutators of $S_j$ with $J_i$. The required relation is
\begin{align}\begin{split}
J_+S_-+J_-S_+&=S_-J_++[J_+,S_-]+S_+J_-+[J_-,S_+]\\
&=S_-J_++[S_+,S_-]+S_+J_-+[S_-,S_+]\\
&=S_-J_++S_+J_-.
\end{split}\end{align}

\begin{align}
\begin{split}
\frac{1}{2}\sum_m(-1)^m(&(\mathcal J\mathcal S)_{-+}+(\mathcal J\mathcal S)_{+-})\\
=\sum_m(-1)^m\Big(
&-\sqrt{j(j+1)-m(m+1)}\braket{\boldsymbol x\boldsymbol a\boldsymbol\alpha|S_-|\phi^{m+1}}\braket{\boldsymbol x\boldsymbol a-\boldsymbol\alpha|\phi^{-m}}\\
&-\sqrt{j(j+1)-m(m-1)}\braket{\boldsymbol x\boldsymbol a\boldsymbol\alpha|S_+|\phi^{m-1}}\braket{\boldsymbol x\boldsymbol a-\boldsymbol\alpha|\phi^{-m}}\\
&-\sqrt{j(j+1)-m(m-1)}\braket{\boldsymbol x\boldsymbol a\boldsymbol\alpha|S_-|\phi^m}\braket{\boldsymbol x\boldsymbol a-\boldsymbol\alpha|\phi^{-(m-1)}}\\
&-\sqrt{j(j+1)-m(m+1)}\braket{\boldsymbol x\boldsymbol a\boldsymbol\alpha|S_+|\phi^m}\braket{\boldsymbol x\boldsymbol a-\boldsymbol\alpha|\phi^{-(m+1)}}\\
&-\sqrt{j(j+1)-m(m+1)}\braket{\boldsymbol x\boldsymbol a\boldsymbol\alpha|\phi^{m+1}}\braket{\boldsymbol x\boldsymbol a-\boldsymbol\alpha|S_-|\phi^{-m}}\\
&-\sqrt{j(j+1)-m(m-1)}\braket{\boldsymbol x\boldsymbol a\boldsymbol\alpha|\phi^{m-1}}\braket{\boldsymbol x\boldsymbol a-\boldsymbol\alpha|S_+|\phi^{-m}}\\
&-\sqrt{j(j+1)-m(m-1)}\braket{\boldsymbol x\boldsymbol a\boldsymbol\alpha|\phi^m}\braket{\boldsymbol x\boldsymbol a-\boldsymbol\alpha|S_-|\phi^{-(m-1)}}\\
&-\sqrt{j(j+1)-m(m+1)}\braket{\boldsymbol x\boldsymbol a\boldsymbol\alpha|\phi^m}\braket{\boldsymbol x\boldsymbol a-\boldsymbol\alpha|S_+|\phi^{-(m+1)}}\Big)\\
=\sum_m(-1)^m\Big(
&+\sqrt{j(j+1)-m(m-1)}\braket{\boldsymbol x\boldsymbol a\boldsymbol\alpha|S_-|\phi^m}\braket{\boldsymbol x\boldsymbol a-\boldsymbol\alpha|\phi^{-(m-1)}}\\
&+\sqrt{j(j+1)-m(m+1)}\braket{\boldsymbol x\boldsymbol a\boldsymbol\alpha|S_+|\phi^m}\braket{\boldsymbol x\boldsymbol a-\boldsymbol\alpha|\phi^{-(m+1)}}\\
&-\sqrt{j(j+1)-m(m-1)}\braket{\boldsymbol x\boldsymbol a\boldsymbol\alpha|S_-|\phi^m}\braket{\boldsymbol x\boldsymbol a-\boldsymbol\alpha|\phi^{-(m-1)}}\\
&-\sqrt{j(j+1)-m(m+1)}\braket{\boldsymbol x\boldsymbol a\boldsymbol\alpha|S_+|\phi^m}\braket{\boldsymbol x\boldsymbol a-\boldsymbol\alpha|\phi^{-(m+1)}}\\
&+\sqrt{j(j+1)-m(m-1)}\braket{\boldsymbol x\boldsymbol a\boldsymbol\alpha|\phi^m}\braket{\boldsymbol x\boldsymbol a-\boldsymbol\alpha|S_-|\phi^{-(m-1)}}\\
&+\sqrt{j(j+1)-m(m+1)}\braket{\boldsymbol x\boldsymbol a\boldsymbol\alpha|\phi^m}\braket{\boldsymbol x\boldsymbol a-\boldsymbol\alpha|S_+|\phi^{-(m+1)}}\\
&-\sqrt{j(j+1)-m(m-1)}\braket{\boldsymbol x\boldsymbol a\boldsymbol\alpha|\phi^m}\braket{\boldsymbol x\boldsymbol a-\boldsymbol\alpha|S_-|\phi^{-(m-1)}}\\
&-\sqrt{j(j+1)-m(m+1)}\braket{\boldsymbol x\boldsymbol a\boldsymbol\alpha|\phi^m}\braket{\boldsymbol x\boldsymbol a-\boldsymbol\alpha|S_+|\phi^{-(m+1)}}\Big)\\
&=0.
\end{split}
\end{align}

\paragraph{\large Terms with double $S_i$.}
In $(\mathcal S\mathcal S)_{33}$ each $S_3$ acts to the left giving a constant $m_{\boldsymbol\alpha}$ or $-m_{\boldsymbol\alpha}$. Therefore,
\begin{align}
(\mathcal S\mathcal S)_{33}=(m_{\boldsymbol\alpha}^2-m_{\boldsymbol\alpha}^2-m_{\boldsymbol\alpha}^2+(-m_{\boldsymbol\alpha})^2)\braket{\boldsymbol x\boldsymbol a\boldsymbol\alpha|\phi^m}\braket{\boldsymbol x\boldsymbol a-\boldsymbol\alpha|\phi^{-m}}=0
\end{align}
Finally,
\begin{align}
\begin{split}
&\sum_{\boldsymbol a\boldsymbol\alpha}(-1)^{m_{\boldsymbol\alpha}}\left((\mathcal S\mathcal S)_{+-}+(\mathcal S\mathcal S)_{-+}\right)\\
&=\sum_{\boldsymbol a\boldsymbol\alpha}(-1)^{m_{\boldsymbol\alpha}}\Big(\\&\quad
+\frac{1}{2}\braket{\boldsymbol x\boldsymbol a\boldsymbol\alpha|S_+^\dagger S_-^\dagger |\phi^m}\braket{\boldsymbol x\boldsymbol a-\boldsymbol\alpha|\phi^{-m}}
+\frac{1}{2}\braket{\boldsymbol x\boldsymbol a\boldsymbol\alpha|S_-^\dagger S_+^\dagger |\phi^m}\braket{\boldsymbol x\boldsymbol a-\boldsymbol\alpha|\phi^{-m}}\\&\quad
+\frac{1}{2}\braket{\boldsymbol x\boldsymbol a\boldsymbol\alpha|\phi^m}\braket{\boldsymbol x\boldsymbol a-\boldsymbol\alpha|S_+^\dagger S_-^\dagger |\phi^{-m}}
+\frac{1}{2}\braket{\boldsymbol x\boldsymbol a\boldsymbol\alpha|\phi^m}\braket{\boldsymbol x\boldsymbol a-\boldsymbol\alpha|S_-^\dagger S_+^\dagger |\phi^{-m}}\\&\quad
+\braket{\boldsymbol x\boldsymbol a\boldsymbol\alpha|S_+^\dagger |\phi^m}\braket{\boldsymbol x\boldsymbol a-\boldsymbol\alpha|S_-^\dagger |\phi^{-m}}
+\braket{\boldsymbol x\boldsymbol a\boldsymbol\alpha|S_-^\dagger |\phi^m}\braket{\boldsymbol x\boldsymbol a-\boldsymbol\alpha|S_+^\dagger |\phi^{-m}}\Big)
\\
&=\sum_{\boldsymbol a\boldsymbol\alpha}(-1)^{m_{\boldsymbol\alpha}}\Big(\\&\quad
+\frac{1}{2}\sum_{kl}\braket{\boldsymbol x\boldsymbol a(\boldsymbol\alpha_+^k)_-^l|\phi^m}\braket{\boldsymbol x\boldsymbol a-\boldsymbol\alpha|\phi^{-m}}
+\frac{1}{2}\sum_{kl}\braket{\boldsymbol x\boldsymbol a(\boldsymbol\alpha_-^k)_+^l|\phi^m}\braket{\boldsymbol x\boldsymbol a-\boldsymbol\alpha|\phi^{-m}}\\&\quad
+\frac{1}{2}\sum_{kl}\braket{\boldsymbol x\boldsymbol a\boldsymbol\alpha|\phi^m}\braket{\boldsymbol x\boldsymbol a-(\boldsymbol\alpha_-^k)_+^l|\phi^{-m}}
+\frac{1}{2}\sum_{kl}\braket{\boldsymbol x\boldsymbol a\boldsymbol\alpha|\phi^m}\braket{\boldsymbol x\boldsymbol a-(\boldsymbol\alpha_+^k)_-^l|\phi^{-m}}\\&\quad
+\sum_{kl}\braket{\boldsymbol x\boldsymbol a\boldsymbol\alpha_+^k|\phi^m}\braket{\boldsymbol x\boldsymbol a-(\boldsymbol\alpha_+^l)|\phi^{-m}}
+\sum_{kl}\braket{\boldsymbol x\boldsymbol a\boldsymbol\alpha_-^k|\phi^m}\braket{\boldsymbol x\boldsymbol a-(\boldsymbol\alpha_-^l)|\phi^{-m}}\Big)
\\
&=\sum_{\boldsymbol a\boldsymbol\alpha}(-1)^{m_{\boldsymbol\alpha}}\Big(\\&\quad
+\frac{1}{2}\sum_{kl}\braket{\boldsymbol x\boldsymbol a(\boldsymbol\alpha_+^k)_-^l|\phi^m}\braket{\boldsymbol x\boldsymbol a-\boldsymbol\alpha|\phi^{-m}}
+\frac{1}{2}\sum_{kl}\braket{\boldsymbol x\boldsymbol a(\boldsymbol\alpha_-^k)_+^l|\phi^m}\braket{\boldsymbol x\boldsymbol a-\boldsymbol\alpha|\phi^{-m}}\\&\quad
+\frac{1}{2}\sum_{kl}\braket{\boldsymbol x\boldsymbol a(\boldsymbol\alpha_+^k)_-^l|\phi^m}\braket{\boldsymbol x\boldsymbol a-\boldsymbol\alpha|\phi^{-m}}
+\frac{1}{2}\sum_{kl}\braket{\boldsymbol x\boldsymbol a(\boldsymbol\alpha_-^k)_+^l|\phi^m}\braket{\boldsymbol x\boldsymbol a-\boldsymbol\alpha|\phi^{-m}}\\&\quad
-\sum_{kl}\braket{\boldsymbol x\boldsymbol a(\boldsymbol\alpha_-^l)_+^k|\phi^m}\braket{\boldsymbol x\boldsymbol a-\boldsymbol\alpha|\phi^{-m}}
-\sum_{kl}\braket{\boldsymbol x\boldsymbol a(\boldsymbol\alpha_+^l)_-^k|\phi^m}\braket{\boldsymbol x\boldsymbol a-\boldsymbol\alpha|\phi^{-m}}\Big)
\\
&=0.
\end{split}
\end{align}
In the third equality the sum over $\boldsymbol a$ and $\boldsymbol\alpha$ was reorganized so that there is an identical bracket $\braket{\boldsymbol x\boldsymbol a-\boldsymbol\alpha|\phi^{-m}}$ in each term. To this end the variables were changed in the following way. In the third term
\begin{align}
\boldsymbol\alpha\rightarrow\boldsymbol\alpha'=(\boldsymbol\alpha_+^k)_-^l.
\end{align}
Then,
\begin{align}
(\boldsymbol\alpha'{}_-^k)_+^l=\boldsymbol\alpha,\ m_{\boldsymbol\alpha'}=m_{\boldsymbol\alpha}.
\end{align}
A opposite change was done in the fourth term. In the fifth term
\begin{align}
\boldsymbol\alpha\rightarrow\boldsymbol\alpha''=\boldsymbol\alpha_-^l,
\end{align}
and then
\begin{align}
\boldsymbol\alpha''{}_+^l=\boldsymbol\alpha.
\end{align}
An opposite change was performed in the sixth component. The ket $\ket{\boldsymbol x\boldsymbol a\boldsymbol\alpha''}$ is created by one less operator $f_+^\dagger$ and one more operator $f_-^\dagger$ than $\ket{\boldsymbol x\boldsymbol a\boldsymbol \alpha}$. Therefore, $\ m_{\boldsymbol\alpha''}=m_{\boldsymbol\alpha}-1$ and $\ (-1)^{m_{\boldsymbol\alpha'}}=-(-1)^{m_{\boldsymbol\alpha}}$, hence additional minus sign.

We proved than all terms in (\ref{eq:LiLj_big_expression}) vanish and therefore $L^2\rho(\boldsymbol x)=0$. The proof is now complete.

\chapter{Gauge fixing for $SU(3)$}\label{ap:fixing_gauge}
\newcommand*\mycirc[1]{%
\begin{minipage}[m]{20pt}
  \begin{tikzpicture}
    \node[circle, draw, text centered, font=\small] {#1};
  \end{tikzpicture}
  \end{minipage}
}

In \ref{sec:fixing_gauge} we consider gauge fixing for a set of three bosonic variables $X_i$. Each $X_i$ is a hermitean, traceless $3\times3$ matrix. The gauge transformation is $X_i\rightarrow UX_iU^\dagger$, where $U$ is an element of $SU(3)$ in the fundamental representation. In \ref{sec:fixing_gauge} we shown how to completely fix the gauge in the generic case. Gauge fixing can be represented by conditions coordinates $x_i^a$ (cf. (\ref{eq:gauge_constraints})). There are however certain configurations $X_i$ which have to be addressed separately. This is because in these cases there is come residual gauge freedom left. This freedom leads e.g. to identification of false flat valleys of the potential (c.f. (\ref{eq:false_valley})). In this appendix all such cases are discussed.

The procedure of gauge fixing is algorithmic. That is, it consists several steps. At each step there is some gauge symmetry available. It is then used to transform $X_i$ so that they satisfy some additional constraints. The gauge freedom at each step depends on particular configuration of $X_i$. For example, in the first step $X_1$ is diagonalized. In the generic case it is then invariant under gauge transformation only with $U=\mathrm{diag}(e^{i\phi},e^{i\psi},e^{-i\phi-i\psi})$. However, if $X_1=0$, then it is invariant under gauge transformation with general $U\in SU(3)$. In Tab. \ref{tab:gauge_freedom_notation} we introduce notation for residual gauge symmetries which will be used in what follows.

\setlength\bigstrutjot{15pt}
\begin{table}[h]
\centering
\begin{tabular}{|c|c||c|c|}
\hline\bigstrut\mycirc{A}&$U$ -- general $SU(3)$ matrix&\mycirc{E}&\(\displaystyle U=diag(e^{i\phi},e^{i\phi},e^{-2i\phi})\)\\
\hline\bigstrut\mycirc{B}&\(\displaystyle U=\left(\begin{array}{cc}U_2&0\\0&(detU_2)^*\end{array}\right)\)&\mycirc{F}&\(\displaystyle U=diag(e^{-2i\phi},e^{i\phi},e^{i\phi})\)\\
\hline\bigstrut\mycirc{C}&\(\displaystyle U=\left(\begin{array}{cc}(detU_2)^*&0\\0&U_2\end{array}\right)\)&\mycirc{G}&\(\displaystyle U=diag(e^{i\phi},e^{-2i\phi},e^{i\phi})\)\\
\hline\bigstrut\mycirc{D}&\(\displaystyle U=diag(e^{i\phi},e^{i\psi},e^{-i\phi-i\psi})\)&&\\
\hline
\end{tabular}
\caption{Notations for residual gauge freedom in midsteps in the procedure of gauge fixing. $U_2$ stands for a general, $SU(2)$ matrix.}
\label{tab:gauge_freedom_notation}
\end{table}

Recall how the gauge is fixed in the generic case. First, one diagonalizes $X_1$ and sets the diagonal elements in increasing order. The residual gauge freedom is  \mycirc{D}. The diagonal matrix $X_1$ is invariant under transformations with such $U$. The gauge transformation is then used to set two matrix elements $(X_2)_{12}$ and $(X_2)_{23}$ real and positive. This fixes the gauge completely.

In \ref{sec:fixing_gauge} we considered three basic degenerate cases. They are summarized in Tab. \ref{tab:special_cases} together with the residual gauge freedom and how one can use it to impose further constraints. There are certainly many more different cases. Nevertheless, there are usually similar to one of the three basic cases.
\setlength\bigstrutjot{5pt}
\begin{table}[h]
\centering
\begin{tabularx}{.9\textwidth}{|X|c|X|}
\hline
special case&residual gauge freedom&action\\\hline
$X_1=0$&\mycirc{A}&use $U$ to diagonalize $X_2$ and then treat $X_3$ as $X_2$ in the generic case\\\hline
after diagonalization of $X_1$ two eigenvalues are equal, e.g. $(X_1)_{11}=(X_1)_{22}$&\mycirc{B}&diagonalize upper--left 2x2 submatrix of $X_2$ and set $(X_2)_{13},\ (X_2)_{23}$ positive\\\hline
after diagonalization of $X_1$ matrix element $(X_2)_{12}$ vanishes&\mycirc{D}&set $(X_2)_{13}$ and $(X_2)_{23}$ positive\\\hline
\end{tabularx}
\caption{The basic degenerate cases in gauge fixing procedure.}
\label{tab:special_cases}
\end{table}

Finally, we proceed to describing the full algorithm. At each step $X_i$ obey some constraints. There is a residual gauge symmetry that does not violate these constraints. The residual symmetry depends on both, constraints and particular configuration $X_i$. The gauge symmetry is then used to perform some action (such as to diagonalize a matrix) so that $X_i$ are subject to stricter constraints. This leaves a smaller symmetry group which is then used in next step. Different residual gauge symmetries allow for different sets of actions. In each case an action with highest priority is chosen. The list of possible actions is
\begin{enumerate}
\item diagonalize $X_i$,
\item diagonalize a $2\times2$ upper--left or lower--right submatrix of $X_i$,
\item set an off--diagonal matrix element to zero; $(X_i)_{12}$ has priority over $(X_i)_{23}$ and $(X_i)_{23}$ has priority over $(X_i)_{13}$,
\item set an off--diagonal matrix element positive, $(X_i)_{12}$ has priority over $(X_i)_{23}$ and $(X_i)_{23}$ has priority over $(X_i)_{13}$.
\end{enumerate}
First action has highest priority and the last has lowest. Each action concerns only one of the three matrices $X_i$. We say that $X_1$ has always priority over $X_2$ and $X_2$ has priority over $X_3$. That is, if the same action is possible for $X_1$ and $X_2$ then it is done for $X_1$. By diagonalization we mean diagonalization with setting eigenvalues on the diagonal in increasing order.

The full procedure of gauge fixing is presented in a flowchart, Figs \ref{fig:flowchart1} and \ref{fig:flowchart2}. Interpretation of the flowchart is following. The pink rectangle is the first step. Rectangles represent new constraints that are imposed on matrices $X_i$. These conditions can be always fulfilled by performing an action on $X_i$ which is allowed by the residual symmetry group. Each diamond represents a question. If the condition in a diamond node is fulfilled, then there is a degeneracy and the 'yes' arrow has to be followed. Otherwise, one has to follow the 'no' arrow. The blue triangle in Fig. \ref{fig:flowchart1} means that one should move to the triangle in Fig. \ref{fig:flowchart2}. Some arrows have additional symbols, \mycirc{A},\ldots,\mycirc{G}. These denote the residual gauge transformations at given stage. For brevity we use $X=X_1,\ Y=X_2,\ Z=X_3$. Moreover, 'diagonalize UL of M' means: 'diagonalize $2\times2$ submatrix in the upper-left corner of M'. Similarly, LR stands for 'the lower-right submatrix'.

\newlength{\horsep}
\setlength{\horsep}{\textwidth/4}
\newlength{\versep}
\setlength{\versep}{.6\horsep}
\newlength{\elwid}
\setlength{\elwid}{.7\horsep}
\newlength{\texwid}
\setlength{\texwid}{.7\elwid}
\newlength{\ssep}
\setlength{\ssep}{.05\versep}
\newlength{\bulhsep}
\setlength{\bulhsep}{.5\horsep}
\newlength{\bulvsep}
\setlength{\bulvsep}{.5\versep}

\tikzstyle{proc} = [rectangle, draw, text width = \texwid, minimum width = \elwid, text centered, inner sep = 3pt, font=\normalsize]
\tikzstyle{ques} = [diamond, draw, text width = \texwid, minimum width = \elwid, text centered, aspect=1.5, inner sep = 1pt, font=\normalsize]
\tikzstyle{opt} = [circle, draw, text centered, font=\footnotesize]
\tikzstyle{tri} = [draw, shape border rotate=180, regular polygon, regular polygon sides=3, minimum height=.7\elwid, font=\normalsize, fill=blue!20]

\begin{sidewaysfigure}
\begin{center}
\begin{tikzpicture}[xscale=1, line/.style={draw, ->, >=latex, very thick}, end/.style={draw, -*, very thick}]
\node[proc, thick, fill=pink](start){diagonalize X};
\node[ques, below = \versep of start.center, anchor = center] (x1){$X=0$};
\node[proc, right = \horsep of x1.center, anchor = center]    (x1Y){diagonalize $Y$};
\node[ques, right = \horsep of x1Y.center, anchor = center]   (x1Yy){$Y=0$};
\node[proc, above = \versep of x1Yy.center, anchor = center]  (x1YyZ){diagonalize $Z$};
\node[ques, right = \horsep of x1Yy.center, anchor = center]  (y1){$Y_{11}=Y_{22}$};
\node[proc, above = \versep of y1.center, anchor = center]    (y1Z){diagonalize UL of $Z$};
\node[proc, right = \horsep of y1Z.center, anchor = center]   (y1ZZ){$Z_{13}\geq0$ $Z_{23}\geq0$};
\node[ques, below = \versep of y1.center, anchor = center]    (y1y){$Y_{22}=Y_{33}$};
\node[proc, right = \horsep of y1y.center, anchor = center]   (y1yZ1){diagonalize LR of $Z$};
\node[proc, above = \versep of y1yZ1.center, anchor = center] (y1yZ1Z){$Z_{12}\geq0$ $Z_{13}\geq0$};
\node[proc, below = \versep of y1y.center, anchor = center]   (y1yZ2){$Z_{12}\geq0$ $Z_{23}\geq0$};
\node[ques, right = \horsep of y1yZ2.center, anchor = center] (y1yZ2z){$Z_{12}=0$ or $Z_{23}=0$};
\node[proc, below = \versep of y1yZ2z.center, anchor = center](y1yZ2zZ){$Z_{13}\geq0$};
\node[ques, below = \versep of x1.center, anchor = center]    (x2){$X_{11}=X_{22}$};
\node[tri, below = \versep of x2.center, anchor = center]     (portal){};
\node[proc, right = \horsep of x2.center, anchor = center]    (x2Y){diagonalize UL of $Y$};
\node[ques, below = \versep of x2Y.center, anchor = center]   (x2Yy){$Y_{11}=Y_{22}$};
\node[proc, below = \versep of x2Yy.center, anchor = center]  (x2YyZ){diagonalize UL of $Z$};
\node[ques, right = \horsep of x2YyZ.center, anchor = center] (z1){$Z_{11}=Z_{22}$};
\node[proc, below = \versep of z1.center, anchor = center]    (z1Y){$Y_{23}=0$};
\node[ques, left  = \horsep of z1Y.center, anchor = center]   (z1Yy){$Y_{13}=0$};
\node[proc, left  = \horsep of z1Yy.center, anchor = center]  (z1YyZ){$Z_{23}=0$ $Z_{12}\geq0$};
\node[proc, below = \versep of z1Yy.center, anchor = center]  (z1YyY){$Y_{13}\geq0$ $Z_{23}\geq0$};
\node[proc, right = \horsep of z1.center, anchor = center]    (z1Y2){$Y_{13}\geq0$ $Y_{23}\geq0$};
\node[ques, below = \versep of z1Y2.center, anchor = center]  (y2){$Y_{23}=0$};
\node[proc, below = \versep of y2.center, anchor = center]    (y2Z){$Z_{23}\geq0$};
\node[ques, right = \horsep of y2Z.center, anchor = center]   (y2y){$Y_{13}=0$};
\node[proc, above = \versep of y2y.center, anchor = center]   (y2yZ){$Z_{13}\geq0$};
\path[line] (start)--(x1);
\path[line] (x1)--(x1Y) node [midway, above]{yes} node[midway, below = \ssep, opt]{A};
\path[line] (x1Y)--(x1Yy);
\path[line] (x1Yy)--(x1YyZ) node [midway, right]{yes} node[midway, left = \ssep, opt]{A};
\path[end]  (x1YyZ)--++(\horsep/2,0);
\path[line] (x1Yy)--(y1) node [midway, above]{no};
\path[line] (y1)--(y1Z) node [midway, right]{yes} node[midway, left = \ssep, opt]{B};
\path[line] (y1Z)--(y1ZZ) node[midway, below = \ssep, opt]{D};
\path[end]  (y1ZZ)--++(\bulhsep,0);
\path[line] (y1)--(y1y) node [midway, right]{no};
\path[line] (y1y)--(y1yZ1) node [midway, above]{yes} node[midway, below = \ssep, opt]{C};
\path[line] (y1yZ1)--(y1yZ1Z) node[midway, left = \ssep, opt]{D};
\path[end]  (y1yZ1Z)--++(\bulhsep,0);
\path[line] (y1y)--(y1yZ2) node [midway, right]{no} node[midway, left = \ssep, opt]{D};
\path[line] (y1yZ2)--(y1yZ2z);
\path[end]  (y1yZ2z)--++(\horsep/2,0) node [midway, above]{no};
\path[line] (y1yZ2z)--(y1yZ2zZ) node [midway, right]{yes};
\path[end]  (y1yZ2zZ)--++(\horsep/2,0);
\path[line] (x1)--(x2) node [midway, right]{no};
\path[line] (x2)--(portal) node [midway, right]{no};
\path[line] (x2)--(x2Y) node [midway, above]{yes} node[midway, below = \ssep, opt]{B};
\path[line] (x2Y)--(x2Yy);
\path[line] (x2Yy)--(x2YyZ) node [midway, right]{yes} node[midway, left = \ssep, opt]{B};
\path[line] (x2YyZ)--(z1);
\path[line] (z1)--(z1Y) node [midway, right]{yes} node[midway, left = \ssep, opt]{B};
\path[line] (z1Y)--(z1Yy);
\path[line] (z1Yy)--(z1YyY) node [midway, right]{no} node[midway, left = \ssep, opt]{G};
\path[line] (z1Yy)--(z1YyZ) node [midway, above]{yes} node[midway, below = \ssep, opt]{B};
\path[end]  (z1YyY)--++(0,-\bulvsep);
\path[end]  (z1YyZ)--++(0,-\bulvsep);
\path[line] (z1)--(z1Y2) node [midway, above](no){no} node[midway, below = \ssep, opt]{D};
\path[line] (x2Yy) -- node [midway, above]{no} ++(\horsep+\bulhsep,0)--(no);
\path[line] (z1Y2)--(y2);
\path[line] (y2)--(y2Z) node [midway, right]{yes};
\path[line] (y2Z)--(y2y) node [midway](here){};
\path[line] (y2)--++(\bulhsep,0)-- node [midway,right]{no} ++(0,-\versep);
\path[line] (y2y)--(y2yZ) node [midway, right]{yes};
\path[end] (y2y)--++(\bulhsep,0) node [midway, above]{no};
\path[end] (y2yZ)--++(\bulhsep,0);
\end{tikzpicture}
\end{center}
\caption{Flowchart of procedure of fixing the gauge, part1.}
\label{fig:flowchart1}
\end{sidewaysfigure}

\begin{sidewaysfigure}
\begin{center}
\begin{tikzpicture}[xscale=1, line/.style={draw, ->, >=latex, very thick}, end/.style={draw, -*, very thick}]
\node[tri](portal){};
\node[ques, below = \versep of portal.center, anchor = center] (x){$X_{22}=X_{33}$};
\node[proc, right = \horsep of x.center, anchor = center] (xY){diagonalize LR of $Y$};
\node[ques, below = \versep of xY.center, anchor = center] (xYy){$Y_{22}=Y_{33}$};
\node[proc, right = \horsep of xYy.center, anchor = center] (xYyZ){diagonalize LR of $Z$};
\node[ques, above = \versep of xYyZ.center, anchor = center] (z){$Z_{22}=Z_{33}$};
\node[proc, above = \versep of z.center, anchor = center] (A){$Y_{12}=0$};
\node[ques, right = \horsep of A.center, anchor = center] (Ay){$Y_{13}=0$};
\node[proc, right = \horsep of Ay.center, anchor = center] (AyZ){$Z_{12}=0$ $Z_{23}\geq0$};
\node[proc, below = \versep of AyZ.center, anchor = center] (AyY){$Y_{13}\geq0$ $Z_{12}\geq0$};
\node[proc, right = \horsep of z.center, anchor = center] (B){$Y_{12}\geq0$ $Y_{13}\geq0$};
\node[ques, below = \versep of B.center, anchor = center] (By){$Y_{12}=0$};
\node[proc, right = \horsep of By.center, anchor = center] (ByZ){$Z_{12}\geq0$};
\node[ques, below = \versep of ByZ.center, anchor = center] (Byy){$Y_{13}=0$};
\node[proc, left  = \horsep of Byy.center, anchor = center] (ByyZ){$Z_{13}\geq0$};
\node[proc, below = \versep of xYy.center, anchor = center] (xYyY){$Y_{12}\geq0$ $Y_{13}\geq0$};
\node[ques, right = \horsep of xYyY.center, anchor = center] (xYyYy){$Y_{12}=0$};
\node[proc, below = 2\versep of x.center, anchor = center] (xC){$Y_{12}\geq0$ $Y_{23}\geq0$};
\node[ques, below = \versep of xC.center, anchor = center] (xCy){$Y_{12}=0$};
\node[proc, right = \horsep of xCy.center, anchor = center] (xCyY){$Y_{13}\geq0$};
\node[ques, right = \horsep of xCyY.center, anchor = center] (y){$Y_{23}=0$};
\node[proc, right = \horsep of y.center, anchor = center] (yZ){$Z_{12}\geq0$};
\node[ques, right = \horsep of yZ.center, anchor = center] (yZy){$Y_{13}=0$};
\node[proc, below = \versep of yZy.center, anchor = center] (yZyZ){$Z_{23}\geq0$};
\node[ques, left =  \horsep of yZyZ.center, anchor = center] (z2){$Z_{12}=0$ or $Z_{23}=0$};
\node[proc, left =  \horsep of z2.center, anchor = center] (z2Z){$Z_{13}\geq0$};
\node[ques, below = \versep of xCy.center, anchor = center] (xCyy){$Y_{23}=0$};
\node[proc, below = \versep of xCyy.center, anchor = center] (Y){$Y_{13}\geq0$};
\node[ques, right = \horsep of Y.center, anchor = center] (Yy){$Y_{13}=0$};
\node[proc, right = \horsep of Yy.center, anchor = center] (YyZ){$Z_{23}\geq0$};
\node[ques, right = \horsep of YyZ.center, anchor = center] (z3){$Z_{23}=0$};
\node[proc, right = \horsep of z3.center, anchor = center] (z3Z){$Z_{13}\geq0$};
\path[line] (portal)--(x);
\path[line] (x)--(xY) node [midway, above]{yes} node[midway, below = \ssep, opt]{C};
\path[line] (xY)--(xYy);
\path[line] (xYy)--(xYyZ) node [midway, above]{yes};
\path[line] (xYyZ)--(z);
\path[line] (z)--(A) node [midway, right]{yes} node[midway, left = \ssep, opt]{C};
\path[line] (A)--(Ay);
\path[line] (Ay)--(AyZ) node [midway, above]{yes} node[midway, below = \ssep, opt]{C};
\path[line] (z)--(B) node [midway, above]{no} node[midway, below = \ssep, opt]{D};
\path[line] (B)--(By);
\path[line] (By)--(ByZ) node [midway, above]{yes};
\path[line] (By)--++(0,-\bulvsep)--node [midway,above]{no}++(\horsep,0);
\path[line] (ByZ)--(Byy);
\path[line] (Byy)--(ByyZ) node [midway, above]{yes} node[midway, below = \ssep, opt]{E};
\path[line] (xYy)--(xYyY) node [midway, right]{no} node[midway, left = \ssep, opt]{D};
\path[line] (xYyY)--(xYyYy);
\path[line] (x)--(xC) node [midway, right]{no} node[midway, left = \ssep, opt]{D};
\path[line] (xC)--(xCy);
\path[line] (xCy)--(xCyY) node [midway, above]{yes};
\path[line] (xCyY)--(y);
\path[line] (y)--(yZ) node [midway, above](yesnode){yes};
\path[line] (yZ)--(yZy);
\path[line] (yZy)--(yZyZ) node [midway, right]{yes} node[midway, left = \ssep, opt]{D};
\path[line] (yZyZ)--(z2);
\path[line] (z2)--(z2Z) node [midway, above]{yes};
\path[line] (xCy)--(xCyy) node [midway, right]{no};
\path[line] (xCyy)--(Y) node [midway, right]{yes};
\path[line] (Y)--(Yy);
\path[line] (Yy)--(YyZ) node [midway, above]{yes} node[midway, below = \ssep, opt]{E};
\path[line] (YyZ)--(z3);
\path[line] (z3)--(z3Z) node [midway, above]{yes} node[midway, below = \ssep, opt]{E};
\path[line] (Ay)--++(0,-\bulvsep)--node [midway, above]{no} node[midway, below = .4\ssep, opt]{G} ++(\bulhsep,0)--++(0,-\bulvsep)--(AyY);
\path[line] (xYyYy)--++(\bulhsep,0)--node [midway, right]{yes} (yesnode);
\path[line] (xYyYy)--++(0,-\bulvsep)-- node [midway, above]{no}++(-\horsep-\bulhsep,0)-- ++(0,-0.55\bulvsep) arc (-90:90:-0.45\bulvsep) -- ++(0,-3.55\bulvsep);
\path[end] (AyZ)--++(\bulhsep,0);
\path[end] (AyY)--++(\bulhsep,0);
\path[end] (Byy)--++(\bulhsep,0) node [midway, above]{no};
\path[end] (yZy)--++(\bulhsep,0) node [midway, above]{no};
\path[end] (z3Z)--++(\bulhsep,0);
\path[end] (ByyZ)--++(0,-\bulvsep);
\path[end] (z2)--++(0,1.3\bulvsep) node [midway, right]{no};
\path[end] (z3)--++(0,-1.2\bulvsep) node [midway, right]{no};
\path[end] (Yy)--++(0,-1.2\bulvsep) node [midway, right]{no};
\path[end] (xCyy)--++(-\bulhsep,0) node [midway, above]{no};
\path[end] (y)--++(0,-1.2\bulvsep) node [midway, right]{no};
\end{tikzpicture}
\end{center}
\caption{Flowchart of procedure of fixing the gauge, part2.}
\label{fig:flowchart2}
\end{sidewaysfigure}


\begin{thebibliography}{10}

\bibitem{Weinberg_hierarchy}
S.~Weinberg.
\newblock {Implications of dynamical symmetry breaking}.
\newblock {\em Phys.Rev.}, D13:974--996, 1976.

\bibitem{Dimopoulos}
S.~Dimopoulos and S.~Raby.
\newblock {Supercolor}.
\newblock {\em Nucl.Phys.}, B192:353, 1981.

\bibitem{Raby}
S.~Dimopoulos, S.~Raby, and Frank Wilczek.
\newblock {Supersymmetry and the scale of unification}.
\newblock {\em Phys.Rev.}, D24:1681--1683, 1981.

\bibitem{Ellis}
J.~R. Ellis, J.S. Hagelin, D.~V. Nanopoulos, K.~A. Olive, and M.~Srednicki.
\newblock {Supersymmetric relics from the Big Bang}.
\newblock {\em Nucl.Phys.}, B238:453--476, 1984.

\bibitem{Ferrara}
S.~Ferrara and B.~Zumino.
\newblock {Supergauge invariant Yang-Mills theories}.
\newblock {\em Nucl.Phys.}, B79:413, 1974.

\bibitem{Brink}
L.~Brink, J.~H. Schwarz, and J.~Scherk.
\newblock {Supersymmetric Yang-Mills t heories}.
\newblock {\em Nucl.Phys.}, B121:77, 1977.

\bibitem{Claudson}
M.~Claudson and M.~B. Halpern.
\newblock {Supersymmetric ground state wave functions}.
\newblock {\em Nucl.Phys.}, B250:689, 1985.

\bibitem{Samuel}
S.~Samuel.
\newblock {Solutions of extended supersymmetric matrix models for arbitrary
  gauge groups}.
\newblock {\em Phys.Lett.}, B411:268--273, 1997, hep-th/9705167.

\bibitem{Hoppe}
J.~Hoppe.
\newblock {Quantum theory of a massless relativistic surface and a
  two-dimensional bound state problem}.
\newblock {\em unpublished}, http://hdl.handle.net/1721.1/15717.

\bibitem{Bergshoeff}
E.~Bergshoeff, E.~Sezgin, and P.K. Townsend.
\newblock {Supermembranes and eleven-dimensional supergravity}.
\newblock {\em Phys.Lett.}, B189:75--78, 1987.

\bibitem{deWit}
B.~de~Wit, J.~Hoppe, and H.~Nicolai.
\newblock {On the quantum mechanics of supermembranes}.
\newblock {\em Nucl.Phys.}, B305:545, 1988.

\bibitem{Nicolai}
B.~de~Wit, M.~Luscher, and H.~Nicolai.
\newblock {The supermembrane is unstable}.
\newblock {\em Nucl.Phys.}, B320:135, 1989.

\bibitem{Helling}
H.~Nicolai and R.~Helling.
\newblock {Supermembranes and M(atrix) theory}.
\newblock 1998, hep-th/9809103.

\bibitem{BFSS}
T.~Banks, W.~Fischler, S.H. Shenker, and L.~Susskind.
\newblock {M theory as a matrix model: A Conjecture}.
\newblock {\em Phys.Rev.}, D55:5112--5128, 1997, hep-th/9610043.

\bibitem{Kogut}
John~B. Kogut and Leonard Susskind.
\newblock {The Parton picture of elementary particles}.
\newblock {\em Phys.Rept.}, 8:75--172, 1973.

\bibitem{Moore}
G.~W. Moore, N.~Nekrasov, and S.~Shatashvili.
\newblock {D particle bound states and generalized instantons}.
\newblock {\em Commun.Math.Phys.}, 209:77--95, 2000, hep-th/9803265.

\bibitem{Staudacher}
M.~Staudacher.
\newblock {Bulk Witten indices and the number of normalizable ground states in
  supersymmetric quantum mechanics of orthogonal, symplectic and exceptional
  groups}.
\newblock {\em Phys.Lett.}, B488:194--198, 2000, hep-th/0006234.

\bibitem{Fischbacher}
T.~Fischbacher.
\newblock {Bulk Witten indices from D = 10 Yang-Mills integrals}.
\newblock {\em Nucl.Phys.}, B694:525--535, 2004, hep-th/0312262.

\bibitem{Taylor}
W.~Taylor.
\newblock {M(atrix) theory: Matrix quantum mechanics as a fundamental theory}.
\newblock {\em Rev.Mod.Phys.}, 73:419--462, 2001, hep-th/0101126.

\bibitem{Becker}
Katrin Becker and Melanie Becker.
\newblock {A two loop test of M(atrix) theory}.
\newblock {\em Nucl.Phys.}, B506:48--60, 1997, hep-th/9705091.

\bibitem{Porrati}
M.~Porrati and A.~Rozenberg.
\newblock {Bound states at threshold in supersymmetric quantum mechanics}.
\newblock {\em Nucl.Phys.}, B515:184--202, 1998, hep-th/9708119.

\bibitem{Danielsson}
U.~H. Danielsson, G.~Ferretti, and B.~Sundborg.
\newblock {D particle dynamics and bound states}.
\newblock {\em Int.J.Mod.Phys.}, A11:5463--5478, 1996, hep-th/9603081.

\bibitem{Halpern}
M.B. Halpern and C.~Schwartz.
\newblock {Asymptotic search for ground states of SU(2) matrix theory}.
\newblock {\em Int.J.Mod.Phys.}, A13:4367--4408, 1998, hep-th/9712133.

\bibitem{Catterall}
S.~Catterall and T.~Wiseman.
\newblock {Black hole thermodynamics from simulations of lattice Yang-Mills
  theory}.
\newblock {\em Phys.Rev.}, D78:041502, 2008, 0803.4273.

\bibitem{Wiseman}
S.~Catterall and T.~Wiseman.
\newblock {Extracting black hole physics from the lattice}.
\newblock {\em JHEP}, 1004:077, 2010, 0909.4947.

\bibitem{Anagnostopoulos}
K.~N. Anagnostopoulos, M.~Hanada, J.~Nishimura, and S.~Takeuchi.
\newblock {Monte Carlo studies of supersymmetric matrix quantum mechanics with
  sixteen supercharges at finite temperature}.
\newblock {\em Phys.Rev.Lett.}, 100:021601, 2008, 0707.4454.

\bibitem{Hanada}
M.~Hanada, Y.~Hyakutake, G.~Ishiki, and J.~Nishimura.
\newblock {Holographic description of quantum black hole on a computer}.
\newblock 2013, 1311.5607.

\bibitem{Nishimura}
M.~Hanada, J.~Nishimura, and S.~Takeuchi.
\newblock {Non-lattice simulation for supersymmetric gauge theories in one
  dimension}.
\newblock {\em Phys.Rev.Lett.}, 99:161602, 2007, 0706.1647.

\bibitem{Bjorken}
J.D. Bjorken.
\newblock {Elements of Quantum Chronodynamics}.
\newblock 1979.

\bibitem{Luscher}
M.~Luscher.
\newblock {Some analytic results concerning the mass spectrum of Yang-Mills
  gauge theories on a torus}.
\newblock {\em Nucl.Phys.}, B219:233--261, 1983.

\bibitem{Munster}
M.~Luscher and G.~Munster.
\newblock {Weak coupling expansion of the low lying energy values in the SU(2)
  gauge theory on a torus}.
\newblock {\em Nucl.Phys.}, B232:445, 1984.

\bibitem{Weisz}
P.~Weisz and V.~Ziemann.
\newblock {Weak coupling expansions of the low lying energy values in SU(3)
  gauge theory on a torus}.
\newblock {\em Nucl.Phys.}, B284:157, 1987.

\bibitem{Ziemann_phd}
V.~Ziemann.
\newblock {\em {Qualitative Untersuchung des niedrig liegenden Spektrums reiner
  Yang-Mills Theorien im endlichen Volumen mit besonderer Ber\"ucksichtigung
  von SU(3)}}.
\newblock PhD thesis, {Universit\"at Hamburg}, June 1986.

\bibitem{Koller1}
J.~Koller and P.~van Baal.
\newblock {A rigorous nonperturbative result for the glueball mass and electric
  flux energy in a finite volume}.
\newblock {\em Nucl.Phys.}, B273:387, 1986.

\bibitem{Koller2}
J.~Koller and P.~van Baal.
\newblock {A nonperturbative analysis in finite volume gauge theory}.
\newblock {\em Nucl.Phys.}, B302:1, 1988.

\bibitem{Janik}
R.A. Janik and J.~Wosiek.
\newblock {Towards the matrix model of M theory on a lattice}.
\newblock {\em Acta Phys.Polon.}, B32:2143--2154, 2001, hep-th/0003121.

\bibitem{Wosiek}
J.~Wosiek.
\newblock {Spectra of supersymmetric Yang-Mills quantum mechanics}.
\newblock {\em Nucl.Phys.}, B644:85--112, 2002, hep-th/0203116.

\bibitem{Campostrini}
M.~Campostrini and J.~Wosiek.
\newblock {High precision study of the structure of D=4 supersymmetric
  Yang-Mills quantum mechanics}.
\newblock {\em Nucl.Phys.}, B703:454--498, 2004, hep-th/0407021.

\bibitem{Trzetrzelewski_susyd2}
M.~Trzetrzelewski.
\newblock {Large N behavior of two dimensional supersymmetric Yang-Mills
  quantum mechanics}.
\newblock {\em J.Math.Phys.}, 48:012302, 2007, hep-th/0608147.

\bibitem{Korcyl}
P.~Korcyl.
\newblock {Exact solutions to D=2 Supersymmetric Yang-Mills Quantum Mechanics
  with SU(3) gauge group}.
\newblock {\em Acta Phys.Polon.Supp.}, 2:623, 2009, 0911.2152.

\bibitem{KorcylN}
P.~Korcyl.
\newblock {Solutions of D=2 supersymmetric Yang-Mills quantum mechanics with
  SU(N) gauge group}.
\newblock {\em J.Math.Phys.}, 52:052105, 2011, 1101.0591.

\bibitem{vanBaal}
P.~van Baal.
\newblock {The Witten index beyond the adiabatic approximation}.
\newblock 2001, hep-th/0112072.

\bibitem{Kotanski}
J.~Kotanski.
\newblock {Energy spectrum and wave-functions of four-dimensional
  Supersymmetric Yang-Mills Quantum Mechanics for very high cut-offs}.
\newblock {\em Acta Phys.Polon.}, B37:2813--2838, 2006, hep-th/0607012.

\bibitem{Kotanski2}
Jan Kotanski.
\newblock {Virial theorem for four-dimensional supersymmetric Yang-Mills
  quantum mechanics with SU(2) gauge group}.
\newblock {\em Acta Phys.Polon.}, B37:3659--3666, 2006, hep-th/0610091.

\bibitem{Trzetrzelewski_number}
M.~Trzetrzelewski.
\newblock {The Number of gauge singlets in supersymmetric Yang-Mills quantum
  mechanics}.
\newblock {\em Phys.Rev.}, D76:085012, 2007, 0708.2946.

\bibitem{Itzykson}
C.~Itzykson and J.B. Zuber.
\newblock {\em {Quantum Field Theory}}.
\newblock {McGraw-Hill}, 1980.

\bibitem{Weinberg}
S.~Weinberg.
\newblock {\em {The quantum theory of fields. Vol. 3: Supersymmetry}}.
\newblock Cambridge University Press, 2000.

\bibitem{Dancoff}
S.M. Dancoff.
\newblock {Nonadiabatic meson theory of nuclear forces}.
\newblock {\em Phys.Rev.}, 78:382--385, 1950.

\bibitem{Trzetrzelewski_spectra}
M.~Trzetrzelewski and J.~Wosiek.
\newblock {Quantum systems in a cut Fock space}.
\newblock {\em Acta Phys.Polon.}, B35:1615--1624, 2004, hep-th/0308007.

\bibitem{Ambrozinski}
Z.~Ambrozinski and J.~Wosiek.
\newblock {Resumming not summable perturbative series}.
\newblock {\em Acta Phys.Polon.}, B44(1):49--58, 2013.

\bibitem{Ambrozinski_cosine}
Z.~Ambrozinski.
\newblock {Tunneling in cosine potential with periodic boundary conditions}.
\newblock {\em Acta Phys.Polon.}, B44:1261--1272, 2013, arXiv: 1303.0708.

\bibitem{Trzetrzelewski_trees}
M.~Trzetrzelewski.
\newblock {Reduction of su(N) loop tensors to trees}.
\newblock {\em J.Math.Phys.}, 46:103512, 2005, math-ph/0505084.

\bibitem{Procesi}
M.~Bresan, C.~Procesi, and S.~Spenko.
\newblock {Quasi-identities on matrices and the Cayley-Hamilton polynomial}.
\newblock arXiv: 1212.4597.

\bibitem{Macfarlane}
A.J. Macfarlane, A.~Sudbery, and P.~Weisz.
\newblock {Explicit representations of chiral invariant lagrangian theories of
  hadron dynamics}.
\newblock {\em Proc.Roy.Soc.Lond.}, 314:217--250, 1970.

\bibitem{DiFrancesco}
P.~Di~Francesco, P.~Mathieu, and D.~Senechal.
\newblock {\em {Conformal field theory}}.
\newblock Springer-Verlag New York, Inc.

\bibitem{Hamermesh}
M.~Hamermesh.
\newblock {\em {Teoria grup w zastosowaniu do zagadnie\'n fizycznych}}.
\newblock Pa\'nstwowe Wydawnictwo Naukowe, 1968.

\bibitem{Drouffe}
J.M. Drouffe and C.~Itzykson.
\newblock {Lattice gauge fields}.
\newblock {\em Phys.Rept.}, 38:133--175, 1978.

\bibitem{Weyl}
H.~Weyl.
\newblock {\em {The Theory of Groups and Quantum Mechanics}}.
\newblock {Methuen \& Co. Ltd}, 1931.

\bibitem{Trzetrzelewski1}
M.~Trzetrzelewski.
\newblock {Quantum mechanics in a cut Fock space}.
\newblock {\em Acta Phys.Polon.}, B35:2393--2416, 2004, hep-th/0407059.

\bibitem{Simon}
B.~Simon.
\newblock {Some quantum operators with discrete spectrum but clasically
  continuous spectrum}.
\newblock {\em Annals Phys.}, 146:209--220, 1983.

\bibitem{Korcyl_effective}
P.~Korcyl.
\newblock {Classical trajectories and quantum supersymmetry}.
\newblock {\em Phys.Rev.}, D74:115012, 2006, hep-th/0610105.

\bibitem{Wess}
J.~Wess and J.~Bagger.
\newblock {\em {Supersymmetry and supergravity}}.
\newblock Princeton University Press, 1983.

\bibitem{Salomonson}
P.~Salomonson and J.W. van Holten.
\newblock {Fermionic coordinates and supersymmetry in quantum mechanics}.
\newblock {\em Nucl.Phys.}, B196:509, 1982.

\bibitem{Goldstone}
J.~Goldstone and R.~Jackiw.
\newblock {Unconstrained temporal gauge for Yang-Mills theory}.
\newblock {\em Phys.Lett.}, B74:81, 1978.

\bibitem{Savvidy}
G.K. Savvidy.
\newblock {Yang-Mills Quantum Mechanics}.
\newblock {\em Phys.Lett.}, B159:325, 1985.

\bibitem{Sakurai}
J.~J. Sakurai.
\newblock {\em {Modern Quantum Mechanics}}.
\newblock {Addison-Wesley Publishing Company}, 1994.

\end{thebibliography}

\end{document}